\pdfoutput=1
\documentclass[electronic]{mythesis}

\begin{document}
\frontmatter
\maketitle
\chapter{Acknowledgments}

A \<PhD> is a journey.
I might not have travelled very far in space, but I have certainly accomplished an intellectual odyssey.
This thesis can be seen as a kind of \emph{{\selectlanguage{french}récit de voyage}}, an after the fact narration of where I have been and what I have seen, but with equations taking the place of descriptions of landscapes.

Many people have played a significant role during these four years of peregrination as a doctoral student, whom I wish to acknowledge.

First and foremost, I would like to thank my advisor Frank Ferrari for being my guide through the land of theoretical physics.
I learned a lot at his side, both from his deep and clear knowledge of physics and from his focus, dedication and enthusiasm as a physicist.
It was often challenging to work with him and I am confident that his strive for rigor and precision will have a lasting influence on my life as a scientist.

I am also very much indebted to my other collaborators and friends, Antonin Rovai and Eduardo Conde (Pena?).
I am grateful for our numerous discussions and for everything they taught me about physics but also various aspects of life such as bikes or Spanish cooking.
Further, Eduardo's detailed reading of the manuscript and his many interesting comments were very valuable.
All together, it was a pleasure to collaborate with them and I hope that we will manage to work together again in the future.

Next I would like to thank my companions in exile on the second floor, Josef Lindman Hörnlund, Marco Fazzi, Andrea Marzolla, Daniele Musso, Diego Redigolo and Antonin Rovai.
It was a pleasure to share an office or a (very thin) wall with them and I will remember fondly our impromptu conversations and evening walks back home.

More generally, I am grateful to all members of the \emph{{\selectlanguage{french}Physique Théorique et Mathématique}} group for providing a welcoming and friendly environment.
I would particularly like to thank Riccardo Argurio and Andrés Collinucci for explanations and guidance as well as Dominique Bogaerts, Fabienne De Neyn, Marie-France Rogge, Isabelle Van Geet and Delphine Vantighem for the administrative support.
I am also grateful to the local representatives of the Italian \<HEP-TH> mafia Riccardo Argurio, Andrea Campoleoni, Marco Fazzi, Davide Forcella, Manuela Kulaxizi, Simone Giacomelli, Andrea Marzolla, Andrea Mezzalira, Daniele Musso, Diego Redigolo and Antonin Rovai for providing a very beneficial foretaste of Italy, to the Hispanic guys Eduardo Conde Pena, Ignacio Cortese Mombelli, Hernan Gonzales and Gustavo Lucena Gómez for trying to fight back, to Pujian Mao, Rakibur Rahman and Amitabh Virmani for some oriental wisdom (and jokes), to David Chow, Semyon Klevtsov, Josef Lindman Hörnlund, Jakob Palmkvist, Christoffer Petersson and Waldemar Schulgin for completing the European picture, to the French speakers Glenn Barnich, Cyril Closset, Geoffrey Compère, François Dehouck, Stéphane Detournay, Laura Donnay, Laure-Anne Douxchamps, Marc Henneaux, Frank Ferrari, Pierre-Henry Lambert, Gustavo Lucena Gómez, Blaža Oblak, Antonin Rovai and Cédric Troessart for still managing to provide a local touch and finally to Andrés Collinucci for ruining this classification.

I also wish to thank my friends who have been accompanying me from the beginning of my studies, Martin, Ann, Gustavo and Florian.
Even though we have gradually parted company in recent times, I hope that we will manage to stay in contact in the future.

I warmly thank my family for their help and support since the beginning.
Even if my work is largely mysterious for them, they have always encouraged me and helped me overcome some more difficult periods.

My final words of gratitude go to my Penelope who fortunately was by my side all along, and without whom I would probably never have managed to survive the rough waters.

\thispagestyle{plain}
\begin{refsection}
  \defbibnote{mypapers}{The original contributions of this thesis are based on the following published papers.}
  \nocite{Ferrari:2013pq,Ferrari:2013hg,Conde:2013wpa}
  \printbibliography[heading=none,prenote=mypapers]
\end{refsection}
\nocite{Ferrari:2013pq,Ferrari:2013hg,Conde:2013wpa}

\listoftodos
\todototoc
\tableofcontents
\mainmatter
\chapter{Introduction}

The twentieth century has witnessed a giant leap in the understanding of the fundamental laws of Nature, culminating in the Standard Model of particle physics on the one hand and in Einstein's theory of gravitation on the other hand.
These two theories are on very different footings.

The Standard Model is described in the framework of Quantum Field Theory (\<QFT>), which looked rather strange in the beginning (and still does to the beginner!), but whose organizing principles \cite{Wilson:1973jj} seem quite well understood by now.
It is extremely accurate and to date perfectly matches the results of experiments performed at the increasingly high energies probed by particle colliders, of which the Large Hadron Collider is the most recent contender.

On the contrary, General Relativity is a very elegant theory, describing gravitation geometrically as the curvature of space-time.
However, it does not make sense as a fundamental quantum theory, so one might be tempted to view it as an effective theory that breaks down and must be replaced by something else at about the Planck scale, where it becomes strongly coupled.
However, the problematic high energy behavior of classical gravity viewed as a quantum theory is only part of the story and there are other much more subtle issues that cannot be directly ascribed to it.
A well-known example in this class is the information paradox \cite{Hawking:1976ra}, which can be explained as follows.
Black holes can be formed by gravitational collapse of matter.
A semi-classical reasoning that should be valid according to effective field theory arguments shows that the black hole evaporates by emitting thermal radiation.
If the black hole completely evaporates, any initial quantum state, possibly pure, then evolves into a mixed state, destroying all information of the original state.
This is in sharp tension with unitarity, one of the cornerstones of quantum mechanics, hence the paradox, suggesting that either local effective field theory or quantum mechanics must fail in gravitational theories.

Different approaches to tame the behavior of gravity at high energies have been attempted,
of which the most popular is String Theory.
String Theory replaces the point particles of ordinary \<QFT> by extended one-dimensional objects, strings, which has the effect of adding an infinite number of new massive degrees of freedom, drastically altering the ultraviolet (\<UV>) behavior of Einstein gravity.
Whether it solves also the black hole evaporation issue remained unclear until the discovery of the gauge/gravity correspondence~\cite{Maldacena:1997re}.
This correspondence provides a radically different perspective on gravity with negative cosmological constant, stating that it is in fact an ordinary~\<QFT> in disguise.
I will now briefly review these developments.

\section{The context: the gauge/gravity correspondence}

String Theory contains other extended objects besides strings, the most important for the purposes of this thesis being D-branes.
Looking at the low-energy behavior of a large number of D-branes with three spatial dimensions (D3-branes for short) in two different descriptions, Maldacena conjectured in~\cite{Maldacena:1997re} that two seemingly different theories are actually related.
These two theories are a maximally supersymmetric gauge theory in four dimensions and string theory on the asymptotically $\AdSS$ space-time respectively.
The first theory is a particularly well-behaved cousin of the usual gauge theories also used in the Standard Model.
Because of the large amount of supersymmetry, it is actually a Conformal Field Theory (\<CFT>) which implies that its coupling constant is not renormalized.
The second theory is a full-blown string theory in ten dimensions, that in particular also describes gravity.
While they look extremely dissimilar, e.g.\ not even the number of dimensions is the same, Maldacena claimed that these two theories are \emph{dual} in the sense that they describe the same physics but in two different languages, more natural in different regimes.

The perturbative regime, where the \<CFT> can be described with the usual Feynman diagrams, is valid, when the gauge group has a large rank, for small ('t~Hooft) coupling.
In this range of parameters, the string theory on $\AdSS$ is classical, however the space-time is highly curved and all the massive stringy modes are important.
In order to decouple them and have a description where Einstein's equations are valid, one needs the curvature to be very small which translates into very large 't~Hooft coupling on the \<CFT> side.

By now, Maldacena's correspondence is very well tested and has been generalized to much more general gauge/gravity dualities.
It constitutes an invaluable tool to study strongly coupled systems by mapping hard questions in the \<QFT> to tractable problems in classical gravity.
Using the correspondence in the opposite direction, it is in principle also possible to gain insights into the mysteries of quantum gravity.

For instance, the duality realizes in a precise way the idea, pioneered by \textcite{tHooft:1993gx} and \textcite{Susskind:1994vu}, that gravity is \emph{holographic}, in the sense that its number of degrees of freedom scales like the surface area rather than the volume of space: the microscopic theory is a field theory defined on a rigid space-time with less dimensions.
The extra dimensions then emerge from the strongly coupled dynamics of the gauge theory.
It also explains the laws of black hole thermodynamics as arising from the genuine thermodynamics of the dual gauge theory.
Finally, the \<QFT> is manifestly unitary and the \<AdS> dual contains black holes, hence the \<AdS>/\<CFT> correspondence can teach us a lot about black holes and the information paradox.
For example, one might want to track in the \<CFT> the evolution of a collapsing shell of matter that forms a black hole, which by construction yields a pure black hole microstate.

How to do this explicitly is however far from clear.
This can be blamed on the difficulty of tackling strongly coupled gauge theories directly.
A second hurdle is that the standard \<AdS>/\<CFT> dictionary \cite{Witten:1998qj,Gubser:1998bc} is far from transparent.
While on the field theory side it deals with quite natural objects, namely correlators of local gauge invariant operators, it maps them on the gravity side to coefficients in an asymptotic expansion of the fields near the boundary.
It is then highly non-trivial to recover the full bulk fields from this asymptotic data.
Other holographic observables, such as entanglement entropy \cite{Ryu:2006bv,Lewkowycz:2013nqa}, are extremely difficult to construct directly in the field theory and while more geometric in nature, still do not map directly to the bulk fields.

These two difficulties of strongly coupled \<QFT> and observables explain why comparatively little work has focused on the study of gravitational physics starting from the field theory side.
A basic question that one might want to ask in this context is the following:

\emph{Given a strongly coupled \<CFT> (or some deformation thereof) and some state in this theory, what is its holographic dual? In other words, can one find explicitly the bulk fields in the dual theory directly from the gauge theory, without assuming, let alone solving, any supergravity equation of motion?}

While answering this question is in general very hard, it turns out that it can be done in some specific and non-trivial cases.
A good strategy to address this question was proposed in~\cite{Ferrari:2012nw} (see also \cite{Ferrari:2013aba} for an extension to more general setups).
The idea is to consider a finite number of \enquote{probe} branes in addition to the large number of \enquote{background} branes whose strong coupling dynamics generate the background.
In the open string picture, this D-brane system admits different sectors of open strings, background/background, background/probe and probe/probe.
Integrating out the background/probe and background/background strings in a suitable low-energy field theory limit corresponding to Maldacena's decoupling limit, one obtains an effective theory for the light degrees of freedom of the probe.
This effective action can then be compared to the D-brane probe action in an arbitrary supergravity background (the Dirac-Born-Infeld plus Wess-Zumino action in the case of a single probe or a generalization thereof \cite{Myers:1999ps} in the case of multiple probes) and the background can be read-off by matching the two actions.

The case of the $\AdSS$ background, that is its metric and all the fluxes, was derived in this way by adding \Dmi{}-brane probes to the D3-branes in the original paper~\cite{Ferrari:2012nw}.
The purpose of the present thesis is to show that one can also successfully apply this approach to less simple theories and convey the message that it could be a very useful tool for future investigations of the gauge/gravity correspondence.

\section{Outline of the thesis}

This thesis is divided into two main parts.
The first part introduces the background material that is then exploited in the second part presenting the original results of this thesis, which have already appeared in the published papers \cite{Ferrari:2013pq,Ferrari:2013hg,Conde:2013wpa}.
A discussion of the same \cite{Ferrari:2013pq} or similar \cite{Ferrari:2013wla} results from a different and complementary point of view can be found in A.~Rovai's thesis \cite{antothese}.

\subsection*{The first part}
It contains no new results.
Instead, the goal is to collect and present the context and the technical tools that are required for a proper understanding of the work presented in the second part.
The presentation aims at being concise and focused on this purpose rather than exhaustive.
Most of the material is already quite well covered in textbooks or reviews, and this approach will hopefully allow the reader not to lose track.
The same objective also explains some unorthodox presentation choices compared to the standard references.

The first part starts in \cref{chap:instantons} with a presentation of some aspects of instantons in four-dimensional gauge theories.
Instantons in pure Yang-Mills theory are introduced as topologically non-trivial solutions of the equations of motion.
The one-instanton \<BPST>~solution is given and the role of the moduli is emphasized.
The \<ADHM> construction as an ansatz to construct multi-instanton solutions is presented next.
The most important part of this contruction is the identification of the moduli as a set of variables subject to constraints.
The instanton effective action, describing the dynamics of a subset of moduli is the last step in the bosonic theory, which is finally generalized to the supersymmetric setting.

After the field theory \emph{mise en bouche}, the relevant parts of type~\<II> string theory are introduced in \cref{chap:D-branes}.
First, the closed string spectrum on flat space-time and on orbifolds is introduced.
Then, D-branes are added to the game, first from the point of view of Ramond-Ramond charges and subsequently as boundary conditions for open strings.
Different aspects of their dynamics are discussed: supersymmetry, single D-branes in curved background, multiple D-branes in flat and in curved space-times.
Instantons are revisited in this context as a specific decoupling limit of \Dmi{}-branes within D3-branes.
Finally, the chapter closes by extending the discussion to D-branes on orbifolds.

The gauge/gravity (or more properly the gauge/string) correspondence is discussed in \cref{chap:holo}.
It is presented first through a similar argument to the one used by Maldacena in his original paper.
This argument is shown to still hold when adding \Dmi{}-brane probes, providing along the way a justification for the decoupling limit of the \Dmi{}/D3 theory, and it is explained how one can construct the \Dmi{}-brane effective action and recover the supergravity background from it.

\subsection*{The second part}
It contains the original contributions of the author and his collaborators.

\Cref{chap:adsdeformed} shows that probe \Dmi{}-branes can be used to reconstruct supergravity duals to deformations of the $\nn=4$ theory corresponding to $\AdSS$ \cite{Ferrari:2013pq}.
Three such deformations are considered.
The first is the Coulomb branch deformation, corresponding to turning on vacuum expectation values (\<VEV>s) for the scalar fields in the field theory.
The second is the non-commutative deformation, corresponding to placing the gauge theory on a non-commutative space-time, breaking conformality but preserving supersymmetry.
The third is the $\beta$-deformation, which is an exactly marginal\footnote{see \cref{chap:adsdeformed} for caveats} deformation breaking all supersymmetry in the most general case.
In the first case, we can recover the full dual background, whereas in the other two cases, one can find the background for small deformations and we explain the reason for this limitation.

\Cref{chap:enhancon} is based on \cite{Conde:2013wpa} and shows that one can \emph{derive} a stringy resolution of a classical singularity in supergravity, namely the enhançon mechanism, from tractable non-perturbative effects in an $\nn=2$ supersymmetric gauge theory.
The specific theory under consideration is the $\nn=2$ quiver gauge theory corresponding to D3-branes on the $\CC^2/\ZZ_2$ orbifold, taken on its Coulomb branch.
One can use a fractional \Dmi{}-brane as a probe to relate a twisted supergravity scalar to a chiral correlator in the field theory.
Exploiting recent results in instanton technology for $\nn=2$ quivers, this correlator can be computed explicitly for any value of the coupling and at any point on the Coulomb branch.
By carefully analyzing the large~$N$ limit, one can then show explicitly the enhançon mechanism at work.

\Cref{chap:D4brane} shows that this approach is not limited to D3-branes probed by \Dmi{}-branes, but that one can also recover the near-horizon D4-brane background by using D0-branes as a probe.
This chapter is based on~\cite{Ferrari:2013hg}.

There are also a number of appendices provided to cover the more technical aspects of this work.
\Cref{chap:notations} collects the conventions and notations in use as well as various useful algebraic formulas.
\Cref{chap:Myers} contains the expansion of the \Dmi{} probe action for arbitrary backgrounds that is compared with the explicit actions obtained in the main text in order to deduce the holographic duals to the considered field theory setups.
\Cref{chap:sugrasols} reviews the full supergravity solutions for the non-commutative and $\beta$-deformations of \cref{chap:adsdeformed}.
Finally, in \cref{chap:correlator}, the gauge theory correlator that is relevant for \cref{chap:enhancon} is derived through some gymnastics with elliptic functions.

\part{Background material}
\label{part1}
\chapter{Instantons as gauge theory phenomena}
\label{chap:instantons}

Instantons play a central role in this thesis and will be considered from two different perspectives, namely as non-perturbative objects in field theory and subsequently as D-brane configurations in string theory.
In this chapter, instantons are presented from the field theory point of view.
Starting with the non-supersymmetric pure Yang-Mills case in \cref{sec:YMinst}, the general features of instantons are presented in \cref{sec:YMinstgen}, the one-instanton example is discussed in \cref{sec:BPST} and subsequently generalized to the multi-instanton case thanks to the \<ADHM> construction in \cref{sec:ADHM}.
The instanton effective action, which will play a major role in the following, is introduced in \cref{sec:YMinstaction}.
After a brief reminder about supersymmetric gauge theories in \cref{sec:SUSYreview}, these topics are finally extended to the supersymmetric setting in \cref{sec:SUSYinst}, which will be of main relevance.

\section{Instantons in the pure Yang-Mills theory}
\label{sec:YMinst}

\subsection{Instantons as anti-self-dual solutions}
\label{sec:YMinstgen}

Let us start with a Euclidean $\SU(N)$ Yang-Mills theory, with action
\begin{equation}
  S_{\text{\<YM>}}=\frac{1}{2g^2} \int \d^4x \tr F_{\mu\nu} F_{\mu\nu} \, ,
  \label{eq:YMaction}
\end{equation}
where the field strength is given by $F_{\mu\nu}=\partial_\mu A_\nu - \partial_\mu A_\mu + i [A_\mu, A_\nu]$ and the trace is taken in the fundamental representation $\mathbf{N}$ of $\SU(N)$.
This is not the most general action one may write in this theory.
One may supplement it with the following $\vartheta$-term:
\begin{equation}
  \frac{i\vartheta}{8\pi^2}\int \tr F \wedge F = \frac{i\vartheta}{16\pi^2}\int \d^4x \tr F_{\mu\nu} \hodge{F}_{\mu\nu} = -i\vartheta K \, , \quad K \in \ZZ \, ,
  \label{eq:YMtheta}
\end{equation}
where the Hodge dual of the field strength is given by $\hodge{F}_{\mu\nu} = \frac{1}{2}\epsilon_{\mu\nu\rho\sigma}F_{\rho\sigma}$.
The integrand of \cref{eq:YMtheta} can be written as a total derivative, but the integral can be non-vanishing for field configurations that fall off sufficiently slowly at infinity to contribute to this boundary term.
We will be looking for such configurations with the additional requirement that this contribution is finite.
One can then show that $K$ must be an integer, called \define{instanton number} for reasons that will become clear shortly.

Since
\begin{equation}
  0\le \tr (F \pm \hodge{F})_{\mu\nu} (F \pm \hodge{F})_{\mu\nu} = 2 \tr (F_{\mu\nu} F_{\mu\nu} \pm F_{\mu\nu} \hodge{F}_{\mu\nu}) \, ,
  \label{eq:FpmstarF}
\end{equation}
one can rewrite the total action as
\begin{equation}
  S_{\text{\<YM>}}-i\vartheta K = - (i\vartheta \mp \frac{8\pi^2}{g^2})K+ \frac{1}{4g^2} \int \d^4x \tr (F_{\mu\nu}\pm \hodge{F}_{\mu\nu})^2 \, .
  \label{eq:YMsquare}
\end{equation}
By positivity of the integral on the \RHS, this action is minimized at fixed $K$ for field configurations satisfying $F\pm *F=0$, i.e.\ configurations with (anti-)self-dual field strengths.
But variations $A\to A+\delta A$ that vanish sufficiently fast at infinity cannot alter the value of $K$, hence such configurations automatically satisfy the equations of motion.
This can also be easily checked by noticing that the Bianchi identity and (anti-)self-duality imply the equations of motion.

As a consequence of the positivity of \cref{eq:YMaction}, the real part of \cref{eq:YMsquare} must be positive and the sign of $K$ is thus correlated with the sign of the $\pm$.
We will focus on the solutions with $K>0$ satisfying the anti-self-duality equation that can be written in any of the equivalent ways
\begin{equation}
  F=-\hodge{F}  \Leftrightarrow F=F^{-}=\frac{1}{2}(F-\hodge{F})\Leftrightarrow F^{+} = \frac{1}{2} (F + \hodge{F}) = 0 \, .
  \label{eq:inst}
\end{equation}
We will call the solutions to this equation \define{instantons}\footnote{Often the opposite convention of defining $F=\hodge{F}$ to be an instanton is used, and what we call an instanton would then be called anti-instanton.
It is of course completely equivalent to work with self- and anti-self-dual solutions, but this choice turns out to be quite natural as some important objects appearing in the instanton calculus will then be self-dual.} of charge $K$.
From \cref{eq:YMsquare}, we see that their action is equal to
\begin{equation}
  S_K = -2\pi i \tau K \, , \quad \text{where} \quad \tau=\frac{\vartheta}{2\pi}+\frac{4\pi i}{g^2}
  \label{eq:deftau}
\end{equation}
is the \define{holomorphic gauge coupling}.

We will solve the equation \cref{eq:inst} in \cref{sec:ADHM} but a few comments are in order before that.
Firstly, note that it is crucial to work in Euclidean signature, since under Wick rotation the Hodge~$*$-operator receives an extra $i$~factor and the Minkowskian version of \cref{eq:inst} does not admit solutions for real gauge fields.
Secondly, it is useful to rewrite this equation in terms of spinor indices of $\text{Spin}(4)\simeq \SU(2)_+\times \SU(2)_-$.
It is straightforward to check that the matrices $\sigma_{\mu\nu}$ (defined in \cref{sigma2def}) are self-dual while $\bar\sigma_{\mu\nu}$ are anti-self-dual.
Hence we can equivalently write \cref{eq:inst} as
\begin{equation}
  F\indices{_{\alpha}^{\beta}}=\tfrac{1}{2}F_{\mu\nu}\sigma\indices{_\mu_\nu_{\alpha}^{\beta}}=0 \, .
  \label{eq:instspinor}
\end{equation}
This means that an instanton is required to be invariant under $\SU(2)_+$ only, unlike the true vacuum with $K=0$ which is invariant under the full Lorentz group.

Let us now comment on the relevance of \emph{classical} solutions in a \emph{quantum} theory, especially a strongly coupled one.
Let us first discuss this issue at the level of perturbation theory.
Ordinary perturbation theory involves fluctuations of the fields about the vacuum and yields expressions for correlators which are power series in the coupling constant $g$ of the general form
\begin{equation}
  \corr*{ \mathcal O_1(x_1) \cdots \mathcal O_n(x_n) }_{\text{pert}} \equiv F_{0}(x_1,\cdots,x_n) = \sum_j g^{2j} c_{0,j}(x_1, \cdots x_n) \, .
  \label{eq:pertseries}
\end{equation}
Because the action for an instanton of charge $K$ does not vanish, the same correlator evaluated in the instanton background, i.e.\ on a connection satisfying the equation \cref{eq:inst}, will be weighted  by the factor
\begin{equation}
    \qq^K = e^{2\pi i \tau K} = e^{-\frac{8\pi^2K}{g^2} + i\theta K} \, .
  \label{eq:defq}
\end{equation}
The series expansion of $\qq$ in powers of $g$ vanishes at all orders, hence instantons do not contribute to the perturbative series \cref{eq:pertseries} and constitute new \emph{non-perturbative} contributions.
They are the starting point for a new perturbative series for each value of $K$, computed by the usual rules of perturbation theory, but expanding the gauge field around its instanton value rather than the vacuum.
This procedure then yields an expression of the correlator of the schematic form
\begin{multline}
  \corr*{ \mathcal O_1(x_1) \cdots \mathcal O_n(x_n) } = F_{0}(x_1,\cdots,x_n) \\
  + \sum_{K>0} \qq^K F_{K}(x_1,\cdots,x_n) + \sum_{K<0} \bar\qq^{-K} F_{K}(x_1,\cdots,x_n) \, ,
  \label{eq:npseries}
\end{multline}
where each function $F_K$ is the perturbation series in the sector of instanton charge~$K$ and has a similar expansion to $F_0$ in \cref{eq:pertseries}.
As is well known, the asymptotic series in \cref{eq:npseries} is divergent in quantum field theory for any positive value of $g$.
On the other hand, if one only cares about the relative size of the different terms, as is the case in the perturbative point of view, instanton effects are exponentially suppressed, and one might think that they are completely useless.
This is not true however if for a given correlator, the perturbative contribution $F_0$ identically vanishes, making the one-instanton contribution the leading term, as can happen when the correlator vanishes by a classical symmetry which turns out to be anomalous in the quantum theory \cite{tHooft:1976fv}.
Moreover, in supersymmetric theories, for correlators of chiral operators, the series is an analytic series in $\qq$ only which does not depend on $\bar\qq$ nor $g$ directly and converges for weakly coupled vacua.
Analytic continuation can then be used to gain information on the strong coupling regime of the theory.
The main focus of this thesis will be on supersymmetric (or near-supersymmetric) theories.
The pure Yang-Mills case serves as a useful toy model that already captures some important properties of the generalizations we will consider later.

\subsection{A one-instanton example: the \<BPST>\ solution}
\label{sec:BPST}

Let us now look at a solution of \cref{eq:inst} with $K=1$ and gauge group $\SU(2)$, found by \textcite{Belavin:1975fg}.
It will be sufficient to have the field strength, which reads (in regular gauge),
\begin{equation}
  F^{\textsc{BPST}}_{\mu\nu}(\rho) = \frac{4i\rho^2\bar\sigma_{\mu\nu}}{(x^2+\rho^2)^2} \, ,
  \label{eq:FBPST}
\end{equation}
where $\rho$ is an arbitrary length scale that we will discuss shortly.
Note that in this gauge, the $\SU(2)$ gauge group indices are identified with the $\SU(2)_-$ indices of the Lorentz group, hence the instanton breaks the product of these gauge groups down to their diagonal subgroup.
The field strength \cref{eq:FBPST} is automatically anti-self-dual by being proportional to $\bar\sigma_{\mu\nu}$, and it can be checked straightforwardly that $K=1$ by plugging this expression for $F$ into \cref{eq:YMtheta} and using $\tr \bar\sigma_{\mu\nu} \bar\sigma_{\mu\nu} = -6$.

An important property of the \<BPST> solution, which will have far-reaching generalizations at any~$K$ and for any gauge group, is that the solution depends on an arbitrary parameter~$\rho$, which sets the characteristic size of the instanton: from \cref{eq:FBPST}, we see that the radius of the instanton is $\sim \rho$.
We can take $\rho>0$, since only $\rho^2$ enters \cref{eq:FBPST} and the solution is singular in the limit $\rho\to0$.
In this limit, the instanton shrinks to zero size and becomes point-like.
This singularity manifests itself as a \<UV> divergence in the instanton calculus and will have an interesting interpretation in the string theory \<UV>-completion of the field theory.

Every value of $\rho>0$ corresponds to a different solution of the equation \cref{eq:inst}, as can be understood by a classical version of the Goldstone theorem: classical Yang-Mills theory is scale invariant (and even conformally invariant), the scale invariance is broken by the instanton solution and acting with a scale transformation on a solution hence produces a family of inequivalent solutions.
Such a continuous parameter on which the solution depends is called a \define{modulus} and the set\footnote{We will see in the following that the moduli space carries a lot of structure, it is actually a hyperkähler manifold (or an orbifold when compactifying the moduli space by including the point-like instantons).} of all parameters on which the general K-instanton solution depends is called the \define{moduli space} of $K$-instantons.

Actually, $\rho$ is not the only modulus of the \<BPST> instanton.
The solution \cref{eq:FBPST} breaks translational invariance and is centered at the origin, hence by acting with translations $x\to x-X$ we can generate a solution that is centered at any $X_\mu \in \RR^4$.
We can also act with global gauge transformations (or equivalently $\SU(2)_-$ rotations) parameterized by a constant $\SU(2)$ matrix~$V$, to finally obtain the following form
\begin{equation}
  F^{\textsc{BPST}}_{\mu\nu}(\rho,X_\mu,V) = \frac{4i\rho^2V\bar\sigma_{\mu\nu}V^\dagger}{((x-X)^2+\rho^2)^2} \, .
  \label{eq:FBPSTgen}
\end{equation}
This solution depends on 8 moduli: 1 scale~$\rho$, 4 coordinates~$X_\mu$ for the position of the center and 3 parameters for the gauge orientation~$V$.
It can be shown\footnote{An index theorem \cite{Atiyah:1968mp} implies that there are locally 8 parameters, see \cite{Vandoren:2008xg} for a pedagogical review. Globally, it is a consequence of the completeness of the \<ADHM> construction \cite{Atiyah:1977pw,Atiyah:1978ri}.} that \cref{eq:FBPSTgen} is the most general one-instanton solution (up to gauge transformations) and the one-instanton moduli space for gauge group $\SU(2)$ has hence dimension~8.

The 8-parameter \<BPST> solution \cref{eq:FBPSTgen} for $\SU(2)$ immediately yields a one-instanton solution for any $\SU(N)$ gauge group by embedding $\SU(2) \hookrightarrow \SU(N)$ through the map that sends a $2\times 2$ matrix to the top-left corner of a $N\times N$ matrix.
The chosen embedding is of course completely arbitrary, and we can generate inequivalent solutions by conjugation with an $\SU(N)$ matrix~$U$,
\begin{equation}
  F_{\mu\nu}(\rho,X_\mu,V,U) = U^\dagger \begin{pmatrix}
    F_{\mu\nu}^{\textsc{BPST}}(\rho,X_\mu,V) & 0 \\
      0 & 0
  \end{pmatrix} U \, .
  \label{eq:BPSTN}
\end{equation}
Actually, not all choices of~$U \in \SU(N)$ correspond to inequivalent solutions.
Indeed, the action of $U$ will be trivial whenever the top-left $2\times 2$ block of $U$ is the identity or when $U$ is of the form
\begin{equation}
  U(\theta) = \left( \begin{smallmatrix}
    e^{i(1-N/2)\theta} & & & & \\
    & e^{i(1-N/2)\theta} & & & \\
    & & e^{i\theta} & & \\
    & & & \ddots & \\
    & & & & e^{i\theta}
  \end{smallmatrix} \right) \, .
  \label{eq:Utrivial}
\end{equation}
Also, we can view any $V \in \SU(2)$ as a specific $U$ where only the $2\times 2$ top-left corner is non-trivial.
Hence, the subspace of the moduli space that parameterizes the orientation of the instanton inside $\SU(N)$ is given by the coset\footnote{
  The denominator is not simply the direct product $\SU(N-2)\times \U(1)$ because $U(\theta) \in \SU(N-2)$ if $\theta=\frac{2k\pi}{N-2}$ with $k\in \ZZ$, hence we need to identify this subgroup of $\U(1)$ with the $\ZZ_{N-2}$ center of $\SU(N-2)$.
}
\begin{equation}
  \frac{\SU(N)}{(\SU(N-2)\times \U(1))/\ZZ_{N-2}} \ni U \, .
  \label{eq:instorient}
\end{equation}
This manifold has dimension $N^2-1-[(N-2)^2-1+1]=4N-5$, so altogether the one-instanton moduli space for gauge group $\SU(N)$ has at least dimension $4N$.
It can be shown that this is actually the full story for $K=1$ and that the moduli space has exactly dimension $4N$.

What about the case $K>1$?
One can construct multi-instanton solutions of $\SU(N)$ by considering an $\cramped{\SU(2)^{\floor{N/2}}}$ subgroup (where $\floor\cdot$ is the integer part) and embedding at most one \<BPST> instanton in each $\SU(2)$ factor.
This yields a solution with $K\le\floor{\frac{N}{2}}$ which cannot be the most general case given that $K$ is unbounded for any fixed~$N$, as we now argue.
One could also try to embed several copies of a \<BPST> instanton in the same $\SU(2)$ factor but this does not correspond to a solution of the anti-self-duality equation because of its non-linearity.
While at this stage we cannot construct exact solutions for any $K$, we can nevertheless construct approximate solutions.
Indeed, let us take a family of $K$ copies of the one-instanton solution parameterized by $(\rho^{(i)},X_\mu^{(i)},U^{(i)})$ with $i=1,\ldots,K$.
If the different instantons are well separated, that is if
\begin{equation}
  (X_\mu^{(i)}-X_\mu^{(j)})^2\gg \rho^{(i)}\rho^{(j)} \qquad \forall \, 1\le i\neq j \le K \, ,
  \label{eq:instgas}
\end{equation}
the different instantons will have a very small overlap and the total field strength $F_{\mu\nu}=\sum_i F_{\mu\nu}^{(i)}$ will approximately solve equation \cref{eq:inst}.
This approximation is known as the \define{dilute instanton gas} limit since a $K$-instanton is considered as a gas made of $K$ copies of non-interacting identical one-instantons.
This argument shows that the full instanton moduli space has an asymptotic region in which the instanton gas picture is valid and hence looks like $K$ copies of the one-instanton moduli space.
This argument also shows that a $K$-instanton solution has $4KN$ moduli in the dilute limit, but it turns out that the equation \cref{eq:inst} is sufficiently well-behaved for the moduli space to be a manifold and hence have constant dimension~$4KN$.

\subsection{Solving the anti-self-duality equation: the \<ADHM>\ construction}
\label{sec:ADHM}

We are now going to discuss the general solution to the anti-self-duality equation \cref{eq:inst} for gauge group~$\SU(N)$ at any value of~$K$.
The construction of general instanton solutions was first achieved by \textcite{Atiyah:1978ri} and is called the \define{\<ADHM> construction} after them, but we will present here a slight variation thereof \cite{Khoze:1998gy,Dorey:2002ik} that is more adapted to $\SU(N)$ gauge group (as opposed to the $\operatorname{Sp}(n)$ formalism of the original paper).

The starting point of the construction is to take an ansatz for the gauge field of the form
\begin{equation}
  A = - i\, \bar\cU \d \cU \, ,
  \label{eq:AADHM}
\end{equation}
where we have used a form notation for conciseness, i.e.\ $A=A_\mu \d x_\mu$, and also adopted the notation $\bar\cU = \cU^\dagger$ that will prove convenient when dealing with explicit indices.
If $\cU$ were a (position-dependent) $\SU(N)$ matrix, this ansatz would be pure gauge, and we would get a trivial solution $F=0$ corresponding to $K=0$.
The clever idea is to relax this requirement, and take $\cU$ to be a \emph{non-square} matrix, of dimensions $(N+2K)\times N$.
By imposing further constraints on $\cU$, \cref{eq:AADHM} will then correspond to a solution of instanton number~$K$.

To this end, let us start by defining a $(N+2K) \times 2K$ complex-valued matrix $\Delta$ that depends linearly on~$x$,
\begin{subequations}
  \begin{align}
    \Delta\indices{_I^\alpha^j}(x) &=a\indices{_I^\alpha^j}+ i b\indices{_I_{\dot\alpha}^j}\bar\sigma\indices{_\mu^{\dot\alpha\alpha}}x_\mu \, , \label{eq:DeltaADHM} \\
  \Rightarrow \quad \bar\Delta\indices{_\alpha_i^J}(x) &=\bar a\indices{_\alpha_i^J} + i x_\mu \sigma_{\mu\alpha\dot\alpha}\bar b\indices{^{\dot\alpha}_i^J} \, ,
  \label{eq:DeltabarADHM}
\end{align}
\end{subequations}
where $I,J=1,\ldots N+2K$ and we have already anticipated the fact that $\alpha$ is an $\SU(2)_+$ index.
This matrix will eventually depend on the instanton moduli, but at this stage $a$ and $b$ contain much more than $4KN$ free parameters and we need to impose additional constraints on $\Delta$ to reduce their number and eventually be able to describe a $K$-instanton.
We will first impose a non-degeneracy condition on $\Delta$, requiring it to be of maximal rank for any value of $x$.
This amounts to asking that the family of linear maps $\Delta(x)\from \CC^{2K} \to \CC^{N+2K}$ be injective; its Hermitian conjugate $\bar\Delta(x)\from \CC^{N+2K} \to \CC^{2K}$ will then be surjective.
This non-degeneracy condition amounts to excluding the singular point-like instanton configurations~\cite{Dorey:2002ik}.
The matrix $\cU$ is then defined by demanding that the following short sequences be exact\footnote{Exactness of the second one follows automatically from exactness of the first one by taking the Hermitian conjugate.},
\begin{gather}
  0 \maprarrow{} \CC^{N} \maprarrow{\cU} \CC^{N+2K} \maprarrow{\bar\Delta} \CC^{2K} \maprarrow{} 0 \, , \label{eq:ADHMses1} \\
  0 \maplarrow{} \CC^{N} \maplarrow{\bar\cU} \CC^{N+2K} \maplarrow{\Delta} \CC^{2K} \maplarrow{} 0 \, . \label{eq:ADHMses2}
\end{gather}
The reader unfamiliar with exact sequences need not be scared, these diagrams only encode in a more visual way the additional requirements that $\cU$ and $\bar\cU$ are injective and surjective respectively, and that
\begin{equation}
  \bar\Delta\cU = 0 = \bar \cU \Delta \, .
  \label{eq:DeltabarUADHM}
\end{equation}
The matrix $\bar \cU \cU$ is an invertible $N\times N$ Hermitian matrix, and we can choose the basis in $\CC^N$ such that $\cU$ is a \enquote{generalized unitary matrix},
\begin{equation}
  \bar \cU \cU = \1_{N} \, .
  \label{eq:barUUADHM}
\end{equation}
Also $\bar\Delta \Delta$ is an invertible Hermitian $2K \times 2K$ matrix.
Its form is constrained by the ansatz \cref{eq:DeltaADHM,eq:DeltabarADHM} and we write
\begin{equation}
  \bar\Delta \Delta = \tilde f^{-1} \, .
  \label{eq:tildefADHM}
\end{equation}
Let us now consider the $(N+2K)\times (N+2K)$ matrix
\begin{equation}
  P = \cU \bar \cU \, .
  \label{eq:PADHM}
\end{equation}
By \cref{eq:barUUADHM}, $P = P^2$ is a projector onto the $N$-dimensional subspace $\im \cU = \ker \bar \Delta \subset \CC^{N+2K}$.
With a bit of linear algebra, one can see that the middle vector space splits as $\CC^{N+2K}=\im \Delta \oplus \im \cU$, which allows to show that
\begin{equation}
  P + \Delta \tilde f \bar \Delta = \1_{N+2K}
  \label{eq:PDeltaADHM}
\end{equation}
by acting with both sides on $\cU$ and $\Delta$.

We are now well-equipped to compute the field strength from \cref{eq:AADHM},
\begin{subequations}
  \begin{align}
  F = \d A + i A \wedge A &= -i \d ( \bar\cU \d \cU ) - i \,\bar\cU \d \cU \wedge \bar\cU \d \cU \label{eq:FADHM1} \\
  &= -i \d \bar\cU \wedge (\1 - \cU \bar \cU )\d \cU \label{eq:FADHM2} \\
  &= -i \d \bar\cU \wedge \Delta \tilde f \bar\Delta \d \cU \label{eq:FADHM3} \\
  &= -i \bar \cU \d\Delta \wedge \tilde f \d\bar\Delta \, \cU \, . \label{eq:FADHM4}
\end{align}
\end{subequations}
To get \cref{eq:FADHM2} one uses the Leibniz law as well as \cref{eq:barUUADHM}, to get \cref{eq:FADHM3} one uses \cref{eq:PDeltaADHM} and to get \cref{eq:FADHM4} one uses again the Leibniz law as well as \cref{eq:DeltabarUADHM} to see that all but the written term vanish.
From \cref{eq:DeltaADHM,eq:DeltabarADHM}, we see that $\d \Delta = i b \bar\sigma_\mu \d x_\mu$ and $\d \bar \Delta = i\sigma_\mu \bar b \d x_\mu$.
Taking into account the antisymmetry of the wedge product, we almost have $F_{\mu\nu} \propto \bar \sigma_{\mu\nu}$ which would imply anti-self-duality, but this is not quite the case since the matrix $\tilde f$ sits in the middle, $F_{\mu\nu} \propto \bar\sigma_{[\mu} \tilde f \sigma_{\nu]}$.
The solution to this problem is to force $\tilde f$ to commute with $\sigma$ by replacing the generic expression \cref{eq:tildefADHM} with the constraint
\begin{equation}
  \bar\Delta\indices{_\alpha_i^I} \Delta\indices{_I^\beta^j}= \delta_\alpha^\beta (f^{-1})\indices{_i^j} \, .
  \label{eq:ADHMconstraint}
\end{equation}
The field strength then takes the manifestly anti-self-dual form
\begin{equation}
  F_{\mu\nu} = 4 i \bar \cU b \bar\sigma_{\mu\nu} f \bar b\,  \cU \, .
  \label{eq:FADHM}
\end{equation}
While we will not have much use for this expression, the constraint \cref{eq:ADHMconstraint} plays an important role in the \<ADHM> construction and hence in this thesis.
We are thus going to make it more explicit.
First, note that by tracing on the $\SU(2)_+$ indices, we obtain $f^{-1} = \frac{1}{2}\tr_2 \bar\Delta \Delta$.
Since \cref{eq:ADHMconstraint} has to hold for any value of $x$, we can rewrite it as three different equations by expanding the definitions \cref{eq:DeltaADHM,eq:DeltabarADHM},
\begin{subequations}
  \begin{align}
    \bar a\indices{_\alpha_i^I} a\indices{_I^\beta^j} &= \tfrac{1}{2} (\tr_2 \bar a a)\indices{_i^j} \delta_\alpha^\beta \label{eq:ADHMconst1} \, , \\
    \bar a\indices{^\alpha_i^I} b\indices{_I^{\dot\beta j}} &= \bar b\indices{^{\dot\beta}_i^I} a\indices{_I^{\alpha j}} \label{eq:ADHMconst2} \, , \\
    \bar b\indices{^{\dot\alpha}_i^I} b\indices{_{I\dot\beta}^j} &= \tfrac{1}{2} (\tr_2 \bar b b)\indices{_i^j} \delta^{\dot\alpha}_{\dot\beta} \label{eq:ADHMconst3} \, .
  \end{align}
\end{subequations}
To obtain \cref{eq:ADHMconst2}, one can write $\bar ab = v_\mu \sigma_\mu$, $\bar b a = \bar v_\mu \sigma_\mu$ and realize that the only way to satisfy the constraint $v_\mu \sigma_\mu \bar \sigma_\nu + \bar v_\mu \sigma_\nu \bar \sigma_\mu \propto \I_2$ is $v_\mu = \bar v_\mu$ which can be massaged into \cref{eq:ADHMconst2}.
To obtain \cref{eq:ADHMconst3}, one can similarly expand $\bar b b = u_0 \1_2 + u_{\mu\nu} \sigma_{\mu\nu}$, where the constraint now implies that $w_{\mu\nu}=0$.

We can solve the constraints \cref{eq:ADHMconst2,eq:ADHMconst3} by exploiting the large gauge freedom of the \<ADHM> construction.
Indeed, we have not specified any basis for the vector space $\CC^{N+2K}$ nor $\CC^{2K}$.
In order to preserve \cref{eq:barUUADHM,eq:FADHM}, the most general transformations  we can do on these spaces are parameterized by constant matrices $\Lambda \in \operatorname{U(N+2K)}$ and $M \in \operatorname{GL}(K,\CC)$ respectively\footnote{We cannot act on the $\CC^{N}$ factor since this would correspond to a large gauge transformation.}.
They act as
\begin{equation}
  \Delta \to \Lambda \Delta M^{-1} \, , \quad \cU \to \Lambda \cU \, , \quad f \to M f M^{\dagger} \, ,
  \label{eq:gaugeADHM}
\end{equation}
and allow to gauge fix $b$ completely \cite{Corrigan:1978ce} ,
\begin{equation}
  b = \begin{pmatrix}
   0_{N\times 2K} \\
   \1_{2K\times 2K}
  \end{pmatrix} \, ,
  \label{eq:gaugefixADHM}
\end{equation}
which has the effect of splitting the lower index $I$ into an $\SU(N)$-fundamental (lower) index $f$ for its $N$ first values and a pair $(\dot\alpha,i)$ of (upper,lower) indices for the $2K$ next values.
Correspondingly, the matrices $a$ and $\bar a$ decompose as
\begin{equation}
  a\indices{_I^\alpha^j}= \begin{pmatrix}
    \frac{g}{4\pi} \tilde q\indices{^\alpha_f^j} \\
    \bar X\indices{^{\dot\alpha}^\alpha_i^j}
  \end{pmatrix} \, , \quad
  \bar a\indices{_{\alpha i}^J} = \left( \tfrac{g}{4\pi} q\indices{_{\alpha i}^f} \quad X\indices{_\alpha_{\dot\alpha}_i^j} \right) \, ,
  \label{eq:asplitADHM}
\end{equation}
where the $\frac{g}{4\pi}$ factor has been added for later convenience.
The form \cref{eq:gaugefixADHM} for $b$ does not fix the gauge completely, we are still allowed to act with $\U(K)$ transformations embedded in $\U(N+2K)$ and $\operatorname{GL}(K,\CC)$ in the obvious way.
It is most convenient not to fix the gauge completely and to leave $\U(K)$ as the residual gauge freedom of the construction.
It is called the \define{instanton gauge group}.
With $b$ given by \cref{eq:gaugefixADHM}, the constraint \cref{eq:ADHMconst3} is trivially solved, while \cref{eq:ADHMconst2} becomes the reality condition
\begin{equation}
  X\indices{^{\alpha\dot\alpha}_i^j} = \bar X\indices{^{\dot\alpha \alpha}_i^j} \, ,
  \label{eq:realXADHM}
\end{equation}
which means that we can take
\begin{equation}
  \bar X\indices{^{\dot\alpha \alpha}_i^j} = i X\indices{_{\mu i}^j}\bar\sigma\indices{_{\mu}^{\dot\alpha \alpha}} \, , \quad X\indices{_{\alpha\dot\alpha i}^j} = i X\indices{_\mu_i^j}\sigma_{\mu\alpha\dot\alpha} \, ,  \quad \text{with} \quad X_{\mu}=X_{\mu}^\dagger \, .
  \label{eq:XHermitianADHM}
\end{equation}
To take care of the remaining constraint \cref{eq:ADHMconst1}, we rewrite it in the form $\tr_2 \sigma_{\mu\nu} \bar a a = 0$, which combined with \cref{eq:asplitADHM,eq:XHermitianADHM,2sigmaid} yields
\begin{equation}
  \bmu_{\mu\nu} \equiv \frac{g^2}{16\pi^2} q_\alpha \tilde q^\beta \sigma\indices{_{\mu\nu}_\beta^\alpha} + [X_\mu,X_\nu]^+ = 0 \, .
  \label{eq:ADHMconst}
\end{equation}
This remaining constraint is called \define{the \<ADHM> constraint}.
It is an $\SU(2)_+$-triplet of $K\times K$ equations (the corresponding indices have been suppressed in \cref{eq:ADHMconst}) which cannot be solved in general for arbitrary $K$ and $N$.
This is a major technical hurdle in the practical use of instantons, but we will see later that one can manage to enforce this constraint in the $N\to\infty$ limit.

Let us now check that by taking the \<ADHM> constraint into account, this construction indeed describes $4KN$ independent moduli.
Counting the number of parameters in $X_\mu$ and $q_\alpha$ and subtracting the number of constraints in \cref{eq:ADHMconst} as well as the gauge redundancy parameterized by a $\U(K)$ matrix, we obtain
\begin{equation}
  4 K^2 + 4KN - 3K^2 - K^2 = 4 KN \, .
  \label{eq:ADHMcount}
\end{equation}
This is a strong indication that the parameter $K$ in this construction indeed coincides with the instanton number, though this can be checked explicitly by computing \cref{eq:YMtheta} from \cref{eq:FADHM} with the use of the \<ADHM> constraint \cref{eq:ADHMconst} (this is done in detail in \cite{Dorey:2002ik}).

Note that the formula \cref{eq:ADHMcount} is more than simple numerology and has an interesting geometric origin.
Indeed, one can view the space $\RR^{4K(K+N)}$ with coordinates $(X_\mu,\tilde q^\alpha, q_\alpha)$ as a hyperkähler manifold\footnote{A manifold is hyperkähler if it has $\operatorname{Sp}(n)$ holonomy. The space $\RR^{p}$ with the flat metric has trivial holonomy, hence it is hyperkähler when $p$ is a multiple of four.}.
The action of $\U(K)$ on the coordinates preserves the hyperkähler structure and the $\bmu_{\mu\nu}$ in \cref{eq:ADHMconst} is then the triplet of moment maps corresponding to this action.
The procedure of restricting to the zero-level set of the moment maps by imposing the \<ADHM> constraint \cref{eq:ADHMconst} and identifying the points in the same orbit of $\U(K)$ is then simply the \define{hyperkähler quotient}\footnote{The reader might be more familiar with the Kähler quotient construction that arises in supersymmetric theories when imposing the $D$-flatness constraints and identifying the vacua that differ by a gauge transformation.
The hyperkähler quotient is the quaternionic version of this.} of $\RR^{4K(K+N)}$ by $\U(K)$.
Many expressions that arise in the instanton calculus have an interpretation in the context of hyperkähler geometry, which explains the emphasis on the hyperkähler construction e.g.\ in \cite{Dorey:2002ik}.
In this thesis however, we will only scratch the surface of the instanton calculus in field theory, opting instead for the more straightforward approach of embedding it into brane constructions in string theory.

To close this section, let us summarize the most important point, namely that a general $K$-instanton solution in pure gauge theory can be constructed from the data in \cref{tab:ADHMdata}: the moduli $X_\mu$, $q_\alpha$, $\tilde q^\alpha$, subject to the \<ADHM> constraint \cref{eq:ADHMconst}.
We can make the further distinction between the \define{neutral moduli}, which are not charged under the $\SU(N)$ Yang-Mills gauge group, and the \define{charged moduli}, which are.
We will also write from now on $\cX$ collectively for the $4KN$ moduli satisfying the constraints.
\begin{table}
  \caption{The \<ADHM> data. The table summarizes the data (moduli subject to the constraints) used in the \<ADHM> construction of $K$-instanton solutions in Yang-Mills theory, as well as their representations under the Lorentz group~$\SU(2)_+\times \SU(2)_-$, Yang-Mills gauge group~$\SU(N)$ and instanton gauge group~$\U(K)$.}
  \centering
  \begin{tabular}{@{}lM{c}M{c}M{c}M{c}M{c}@{}}
    \toprule
    Type & \text{Name} & \SU(2)_+ & \SU(2)_- & \SU(N) & \U(K) \\
    \midrule
    Moduli (neutral)& X\indices{_\mu_i^j} & \mathbf{2} & \mathbf{2} & \mathbf{1} & \mathbf{Adj} \\
    \phantom{Moduli} (charged)& q\indices{_\alpha_i^f} & \mathbf{2} & \mathbf{1} & \mathbf{\bar N} & \mathbf{K} \\
    \phantom{Moduli} (charged)& \tilde q\indices{^\alpha_f^i} & \mathbf{2} & \mathbf{1} & \mathbf{N} & \mathbf{\bar K} \\
    Constraints & \bmu\indices{_{\mu\nu}_i^j} & \mathbf{3} & \mathbf{1} & \mathbf{1} & \mathbf{Adj} \\
    \bottomrule
  \end{tabular}
  \label{tab:ADHMdata}
\end{table}

\subsection{The instanton effective action}
\label{sec:YMinstaction}

In the previous section, we have constructed instanton solutions, that is \emph{classical} solutions to the anti-self-duality equations \cref{eq:inst}.
We are now going to look at the fate of instantons in the quantum theory.
As explained at the end of \cref{sec:YMinstgen}, instantons in Yang-Mills theory only make sense semi-classically, in a formal expansion in the coupling $g$.
The starting point for this is an expansion of the gauge field $A_\mu$ around a $K$-instanton solution,
\begin{equation}
  A_\mu(x) = A_\mu^K(x;\cX) + g \delta A_\mu(x) \, ,
  \label{eq:instpert}
\end{equation}
the gauge field being split into the solution (which depends on the moduli) and fluctuations around it.
Plugging this decomposition into the action \cref{eq:YMaction} plus \cref{eq:YMtheta} and treating $A^K$ as a background field and $\delta A_\mu$ as a quantum field then allows to compute correlators by following the same rules as in textbook QFT.
The only subtlety is that the field configuration $A^K_\mu$ around which we expand is not an isolated extremum of the action, but belongs to a moduli space of solutions.
As a consequence, we need to separate the fluctuations $\delta A_\mu$ into three categories.
\begin{description}
  \item[The pure gauge modes] $\delta_g A_\mu$ correspond to infinitesimal gauge transformations of the instanton gauge field
    \begin{equation}
      \delta_g A_\mu(x) = - \nabla^K_\mu \epsilon(x)
      \label{eq:puregaugemodes}
    \end{equation}
    where $\nabla^K= \d + i [A^K,\cdot]$ is the gauge-covariant derivative evaluated on the instanton.
    Pure gauge modes are always present in gauge theories, also when doing perturbation theory around the vacuum.
    Two field configurations differing by a gauge transformation are physically equivalent.
    In order to account for this, we need to impose a gauge condition that selects only one field configuration in each gauge orbit.
    A natural and convenient choice of gauge in this context is the background field gauge
    \begin{equation}
      \nabla^K_\mu \delta A_\mu = 0 \, .
      \label{eq:gaugefix}
    \end{equation}
    In the rest of this section, we will always assume that this gauge has been imposed.

  \item[The \define{zero modes}] $\delta_0 A_\mu$ correspond to fluctuations induced by varying the moduli\footnote{One needs to supplement this with a gauge transformation in order to preserve the gauge condition \cref{eq:gaugefix}, but such technicalities are not important at this level of discussion.},
    \begin{equation}
      g \delta_0 A_\mu(x) = A_\mu^K(x;\cX + \delta\cX) - A_\mu^K(x;\cX) + O(\delta \cX^2)\, .
      \label{eq:zeromodes}
    \end{equation}
    By construction, $A_\mu^K$ and $A_\mu^K+ g\delta_0 A_\mu$ are both $K$-instanton solutions and hence have the same action \cref{eq:deftau}.

  \item[The non-zero modes] $\deltap  A_\mu$ are all the remaining physical modes which do not correspond to variations of the moduli.
\end{description}

Geometrically, we can see small fluctuations $\delta A_\mu(x)$ as belonging to the tangent space at the point $A_\mu^K(x;\cX)$ of the (infinite-dimensional) field configuration manifold.
The moduli space of $K$-instantons is a finite-dimensional submanifold of this and the distinction between zero and non-zero modes then amounts to splitting this infinite-dimensional tangent space into the tangent and the normal space to the moduli space.
The normal space is non-canonical: it is isomorphic to a quotient of the full tangent space by the tangent space to the submanifold, but to view it in a unique way as a subspace requires a choice of metric.
This general statement has a simple realization also in our case: if we add a zero mode to any non-zero mode, we still have a non-zero mode, hence non-zero modes are a priori only defined up to the addition of zero modes.
This ambiguity can be fixed by introducing the following inner-product between fluctuations,
\begin{equation}
  \left(\delta_1 A, \delta_2 A\right) = \int \d^4x \tr \delta_1 A_\mu(x) \delta_2 A_\mu(x) \, ,
  \label{eq:instmetric}
\end{equation}
and demanding that the non-zero modes be orthogonal to the zero modes.

Plugging \cref{eq:instpert} into the action, we obtain an expansion of the form
\begin{equation}
  S[A^K + g\delta A] = - 2\pi i \tau K + \int \d^4x \tr \deltap  A_\mu \Delta_{\mu\nu}(x;\cX+\delta\cX) \deltap  A_\nu + O(g) \, .
  \label{eq:instpertaction}
\end{equation}
To quadratic order in the fluctuations, only the non-zero modes can contribute given that both $A^K$ and $A^K + g\delta_0 A$ solve the equations of motion but we can more generally view the addition of a zero mode as a shift of the moduli.
Hence what we called zero modes can be seen alternatively as zero modes of the quadratic operator $\Delta_{\mu\nu}$ appearing when expanding the action around an instanton background.

Let us now focus on the dynamics of the zero modes, or more precisely, of the moduli, induced by the dynamics of the full gauge theory.
The partition function $\mathcal Z_K$ in the $K$-instanton sector is given by the functional integral
\begin{equation}
  \mathcal Z_K = \int \mathcal DA \mathcal D\omega \, e^{i\vartheta K-S_{\text{\<YM>}}[A]-S_{\text{g.f.}}[A,\omega]} \, ,
  \label{eq:Kinstpart}
\end{equation}
where we have added the ghost sector $\omega$ and a gauge fixing functional, and suitable asymptotic conditions have been imposed on the gauge fields for the instanton number to stay fixed at $K$.
We can formally rewrite this partition function equivalently as
\begin{equation}
  \mathcal Z_K = \int \d\cX e^{-S_{\text{eff}}(\cX)} \, ,
  \label{eq:Kinstparteff}
\end{equation}
with the \define{instanton effective action} $S_{\text{eff}}$ given by
\begin{equation}
  e^{-S_{\text{eff}}(\cX)} = \int \mathcal D\deltap A \mathcal J(\cX) \det (\nabla^K)^2 \, e^{-i\theta K - S_{\text{\<YM>}}[A^K(\mathcal X)+\deltap  A]} \, .
  \label{eq:Kinsteffact}
\end{equation}
Here we integrate only over the non-zero modes, and in addition to the Fadeev-Popov determinant $\det (\nabla^K)^2$, there is a Jacobian~$\mathcal J$ for the change of variables from the zero modes $\delta_0 A_\mu$ to the moduli $\cX$.
The computation of \cref{eq:Kinstparteff} is technically quite involved in practice, even when truncating the perturbative series at one loop, because it involves computing the determinant of the operator in \cref{eq:instpertaction} over all the non-zero modes and also because the moduli space of instantons cannot be described completely explicitly due to the impossibility of solving the \<ADHM> constraint in general.
It has nevertheless been done for $K=1,2$, starting with the seminal paper \cite{tHooft:1976fv}.
The interested reader can find more details in \cite{Osborn:1981yf}.

In the supersymmetric setting, the first problem is absent because there are also fermionic fluctuations that exactly cancel the bosonic ones.
As will be shown later, the second problem can be dealt with in the $N\to\infty$ limit.
It will turn out that the effective action that is most relevant for applications to holography is not quite the same as the one appearing in \cref{eq:Kinstparteff}.
Instead of depending on the $4KN$ moduli satisfying the constraints, it depends on the $4K^2$ moduli $X_\mu$ and hence \cref{eq:Kinstparteff,eq:Kinsteffact} get replaced by
\begin{equation}
  \mathcal Z_K = \int \d X e^{-S_{\text{eff}}(X)}
  \label{eq:KinstparteffX}
\end{equation}
with
\begin{equation}
 e^{-S_{\text{eff}}(X)} = \int \d q \d\tilde q \d D\, \mathcal D \deltap A \, e^{- S_{\text{inst}}(X,q,\tilde q, D)-S_{\text{gauge}}[\deltap A;X,q,\tilde q]} \, .
  \label{eq:KinsteffactX}
\end{equation}
The main difference with \cref{eq:Kinsteffact} is that we now need to integrate over the remaining moduli $q^\alpha$ and $\tilde q_\alpha$ to compute the effective action and that we also need to enforce the \<ADHM> constraint \cref{eq:ADHMconst}.
This is done by adding a Lagrange multiplier $D$ for the constraint.
In order to be able to write a linear term of the form $D\cdot \bmu$ that is a scalar, it needs to be a self-dual antisymmetric tensor $D_{\mu\nu}=D^+_{\mu\nu}$ in the adjoint of $\U(K)$.
The instanton action $S_{\text{inst}}$ then contains the terms that only depend on the moduli,
\begin{multline}
  S_{\text{inst}}(X,q,\tilde q,D) = -2\pi i\tau K + \frac{8\pi^2}{g^2} \tr \left( D_{\mu\nu}[X_\mu,X_\nu] \right) + \frac{1}{2}\tilde q\indices{^\alpha_f} D_{\mu\nu}\sigma\indices{_{\mu\nu\alpha}^\beta} q\indices{_\beta^f} \\ +  \log \mathcal J(X,q,\tilde q) \, ,
  \label{eq:SinstYM}
\end{multline}
namely the constant term coming from the action evaluated on the instanton solution, the Lagrange multiplier term that we have written in a way convenient for later use, as well as the term involving the Jacobian $\mathcal J$ of the change of variables $\delta_0 A \to (X,q,\tilde q)$.
The gauge action $S_{\text{gauge}}$ then contains all the other terms in \cref{eq:Kinsteffact}, that is the Yang-Mills action evaluated on the non-zero modes (with the constant piece subtracted) and the gauge fixing term.
In the supersymmetric case, we will have very similar formulas with additional fermionic moduli to serve as the superpartners of the bosonic moduli.

\section{Instantons in supersymmetric theories}

In this section, we are going to extend the previous discussion of instantons in the pure Yang-Mills theory to their supersymmetric siblings in supersymmetric gauge theories.

\subsection{A quick review of supersymmetric gauge theories}
\label{sec:SUSYreview}

Supersymmetric theories have the defining property of possessing conserved charges that are \emph{fermionic} and extend the bosonic Poincaré algebra to a superalgebra \cite{Haag:1974qh}, see e.g.\ \cite{Sohnius:1985qm,Bilal:2001nv} for reviews.
Recall that these additional fermionic charges, the \define{supercharges} $Q$ and $\bar Q$, obey the following anti-commutation relations:
\begin{equation}
  \!\!\left\{ Q\indices{_\alpha^a},Q\indices{_\beta^b} \right\} = 2\epsilon_{\alpha\beta} Z^{ab} \, , 
  \left\{ \bar Q\indices{^{\dot\alpha}_a},\bar Q\indices{^{\dot\beta}_b} \right\} = 2 \epsilon^{\dot\alpha\dot\beta} \bar Z_{ab} \, , 
  \left\{ Q\indices{_\alpha^a},\bar Q\indices{_{\dot\alpha}_b} \right\} = 2 i \sigma_{\mu\alpha\dot\alpha} \delta^a_b P_\mu \, ,
  \label{eq:susyalg}
\end{equation}
where $a=1,\cdots,\nn$ is an R-symmetry index and $Z$ and $\bar Z$ are central charges (they commute with any element of the algebra).

The case we will be mostly interested in is $\nn=4$, as all the examples we will consider in this thesis can be obtained from this theory by small deformations, orbifolding or T-duality.
Four is the highest possible value of $\nn$ that can be realized by an ordinary quantum field theory with fields whose spin is at most one.
This field theory is almost unique, depending only on a gauge group, which we will take to be $\SU(N)$, and one holomorphic coupling $\tau$ as in \cref{eq:deftau}.
It is an extension of the bosonic Yang-Mills theory discussed in \cref{sec:YMinst}, supplementing the gauge field with an $\operatorname{SO}(6)$-vector of Hermitian scalar fields $\varphi_A$ and two spinors of $\operatorname{Spin}(6)\simeq \SU(4)$, $\lambda_a$ and $\bar\lambda^a$, all transforming in the adjoint of $\SU(N)$.
Its action can be conveniently found by dimensional reduction of the ten-dimensional $\nn=1$ theory \cite{Brink:1976bc} and reads
\begin{multline}
  S_{\nn=4} =\frac{1}{g^2}\int \d^4x \tr \left\{ \frac{1}{2}F_{\mu\nu}F_{\mu\nu} + \frac{i\vartheta g^2}{16\pi^2} F_{\mu\nu}\hodge{F}_{\mu\nu} + \nabla_\mu \varphi_A  \nabla_\mu \varphi_A  \right. \\
    \left. -\frac{1}{2}[\varphi_A,\varphi_B][\varphi_A,\varphi_B] + 2i \lambda\indices{^\alpha_a}\sigma_{\mu\alpha\dot\alpha}\nabla_\mu\bar\lambda^{\dot\alpha a} \right. \\
    \left. - \lambda\indices{^\alpha_a}\bar\Sigma_A^{ab}[\varphi_A,\lambda_{\alpha b}] - \bar\lambda\indices{_{\dot\alpha}^a}\Sigma_{Aab}[\varphi_A,\bar\lambda^{\dot\alpha b}] \vphantom{\frac{1}{2}} \right\} \, .
  \label{eq:actionN4}
\end{multline}
The definitions of the $\Sigma_A$ and $\bar\Sigma_A$ matrices can be found in \cref{Sigmadef6D,Sigmabardef6D}.
This theory has the remarkable property of being conformal for any value of the coupling \cite{Mandelstam:1982cb}.
At the bosonic level, the Poincaré algebra is enhanced to the conformal algebra, but since we are dealing with a supersymmetric theory, the super-Poincaré algebra is enhanced to a superconformal algebra that contains 32 supercharges: the 16 super-Poincaré charges in \cref{eq:susyalg} and 16 additional superconformal charges $S$ and $\bar S$ whose anticommutator closes on the special conformal transformations.

The supersymmetry transformations of the fields read
\begin{subequations}
  \begin{align}
  \delta A_\mu &= -i \xi\indices{^\alpha_a}\sigma_{\mu\alpha\dot\alpha}\bar\lambda^{\dot\alpha a} - i \bar\xi\indices{_{\dot\alpha}^a}\bar\sigma_\mu^{\dot\alpha\alpha}\lambda_{\alpha a} \, , \label{eq:susyN4A} \\
  \delta \varphi_A &= -i \xi\indices{^\alpha_a}\bar\Sigma_A^{ab}\lambda_{\alpha b} - i \bar\xi\indices{_{\dot\alpha}^a}\Sigma_{Aab}\bar\lambda^{\dot\alpha b} \, , \label{eq:susyN4phi} \\
  \delta \lambda\indices{^\alpha_a} &= \xi\indices{^\beta_a}\sigma\indices{_{\mu\nu\beta}^\alpha} F_{\mu\nu} - i \xi\indices{^\alpha_b}\bar\Sigma\indices{_{AB}^b_a}[\varphi_A,\varphi_B] - \bar\xi\indices{_{\dot\alpha}^b}\bar\sigma_\mu^{\dot\alpha\alpha}\Sigma_{Aba}\nabla_\mu\varphi_A \, , \label{eq:susyN4lambda} \\
  \delta \bar\lambda\indices{_{\dot\alpha}^a} &= \bar\xi\indices{_{\dot\beta}^a}\bar\sigma\indices{_{\mu\nu}^{\dot\beta}_{\dot\alpha}} F_{\mu\nu} - i \bar\xi\indices{_{\dot\alpha}^b}\bar\Sigma\indices{_{ABb}^a}[\varphi_A,\varphi_B] -\xi\indices{^\alpha_b}\sigma_{\mu\alpha\dot\alpha}\bar\Sigma_{A}^{ba}\nabla_\mu\varphi_A \, . \label{eq:susyN4lambdabar}
\end{align}
\end{subequations}
With $\xi_a$ and $\bar\xi^a$ constant spinors, the transformations correspond to the action of the super-Poincaré charges on the fields.
If we take $\xi_a=ix_\mu\sigma_{\mu}\bar\zeta_a$ and $\bar\xi^a = ix_\mu\bar\sigma_{\mu}\zeta^a$ instead, we obtain the action of the superconformal charges on the fields.

By setting some of the fields to zero, one can obtain theories that still preserve some supersymmetry.
To obtain the pure $\nn=2$ super-Yang-Mills theory one needs to set $\varphi_A=0$ for $A=3,\cdots,6$ as well as $\lambda_a = \bar\lambda^a=0$ for $a=3,4$.
To obtain the pure $\nn=1$ super-Yang-Mills theory, one needs to set also the remaining scalars to zero and keep only one pair $(\lambda_1,\bar\lambda^1)$.
If we also set these fermions to zero, we are back at the purely bosonic $\nn=0$ theory considered in \cref{sec:YMinst}.
We will see later that one can give a nice geometric interpretation in string theory for this procedure of setting well-chosen fields to zero.
These less supersymmetric theories are not conformal at the quantum level.
For $\nn=2$, the perturbative $\beta$-function for $g^2$ is one-loop exact, but the theory has a much richer structure non-perturbatively \cite{Seiberg:1994rs}.
For $\nn=1$, the $\beta$-function is non-trivial already at the perturbative level \cite{Novikov:1983uc,ArkaniHamed:1997mj}.

\subsection{The super-instanton}
\label{sec:SUSYinst}

Let us now generalize the discussion of instantons to these supersymmetric theories.
We can go through the same steps as in \cref{sec:YMinstgen}, writing the action \cref{eq:actionN4} as a sum of manifestly positive terms.
For the $F^2$ terms, exactly the same argument applies and yields the \RHS\ of \cref{eq:YMsquare}.
The other bosonic terms are already positive: for the scalar kinetic term this is manifest, while for the quartic terms we need to remember that a commutator of Hermitian matrices is anti-Hermitian, hence $i[\varphi_A,\varphi_B]$ is a Hermitian matrix.
We can thus rewrite the bosonic part of the action \cref{eq:actionN4} as
\begin{multline}
  S_{\nn=4} =  \frac{1}{g^2} \int \d^4x \tr \left\{ \frac{1}{4}(F_{\mu\nu}\pm \hodge{F}_{\mu\nu})^2 + (\nabla_\mu\varphi_A)^2 + \frac{1}{2} (i[\varphi_A,\varphi_B])^2  \right\} \\
  - (i\vartheta \mp \frac{8\pi^2}{g^2})K + \text{fermions} \, .
  \label{eq:N4squares}
\end{multline}
An instanton solution, which minimizes the bosonic part of the action at fixed instanton number $K$, now requires
\begin{subequations}
  \begin{gather}
  F_{\mu\nu}^+ = 0 \, , \label{eq:SinstF} \\
  \nabla_{\mu}\varphi_A = 0 \, , \label{eq:Sinstphi} \\
  [\varphi_A,\varphi_B] = 0 \, , \label{eq:Sinstphi2}
  \intertext{and the full equations of motion taking the fermions into account allow us to set also}
  \lambda = \bar\lambda = 0 \, . \label{eq:Sinstlambda}
\end{gather}
\end{subequations}
Condition \cref{eq:SinstF} is the same as \cref{eq:inst} in the $\nn=0$ theory and its solutions are provided by the \<ADHM> construction.
Condition \cref{eq:Sinstphi2} asks for the scalar fields to be diagonalizable simultaneously, but \cref{eq:Sinstphi} asks for the stronger condition $\varphi_A=0$ since it implies $(\nabla^K)^2\varphi_A=\nabla^K_\mu\nabla^K_\mu\varphi_A=0$ which has no other normalizable solutions given that $-(\nabla^K)^2$ is a positive operator.
We thus conclude that any bosonic instanton solution of the $\nn$-extended super-Yang-Mills theory comes from an instanton solution of the non-supersymmetric Yang-Mills theory.

We have seen in \cref{eq:instspinor} that from the point of view of the preserved symmetries, an instanton in the $\nn=0$ theory preserves the $\SU(2)_+$ subgroup of the Lorentz group.
This fact has an important generalization in the supersymmetric setting.
Indeed, plugging the conditions \cref{eq:SinstF,eq:Sinstphi,eq:Sinstphi2,eq:Sinstlambda} into the supersymmetry transformations \cref{eq:susyN4A,eq:susyN4phi,eq:susyN4lambda,eq:susyN4lambdabar}, we see that an instanton configuration is also invariant under transformations with vanishing $\bar\xi$ but arbitrary $\xi$, that is under the left \emph{super}-charges $Q_\alpha$ (and also $S_\alpha$ for $\nn=4$).
Hence an instanton solution preserves one-half of the supersymmetries of the vacuum; this is called a \define{$\frac{1}{2}$-\<BPS> configuration}.
The preserved supersymmetry entails non-renormalization properties, explaining the relevance of instantons in supersymmetric theories away from the semi-classical regime.

Given that the instanton breaks half the supercharges, acting with the broken supersymmetry transformations on a solution to \cref{eq:susyN4A,eq:susyN4phi,eq:susyN4lambda,eq:susyN4lambdabar} yields a new instanton solution but with non-vanishing right-handed fermions $\bar\lambda(x;\cX) = \bar\xi \bar\sigma_{\mu\nu}F_{\mu\nu}(x;\cX)$.
One might expect to have more generally $4KN\frac{\nn}{2}$ fermionic moduli as superpartners of the $4KN$ bosonic moduli.
However, this is not quite the case, as we now discuss.
Let us expand the action \cref{eq:actionN4} order by order in the fermions.
At zeroth order, only the bosonic terms in \cref{eq:N4squares} appear and the solution is the one considered earlier.
At quadratic order, the kinetic terms contribute and yield the equations of motion
\begin{subequations}
  \begin{align}
    \bar\sigma_{\mu}^{\dot\alpha \alpha} \nabla^K_\mu\lambda_{\alpha a} & = 0 \, , \label{eq:Sinstlambda2} \\
    \sigma_{\mu\alpha\dot\alpha}\nabla^K_\mu\bar\lambda^{\dot\alpha a} & = 0 \, . \label{eq:Sinstlambdabar2}
  \end{align}
\end{subequations}
Acting with $\sigma_\nu\nabla_\nu^K$ on \cref{eq:Sinstlambda2} and $\bar\sigma_\nu\nabla_\nu^K$ on \cref{eq:Sinstlambdabar2}, we obtain by symmetrizing and antisymmetrizing the $\mu,\nu$ indices
\begin{subequations}
  \begin{align}
    [-(\nabla^K)^2+\sigma_{\mu\nu}F_{\mu\nu}]\lambda &= 0 \, , \label{eq:Sinstlambda3} \\
    [-(\nabla^K)^2+\bar\sigma_{\mu\nu}F_{\mu\nu}]\bar\lambda &= 0 \, . \label{eq:Sinstlambdabar3}
  \end{align}
\end{subequations}
Since $F_{\mu\nu}^+=0$, \cref{eq:Sinstlambda3} reduces to $-(\nabla^K)^2\lambda=0$, which has no non-trivial normalizable solutions.
On the other hand, \cref{eq:Sinstlambdabar3} does admit non-trivial solutions.
They can be found \cite{Dorey:2002ik} by going through a construction similar to the bosonic \<ADHM> construction explained in \cref{sec:ADHM}.
The result is that the solutions are given in terms of data similar to the bosonic \<ADHM> data, namely fermionic matrices $\bar\psi$, $\chi$ and $\tilde\chi$, with charges as in \cref{tab:ADHMferm}, and subject to fermionic \<ADHM> constraints
\begin{equation}
  \bmu\indices{_\alpha^a^i_j} \equiv \frac{8\sqrt{2}\pi^2}{g^2}\sigma_{\mu\alpha\dot\alpha}[X_\mu,\bar\psi^{\dot\alpha a}]\indices{_i^j} + \chi\indices{^a_i^f}\tilde q\indices{_\alpha_f^j} +  q\indices{_\alpha_i^f}\tilde\chi\indices{^a_f^j} =0 \, .
  \label{eq:ADHMconstferm}
\end{equation}
\begin{table}
    \caption{The fermionic \<ADHM> data and their representations under the different symmetry and gauge groups. They supplement the bosonic data in \cref{tab:ADHMdata} on \cpageref{tab:ADHMdata} in a theory with $\nn$ adjoint right-handed fermions.}
  \label{tab:ADHMferm}
  \leavevmode
  \begin{minipage}{\textwidth}
    \setfootnoterule{0pt}
    \hspace{-2pt}\begin{tabular*}{\textwidth}{@{}lMcMcMcMcMcMc@{}}
    \toprule
   Type & \text{Name} & \SU(2)_+ & \SU(2)_- & \SU(\nn) & \SU(N) & \U(K)  \\
   \midrule
   Moduli\footnote{As discussed in the text, these are not proper moduli in general beyond linear order.} (neutral) & \bar\psi\indices{^{\dot\alpha a}_i^j} & \mathbf{1} & \mathbf{2} & \boldsymbol{\bar\nn} & \mathbf{1} & \mathbf{Adj} \\
   \phantom{Moduli$^a$} (charged) & \chi\indices{^a_i^f} & \mathbf{1} & \mathbf{1} & \boldsymbol{\bar\nn} & \mathbf{\bar N} & \mathbf{K} \\
   \phantom{Moduli$^a$} (charged) & \tilde\chi\indices{^a_f^i} & \mathbf{1} & \mathbf{1} & \boldsymbol{\bar\nn} & \mathbf{N} & \mathbf{\bar K} \\
   Constraints & \bmu\indices{_\alpha^a^i_j} & \mathbf{2} & \mathbf{1} & \boldsymbol{\bar\nn} & \mathbf{1} & \mathbf{Adj} \\
   \bottomrule
  \end{tabular*}
\end{minipage}
\end{table}
Counting the number of independent components, we arrive at $2KN\nn$ as expected by supersymmetry.
However, so far we have only solved the equations of motion for the fermions at linear order, disregarding their couplings to the other fields.
From the last term of \cref{eq:actionN4}, we see that the scalar fields are sourced by bilinear terms in the right-handed fermions.
Solving the scalar equation of motion with these sources then yields a solution $\varphi_A \propto \bar\lambda\bar\lambda$ that will not be written down explicitly.
Non-vanishing scalars in turn source all the fields in the theory at sufficiently high order\footnote{This is the generic situation. For $\nn=1$ or $\nn=2$ with vanishing \VEV s, this does not happen and the classical solution with the fermionic moduli turned on is exact.} in $\bar\lambda$.
While one could attempt to solve these coupled equations of motion perturbatively, this is not so useful in practice, as this classical perturbative expansion is quantum-corrected anyway.
Rather, we give up the requirement of having an exact solution, and consider a super-instanton to be an off-shell object, parameterized by the collection of the bosonic and fermionic \<ADHM> data subject to their respective constraints, but which only solves the equations of motion up to quadratic order in $\bar\lambda$.
At this point, since we are not looking for an exact solution, we may as well relax the requirement $\varphi_A=0$ coming from \cref{eq:Sinstphi} and only enforce \cref{eq:Sinstphi2}.
Evaluating the action \cref{eq:actionN4} on the \<ADHM> data then yields a potential already in the classical theory:
the moduli space is lifted and we are dealing with \define{quasi-moduli} rather than moduli for $\bar\lambda\neq0$, though we will keep calling them moduli for brevity.

With this additional subtletly compared to the bosonic case, one can quantize the theory around the super-instanton as in \cref{sec:YMinstaction}.
Our goal is to write supersymmetric versions of \cref{eq:KinstparteffX,eq:KinsteffactX,eq:SinstYM}.
It is again convenient to add a Lagrange multiplier $\Lambda^\alpha_a$ in the adjoint of $\U(K)$ to enforce the fermionic \<ADHM> constraint \cref{eq:ADHMconstferm}.
It turns out to be also quite natural to introduce further auxiliary moduli $\phi_A$ in the adjoint of $\U(K)$, which produce both the Jacobian determinants for the change of variables from zero-modes to moduli as well as the potential for the moduli when put on-shell.
The analogue of the partition function \cref{eq:KinstparteffX} that will be relevant is
\begin{equation}
  \mathcal Z_K = \int \d X \d\phi \d\bar\psi \d\Lambda \, e^{-S_{\text{eff}}(X,\phi,\lambda,\bar\psi)} \, .
  \label{eq:SKinstparteff}
\end{equation}
In addition to the minimal extension of the bosonic case by including the dependence on the fermionic moduli $\bar\psi$, also the moduli $\phi$ and $\Lambda$ have been kept because they will enter crucially in the holographic interpretation later on.
The effective action is given by integrating out all other moduli and non-zero modes of the gauge-fixed theory, generalizing \cref{eq:KinsteffactX}, with an action $S_{\text{inst}} + S_{\text{gauge}}$.
The former, the instanton action, again contains all the terms that do not depend on the non-zero modes but only on the moduli and generalizes \cref{eq:SinstYM},
\begin{multline}\label{eq:SinstN4}
  S_{\text{inst}}(X,\phi,D,q,\tilde q,\bar\psi,\Lambda,\chi,\tilde\chi)=-2\pi i \tau K
 + \frac{4\pi^{2}}{g^2}\tr \Bigl\{
2iD_{\mu\nu}\bigl[X_{\mu},X_{\nu}\bigr] \\
-\bigl[X_{\mu},\phi_{A}\bigr]\bigl[X_{\mu},\phi_{A}\bigr]
-2  \Lambda^{\alpha}_{\ a}\sigma_{\mu\alpha\dot\alpha}
\bigl[X_{\mu},\bar\psi^{\dot\alpha a} \bigr]
- \bar\psi_{\dot\alpha}^{\ a}\Sigma_{Aab}
\bigl[\phi_{A},\bar\psi^{\dot\alpha b}\bigr]\Bigr\}\\
+ \frac{i}{2}
\tilde q^{\alpha}D_{\mu\nu}\sigma_{\mu\nu\alpha}^{\hphantom{\mu\nu\alpha}\beta}q_{\beta}+\frac{1}{2}\tilde q^{\alpha}\phi_{A}\phi_{A}q_{\alpha}-\frac{1}{2}
\tilde\chi^{a}\Sigma_{Aab}\phi_{A}\chi^{b}\\
+ \frac{1}{\sqrt{2}} \tilde q^{\alpha}\Lambda_{\alpha a}\chi^{a} + \frac{1}{\sqrt{2}}
\tilde\chi^{a}\Lambda^{\alpha}_{\ a}q_{\alpha} \, .
\end{multline}
As expected, this action is invariant under supersymmetry transformations generated by the left-moving supercharges.
The supersymmetry variations of the moduli read
\begin{subequations}
  \begin{align}
\delta X_\mu &= -i \xi\indices{^\alpha_a}\sigma_{\mu\alpha\dot\alpha}\bar\psi^{\dot\alpha a}  \, , \label{eq:susyinstX} \\
  \delta \phi_A &= -i \xi\indices{^\alpha_a}\bar\Sigma_A^{ab}\Lambda_{\alpha b} \, , \label{eq:susyinstphi} \\
  \delta \Lambda\indices{^\alpha_a} &= \xi\indices{^\beta_a}\sigma\indices{_{\mu\nu\beta}^\alpha} D_{\mu\nu} - i \xi\indices{^\alpha_b}\bar\Sigma\indices{_{AB}^b_a}[\phi_A,\phi_B] \, , \label{eq:susyinstLambda} \\
  \delta \bar\psi\indices{_{\dot\alpha}^a} &= -i\xi\indices{^\alpha_b}\sigma_{\mu\alpha\dot\alpha}\bar\Sigma_{A}^{ba} [X_\mu,\phi_A] \, , \label{eq:susyinstpsibar} \\
  \delta D_{\mu\nu} &= -\xi\indices{^\alpha_a}\sigma\indices{_{\mu\nu\alpha}^\beta}\bar\Sigma_A^{ab} [\phi_A,\Lambda_{\beta b}] \, , \label{eq:susyinstD} \\
  \delta q_\alpha &= -i\sqrt2 \xi_{\alpha a}\chi^a \, , \label{eq:susyinstq} \\
  \delta \chi^a &= i\sqrt2 \xi\indices{^\alpha_b}\bar\Sigma^{ab}_A\phi_A q_\alpha \, , \label{eq:susyinstchi} \\
  \delta \tilde q^\alpha &= -i\sqrt2\xi\indices{^\alpha_a}\tilde\chi^a \, , \label{eq:susyinstqtilde} \\
  \delta \tilde \chi^a &= -i\sqrt2 \xi_{\alpha b}\bar\Sigma_A^{ab}\phi_A\tilde q^\alpha \, . \label{eq:susyinstchitilde}
  \end{align}
\end{subequations}
Note that the transformations of the neutral moduli \cref{eq:susyinstX,eq:susyinstphi,eq:susyinstLambda,eq:susyinstpsibar} look very similar to the transformations of the fields \cref{eq:susyN4A,eq:susyN4phi,eq:susyN4lambda,eq:susyN4lambdabar}: one only needs to substitute $A \to X$, $\varphi \to \phi$, $\lambda \to \Lambda$, $\bar\lambda \to \bar\psi$, $F\to D$ and forget the partial derivative inside the $\nabla$.
One might argue that they are constrained to be similar by symmetry arguments, but we will see later that there is a better reason for this fact.

The field theoretic derivation of the instanton action \cref{eq:SinstN4} has only been sketched in this chapter, and the interested reader may consult \cite{Dorey:2002ik} for the details.
Indeed, the explicit computations are technically too involved and not enlightening enough to fit in this thesis.
As a consequence, only the main features have been presented in order to convince the reader that instantons exist in (supersymmetric) gauge theories and can be parameterized by the moduli that were discussed.
We will see in the coming chapters that, by using insights from string theory, one may derive \cref{eq:SinstN4} in a much simpler way and also provide an interpretation for the effective action in \cref{eq:SKinstparteff} that will be of paramount importance to this thesis.

\chapter{Strings, D-branes and instantons}
\label{chap:D-branes}

In this chapter, we present the String Theory ingredients that are necessary to provide the groundwork for the rest of the thesis.
After a brief reminder of some basic features of type \II\ superstring theories in \cref{sec:stringgen}, \cref{sec:D-branes} discusses D-branes in the open string picture in various settings and most importantly in \cref{sec:Dinstantons} how to embed the super-instanton of \cref{chap:instantons} in a D-brane setup.
Standard references on String Theory are \cite{Green:1987sp,Green:1987mn,Polchinski:1998rq,Polchinski:1998rr} and a nice review about instantons (and other solitons) in the D-brane context can be found in~\cite{Tong:2005un}.

\section{Elements of type \<II>\ string theory}
\label{sec:stringgen}

Let us recall some basic facts about String Theory, focusing in particular on the type \II\ theories.
Perturbative string theory is constructed by quantizing the embeddings~$X\from \Sigma \to \mathcal M$ of the two-dimensional string worldsheet into spacetime \cite{Polchinski:1998rq}.
We will first assume that $\mathcal M$ is flat $D$-dimensional Euclidean spacetime but we will consider slightly more general target spaces later.
In the Ramond-Neveu-Schwarz approach to superstring theory, the Riemann surface $\Sigma$ is replaced by a super-Riemann surface (see \cite{Green:1987mn,Polchinski:1998rr} for textbook discussions, \cite{Witten:2012bh} for a recent detailed approach).
The bosonic embedding coordinates $X_M$, $M=1,\ldots,D$ now have fermionic superpartners $(\psi_M,\tilde\psi_M)$, and consistency of the theory requires $D=10$.

\subsection{The closed string spectrum}
\label{sec:stringintro}

The simplest boundary condition one can put on the string is to demand that is has no spatial boundaries: this defines the \define{closed string}.
If we take the worldsheet to be a cylinder with Euclidean time $\tau$ and position $\sigma$ along the string, this boundary condition requires the bosons to be periodic
\begin{equation}
  X_M(\tau,\sigma+2\pi)=X_M(\tau,\sigma) \, ,
  \label{eq:Xbdcond}
\end{equation}
while the fermions can be either periodic or anti-periodic\footnote{One can also make the bosons anti-periodic, but this necessarily breaks translation invariance of the target space. This possibility will be considered later.}
\begin{equation}
  \psi_M(\tau,\sigma+2\pi) =
  \begin{cases*}
   + \psi_M(\tau,\sigma) & Ramond (\<R>)\\
   - \psi_M(\tau,\sigma) & Neveu-Schwarz (\<NS>)
  \end{cases*}
  \label{eq:psibdcond}
\end{equation}
and similarly for $\tilde\psi$.
Hence, there are four different closed string sectors differing by the choice of the fermion boundary conditions for $(\psi,\tilde\psi)$: $(\text{\<NS>},\text{\<NS>})$, $(\text{\<R>},\text{\<NS>})$, $(\text{\<NS>},\text{\<R>})$, $(\text{\<R>},\text{\<R>})$.
The zero modes in the Ramond sector are massless and obey a Clifford algebra in ten dimensions.
The mode expansion in the \<NS> sector contains a tachyon, which has to be eliminated by performing a \<GSO> projection.
The action of the \<GSO> projection in the Ramond sector leads to two inequivalent theories: the vacua in the left- and right-moving Ramond sectors are Weyl spinors and one can choose their chiralities to differ or to coincide.
The two resulting theories are called \define{Type \IIA} and \define{Type \IIB} String Theory respectively.

Both theories in flat ten-dimensional space-time are invariant under 32 global supercharges.
In the type \IIA\ case, the supersymmetry transformations are parameterized by two Majorana-Weyl fermions $\epsilon_1$ and $\epsilon_2$ of opposite chiralities, in the type \IIB\ case, they are two Majorana-Weyl fermions of the same chirality.

The massless spectrum of these theories is of particular relevance.
Let us first describe the \IIA\ theory.
From the $(\text{\<NS>},\text{\<NS>})$ sector one gets a dilaton $\Phi$, a graviton $h_{MN}$ and an antisymmetric two-form potential $B_{MN}$.
From the $(\text{\<R>},\text{\<NS>})$ and $(\text{\<NS>},\text{\<R>})$ sectors one gets two Majorana-Weyl gravitinos (with spin $\frac{3}{2}$) and dilatinos (with spin $\frac{1}{2}$) with opposite chirality.
The $(\text{\<R>},\text{\<R>})$ sector is a bi-spinor $\Psi \bar\Psi$, bilinear in two spinors of different chiralities which can be traded for a set of differential forms by contraction with gamma-matrices $\Gamma_{M_1\cdots M_n}$.
The non-vanishing terms correspond to $p$-form fields with even\footnote{This is for the massless type \IIA\ case. The massive case \cite{Romans:1985tz} has also a $F_0$ but will not be considered in this thesis.}~$p$, $F_2$ and $F_4$.
The \IIB\ theory has the same $(\text{\<NS>},\text{\<NS>})$ sector since the \<GSO> projection in the \<NS> sector is unique.
The other sectors differ:
the mixed sector has two gravitinos and dilatinos of same chirality, while the \<RR> $p$-form fields have odd $p$, they are $F_1$, $F_3$ and $F_5$.
The field $F_5$ obeys a self-duality condition, which reads in Euclidean signature
\begin{equation}
  \star F_5 = -i F_5
  \label{eq:F5sd}
\end{equation}
as a consequence of the similar identity obeyed by the 5-index gamma-matrix.

The low-energy limit of type \IIA/\IIB\ string theory is type \IIA/\IIB\ supergravity.
The details of these supergravity theories will not be needed in this thesis, whose focus is on deriving supergravity backgrounds from field theory \emph{without} using the supergravity equations of motion.
Their only feature that will be important is the non-standard Bianchi identities obeyed by the Ramond-Ramond fields $F$.
These fields should be interpreted as $p$-form field strengths rather than potentials because their equations of motion are first order, of the form
\begin{align}
  \d F &= \cdots \, , \label{eq:BianchiRR} \\
  \d{\star F} &= \cdots \, . \label{eq:eomRR}
\end{align}
Equation~\cref{eq:BianchiRR} generalizes the usual Bianchi identity for the Maxwell two-form, \cref{eq:eomRR} generalizes its equation of motion.
The \RHS\ of \cref{eq:BianchiRR,eq:eomRR} contain a generalization of the electric and magnetic sources that we shall discuss later.
However, unlike in electromagnetism, the \RHS\ of the Bianchi identity \cref{eq:BianchiRR} is non-vanishing in the absence of external magnetic sources.
A convenient way to write it down is to define the following field strengths
\begin{align}
  H &= \d B \, , \label{eq:defHdB} \\
  F_{\text{\IIA}} &= F_2 + F_4 \, , \label{eq:FIIA} \\
  F_{\text{\IIB}} &= F_1 + F_3 + F_5 \, . \label{eq:FIIB}
\end{align}
The \define{modified Bianchi identities} for the Ramond-Ramond fields then read
\begin{equation}
  \d F_{\text{\IIA}} = H\wedge F_{\text{\IIA}}\, , \quad
  \d F_{\text{\IIB}} = H\wedge F_{\text{\IIB}}\, ,
  \label{eq:RRmodBianchi}
\end{equation}
in type \IIA\ and \IIB\ supergravity respectively.
In order to solve these equations and make them \emph{bona fide} identities, one introduces $p$-form gauge potentials~$C$ for the \<RR> field strengths.
Because the Bianchi identities are modified, the relation between $F$ and $C$ is not simply $F=\d C$, but rather
\begin{align}
  F_{\text{\IIA}} &= \d C_{\text{\IIA}} - H \wedge C_{\text{\IIA}} \, , \quad C_{\text{\IIA}} = C_1 + C_3 \, , \label{eq:CRRIIA} \\
  F_{\text{\IIB}} &= \d C_{\text{\IIB}} - H \wedge C_{\text{\IIB}} \, , \quad C_{\text{\IIB}} = C_0 + C_2 + C_4 \, , \label{eq:CRRIIB}
\end{align}
for the two theories.
Because of the additional term in the relation between the \<RR> field strengths and potentials, the gauge transformation laws of the potentials need to be modified too.
In order for the \LHS\ of \cref{eq:CRRIIA,eq:CRRIIB} to be gauge invariant under \<RR> gauge transformations, the potentials need to transform as
\begin{align}
  \delta C_{\text{\IIA}} &= \d \lambda_{\text{\IIA}} + H \wedge \lambda_{\text{\IIA}} \, , \quad \lambda_{\text{\IIA}} = \lambda_0 + \lambda_2 \, , \label{eq:gaugeRRIIA} \\
  \delta C_{\text{\IIB}} &= \d \lambda_{\text{\IIB}} + H \wedge \lambda_{\text{\IIB}} \, , \quad \lambda_{\text{\IIB}} = \lambda_1 + \lambda_3 \, . \label{eq:gaugeRRIIB}
\end{align}
In addition to these transformations, the \<RR> field strengths are of course also invariant under the $B$-field gauge transformation
\begin{equation}
  \delta B = \d \zeta \, ,
  \label{eq:gaugeB}
\end{equation}
since they depend on $B$ only through its gauge invariant field strength $H$.

\subsection{Closed strings on orbifolds}
\label{sec:orbifolds}

So far, we have only considered strings in flat space-time which preserves 32 supercharges.
The high amount of supersymmetry can be a drawback because the resulting physics is highly constrained.
One simple way to reduce the supersymmetry is to kill part of the string spectrum by introducing a discrete gauge symmetry on the worldsheet.

The first step is to choose a discrete subgroup of the global symmetry group\footnote{Much of the literature emphasizes toroidal orbifolds, where one first compactifies part of the space-time on a torus and one needs to consider a subgroup of the discrete symmetry group of the lattice used to construct the torus.
Here we will rather be concerned with non-compact orbifolds of $\RR^{10}$.} $\Gamma \subset \SO(10)$.
An element $\gamma \in \Gamma$ then naturally acts on the coordinates of space-time,
\begin{equation}
  (\gamma \cdot x)_M = \gamma\indices{_M^N}x_N \, ,
  \label{eq:orbifoldcoord}
\end{equation}
and hence in the same way on the worldsheet bosons $X_M$ and fermions $\psi_M, \tilde\psi_M$.
The discrete group $\Gamma$ is then gauged, that is to say treated as a gauge group: we declare that the points $x \in \RR^{10}$ and $\gamma\cdot x \in \RR^{10}$ be equivalent.
This amounts to replacing the space-time by $\mathcal M = \RR^{10}/\Gamma$, but it is more convenient (as in the case of usual gauge theories) to work with gauge-dependent quantities and only consider gauge-invariant expressions at the end.
In the present case, this means working in the covering space $\RR^{10}$ rather than $\mathcal M$.
Note that $\gamma$, being a linear transformation, does not act freely since at least the origin is a fixed point.
This implies that the quotient space $\mathcal M$ is singular: it is not a manifold but rather an \define{orbifold}.

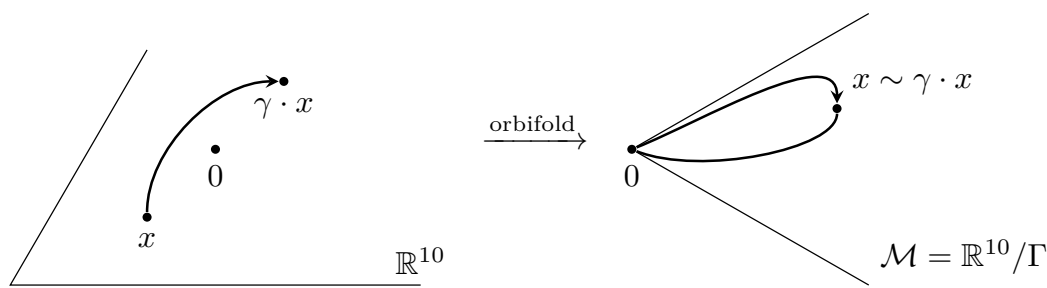
\begin{figure}
  \[ \usetikzlibrary{arrows,positioning}
\begin{tikzpicture}
  [
  scale=1.5,
  point/.style={circle,fill, inner sep=1pt},
  baseline = (o),
  ]
  \draw (3,0) -- (0,0) -- (60:2);
  \node at (3,0) [above] {$\mathbb{R}^{10}$};
  \node[point] (o) at (1.5,1) {} node[below=0 of o] {$0$};
  \node[point] (x) at (1,0.5) {} node[below=0 of x] {$x$};
  \node[point] (gammax) at (2,1.5) {} node[below=0 of gammax] {$\gamma\cdot x$};
  \draw[thick,-stealth] (x) .. controls (1,1) and (1.5,1.5) ..  (gammax);
\end{tikzpicture} \enspace \xrightarrow{\text{orbifold}} \enspace \usetikzlibrary{positioning}
\begin{tikzpicture}
  [
  scale=1.5,
  point/.style={circle,fill, inner sep=1pt},
  baseline = (o),
  ]
  \node[point] (o) {} node[below=0 of o] {$0$};
  \draw (30:2) -- (o) -- (-30:2) node [above right] {$\mathcal M= \mathbb{R}^{10}/\Gamma$};
  \node[point] (x) at (1.5,0.3) {} node[above right=0 of x] {$x \sim \gamma\cdot x$};
  \draw[thick,stealth-] (x) .. controls (1.5,0.8) and (0.5,0.2) ..  (o);
  \draw[thick] (o) .. controls (0.5,-0.2) and (1.5,0) ..  (x);
\end{tikzpicture} \]
\caption{Closed string in the $\gamma$-twisted sector, stretching between the points $x$ and $\gamma\cdot x$ which get identified in the orbifold. The string extends along a non-contractible loop, hence necessarily circles around the orbifold singularity.}
\label{fig:orbifoldstring}
\end{figure}

To see what happens to the mode expansion of the string, we need to consider the boundary conditions that are allowed.
In addition to the \define{untwisted sector} with periodic $X_M$ that was already present in flat space, we can now have more generally for any $\gamma\in \Gamma$,
\begin{equation}
  X_M(\tau,\sigma + 2\pi) = (\gamma\cdot X)_M(\tau,\sigma)
  \label{eq:twistedbdcond}
\end{equation}
and similarly for the \<R> fermions (the \<NS> fermions have an additional factor of $-1$ on the \RHS).
Equation \cref{eq:twistedbdcond} defines the $\gamma$-\define{twisted sector}, which corresponds geometrically to a string stretched along the element $\gamma$ of the fundamental group $\Gamma$ of $\mathcal M$, see \cref{fig:orbifoldstring}.
The fact that these more general boundary conditions exist would not necessarily mean that one is forced to take them into account in the theory.
One can show however that the addition of twisted sectors to the theory is required for modular invariance of the partition function, i.e.\ for consistency of the theory at one-loop~\cite{Green:1987mn}.

The twisted boundary condition forbids zero modes for the $X_M$ that transform non-trivially under the orbifold action:
as the zero modes correspond to the center of mass position and momentum operators, the center of mass is stuck at the orbifold singularity (since $x$ has to satisfy $x = \gamma\cdot x$), and the momenta transverse to the singularity have to vanish.
Geometrically, this arises because the string is embedded along a non-contractible closed loop, which necessarily encircles the orbifold singularity, hence the string is not free to move away from the singularity.

Finally, in each sector that is part of the theory, one must keep only the states invariant under $\Gamma$.

For definiteness, let us focus on an example, with $\Gamma \simeq \ZZ_2$, where the non-trivial element $\gamma$ acts by
\begin{equation}
  (\gamma\cdot x)_M =
  \begin{cases*}
    x_M & for $M = \mu = 1,\ldots,6$ \\
    -x_M & for $M = A = 7,\ldots,10$
  \end{cases*} \, .
  \label{eq:Z2orbaction}
\end{equation}
It is sometimes convenient to use complex coordinates for the directions on which $\Gamma$ acts non-trivially, which we will call the orbifolded directions.
One defines $z_1 = x_7 + i x_8$, $z_2 = x_9 + i x_{10}$ and $\gamma$ then acts as
\begin{equation}
  \gamma\cdot (z_1,z_2) = (-z_1, -z_2)
  \label{eq:Z2orbcomplex}
\end{equation}
which shows that $\Gamma \subset \SU(2) \subset \SO(4) \subset \SO(10)$.
The fact that $\Gamma$ is a subgroup of $\SU(2) \subset \SO(4)$ is equivalent to the fact that the orbifold preserves half the supersymmetry of flat space, namely 16 supercharges.
Indeed, decomposing space-time into orbifolded and non-orbifolded directions breaks the Lorentz group $\SO(10)\to \SO(6)\times \SO(4)$.
But $\operatorname{Spin}(4)\simeq \SU(2)\times\SU(2)$, under which a ten-dimensional Majorana-Weyl spinor transforms as (4 copies of) a Dirac spinor $(\mathbf{2},\mathbf{1}) \oplus (\mathbf{1},\mathbf{2})$.
Requiring invariance of this spinor under the $\SU(2)$ containing the orbifold group amounts to a chirality condition on the Dirac spinor, leaving only half the number of independent components.

Let us now look at the gauge-invariant modes in each sector.
\paragraph{Untwisted sector}
For the zero modes of the $X_M$, which correspond to the string center of mass position and momentum operators, the condition $\gamma\cdot X \sim X$ translates to the fact that $\gamma \cdot x \sim x$, i.e.\ $x \in \mathcal M$, and similarly for the momenta.
    In the $(\text{\<NS>},\text{\<NS>})$ sector, gauge-invariance requires the presence of an even number of modes with indices along the orbifolded directions.
    This kills the massless excitations $h_{\mu A}$ and $B_{\mu A}$ of the metric and $B$-field.
    When considering the Ramond sectors, we need to implement the orbifold projection on the fermionic zero modes.
    By the same argument as for the supercharges, the orbifold action introduces a distinction between left- and right-handed chiral fermions from the four-dimensional point of view, which correspond to the two different eigenvalues $\pm1$ of $\Gamma_{6\cdots{10}}$.
    This kills half the massless states in the $(\text{\<R>},\text{\<NS>})$ and $(\text{\<NS>},\text{\<R>})$ sectors, while in the $(\text{\<R>},\text{\<R>})$ sector it forces the \<RR> field strengths to have an even number of indices in the orbifolded directions like in the $(\text{\<NS>},\text{\<NS>})$ sector.
\paragraph{Twisted sector}
Since $\ZZ_2$ has only two elements, there is only one twisted sector, with $X_A(\tau, \sigma + 2\pi) = -X_A(\tau,\sigma)$.
Looking at the zero modes, this boundary condition forces $x_A = - x_A$ and $p_A = - p_A$: the twisted states are supported on the six-dimensional plane $x_A=0$ and their wavefunctions only depend on the six coordinates $x_\mu$.
To derive the spectrum in the twisted sector, one could again look at the states that are invariant under the orbifold projection.
It will be more convenient however to look at it from a different point of view, following~\cite{Douglas:1996sw}.
The idea is to regard the orbifold as the singular limit of a smooth manifold.
Since a smooth manifold looks locally like flat space, the closed string spectrum is the same as in flat space (ignoring all global issues) and the interactions reduce at low energy to the same supergravity theory but expanded around a different vacuum (one must of course check that the geometry is indeed a solution of the supergravity equations of motion and that the curvature radius is much larger than the string length in order for the supergravity effective description to be valid).
In the case of the $\CC^2/\ZZ_2$ (times $\RR^6$) orbifold under consideration, an explicit description of this smooth Riemannian manifold was found by \textcite{Eguchi:1978gw}.
The details of this solution will not be required, we will only need the fact that the Eguchi-Hanson space has a homologically non-trivial two-cycle that collapses to the orbifold fixed point in the singular limit, called the \define{exceptional cycle}.
This is a general phenomenon: orbifolds are singular complex varieties, which can be smoothened by blowing up their singular points.
Roughly speaking, a blow-up replaces a singular point by a $\mathbb{P}^1$, which is the aforementioned exceptional cycle.
In general, doing one blow-up will not produce a smooth manifold, and one must consider a chain of blow-ups instead, producing several exceptional cycles (this is explained for example in~\cite{Aspinwall:1996mn}).
For example, replacing $\ZZ_2$ by $\ZZ_n$ yields an orbifold with $n-1$ exceptional cycles instead of only one.
One can now consider the periods of the supergravity massless modes on each cycle.
In the singular limit, these will provide new massless modes supported at the singularity: these are precisely the massless states in the twisted sectors and there is one sector for each exceptional cycle.
In the $\ZZ_2$ case, with only one exceptional cycle $\Sigma$ corresponding to the single twisted sector, we get in the $(\text{\<NS>},\text{\<NS>})$ sector
\begin{equation}
  b = \frac{1}{2\pi\ls^2} \int_\Sigma B \, , \quad \zeta_i = \frac{1}{2\pi\ls^2}\int_\Sigma J_i \, , \quad i=1,2,3 \, ,
  \label{eq:NSNStwisted}
\end{equation}
where the $J_i$ are the three Kähler forms that are present in the resolved space.
They are present because the blow-up preserves the $\SU(2)$ holonomy of the orbifold which means that it is a hyper-Kähler manifold and hence admits an $\SU(2)$-triplet of Kähler forms.
In the $(\text{\<R>},\text{\<R>})$ sector, one can construct a twisted $(p-2)$-form by integrating a $p$-form with $p\ge2$ on the exceptional cycle.
For example, in type \IIB, we get the twisted \<RR> potentials
\begin{equation}
  c = \frac{1}{2\pi\ls^2} \int_\Sigma C_2 \, , \quad c_2 = \frac{1}{2\pi\ls^2} \int_\Sigma C_4 \, .
  \label{eq:RRtwisted}
\end{equation}
Similar considerations in type \IIA\ yield a $1$-form potential $c_1$.

\section{D-branes}
\label{sec:D-branes}

Let us now come back to the Ramond-Ramond sector of type \IIA/\IIB\ string theory in flat space $\RR^{10}$.
So far, the discussion of \<RR> fields has not involved any external charged objects\footnote{Because the Bianchi identities (and also the equations of motion) of the Ramond-Ramond fields involve other \<RR> fields the meaning of charge is rather subtle, and one can actually define three distinct but related concepts of charge \cite{Marolf:2000cb}.} which source the \<RR> fields electrically and/or magnetically, i.e.\ which contribute to the \RHS\ of \cref{eq:eomRR} and \cref{eq:BianchiRR} respectively.
These charged objects are \define{D-branes} and we are going to discuss some of their properties.

\subsection{From Ramond-Ramond sources to open strings}
\label{sec:branesintro}

The most immediate property resulting from the fact that D-branes carry \<RR> charges is that they are extended objects.
Indeed, the only source term one can write which involves only a \<RR>-potential yields an action
\begin{equation}
  S_{\text{D}p} = i\mu_p\int_{\Sigma_{p+1}} C_{p+1} \, .
  \label{eq:SDp}
\end{equation}
This generalizes the usual coupling $q\int A$ of the Maxwell field $A$ to a particle of charge $q$ by integrating over the worldline of the particle.
Here, the worldline is replaced by a worldvolume $\Sigma_{p+1}$ of a D$p$-brane, extended in $p$ spatial dimensions in addition to time (taken to be imaginary, explaining the $i$ factor in \cref{eq:SDp}), which couples to a $(p+1)$-form potential.
More precisely, the worldvolume $\Sigma_{p+1}$ is a submanifold $\rho\from\Sigma_{p+1}\hookrightarrow \mathcal M$ of space-time, and the \<RR>-form $C_{p+1}$ defined on $\mathcal M$ needs to be pulled back with the embedding map~$\rho$ to obtain a $(p+1)$-form $\rho^*C_{p+1}$ on $\Sigma_{p+1}$ that is then integrated.
This pullback will often be left implicit, as was done in \cref{eq:SDp}.
The coupling constant $\mu_p$ is the charge of the D$p$-brane and its specific value will be discussed later.

Given that the type \IIA\ theory has only odd \<RR>-potentials, it has charged D$p$-branes with even $p$ only.
In addition to the D0 and D2-branes which couple to the \<RR>-potentials $C_1$ and $C_3$ respectively, there are also D4 and D6-branes which couple to $C_5$ and $C_7$.
These potentials are gauge potentials for the (Hodge) dual field strengths $F_6 = \star F_4$ and $F_8 = \star F_2$ respectively.
If we choose a duality frame where the D0 and D2-branes carry electric charges, the D6 and D4-branes then carry the corresponding magnetic charges.
There also exists a D8-brane which sources magnetically the non-dynamical $F_{0}$ present in the massive type \IIA\ theory.
In type \IIB, there are charged D$p$-branes with odd $p$ only, from $p=-1$ to $p=9$.
The D9-brane is this time special, since the corresponding field strength is an $11$-form which identically vanishes in 10 dimensions.
Also the \Dmi{}-brane is somewhat special: it is not extended in time, but is pointlike in all space-time directions.
This is analogous to the status of instantons in field theory, and for this reason the \Dmi{}-brane is also called \define{D-instanton}.
The D3-brane has the particularity of having both an electric and magnetic charge because of the self-duality of $F_5$, see~\cref{eq:F5sd}.

The coupling \cref{eq:SDp} of a D$p$-brane to the \<RR> sector is not correct as it is and needs to be completed.
Indeed, because of the peculiar form of the \<RR>-gauge transformations \cref{eq:gaugeRRIIA,eq:gaugeRRIIB}, the simple form of the coupling \cref{eq:SDp} is not gauge-invariant, $\delta C_{p+1} \neq \d\,(\cdots)$.
The solution is to couple also to the $B$-field and write instead
\begin{equation}
  S_{\text{D}p} = i\mu_p \int_{\Sigma_{p+1}} [e^B \wedge C]_{p+1} \, ,
  \label{eq:SDPB}
\end{equation}
where $e^B$ is the polyform
\begin{equation}
  e^B = 1 + B + \frac{1}{2} B \wedge B + \cdots + \frac{1}{5!} B \wedge B \wedge B \wedge B \wedge B \, ,
  \label{eq:expB}
\end{equation}
and the integrand in \cref{eq:SDPB} only contains the forms of degree $p+1$, which can be paired with a $(p+1)$-dimensional submanifold of space-time.
Under a \<RR>-gauge transformation,
\begin{equation}
  \delta (e^B \wedge C)  = e^B \wedge (\d \lambda + H \wedge \lambda) = \d\, (e^B \wedge \lambda) \, ,
  \label{eq:deltaexpBC}
\end{equation}
and the integrand is a total derivative.
This is still not sufficient, however, as \cref{eq:SDPB} is not invariant under the $B$-field gauge transformation \cref{eq:gaugeB}.
Except for the $B$-field, the closed string massless sector is inert under that gauge transformation and we thus need to add an extra sector to the string theory in order to be able to compensate for the gauge transformation.
A simple solution, which turns out to be right, is to add an extra \emph{worldvolume} gauge-field $A$.
If it transforms under $B$-field gauge transformations by a shift
\begin{equation}
  \delta A = - \frac{1}{\ls^2} \rho^*\zeta \, ,
  \label{eq:gaugeAzeta}
\end{equation}
its field strength $F= \d A$ (not to be confused with the \<RR> forms $F_p$, which carry an extra label) is a worldvolume two-form that transforms under the $B$-field gauge transformation as
\begin{equation}
  \delta F = - \frac{1}{\ls^2} \rho^*\d \zeta \, .
  \label{eq:gaugeFzeta}
\end{equation}
We can then replace the field $B$ in \cref{eq:SDPB} by the gauge-invariant combination $B + \ls^2 F$,
\begin{equation}
  S_{\text{\<WZ>}} = i\mu_p \int_{\Sigma_{p+1}} [e^{B+\ls^2 F} \wedge C]_{p+1} \, .
  \label{eq:SWZ}
\end{equation}
This action, which is now gauge-invariant, is called the \define{Wess-Zumino action for a D$p$-brane} and we shall see that the fact that it involves a coupling between the \<RR> fields and a worldvolume gauge field has striking consequences.

String theory is a highly constrained theory and we cannot simply postulate the existence of a new gauge field $A$ in order to save the day.
Rather, this gauge field must arise from the dynamics of a consistent string sector that was not taken into account in \cref{sec:stringintro}.
Such a sector does exist: besides the closed string with (anti-)periodic boundary conditions, we can also consider open strings.
To have a well-defined variational principle for the open string, the endpoints, taken to be at $\sigma = 0$ and $\sigma = \pi$, need to satisfy at each end-point suitable boundary conditions.
We can take either
\begin{subequations}
  \begin{align}
  \partial_\sigma X_M(\tau,\sigma_0) &=0 & \text{Neumann boundary condition} \, , \label{eq:Neumann} \\
  X_M(\tau,\sigma_0) &= X_M^0 & \text{Dirichlet boundary condition} \, , \label{eq:Dirichlet}
\end{align}
\end{subequations}
where $\sigma_0 = 0,\pi$ and we can make independent choices for all $X_M$.
For the fermions, we can take $\psi_M(\tau,\pi) = \tilde\psi_M(\tau,\pi)$, and choose one of the two possibilities
\begin{equation}
  \tilde \psi_M(\tau,0) =
  \begin{cases*}
   + \psi_M(\tau,0) & Ramond (\<R>)\\
   - \psi_M(\tau,0) & Neveu-Schwarz (\<NS>)
  \end{cases*}
  \label{eq:psibdopen}
\end{equation}
independently for each value of $M=1,\ldots,10$.
The important point is that imposing a Dirichlet boundary condition for a worldsheet boson at one end or both kills the zero modes in its mode expansion.
Recalling that those zero modes correspond to the position and momentum operators of the center of mass of the string, this means that the string cannot move in the Dirichlet directions.
If we further specialize to the case with $p+1$ bosons $X_{\mu}$, $\mu=1,\ldots,p+1$ having Neumann boundary conditions at both ends and the others having Dirichlet conditions $X_A(\tau,0) = X_A(\tau,\pi) = 0$, $A=p+2,10$, the string is stuck to the $(p+1)$-dimensional surface $\Sigma_{p+1}=\{x_M \mid x_A = 0\}$ in space-time.
This breaks the space-time Lorentz invariance as $\SO(10)\to \SO(p+1)\times\SO(9-p)$.
Because of the reduced number of zero modes, the low-energy description of our string only depends on the $p+1$ coordinates $x_\mu$ and is described by a $(p+1)$-dimensional field theory, for which $\SO(p+1)$ is a Lorentz symmetry and $\SO(9-p)$ an internal global symmetry.
The fields correspond to the massless excitations of the string.
The massless spectrum is, after \<GSO> projection, a gauge field $A_{\mu}$ and (9-p) scalars $\varphi_A$ in the \<NS> sector and the dimensional reduction to $p+1$ dimensions of a ten-dimensional Majorana-Weyl fermion~$\Psi$ in the \<R> sector.
This data contains precisely what is needed for the consistent description of a D$p$-brane, namely a worldvolume gauge field $A_\mu$.
Hence, from the worldsheet perspective, a D$p$-brane corresponds to a specific choice of boundary condition, which also explains its name as the contraction of \enquote{Dirichlet membrane}.
This dual interpretation of D-branes, in terms of sources for closed string modes on the one hand and in terms of endpoints of open strings on the other hand will turn out to be extremely powerful and will be used later to derive closed string backgrounds from open string computations.

For now, let us only give an open string justification of the transformation law \cref{eq:gaugeFzeta} under $B$-field gauge transformations.
To this end, let us consider an open string in a non-trivial $B$-field background.
The open string then couples to the $B$-field through a term
\begin{equation}
  S_{B} = \frac{i}{\ls^2} \int_\Sigma B \, ,
  \label{eq:SBstring}
\end{equation}
where again the pull back by $X\from \Sigma \to \mathcal M$ is left implicit.
Under a $B$-field gauge transformation \cref{eq:gaugeB},
\begin{equation}
  \delta S_B = \frac{i}{\ls^2} \int_\Sigma \d \zeta = \frac{i}{\ls^2} \int_{\partial \Sigma} \zeta
  \label{eq:deltaSBstring}
\end{equation}
which does not vanish because the open string has a boundary $\partial \Sigma$.
The solution to the non-invariance of the action is to add the gauge field $A$ on the boundary,
\begin{equation}
  S_{B+A} = \frac{i}{\ls^2} \int_\Sigma B + i\int_{\partial \Sigma} A \, .
  \label{eq:SBAstring}
\end{equation}
Gauge invariance can now be implemented by having $A$ transform precisely as in \cref{eq:gaugeAzeta}.
Notice that we have also shown that open strings carry \enquote{ordinary} gauge degrees of freedom at their endpoints.

The gauge field $A$ is only part of the open string massless spectrum, which as we have seen is required for consistent interactions of the brane with the \<RR> fields.
Also the other fields can be understood as being required by consistency, more precisely by the Goldstone theorem.
Indeed, since part of the translational invariance of the Minkowski vacuum is broken by the presence of a D-brane, there should be Goldstone bosons (at least for $p>2$) on the worldvolume that account for fluctuations of the D-brane about its position $x_A = 0$.
These are precisely the fields $\varphi_A$, whose interactions need to preserve the shift symmetry $\delta \varphi_A = \ls^2 c_A$.
Similarly, the presence of the D-brane breaks part of the supersymmetry of the vacuum:
the boundary condition on the world-sheet relates the two Killing spinors $\epsilon_1$, $\epsilon_2$ such that the unbroken supersymmetries satisfy
\begin{equation}
  \epsilon_p \equiv \epsilon_1  - \Gamma_{1\cdots (p+1)} \epsilon_2 = 0
  \label{eq:susyDbrane}
\end{equation}
for a D$p$-brane extended along the $p+1$ first directions.
This equation implies that only supersymmetric D$p$-branes with even (resp.\@ odd) $p$ exist in type \IIA\ (resp.\@ \IIB), due to the fact that $\Gamma_{11}\epsilon_p$ also has to vanish, in agreement with the argument using \<RR> charges.

The single constraint~\cref{eq:susyDbrane} allows to express one Killing spinor in terms of the other, hence a D-brane preserves half the supercharges: it is a 1/2-\<BPS> object.
The other supersymmetries are broken, so the fermionic version of the Goldstone theorem requires the presence of a goldstino $\Psi$, invariant under the shift symmetry $\delta\Psi = \epsilon_p$.
Because a D$p$-brane preserves 16 supercharges, the low-energy field theory governing the dynamics of the massless open string modes is highly constrained: it is the $(p+1)$-dimensional maximally supersymmetric $\U(1)$ gauge theory, i.e.\ the dimensional reduction to $p+1$ dimensions of the $\nn=1$ theory in ten dimensions.
Note that the $SO(9-p)$ internal global symmetry group on the D$p$-brane worldvolume is an R-symmetry because the supercharges transform non-trivially under it.

\subsection{D-brane dynamics: the \<DBI>\ action}
\label{sec:DBI}

If the D-brane is placed in a background where closed string modes are turned on, supersymmetry will generically be broken, and the low-energy theory describing the massless open string modes can have more general couplings.
The purely bosonic couplings induced by placing the D-brane in a non-trivial \<RR> background are given by the \<WZ> action \cref{eq:SWZ}.
In addition, there are also couplings to the closed string states from the \<NSNS> sector.
The full supersymmetric action, including the fermions, is also known~\cite{Cederwall:1996ri,Bergshoeff:1996tu} (see also \cite{Aganagic:1996pe}), but we will have no use for it in this thesis since the bosonic terms will be sufficient to determine the holographic backgrounds.
The action for a D$p$-brane then takes the form $S_{\text{D}p} = S_{\text{\<DBI>}} + S_{\text{\<WZ>}} $ where $S_{\text{\<DBI>}}$ is the Dirac-Born-Infeld action
\begin{equation}
  S_{\text{\<DBI>}} = T_p \int_{\Sigma_{p+1}} \d^{p+1}\xi\, e^{-\Phi} \sqrt{\det[\rho^*(G + B) + \ls^2 F]} \, .
  \label{eq:SDBI}
\end{equation}
There is again a pullback of the ten-dimensional closed string background fields $\Phi$, $G$ and $B$ on the worldvolume of the brane parameterized by $\xi^\mu$, $\mu = 1,\ldots,p+1$ and the determinant is taken on the worldvolume indices.

The reader might wonder where the $9-p$ scalar fields $\varphi_A$, that were supposed to describe the transverse fluctuations of the D-brane, hide in this description.
To see them explicitly, one must first unpack the definition of the pullback of the fields under $\rho\from \Sigma_{p+1}\to\mathcal M\from \xi^\mu \mapsto x^M(\xi^\mu)$,
\begin{subequations}
  \begin{align}
  (\rho^*\Phi)(\xi^\mu) &= \Phi(x^M(\xi^\mu)) \label{eq:PBPhi} \, , \\
  (\rho^*G)(\xi^\lambda)_{\mu\nu} &= \frac{\partial x^M}{\partial \xi^\mu} \frac{\partial x^N}{\partial \xi^\nu}G_{MN}(x^P(\xi^\lambda)) \, , \label{eq:PBG} \\
  (\rho^*B)(\xi^\lambda)_{\mu\nu} &= \frac{\partial x^M}{\partial \xi^\mu} \frac{\partial x^N}{\partial \xi^\nu}B_{MN}(x^P(\xi^\lambda)) \, . \label{eq:PBB}
\end{align}
\end{subequations}
This looks like the embedding is determined by the ten functions $x^M(\xi^\mu)$ rather than the $9-p$ scalars $\varphi_A$.
However, all the expressions involved are covariant under worldvolume diffeomorphisms, and one needs to fix this gauge freedom to see the physical worldvolume scalars.
A natural choice is to impose the \define{static gauge}, whereby one identifies the worldvolume coordinates $\xi$ with the space-time coordinates $x^\mu$ that lie along the D-brane,
\begin{equation}
  x^\mu (\xi) = \xi^\mu \, , \quad \mu = 1,\ldots,p+1 \, .
  \label{eq:staticgauge}
\end{equation}
The remaining $9-p$ embedding functions $x^A$ then yield the dynamical scalars $\varphi^A$ when expanded about a reference point $x_0$,
\begin{equation}
  x^A(x^\mu) = x_0^A + \ls^2\varphi^A(x^\mu) \, , \quad A = p+2,\ldots,10 \, .
\label{eq:staticgaugephi}
\end{equation}
The pullback of the metric \cref{eq:PBG} provides the kinetic terms for the scalars,
\begin{equation}
  (\rho^* G)(x^\lambda)_{\mu\nu} = \delta_{\mu\nu} + 2\ls^2 \partial_{\mu} \varphi^A G_{\mu A}(x_0 + \ls^2\varphi) + \ls^4\partial_\mu\varphi^A\partial_\nu\varphi^B G_{AB}(x_0 + \ls^2\varphi) \, .
  \label{eq:PBGstatic}
\end{equation}
If the background is invariant under translations $x^A \to x^A + x_0^A$, the $\varphi_A$ only enter the action through their derivatives, in agreement with the Goldstone theorem.

The \<DBI> action is weighted by the tension $T_p$ of a D$p$-brane, while the \<WZ> action is weighted by the charge $\mu_p$.
These two constants are equal as a consequence of supersymmetry \cite{Polchinski:1995mt} and are given by
\begin{equation}
  T_p = \mu_p = \frac{1}{(2\pi)^{\frac{p-1}{2}}\ls^{p+1}} \, .
  \label{eq:branecharge}
\end{equation}
This uses the convention where the string coupling $g_s$ is not written explicitly but is given by the asymptotic value of the dilaton,
\begin{equation}
  e^{\Phi(x)} \to g_s \qquad \text{for $\abs{x} \to \infty$} \, .
  \label{eq:Phigs}
\end{equation}

A few comments on the validity of the \<DBI> action \cref{eq:SDBI} are in order.
It is an exact action when the fields $(\Phi,G,B,F)$ are constant, in the sense that the equations of motion found by varying $A_\mu$ and $\varphi_A$ are equivalent to conformality of the open string sigma-model in the background specified by these fields~\cite{Leigh:1989jq}.
This regime of constant fields is more general than the low-energy limit $\ls^2\to0$, as the $\ls^2F$ term in~\cref{eq:SDBI} yields an infinite power series in $\ls^2$ that corrects the strict low-energy limit.
Still, there are corrections to the \<DBI> action involving derivatives of the fields, that will become important when the (gauge-invariant combinations of) derivatives are large with respect to the string scale $\ls^2$.

\subsection{Multiple D-branes}

Since an open string has two ends, we can choose different boundary conditions at the two ends of the string.
Let us start by considering the case of two parallel D$p$-branes at different positions in the transverse space, say $x_A^1$ and $x_A^2$.
There are now four different open string sectors: with both ends on the brane at $x_A^1$, with both ends on the brane at $x_A^2$, with the starting point on the brane at $x_A^1$ and the endpoint on the brane at $x_A^2$ and finally the same strings but with the roles of $x_A^1$ and $x_A^2$ reversed.
For bookkeeping purposes, it is convenient to introduce labels for the two D-branes, $i=1,2$, sitting at $x_A^i$.
The different sectors are then labelled by pairs $(i,j)$ for strings going from the brane $i$ to the brane $j$.
These indices are called Chan-Paton factors.

Because the two branes are parallel, they preserve the same supercharges \cref{eq:susyDbrane}.
They are mutually \<BPS>, which means that the central charge behaves additively and the D-branes exert no force on one another: the gravitational attraction exactly balances the electrostatic repulsion of the two like \<RR> charges.
This explains the relation \cref{eq:branecharge} between the tension and charge of a D-brane.

The diagonal sectors $(1,1)$ and $(2,2)$ contain the massless states discussed previously.
The off-diagonal strings (with $i\neq j$) are stretched, hence they have a contribution to the mass proportional to their length
\begin{equation}
  m_{12}^2 \sim \frac{1}{\ls^4}(x_A^1 - x_A^2)^2 \, .
  \label{eq:mstretched}
\end{equation}
This makes the lightest excitations of these strings very massive for separated branes, and the low-energy field theory is a free $(p+1)$-dimensional maximally supersymmetric $\U(1)\times\U(1)$ gauge theory.
However, when taking the D-branes to sit on top of one another, these off-diagonal strings become light, and the D-branes can interact through open strings.
It turns out that the limit of having massless off-diagonal strings corresponds to the gauge symmetry enhancement $\U(1)^2\to \U(2)$, meaning that multiple coinciding D-branes interact through a supersymmetric non-Abelian gauge theory~\cite{Witten:1995im}.
This description makes sense because one can view each massless mode in the sector $(i,j)$ as the element $(i,j)$ of a $2\times2$ matrix.
In the non-Abelian theory, there is a potential for the scalars $V \sim \tr [\varphi_A,\varphi_B]^2$.
This potential is minimized by configurations with commuting constant scalars, which can thus be diagonalized simultaneously.
Remembering that the scalars are interpreted as the transverse fluctuations of the branes, we can then write
\begin{equation}
  \corr{\varphi_A} = \frac{1}{\ls^2} \begin{pmatrix}
    x_A^1 & 0 \\
    0 & x_A^2
  \end{pmatrix} \, .
  \label{eq:vevCBbrane}
\end{equation}
This corresponds to the situation where one D-brane sits at $x_A^1$ and the other at $x_A^2$, i.e.\ the starting point of our discussion.
This vacuum breaks the gauge group $\U(2)\to \U(1)^2$, giving masses to the off-diagonal modes that agree with~\cref{eq:mstretched}.
In the limit where the off-diagonal modes are very heavy, that is in the limit where the D-branes are taken to be very far apart, one can integrate them out by setting them to zero and one recovers in this way the $\U(1)\times \U(1)$ theory describing two non-interacting D-branes.
The overall $\U(1)\subset \U(2)$ factor, which is always decoupled because all the fields are in the adjoint, describes in the geometric picture the free dynamics of the center of mass of the pair of D-branes.
The gauge-coupling $g_{p+1}$ for the $(p+1)$-dimensional theory realized on a stack of D$p$-branes is related to the tension~$T_p$ in \cref{eq:branecharge},
\begin{equation}
  \frac{1}{g_{p+1}^2} = \frac{T_p \ls^4}{2g_s} = \frac{1}{2(2\pi)^\frac{p-1}{2}\ls^{p-3}g_s} \, .
  \label{eq:gcoupling}
\end{equation}
As a quick check, notice that in four dimensions, one gets a dimensionless coupling $g_{4}^2 = 4\pi g_s$ as expected.

To summarize, separating the D-branes corresponds from the worldvolume perspective to going on the Coulomb branch.
The off-diagonal strings are then W-boson multiplets that become massive through the Higgs mechanism.
Of course, the discussion readily generalizes to the case of $N$ parallel D-branes which realize a $\U(N)$ maximally supersymmetric gauge theory on their worldvolume when they are stacked on top of each other.
For instance, in the case of the D3-brane, this is the four-dimensional $\nn=4$ $\U(N)$ gauge theory reviewed in \cref{sec:SUSYreview}.

How does the action for a single D-brane generalize to the setting of multiple D-branes?
When the D-branes are far apart, one simply sums over the contributions of all D-branes.
In the opposite limit, where one deals with a stack of D-branes and a $\U(N)$ gauge group, simply summing over the $N$ D-branes is not gauge-invariant.
It is easy to see that the action
\begin{equation}
  S_{\text{\<WZ>},N} = i\mu_p \int_{\Sigma_{p+1}} \truN [e^{B+\ls^2 F} \wedge C]_{p+1}
  \label{eq:SWZUN}
\end{equation}
provides a gauge invariant generalization of the \<WZ> action \cref{eq:SWZ} for a single D-brane.
This action cannot be the full story in the non-Abelian case, however, because it does not yield the required sum over D-branes upon neglecting the off-diagonal terms.
The reason is that one also needs to generalize the (implicit) pullback to a \enquote{non-Abelian pullback} that depends on the fields $\varphi_A$ that can be seen as some kind of matrix-valued D-brane coordinates.
Further, also the \<DBI> action \cref{eq:SDBI} needs to be generalized to the case of several D-branes.
We will come back to these issues in \cref{sec:Myers}, but for now \cref{eq:SWZUN} will be sufficient.

\subsection{Instantons as branes within branes}
\label{sec:Dinstantons}

We now have all the tools to embed instantons in string theory constructions.
Let us start with a stack of $N$ D3-branes extended along the four directions $x_\mu$, $\mu=1,\ldots,4$, whose low-energy dynamics is described by a four-dimensional $\U(N)$ field theory.
Consider the non-Abelian \<WZ> action \cref{eq:SWZUN} and take the $B$-field and the adjoint scalars to vanish.
Expanding the exponential and selecting the 4-form terms, one obtains
\begin{equation}
  S_{\text{\<WZ>},N} = i\mu_3 \int_{\RR^4} [N C_4 + \ls^2 C_2 \wedge \tr F + \frac{\ls^4}{2} C_0 \tr (F \wedge F) ] \, .
  \label{eq:SWZD3}
\end{equation}
If we take $\tr F = 0$, which we may do consistently as the overall $\U(1)$ factor does not couple to the other worldvolume fields, and also assume that $C_0$ is constant along the D3-branes,
\begin{equation}
  S_{\text{\<WZ>},N} = i\mu_3 N \int_{\RR^4} C_4 -  4\pi^2 \ls^4 \mu_3 i K C_0 \,
  \label{eq:SWZD3inst}
\end{equation}
where $K$ is the instanton number defined in \cref{eq:YMtheta}.
The first term is $N$ times the coupling of a single D3-brane to the \<RR> 4-form potential, which is expected given that we are dealing with $N$ D3-branes.
The second term, however, is similar to the coupling of a \Dmi{}-brane to the \<RR> 0-form potential because the integral over a zero-dimensional \enquote{worldvolume} is simply the evaluation at the position of the \Dmi{}-brane.
Comparing with the normalization of the \Dmi{}-brane charge \cref{eq:branecharge}, we can rewrite it as $\mu_{-1}K C_0$, which means that the configuration we are dealing with behaves like $K$ \Dmi{}-branes\footnote{Because of the minus sign in~\cref{eq:SWZD3inst}, our instantons are more precisely anti-\Dmi{}-branes. This can be traced back to the fact that we are dealing with anti-self-dual solutions rather than self-dual and will be responsible for various additional minus signs in the rest of this thesis.} dissolved in the worldvolume of $N$ D3-branes as far as the charges are concerned, a configuration called \enquote{branes within branes} by \textcite{Douglas:1995bn}.
The \Dmi{}-brane charge is dissolved because the instanton charge, being a topological quantity, is not well-localized but associated with the global behavior of the worldvolume gauge field.

This configuration looks like a bound state of D3 and \Dmi{}-branes not only with respect to the charges but also with respect to the preserved supersymmetry.
We know from \cref{sec:SUSYinst} that an instanton in the $\nn=4$ theory preserves $8$ super-Poincaré charges and precisely the same amount is preserved by a system of \Dmi{} and D3-branes.
Indeed, we can read from \cref{eq:susyDbrane} that the supercharges preserved by such a configuration satisfy the simultaneous equations
\begin{equation}
  \epsilon_1 = \Gamma_{1\cdots4}\epsilon_2 \quad \text{and} \quad \epsilon_1 = - \epsilon_2 \, ,
  \label{eq:SUSYDm1D3}
\end{equation}
which together imply that the unbroken supercharges are chiral from the four-dimensional point of view, in agreement with the field theory considerations of \cref{sec:SUSYinst}.

Both the \<RR> charges and the supersymmetry are strong indications that one can realize instantons in field theory as  bound states of \Dmi{} and D3-branes in a suitable low-energy limit.
Taking this idea seriously, one should be able to obtain the super-\<ADHM> construction and the action \cref{eq:SinstN4} from computations in String Theory.
This is indeed the case and was done in \cite{Green:2000ke,Billo:2002hm} by computing open string amplitudes.
With two different sets of D-branes, there are now four different kinds of open strings, because they can have both their endpoints on the D3-branes, both on the \Dmi{}-branes, or one endpoint on the D3-branes and the other on the \Dmi{}-branes (with two possible orientations).
The amplitudes of the two first sectors correspond, in the low-energy limit, to the maximally supersymmetric Yang-Mills theories in four dimensions (with gauge group $\U(N)$) and in zero dimensions (with gauge group $\U(K)$) respectively.
The massless states are respectively the gauge theory fields $(A_\mu, \varphi_A, \lambda, \bar \lambda)$ and the neutral \<ADHM> moduli $(X_\mu,\phi_A,\bar\psi,\Lambda)$.
The interactions of the strings with mixed boundaries are governed by very different amplitudes, because there are now four dimensions with Dirichlet-Neumann mixed boundary conditions.
All the directions have at least one end with Dirichlet conditions, forbidding the zero modes for all $X_M$, hence the massless states are (coordinate independent) moduli rather than four-dimensional fields.
They turn out to be precisely the charged moduli $(q,\tilde q, \chi, \tilde\chi)$.
By computing disk amplitudes with insertions of these massless modes, the authors of \cite{Green:2000ke,Billo:2002hm} then managed to build the action \cref{eq:SinstN4}, as well as the couplings of the moduli to the D3-brane fields that were hidden in $S_{\text{gauge}}$.

The full derivation involves rather technical string theory machinery, but we can deduce the action \cref{eq:SinstN4} (though not all the couplings to the D3-brane fields) by exploiting supersymmetry and some facts about the D-branes.
First, by using T-duality on the six coordinates transverse to the D3-branes, we can transform the \Dmi{}/D3 system into a D5/D9 system\footnote{Consistency of the type \IIB\ theory in the presence of $N$ D9-branes requires the addition of an O9 orientifold plane as well as the value $N=32$, which would bring us to the type I theory. However, we will use the D5/D9 theory only as a tool to obtain the \Dmi{}/D3 system, making this issue irrelevant.}.
If we only consider $N$ D9-branes with no D5-branes present, the low-energy limit is the ten-dimensional $\nn=1$ Yang-Mills theory with gauge group $\U(N)$ which describes the self-interactions of the D9-D9 strings.
Similarly, we can consider only $K$ D5-branes with no D9-branes present.
The low-energy limit is an $\nn=(1,1)$ six-dimensional theory that can be obtained by the dimensional reduction of the ten-dimensional $\nn=1$ Yang-Mills theory down to six dimensions and has gauge group $\U(K)$.
With the two types of branes present, supersymmetry on the D5 worldvolume is broken from 16 to 8 supercharges.
The $\nn=(1,1)$ vector multiplet then splits into irreducible representations of the reduced supersymmetry algebra.
It yields a vector multiplet $(\phi, \Lambda, D)$ and an adjoint hypermultiplet $(X, \bar\psi)$ with respect to the unbroken subalgebra.
The next step is to add the two extra open string sectors, with one end on the D5 and the other on the D9.
Because of these boundary conditions, they carry (anti-)fundamental charges under both gauge groups, they are \emph{bifundamentals} $(\mathbf{K},\mathbf{\bar N})$ (for the D5$\rightarrow$D9 sector) or $(\mathbf{\bar K}, \mathbf{N})$ (for the D9$\rightarrow$D5 sector) of $\U(K)\times \U(N)$.
Given that the massless states of these strings only depend on six coordinates, they have an interpretation in the six-dimensional theory.
The amount of supersymmetry and the charges require them to fit into a bifundamental hypermultiplet $(q,\chi)$ whose complex conjugate is $(\tilde q, \tilde \chi)$.
From the point of view of the D5-brane worldvolume theory, the D9-brane $\U(N)$ gauge group is a global symmetry group, hence this theory is simply six-dimensional supersymmetric \<QCD> with gauge group $\U(K)$, one adjoint hypermultiplet and $N$ fundamental flavors.
The global symmetries are $\SO(6)\times\SO(4)$ because of the presence of the D5-branes which break part of the ten-dimensional Lorentz group.
The first factor is the Lorentz group along the D5, while the $\SO(4)$ corresponds to transverse rotations.
Under the partial supersymmetry breaking due to the presence of the D9-branes, this $\SO(4)$ group further splits into an $\SU(2)$ R-symmetry and an $\SU(2)$ flavor symmetry.
The action for this theory with canonical kinetic terms is unique up to the value of the gauge coupling when all mass terms are set to zero and is given by the sum of the two contributions,
\begin{multline}
  S_{\text{D5-D5}} = \frac{1}{g_6^2} \int \d^6 x \truK \biggl\{ \frac{1}{2}F_{AB} F_{AB} + \frac{1}{\ls^4} \nabla_A X_\mu \nabla_A X_\mu + \frac{2i}{\ls^{4}} [X_\mu,X_\nu] D_{\mu\nu} \\
   + i\Lambda\indices{^\alpha_a}\bar\Sigma_A^{ab}\nabla_A\Lambda_{\alpha b} + \frac{i}{\ls^4}\bar\psi\indices{_{\dot\alpha}^a}\Sigma_{Aab}\nabla_A\bar\psi^{\dot\alpha b} - \frac{2}{\ls^4}\sigma_{\mu\alpha\dot\alpha}\Lambda\indices{^\alpha_a}[X_\mu,\bar\psi^{\dot\alpha a}] - D_{\mu\nu} D_{\mu\nu} \biggr\}  \label{eq:SD5D5}
\end{multline}
for the massless modes of the D5-D5 strings, and
\begin{multline}
S_{\text{D5-D9}} = \int \d^6x \left\{ \frac{1}{2} \nabla_A \tilde q^\alpha \nabla_A q_\alpha + \frac{i}{2} \tilde\chi^a \Sigma_{Aab} \nabla_A \chi^b  \right. \\
\left. + \frac{1}{\sqrt2} \tilde q^\alpha \Lambda_{\alpha a}\chi^a + \frac{1}{\sqrt2} \tilde \chi^a \Lambda\indices{^\alpha_a} q_\alpha + \frac{i}{2} \tilde q^\alpha D_{\mu\nu}\sigma\indices{_\mu_\nu_\alpha^\beta}q_\beta \right\} + \cdots \label{eq:SD5D9}
\end{multline}
for the massless modes of the D5-D9 strings.
In~\cref{eq:SD5D5}, $F_{AB} = \partial_A \phi_B - \partial_B \phi_A  + i[\phi_A,\phi_B]$ is the field strength of the $\U(K)$ gauge field $\phi_A$ and $\nabla_A = \partial_A + i[\phi_A,\cdot]$ is the gauge-covariant derivative in the adjoint representation.
The explicit factors of $\ls$ are due to the fact that the adjoint hypermultiplet has been rescaled with respect to the canonical normalization by $\ls^{-2}$ in order for the scalars $X_\mu$ to have dimension of length (recall that these fields are interpreted geometrically as the positions of the D5-branes in their transverse space).
The field $D_{\mu\nu}$ is an antisymmetric and self-dual auxiliary field that has been added for the off-shell closure of the super-algebra on the vectormultiplet\footnote{Reducing this theory to 4D and splitting the (now $\nn=2$) vector into an $\nn=1$ vector and an adjoint chiral, the three independent fields in $D_{\mu\nu}$ correspond to the real and imaginary parts of the $F$-component of the chiral multiplet and to the $D$-component of the usual $\nn=1$ vector multiplet.}.
In~\cref{eq:SD5D9}, the covariant derivatives act on the fields according to their charges, i.e.\ $\nabla_A q = \partial_A q + i \phi_A q$, $\nabla_A \tilde q = \partial_A \tilde q - i \tilde q \phi_A$ and similarly for $\chi$ and $\tilde \chi$.
There are extra couplings of the D5-D9 modes to the D9 fields, but most of them will not be needed in this thesis and they have not been written down.

Now that we know the D5 worldvolume action in the D5/D9 system, we can straightforwardly obtain the action for the \Dmi{}-branes in the \Dmi{}/D3 system by T-dualizing back.
T-duality simply amounts to the dimensional reduction of the worldvolume theory down to zero dimensions, which can be implemented very simply by dropping all the partial derivatives in the action as well as the overall integration and replacing the 6D gauge coupling $g_6$ by its 0D counterpart $g_0$.
One obtains
\begin{multline}
  S_{\text{\Dmi{}-\Dmi{}}} = \frac{1}{g_0^2} \truK \left\{ - \frac{1}{\ls^4} [\phi_A, X_\mu] [\phi_A,X_\mu] + \frac{2i}{\ls^{4}} [X_\mu,X_\nu] D_{\mu\nu} \right. \\
  - \frac{1}{\ls^4}\bar\psi\indices{_{\dot\alpha}^a}\Sigma_{Aab}[\phi_A,\bar\psi^{\dot\alpha b}] - \frac{2i}{\ls^4}\sigma_{\mu\alpha\dot\alpha}\Lambda\indices{^\alpha_a}[X_\mu,\bar\psi^{\dot\alpha a}] \\
    \left. -\frac{1}{2}[\phi_A\phi_B][\phi_A,\phi_B] - \Lambda\indices{^\alpha_a}\bar\Sigma_A^{ab}[\phi_A,\Lambda_{\alpha b}] - D_{\mu\nu} D_{\mu\nu} \right\}  \label{eq:SDm1Dm1}
\end{multline}
for the dimensional reduction of~\cref{eq:SD5D5} and
\begin{multline}
S_{\text{\Dmi{}-D3}} = \frac{1}{2} \tilde q^\alpha \phi_A\phi_A q_\alpha - \frac{1}{2} \tilde\chi^\alpha \Sigma_{Aab} \phi_A \chi^b  \\
+ \frac{1}{\sqrt2} \tilde q^\alpha \Lambda_{\alpha a}\chi^a + \frac{1}{\sqrt2} \tilde \chi^a \Lambda\indices{^\alpha_a} q_\alpha + \frac{i}{2} \tilde q^\alpha D_{\mu\nu}\sigma\indices{_\mu_\nu_\alpha^\beta}q_\beta + \cdots \label{eq:SDm1D3}
\end{multline}
for the dimensional reduction of~\cref{eq:SD5D9}.

By using~\cref{eq:gcoupling}, one can relate the zero-dimensional coupling $g_0$ to the four-dimensional coupling $g_4$ (that was simply called $g$ in \cref{chap:instantons}),
\begin{equation}
  \frac{1}{g_0^2} = \frac{4\pi^2 \ls^4}{g_4^2} \, .
  \label{eq:g0g4}
\end{equation}
The sum of \cref{eq:SDm1Dm1} and \cref{eq:SDm1D3} is then close to the instanton action~\cref{eq:SinstN4}, but not exactly the same.
The main difference is the absence in~\cref{eq:SinstN4} of the three terms on the last line of \cref{eq:SDm1Dm1}, hence we are not exactly describing instantons in gauge theory.
To understand the difference, let us look at the moduli space of vacua.
The equation of motion for $D$ reads, by using~\cref{eq:g0g4},
\begin{equation}
  -i \ls^4 D_{\mu\nu} = [X_\mu,X_\nu]^+  + \frac{g_4^2}{16\pi^2} q_\beta\tilde q^\alpha \sigma\indices{_\mu_\nu_\alpha^\beta} \, .
  \label{eq:DbraneADHM}
\end{equation}
The \RHS\ is the same as the \<ADHM> constraint \cref{eq:ADHMconst}, but because the action now contains a term quadratic in $D$, the \LHS\ is no longer zero, and we obtain an equation for $D$ rather than a constraint.
Plugging this value of $D$ back in the action and setting all fermions to zero, we see that there are two types of solutions:
\begin{description}
  \item[The \enquote{Coulomb branch}] where $q= \tilde q = 0$  and the adjoint moduli $X_\mu$, $\Phi_A$ commute.
    The equation of motion for $D$ is then trivially satisfied.
    The gauge group is generically broken as $\U(K)\to\U(1)^K$, explaining the name of this branch in analogy with the higher-dimensional case even if there is no Coulomb potential in zero dimensions.
    These vacua correspond to \Dmi{}-branes that can be placed anywhere in the ten-dimensional space-time.
  \item[The \enquote{Higgs branch}] where $\phi_A = 0$, and the other bosonic moduli $X_\mu$, $q$, $\tilde q$ satisfy the equation \cref{eq:DbraneADHM} with $D=0$ and hence the \<ADHM> constraint.
    The non-trivial values of $q$ and $\tilde q$ generically break the gauge group completely.
    Because $\phi_A=0$, the \Dmi{}-branes sit inside the D3-branes.
\end{description}
Of these two branches, only the Higgs branch enjoys the properties of gauge theory instantons.
However, the two branches are connected in the \Dmi{}/D3 system.
They meet at the points where $q = \tilde q = \phi_A = 0$ and the $X_\mu$ commute.
In the field theory, these points are associated with \<UV> divergences where the instanton shrinks to zero size, and we thus see that String Theory provides a \<UV> completion by allowing such point-like instantons to leave the D3-brane as \Dmi{}-branes.
Since we are interested in genuine field theory instantons, we will need to decouple the Coulomb branch from the Higgs branch by dropping the three terms in the last line of \cref{eq:SDm1Dm1}.
This can be done consistently and has a very nice interpretation in the \<AdS>/\<CFT> context as we will see in the next chapter.

Notice that, starting from the D5/D9 system, one can arrive not only at the \Dmi{}/D3 system, but more generally at all the D$p$/D$(p+4)$ systems for $p=-1,\ldots,5$ by T-duality.
By similar arguments as for $p=-1$, one arrives at the conclusion that in a suitable decoupling limit, the D$p$-branes describe supersymmetric codimension 4 solitons in the D$(p+4)$-brane gauge theory.
This fact will be exploited in \cref{chap:D4brane}.

\subsection{Multiple D-brane effective action}
\label{sec:Myers}

Let us now come back to the issue of generalizing the effective action for a single D$p$-brane in a non-trivial background to the case of $N$ coinciding D$p$-branes.
Recall that for $N$ parallel D$p$-branes, the worldvolume fields are $N\times N$ matrices, and one needs to find the correct way to generalize from commuting variables to matrices.

Since the scalar fields $\varphi_A$ correspond to the transverse fluctuations of the D-brane, they enter the pullback of the bulk fields to the brane, as explained in detail in \cref{sec:DBI}.
The first question is then
\begin{enumerate}
  \item \label{it:pullback} How to define the non-Abelian pullback?
    \suspend{enumerate}
A second issue is that contrarily to the fields on a single D-brane, the fields on a stack of several D-branes are matrices, and one needs to answer
    \resume{enumerate}
  \item \label{it:ordering} How to order the matrices in the action?
\end{enumerate}
The answers to these questions provide the required ingredients to build the non-Abelian D-brane actions.
It is in principle possible to obtain the answers by computing the relevant String Theory amplitudes.
However, this \enquote{shut up and calculate!} approach is very involved in practice, and it has been attempted to find the non-Abelian generalization by indirect methods.

A good strategy to solve the first question is to use T-duality \cite{Myers:1999ps} (see also \cite{Taylor:1999pr}).
Indeed, since all configurations of $N$ D$p$-branes with $p = -1,\ldots,9$ are T-dual to each other, one can deduce the action for all $p$ from the D9-brane action by applying the known T-duality transformation rules both to the closed and open string massless fields.
But since the D9-brane is space-filling, it does not have transverse directions and hence no scalars, the only bosonic worldvolume field being the non-Abelian gauge field.
In static gauge~\cref{eq:staticgauge}, the worldvolume coordinates are simply identified with the ambient space coordinates, and the pullback is trivial.
Implementing the T-duality then yields an effective action $S_{\text{eff}} = S_{\text{\<DBI>}} + S_{\text{\<WZ>}}$ that we will call Myers' action \cite{Myers:1999ps},
\begin{align}
  S_{\text{\<DBI>}} &= T_p \int_{\Sigma_{\mathrlap{p+1}}} \d^{p+1}x \tr \left\{ \pbP[e^{-\Phi}] \sqrt{\det(\pbP[\mathcal E]_{\mu\nu} + \ls^2 F_{\mu\nu})\det(Q\indices{^A_B})}  \right\} \, , \label{eq:MyersDBI} \\
  S_{\text{\<WZ>}} &= i\mu_p \int_{\Sigma_{\mathrlap{p+1}}} \tr \left\{ \pbP[e^{i\ls^2\inner\phi \inner\phi}(C \wedge e^B)]\wedge e^{\ls^2 F} \right\}_{p+1} \, . \label{eq:MyersWZ}
\end{align}
These actions describe the bosonic worldvolume fields, which are a non-Abelian gauge field $A_\mu$, $\mu = 1,\ldots,p+1$ and $9-p$ scalars\footnote{In this section, we use $\phi^A$ instead of $\varphi^A$ for the worldvolume scalars in order to make contact with the notations used for probe branes in the next chapters.} $\phi^A$, $A=1,\ldots,9-p$, in a non-trivial background specified by a dilaton~$\Phi$, a metric~$G_{MN}$, an antisymmetric 2-form potential~$B_{MN}$ and a \<RR> polyform~$C$ as in \cref{eq:CRRIIA} (resp.\ \cref{eq:CRRIIB}) for even (resp.\ odd) $p$.
The tensor $\mathcal E_{MN}$ is defined as
\begin{equation}
  \mathcal E_{MN} = E_{MN} + E_{MA}(Q^{-1} - \delta)\indices{^A_B}E^{BC}E_{CN}
  \label{eq:defcalE}
\end{equation}
in terms of the combination $E_{MN} = G_{MN} + B_{MN}$ of the metric and $B$-field, the matrix $Q\indices{^A_B} =\delta^A_B + i \ls^2[\phi^A,\phi^C]E_{CA}$ and the inverse $E^{AB}$ of $E_{AB}$, i.e.\ $E_{AB}E^{BC} = \delta^C_A$.
The symbol~$\pbP$ stands for the non-Abelian pullback depending on the worldvolume scalars.
It is defined in the following way.
First take the ordinary pullback in static gauge, and replace the partial derivatives of the scalar by gauge-covariant derivatives.
For example on the two-index tensor $\mathcal E$, this yields (compare with \cref{eq:PBGstatic}),
\begin{equation}
  \pbP(\mathcal E)_{\mu\nu} = \mathcal E_{\mu\nu} + \ls^2 \nabla_\mu \phi^A \mathcal E_{A\nu} + \ls^2 \nabla_\nu \phi^A \mathcal E_{\mu A} + \ls^4 \nabla_\mu\phi^A \nabla_\nu\phi^B \mathcal E_{AB} \, .
  \label{eq:MyersPB}
\end{equation}
All background fields still depend on the transverse space-time coordinates that need to be translated into worldvolume coordinates.
In the Abelian setting, this was done by choosing a reference point and identifying the displacement with respect to this point with the worldvolume scalars, as in~\cref{eq:staticgaugephi}.
The second step is to replace in the background fields the transverse space-time coordinates with the matrix-valued coordinates
\begin{equation}
  x^A \to X^A = x^A \1_{N\times N} + \ls^2 \phi^A(x_\mu) \, ,
  \label{eq:matrixcoord}
\end{equation}
which corresponds to expanding around the configuration where all the D-branes are sitting at $x^A$.
This has to be understood in the sense of formal Taylor series:
for example, the non-Abelian pullback of the dilaton is
\begin{equation}
  \pbP(\Phi) = \Phi(x^A \1 + \ls^2\phi^A) = \sum_{n=0}^{\infty} \frac{\ls^{2n}}{n!} \phi^{A_1}\cdots\phi^{A_n} \partial_{x^{A_1}}\cdots\partial_{x^{A_n}}\Phi(x^B) \, .
  \label{eq:MyersPBPhi}
\end{equation}
The other non-standard operation involving the matrix-valued scalar fields is the non-vanishing squared inner product $\inner\phi\inner\phi$.
Recall that, given a $k$-form $\omega_k$ and a vector $v$, one can form the inner product $\inner v\omega$ which is a $(k-1)$-form.
By antisymmetry, the squared inner product $\inner v\inner v\omega_k$ identically vanishes.
However, this is not the case for the inner product $\inner\phi\inner\phi\omega$ since the anti-commutator of $[\phi^A,\phi^B]$ does not necessarily vanish when the $\phi$s are matrices.
For example, on the \<RR> 2-form $C_2 = \frac{1}{2} {C_2}_{MN}{\d x^M}\wedge{\d x^N}$,
\begin{equation}
  \inner\phi\inner\phi C_2 = \phi^B\phi^A {C_2}_{AB} = -\frac{1}{2}[\phi^A,\phi^B] {C_2}_{AB} \, .
  \label{eq:iPhi2}
\end{equation}
The presence of the inner products has the important consequence that D$p$-branes not only couple to all \<RR> $q$-form potentials with $q\le p+1$ (and parity opposite to $p$), but also to all the other \<RR> forms existing in the theory.
This is at the origin of the Myers effect \cite{Myers:1999ps} where a stack of D$p$-branes polarizes into a D$(p+2)$-brane.
It will also play an important role in the following, as it will allow to recover all the \<RR> fields from \Dmi{}-branes (or D0-branes) only.

By unpacking all the definitions, plugging them into the actions \cref{eq:MyersDBI,eq:MyersWZ} and expanding in power series of $\ls^2$, one can obtain an explicit action for the worldvolume bosons with background-dependent couplings.
However, one still needs to answer question~\labelcref{it:ordering} by finding the correct way to order all the matrix-valued worldvolume fields entering the action.
An ordering procedure for the D9-brane, which goes under the name of \define{symmetrized trace prescription} and is denoted by $\operatorname{Str}$, has been defined by Tseytlin in~\cite{Tseytlin:1997csa,Tseytlin:1999dj}.
It amounts in this context to treating the combinations of fields $F_{\mu\nu}$, $\phi^A$, $\nabla_\mu\phi^A$ and $[\phi^A,\phi^B]$ that are involved in the construction of Myers' action as basic objects, and then symmetrizing over all their possible orderings before taking the trace over the gauge group.
This prescription, however natural it may seem, is known to be incorrect.
It fails to capture some effects known to be present in String Theory at third order in $(F, [\phi,\phi],\nabla\phi)$ \cite{Hashimoto:1997gm,Bain:1999hu}.
The issue of finding the correct non-Abelian generalization of the \<DBI> action has received a lot of attention in the literature (see \cite{Koerber:2001uu,Koerber:2002zb,Collinucci:2002ac,Movshev:2009ba,Cederwall:2011vy} for different attempts), but to date there is no clear understanding of the ordering prescription adopted by string theory.
This complication will fortunately not matter for the purposes of this thesis.
Indeed, the effective action will be computed by other means and be compared with Myers' action in order to read-off the background fields from it.
Luckily, all the background fields already appear at the lower orders for which the symmetrized trace prescription is thought to be correct so that the higher orders will not be needed.
The approach pursued in this thesis actually provides an independent way to construct well-defined D-brane actions at all orders in particular cases, even when Myers' form is not valid.

\subsection{D-branes on orbifolds}
\label{sec:Dorbifolds}

So far, we have only considered D-branes in flat space or in weakly curved backgrounds where the light closed string modes are the flat space massless modes.
In \cref{sec:orbifolds}, closed strings in more general backgrounds were constructed by the orbifold procedure.
We are now going to discuss the fate of D-branes on orbifolds \cite{Douglas:1996sw}.

Let us take again an orbifold group~$\Gamma \subset \SO(10)$, and consider the orbifold $\mathcal M \simeq \RR^{10}/\Gamma$.
The elements $\gamma \in \Gamma$ act on the worldsheet fields of the open string in the same way as for the closed string, given by \cref{eq:orbifoldcoord}.
The novelty resides in the action on the boundary conditions.
Indeed, consider a D$p$-brane at the position $x\sim\gamma \cdot x$ on $\mathcal M$, and assume to begin with that it is placed away from the orbifold singularity locus.
Under the orbifold projection $\RR^{10}\to \mathcal M$, all the points in the $\Gamma$-orbit of $x$ map to the same point.
Since $x$ is not an orbifold singularity, $\Gamma$ acts freely and there are $|\Gamma|$ points in the orbit.
Hence, a single physical D-brane in the orbifold maps in the covering space $\RR^{10}$ to $|\Gamma|$ different D-branes.
We can index these D-branes by $i = 0,\ldots,|\Gamma|-1$ and write $\ket{i}$ for the D-brane sitting at $\gamma_i\cdot x$, the element $\gamma_0$ being the identity in $\Gamma$.
The equation
\begin{equation}
  \gamma \gamma_i = R(\gamma)\indices{_i^j} \gamma_j
  \label{eq:defregrep}
\end{equation}
defines the $|\Gamma| \times |\Gamma|$ permutation matrix $R(\gamma)$ uniquely.
The orbifold group acts on the D-branes in the orbit in the same way,
\begin{equation}
  \gamma\cdot \ket{i} = R(\gamma)\indices{_i^j}\ket{j} \, .
  \label{eq:orbifoldbrane}
\end{equation}
By viewing the branes as the basis elements of a complex vector space $\CC[\Gamma]$ of dimension $|\Gamma|$, $R(\gamma)$ can be seen as a linear transformation and it is straigthforward to check that $R\from\Gamma\to\operatorname{End}(\CC[\Gamma])$ is a group representation, called the \define{regular representation} of $\Gamma$.
The physical D-brane that one obtains by implementing \cref{eq:orbifoldcoord} on the worldsheet fields, \cref{eq:orbifoldbrane} on the boundary conditions and projecting on invariant states is similarly called a \define{regular brane}.
More generally, one can consider a stack of $N$ regular branes by starting with $|\Gamma|N$ branes in the covering space and taking $N$ copies of the regular representation.

It is a well-known fact of the representation theory of finite groups that the regular representation $\Gamma$ decomposes into the irreducible representations $R_r$ of $\Gamma$, with multiplicities given by the dimensions,
\begin{equation}
  \CC[\Gamma] = \bigoplus_{r} (\dim R_r) R_r \, .
  \label{eq:irrepsreg}
\end{equation}
The matrices $R(\gamma)$ can thus be simultaneously block-diagonalized along the same pattern.
The gauge group has to preserve the block decomposition and hence gets broken by the orbifold projection as
\begin{equation}
  \U(|\Gamma|N)\to \prod_r \U((\dim R_r)^2N) \, .
  \label{eq:orbifoldgauge}
\end{equation}

The open strings stretched between regular branes are the analogues of the closed string untwisted sector: they are well-defined objects in the covering space and the orbifold projection simply amounts to a truncation of their spectrum.
It is then to be expected that, as happens for the closed string with the twisted sectors, having an orbifold allows to choose more general boundary conditions for the open strings, i.e.\ different kinds of D-branes besides the regular ones.
They could be argued to exist by a similar argument to the one pursued in \cref{sec:branesintro} in order to provide sources for the twisted \<RR> fields.
Let us proceed more directly, noticing that we were led to the regular representation from the condition that $\Gamma$ act freely on the D-branes at a non-fixed point in the covering space.
If we place some D-branes at a fixed point instead, the argument is not valid any more and we can take \emph{any} representation of $\Gamma$ instead of the regular representation to construct D-branes.
The basic building blocks are the different irreducible representations~$R_r$ of $\Gamma$.
To each of these, we can associate a different type of D-brane at the singularity.
The D-brane associated to $R_r$ is called \define{fractional brane} of type~$r$, because \cref{eq:irrepsreg} provides a decomposition of a regular brane, placed at the singularity, into its fractional brane constituents.

Let us see more concretely what happens in the specific case where $\Gamma$ is the group $\ZZ_2 \subset \SU(2)\subset SO(4) \subset \SO(10)$ defined by \cref{eq:Z2orbaction}.
We shall first consider $N$~regular D3-branes on this orbifold, hence we need to start with $2N$ D3-branes on the covering space and implement the orbifold projection.
The transverse space to the D3-branes is $\RR^6$, which is more conveniently viewed as $\CC^3$ with coordinates $(z^I) = (z^1,z^2,z^3)$.
The orbifold group $\ZZ_2$ can be viewed as a subgroup of $\SU(3)$, and its non-trivial element then acts on $\CC^3$ as
\begin{equation}
  \gamma\cdot (z^1,z^2,z^3) = (z^1,-z^2,-z^3) \Rightarrow \left(\gamma\indices{^I_J}\right)=\begin{pmatrix}
  1 &    &    \\
  & -1 &    \\
  &    & -1
\end{pmatrix}\, .
\label{eq:Z2actionSU3}
\end{equation}
This action fixes the complex plane parameterized by the coordinate $z^1$ in addition to the $\RR^4$ parallel to the D3-branes, hence the orbifold space is $\RR^4\times \CC\times \CC^2/\ZZ_2$.
We need to take complex combinations of the six scalars $\varphi_A$, producing three complex scalars, in order to match the identification of the transverse space as a $\CC^3$.
It is convenient to define them by
\begin{equation}
  \Phi^I = \frac{1}{2}\epsilon^{IJK}\Sigma_{AJK}\varphi_A \, , \quad \Phi^\dagger_I = \Sigma_{AI0}\varphi_A \, ,
  \label{eq:defphiI}
\end{equation}
decomposing the global symmetry of the transverse space (which is an R-symmetry in the field theory on the D3-branes) $\operatorname{Spin}(6)\simeq\SU(4)_R\to\U(1)\times \SU(3)$ and the $\SU(4)$ index $a\to(0,I=1,2,3)$.
The matrices $\Phi^I$ and $\Phi^\dagger_I$ are Hermitian conjugate of each other thanks to \cref{relSigma,Sigmabardef6D}.
The action of $\gamma$ in the regular representation is given by the matrix
\begin{equation}
  R(\gamma)=\begin{pmatrix}
    1 & 0 \\
    0 & -1
  \end{pmatrix}
  \label{eq:Z2regrep}
\end{equation}
which makes manifest the decomposition \cref{eq:irrepsreg} into the irreducible representations
\begin{equation}
  R_0(\gamma) = 1  \quad \text{and} \quad R_1(\gamma)= -1 \, .
  \label{eq:Z2irreps}
\end{equation}
The low-energy theory on the D3-branes, before orbifolding, is the $\nn=4$~$\U(2N)$ theory.
In $\nn=1$ language, the spectrum of the theory is a vectormultiplet and three adjoint chiral multiplets whose scalar components are the $\Phi^I$.
Because the orbifold group is a subgroup of $\SU(3)$, the action on the scalars coincides with the action on the fermions and some supersymmetry is preserved.
In fact, $\nn=2$ is preserved because the orbifold group is a subgroup of $\SU(2)$, by the same argument as for the closed string.
On the vectormultiplet $A_\mu$, the orbifold only acts on the Chan-Paton factors and the invariance condition reads
\begin{equation}
  A_\mu = \gamma\cdot A_\mu = R(\gamma)^{-1} A_\mu R(\gamma) \Rightarrow A_\mu =  \begin{pmatrix}
  {A_0}_\mu & \\
  & {A_1}_\mu
\end{pmatrix}
\label{eq:Z2A} \, .
\end{equation}
This breaks the gauge group $\U(2N)\to\U(N)_0\times\U(N)_1$.
On the scalars, the orbifold acts both on the Chan-Paton factors and the R-symmetry indices,
\begin{equation}
  \gamma\cdot\Phi^I = R(\gamma)^{-1} \gamma\indices{^I_J} R(\gamma)
  \label{eq:Z2PhiI}
\end{equation}
hence imposing $\Phi^I = \gamma\cdot\Phi^I$ projects onto the components
\begin{equation}
  \Phi^1 = \begin{pmatrix}
    \varphi_0^1 & \\
    & \varphi_1^1 \\
  \end{pmatrix} \, , \quad
  \Phi^2 = \begin{pmatrix}
    & \varphi_{01}^2 \\
    \varphi_{10}^2 &
  \end{pmatrix} \, , \quad
  \Phi^3 = \begin{pmatrix}
    & \varphi_{01}^3 \\
    \varphi_{10}^3 &
  \end{pmatrix} \, .
  \label{eq:Z2Phi123}
\end{equation}
In $\nn=2$ language, $\varphi^1_r$ combines with ${A_r}_\mu$ to form a vectormultiplet and $(\varphi^2_{01},\varphi^3_{01})$, $(\varphi^2_{10},\varphi^3_{10})$ are two bifundamental hypermultiplets, transforming respectively in the representations $(\mathbf{\bar N}_0,\mathbf{N}_1)$ and $(\mathbf{N}_0,\mathbf{\bar N}_1)$ of the gauge group.
This data can be conveniently summarized in a \define{quiver diagram}, writing a node for each gauge group and an arrow for each chiral multiplet, as in \cref{fig:Z2quiver}.
\begin{figure}
  \centering\begin{tikzpicture}
  [
  every path/.style={thick},
  group/.style={draw,fill=gray!30,minimum size=1cm},
  instanton/.style={rectangle,group},
  gauge/.style={circle,group}
  ]
  \node[gauge] (gauge 1)  {$N$} node[above=\lineskip of gauge 1.north] {Type $1$};
  \node[gauge] (gauge 0) [left=4 of gauge 1] {$N$} node[above=\lineskip of gauge 0.north] {Type $0$};
  \begin{scope}
    [decoration={markings,mark=at position 0.5 with {\arrow{<}}}]
    \draw [postaction={decorate}] (gauge 0.175) arc (5:355:0.5);
    \node [left=1 of gauge 0.west] {$\varphi^1_0$};
    \draw [postaction={decorate}] (gauge 1.5) arc (175:-175:0.5);
    \node [right=1 of gauge 1.east] {$\varphi^1_1$};
  \end{scope}
  \begin{scope}
    \draw [decoration={markings,mark=at position 0.5 with {\arrow{<<}}},postaction={decorate}] (gauge 0.north east) to[out=30, in=150] node [pos=0.5, above] {$(\varphi^2_{10},\varphi^3_{10})$} (gauge 1.north west);
    \draw [decoration={markings,mark=at position 0.5 with {\arrow{>>}}},postaction={decorate}] (gauge 0.south east) to[out=-30, in=-150] node [pos=0.5, below] {$(\varphi^2_{01},\varphi^3_{01})$} (gauge 1.south west);
  \end{scope}
\end{tikzpicture}
  \caption{The quiver diagram for the $\CC^2/\ZZ_2$ orbifold}
  \label{fig:Z2quiver}
\end{figure}
The superpotential of the parent $\nn=4$ theory is $W_{\nn=4}=\tr \Phi^1[\Phi^2,\Phi^3]$, which becomes, by implementing the projection \cref{eq:Z2Phi123},
\begin{equation}
  W_{\CC^2/\ZZ_2}=\tr \left[ \varphi^1_0 \left(\varphi^2_{01}\varphi^3_{10}-\varphi^3_{01}\varphi^2_{10}\right) + \varphi^1_1\left(\varphi^2_{10}\varphi^3_{01}-\varphi^3_{10}\varphi^2_{01}\right) \right] \, .
  \label{eq:Z2superpotential}
\end{equation}
The $\nn=2$ gauge theory that we have just constructed can be shown to inherit the property of being conformal from the parent $\nn=4$ theory \cite{Kachru:1998ys,Lawrence:1998ja}.

It is instructive to look at the classical moduli space of vacua of this theory\footnote{Because of the reduced supersymmetry, the classical moduli space is very different from the quantum one. This will be discussed in detail in \cref{chap:enhancon}.}.
Let us take the case $N=1$ of a single regular D3-brane for simplicity.
Imposing the $D$-flatness conditions for the gauge groups $\U(1)_0$ and $\U(1)_1$ and looking at the extrema of the superpotential, it is not hard to see that there are two branches.
\begin{itemize}
  \item A Higgs branch, where one of the bifundamentals is given expectation values, breaking the gauge group to the diagonal subgroup, $\U(1)_0\times\U(1)_1\to \U(1)_{\text{diag}}$.
    The relations satisfied by the scalars precisely match the interpretation of a regular brane moving on the orbifold, because the positions of its fractional constituents $0$ and $1$ are constrained to be orbifold images of each other.
  \item A Coulomb branch, where all bifundamentals vanish and $\varphi^1_0$ and $\varphi^1_1$ are arbitrary.
    The interpretation is that the regular brane, being placed at the orbifold singularity, can split into two independent fractional branes.
\end{itemize}

Until now, we have described only the case where we have $N$ regular branes.
To add, say, $M$ fractional branes of type~$1$ is very easy: all we have to do is change the representation content:
instead of taking $N$ copies of each irreducible representation, we now take $N$ copies of the trivial representation $R_0$ and $N+M$ copies of the irreducible representation $R_1$ in \cref{eq:Z2irreps}.
The effect is to change the gauge group to $\U(N)_0\times \U(N+M)_1$ and the shapes of the blocks in \cref{eq:Z2Phi123} change accordingly.
Note that if we take \emph{only} $M$ fractional branes of type~1 and no regular branes, the theory one obtains has gauge group $\U(M)$ and there are no bifundamentals: this is simply the pure $\nn=2$ super-Yang-Mills theory.

We can also add \Dmi{}-branes to the game, considering the \Dmi{}/D3 system on orbifolds.
The procedure to obtain the spectrum is the same as for the D3-branes, one chooses a representation of the orbifold group on the Chan-Paton indices corresponding to the \Dmi{}-brane (regular for a regular brane, irreducible for a fractional brane) and one lets the orbifold group act both on the gauge indices and on the global symmetry indices.
It is again useful to combine the six Hermitian scalar moduli $\phi_A$ into complex scalars $\phi^I$, $\cramped{\phi^\dagger_I}$ as in \cref{eq:defphiI}.
The moduli then have block decompositions similar to either $\Phi^2$ or $\Phi^3$ in \cref{eq:Z2Phi123} according to the whether they carry an index $I$ with value $2$ or $3$.

Analogously to the closed string twisted spectrum that can be seen as arising from integrating the ordinary closed string massless spectrum on the exceptional two-cycle, the fractional D-branes can be seen as specific configurations of ordinary branes in the resolved space.
For the $\CC^2/\ZZ_2$ orbifold, a D$p$-brane of type~1 is a D$(p+2)$-brane wrapped on the exceptional cycle $\Sigma$.
This implies that wrapped branes are sources for the \<RR> twisted fields.
For example, take a D1-brane, which couples to the \<RR> two-form potential $C_2$ through a term $i\mu_1\int C_2$.
Letting the worldsheet of the D1 be the exceptional cycle, we get by \cref{eq:RRtwisted} a coupling between the fractional \Dmi{} of type~1 and the \<RR> twisted scalar~$c$
\begin{equation}
  S_{\text{\<WZ>,\Dmi{}}_1} \supset i (2\pi\ls^2\mu_1)c = i \mu_{-1} c \, .
  \label{eq:WZDm1frac}
\end{equation}
The fractional branes of type~0 are a bit more complicated from this point of view, as they involve anti-branes with worldvolume flux \cite{Diaconescu:1997br}.
This is reviewed for example in~\cite{Bertolini:2003iv}.

\chapter{The gauge/string correspondence}
\label{chap:holo}

After a review of the required string theory tools in the previous chapter, we will now put them to good use to study the gauge/string correspondence.
This correspondence is presented in~\cref{sec:malda} according to Maldacena's original argument for D3-branes~\cite{Maldacena:1997re}.
It is then shown in \cref{sec:holoprobe} how one can probe this correspondence by adding a small number of \Dmi{}-branes \cite{Ferrari:2012nw}.

\section{Maldacena's original proposal}
\label{sec:malda}

In the previous chapter, we have introduced D-branes as Ramond-Ramond charges, but have then focused on D-branes as boundary conditions for open strings.
That perspective yielded in the low-energy regime a field theory description of the worldvolume dynamics of the D-branes.
We are now going to reconsider D-branes as \<RR> sources, i.e.\ objects coupling to closed strings, which will provide a supergravity point of view on the D-branes.

\subsection{D3-branes as supergravity solutions}
Consider a stack of $N$ D3-branes in type \<IIB> theory on flat Euclidean space-time $\RR^{10}$.
These D3-branes source the RR 5-form field strength as well as the metric.
The strength of the coupling of a single D3-brane to the closed string sector is given by the string coupling $g_s$, hence for $N$ branes the coupling is proportional to $Ng_s$.
When $Ng_s$ is large, it thus makes sense to look for supergravity solutions that carry D3-brane charge.
These solutions are similar to the Reissner-Nordström solution in Einstein-Maxwell theory that carries electric charge.
Since we want to describe planar D3-branes, we need a solution with $\SO(4)\times\SO(6)$ global symmetry.
This solution was found by \textcite{Horowitz:1991cd} and the non-vanishing fields read
\begin{subequations}
  \begin{align}
    e^\Phi &= g_s \, , \label{eq:solD3Phi} \\
    \d s^2 &= H^{-1/2} \delta_{\mu\nu} \d x^\mu \d x^\nu + H^{1/2} \delta_{AB} \d y^A \d y^B \, , \label{eq:solD3G}\\
    F_5 &= \frac{N\ls^4}{\pi R^5}\Bigl( \omega_{\Sfive} - i\omega_x \Bigr)  \, . \label{eq:solD3F5}
  \end{align}
\end{subequations}
The ten dimensions have been decomposed into parallel and transverse directions to the D3, $x^\mu$, $\mu=1,\ldots,4$ and $y^A$, $A=1,\ldots,6$ respectively and the coordinate $r$ is defined by $r^2=\delta_{AB}y^Ay^B$.
Further,
\begin{subequations}
  \begin{align}\label{eq:defHD3}
    H(r) &=1+\frac{R^4}{r^4}\, ,\\\label{eq:omegax}
    \omega_{x}& = \frac{r^2y_{A}}{R^3}\,
    \d x_{1}\wedge\cdots\wedge\d x_{4}\wedge\d y_{A}\, ,\\
    \label{eq:omegaS5}
    \omega_{\Sfive} &= \frac{1}{5!}\frac{R^5 y_{F}}{r^6}\epsilon_{ABCDEF}\,\d y_{A}\wedge\cdots\wedge\d y_{E}\, .
  \end{align}
\end{subequations}
The length scale $R$ entering the solution is proportional to the string length,
\begin{equation}
  R^4 = \frac{N g_s}{\pi}\ls^4 \, .
  \label{eq:RD3}
\end{equation}
$R$ sets the curvature scale of the solution and as expected, it depends only on the product $N g_s$.
At $r=0$, there is a coordinate singularity signalling in this case the presence of a horizon.

\subsection{The decoupling limit}
Let us now focus on an observer at infinity in the transverse space ($r\to\infty$) who can make only low-energy experiments, $\ls E \ll 1$, by scattering closed strings (light open strings need to stay close to the D3-branes).
Since the time component of the metric~\cref{eq:solD3G} vanishes for $r\to0$, closed strings produced in the deep interior will be heavily redshifted, which allows the asymptotic observer to probe physics at a much higher energy scale.
More specifically, excitations produced at radius $r$ with proper energy $E(r)$ will reach her with energy
\begin{equation}
  E(\infty) = E(r) \left(1+\frac{R^4}{r^4}\right)^{-1/4} \, ,
  \label{eq:redshiftD3}
\end{equation}
which will be accessible if $E(\infty)\le E$.
In the limit $\ls E\to 0$, two different sectors can thus be probed.
The first possibility is for the redshift to be negligible, in which case $\ls E(r)\to 0$ and one is probing the massless modes of the closed string in the flat asymptotic region.
The second possibility is for the redshift to be very large, in which case $\ls E(r)$ can be arbitrary and one can probe arbitrarily massive excitations of the closed string.
Seeing the low-energy limit equivalently as a limit with fixed energies but with $\ls\to 0$ and remembering that $R\sim\ls$ by \cref{eq:RD3}, equation \cref{eq:redshiftD3} implies that one needs to take
\begin{equation}
  U=\frac{r}{\ls^2} \quad \text{fixed} \, .
  \label{eq:defU}
\end{equation}
In this limit, one is focusing on the physics near the horizon, hence the name of \define{near-horizon limit}.
One can neglect in this region the leading~1 in \cref{eq:defHD3} to obtain the Euclidean $\AdSS$ metric,
\begin{equation}
  \d s^2 = \frac{r^2}{R^2}\d x_\mu\d x_\mu + \frac{R^2}{r^2}\d r^2 + R^2\d\Omega_5^2 \, ,
  \label{eq:AdSSmetric}
\end{equation}
where $\d\Omega_5^2$ is the round metric on the unit five-sphere.

Let us now go back to the other description of D3-branes in terms of open strings and take the $\ls\to0$ limit from the start.
We saw that in the low-energy $\ls\to0$ limit, the dynamics of the $N$~D3-branes could be described by the $\nn=4$ $\SU(N)$ field theory\footnote{The overall $\U(1)$ factor describing the center of mass of the stack is decoupled and should not be taken into account in the present discussion.}.
In principle, one should still include the closed string sector that interacts with the D3-branes.
However, in the $\ls\to0$ limit, the closed string sector is decoupled because the Planck mass scales like $\ls^{-1}$ and is sent to infinity.
Hence the D3-branes do not backreact, independently of the value of $Ng_s$.
We thus find again two sectors: on the one hand, massless closed strings in flat space and on the other hand, the $\nn=4$ super-Yang-Mills theory.

In both cases we were describing the same theory, the low-energy limit of D3-branes.
In the first case, we first took the backreaction into account and then took the low-energy~$\ls\to0$ limit, whereas in the second case, we took the low-energy limit directly, which allowed to neglect the backreaction.
We have found in each case two decoupled sectors of which one, the massless closed strings in flat space, is common to the two cases.
It is then reasonable to expect that also the two other sectors are identical.
This yields the
\begin{description}
  \item[\<AdS>/\<CFT> correspondence] \emph{Type \<IIB> string theory on $\mathrm{AdS}_5\times\mathrm{S}^5$ is \emph{equivalent} to the $\nn=4$ super-Yang-Mills theory.}
\end{description}

While the two theories look at first sight very different, this is because we are used to considering them in opposite regimes of parameters.
On the gauge theory side, one can reorganize perturbation theory as an expansion in two parameters $N$ and $\lambda = N g^2_4$ \cite{tHooft:1973jz}, where $N$ is the rank of the gauge group and $\lambda$ is the 't~Hooft coupling.
The theory simplifies in the large~$N$ limit with fixed~$\lambda$, since only a subclass of Feynman diagrams, the planar diagrams, can contribute.
The classical limit is then~$\lambda\to0$.
By~\cref{eq:gcoupling}, one can relate the 't~Hooft coupling to the string coupling~$g_s$,
\begin{equation}
  \lambda = 4\pi N g_s \, .
  \label{eq:lambdags}
\end{equation}
The large $N$ limit thus corresponds to a small string coupling limit~$g_s\to0$.
On the $\AdSS$ string theory side, this means that the large $N$ limit corresponds to a classical limit in the sense that it suppresses the contributions from higher genus Riemann surfaces in the string perturbation theory.
However, the relation~\cref{eq:RD3} implies that at fixed \<AdS> radius~$R$, $\ls^2\sim 1/\sqrt\lambda$.
So in order to decouple the massive string states and recover ordinary two-derivative supergravity one needs to take the limit~$\lambda\to\infty$, where the gauge theory is strongly coupled.

\section{Adding D-instanton probes}
\label{sec:holoprobe}

By considering the low-energy limit of a stack of $N$ D3-branes, we arrived at the \<AdS>/\<CFT> correspondence.
We had on the one side only closed strings in \<AdS> and on the other side the low-energy limit of open strings ending on the D3-branes.
Let us now also add a finite number~$K$ of \Dmi{}-branes to the large number~$N$ of D3-branes and repeat the argument.
Because $K$ stays fixed while $N\to\infty$, we can always neglect the backreaction of the \Dmi{}-branes on the massless closed string modes, i.e.\ we treat the \Dmi{}-branes as \define{probes}.
The crucial point is that we can now look at the D3-branes in the two pictures from the point of view of the probe \Dmi{}-branes rather than that of an observer at infinity.

Considering first the regime of large $\lambda$ where the D3-branes backreact, the probe \Dmi{}-branes are described by Myers' action discussed in \cref{sec:Myers} in the background given by~\cref{eq:solD3Phi,eq:solD3G,eq:solD3F5}.
When taking the decoupling limit $\ls\to0$, we want to place the probes in the near-horizon region rather than the flat asymptotic region.
For this, we need to scale the neutral moduli associated to motion in the transverse space as in~\cref{eq:defU}.
This means that we need to take the limit
\begin{equation}
  \ls\to0 \quad \text{with} \quad \phi_A = \ls^{-2}Y_A \quad \text{fixed} \, ,
  \label{eq:NHlimitDm1}
\end{equation}
where $\phi_A$ and $Y_A$ have respectively dimensions of energy and length.
The partition function for the $K$ \Dmi{}-branes can then be written schematically as
\begin{equation}
  \mathcal Z_{\text{eff}} = \int \d Z \d \Psi e^{-S_{\text{eff}}(Z,\Psi; \Phi, G, B, C)} \, ,
  \label{eq:Zeff}
\end{equation}
where $S_{\text{eff}}$ is the non-Abelian probe effective action for the 10 matrix-value coordinates $Z_M$ and their superpartners $\Psi$.
Its couplings depend on the supergravity background (or its stringy correction for finite~$\lambda$) and are given by Myers' action for the bosonic terms at lowest order.

Let us now consider the open string picture, where one takes first the low-energy limit $\ls\to0$.
We are then dealing with the \Dmi{}/D3 system discussed in \cref{sec:Dinstantons}.
In the low-energy limit, the system is described by an action $S = S_{\text{D3-D3}} +S_{\text{\Dmi{}-\Dmi{}}} +S_{\text{\Dmi{}-D3}}$ where the first term is the $\nn=4$ action \cref{eq:actionN4} describing the dynamics of the massless spectrum of the D3-D3 strings and the two other terms are respectively \cref{eq:SDm1Dm1,eq:SDm1D3} describing the moduli on the \Dmi{}-\Dmi{} and \Dmi{}-D3 strings respectively.
Rewriting the part of the action that depends on the moduli in terms of the 't~Hooft coupling $\lambda = Ng^2_4$ with the help of \cref{eq:g0g4}, one has
\begin{multline}
  S_{\text{\Dmi{}-\Dmi{}}} + S_{\text{\Dmi{}-D3}} =  \frac{4\pi^2N}{\lambda} \truK \left\{ - [\phi_A, X_\mu] [\phi_A,X_\mu]  \right. \\
  + 2i [X_\mu,X_\nu] D_{\mu\nu} - \bar\psi\indices{_{\dot\alpha}^a}\Sigma_{Aab}[\phi_A,\bar\psi^{\dot\alpha b}] - 2i\sigma_{\mu\alpha\dot\alpha}\Lambda\indices{^\alpha_a}[X_\mu,\bar\psi^{\dot\alpha a}] \\
    \left. -\frac{\ls^4}{2}[\phi_A\phi_B][\phi_A,\phi_B] - \ls^4\Lambda\indices{^\alpha_a}\bar\Sigma_A^{ab}[\phi_A,\Lambda_{\alpha b}] - \ls^4D_{\mu\nu} D_{\mu\nu} \right\} \\
+ \frac{1}{2} \tilde q^\alpha \phi_A\phi_A q_\alpha - \frac{1}{2} \tilde\chi^a \Sigma_{Aab} \phi_A \chi^b  \\
+ \frac{1}{\sqrt2} \tilde q^\alpha \Lambda_{\alpha a}\chi^a + \frac{1}{\sqrt2} \tilde \chi^a \Lambda\indices{^\alpha_a} q_\alpha + \frac{i}{2} \tilde q^\alpha D_{\mu\nu}\sigma\indices{_\mu_\nu_\alpha^\beta}q_\beta + \cdots
  \label{eq:SDinstls}
\end{multline}

We still need to implement the analogue of the near-horizon limit, in order for the \Dmi{}-branes to probe the D3-brane dynamics and not the decoupled asymptotic region \cite{Ferrari:2012nw,Billo:2002hm}.
By~\cref{eq:NHlimitDm1}, this amounts to taking $\phi_A$ fixed while sending $\ls\to0$.
On the contrary, the distances parallel to the D3-branes do not to scale, hence it is $X_\mu$ rather than $\ls^{-2}X_\mu$ that must kept fixed.
In order to preserve supersymmetry, this scaling must be implemented in a uniform way on the full supermultiplet, which means that all the moduli appearing in \cref{eq:SDinstls} are kept fixed when $\ls\to0$.
The effect of the scaling limit is thus to eliminate the terms on the third line of~\cref{eq:SDinstls}.
As explained in \cref{sec:Dinstantons}, the action~\cref{eq:SDinstls} then reduces to the action~$S_{\text{inst}}$ in \cref{eq:SinstN4} describing the field theoretic instanton.
Additionally, the Coulomb branch of the \Dmi{} theory decouples.
This is consistent with the dual near-horizon limit, as that branch was interpreted as configurations of \Dmi{}-branes that can be far away from the D3-branes and probe the full flat space-time.
The microscopic partition function, capturing the full dynamics of this system, can be written schematically as
\begin{equation}
  \mathcal Z_{\text{mic}} = \int [\mathcal DA] [\d X][\d q] e^{-S_{\text{inst}}[X,q,A] - S_{\nn=4}[A]} \, .
  \label{eq:Zmic}
\end{equation}
The integration measures stand for the functional integration over the D3 fields and the ordinary integrals over the neutral and charged moduli respectively.

Given that the two descriptions of the probe \Dmi{}-branes should be equivalent, we obtain that the two partition functions~\cref{eq:Zeff} and \cref{eq:Zmic} are actually equal.
This provides a strategy to \emph{deduce} the closed string background dual to the field theory on the D3-branes:
one needs to recast $\mathcal Z_{\text{mic}}$ in the same form as $\mathcal Z_{\text{eff}}$ and one can then read-off the background by comparing the effective action~$S_{\text{eff}}$ with Myers' action for a generic background.
Such an effective action can be found by integrating out all the fields and moduli that are not present in the geometric picture, which are the ten $\U(K)$-adjoint scalars~$Z_M$ and their superpartners~$\Psi$.
Let us briefly see how one can do this in the $\AdSS$ case~\cite{Ferrari:2012nw}.
One first needs to identify the moduli that correspond to the geometric scalars~$Z_M$.
The moduli corresponding to the directions parallel to the D3-branes are the $X_\mu$, while for the transverse directions one has by \cref{eq:NHlimitDm1},
\begin{equation}
  Y_A = \ls^2\phi_A \, .
  \label{eq:eq:YPhi}
\end{equation}
The next step is to integrate out the charged moduli $(q,\tilde q,\chi,\tilde\chi)$.
This can be done exactly by Gaussian integration because the instanton action~\cref{eq:SinstN4} is quadratic in the charged moduli%
\footnote{
  From a field theory point of view, this condition can be seen to hold thanks to the presence of the neutral moduli~$\phi_A$: they also appear quadratically in the action and one could integrate them out, at the price of producing cubic and quartic terms for the charged moduli.
  These moduli are nevertheless crucial in order to provide the right non-fluctuating variables to solve the theory on the probe in the large~$N$ limit and this motivates their introduction in cases when they are not present from the start, as in \cite{Ferrari:2013aba}.
}.
This produces a superdeterminant~$\mathscr D$ given by the ratio of fermionic and bosonic \enquote{mass operators}.
The mass operators, and hence the superdeterminant, will depend on the D3-brane fields through couplings that were not written down explicitly in~\cref{eq:SDinstls} and can be found in \cite{Green:2000ke,Billo:2002hm,Ferrari:2012nw}.
Integrating out the D3 fields then amounts to computing the expectation value $\corr{\mathscr D}$ of this superdeterminant in the full-fledged gauge theory.
In general, this  involves an intractable sum over planar diagrams.
However, in some cases, the expectation value can simplify. In particular, for the \Dmi{}/D3 system under study, it corresponds to a one-point function which cannot be quantum corrected if conformal invariance is unbroken \cite{Ferrari:2012nw}, which is the case for a single stack of D3-branes.
More generally, we assume that when eight or more supercharges are preserved (from the probe brane point of view), including when conformal invariance is broken, the expectation value $\corr{\mathscr D}$ is not quantum corrected or, more mildly, that the terms in the effective action $\Seff$ that we use to derive the supergravity background are insensitive to the possible quantum corrections in $\corr{\mathscr D}$.
This will be the case for the theories studied in~\cref{chap:adsdeformed,chap:D4brane} and is a very plausible assumption, which is strongly supported by the consistency of the results obtained.
In \cref{chap:enhancon}, on the other hand, the superdeterminant will turn out to be quantum corrected in a crucial way, but in a way that is computable thanks to a supersymmetric non-renormalization theorem.
In any case, the result is an effective action that only depends on the neutral moduli.
Because we were integrating out $N$ charged moduli, the action is automatically proportional to $N$.
This means that the large~$N$ limit is a classical limit for the probe brane, in agreement with the general statement that one should obtain a classical string theory in the 't~Hooft limit of large~$N$ with fixed~$\lambda$.
The effective action so obtained however still depends on some moduli that do not have a geometric interpretation, such as~$D_{\mu\nu}$.
Those can be integrated out exactly by saddle-point approximation in the large~$N$ limit.
From the field theory point of view, $D_{\mu\nu}$ was a Lagrange multiplier for the \<ADHM> constraint and this step then amounts to enforcing the \<ADHM> constraint in order to obtain honest instantons.
The effective action one obtains can now finally be compared to Myers' and the background can be found by making the two actions coincide.

All the steps were performed explicitly in~\cite{Ferrari:2012nw}, allowing the derivation of the complete $\AdSS$ solution~\cref{eq:solD3Phi,eq:solD3G,eq:solD3F5}, including the full metric and non-trivial \<RR> 5-form from the \Dmi{}-branes probing D3-branes in the conformal phase, without solving any supergravity equation of motion.
In the following chapters, we will see that this procedure can be applied successfully to several different theories.

\part{Original contributions}
\label{part2}
\chapter{Deformations of \texorpdfstring{$\textrm{AdS}_{\textrm 5}\times \textrm{S}^{\textrm 5}$}{AdS⁵×S₅} from D-instanton probes}
\label{chap:adsdeformed}

The aim of this chapter is to start the study, along the lines explained in~\cref{sec:holoprobe}, of holographic duals from field theory thanks to the use of D-brane probes and more specifically of D-instantons.
The case of $\AdSS$ from the conformal vacuum of $\nn=4$ was already studied in~\cite{Ferrari:2012nw} and the focus here is on continuous deformations of this basic case.
The present chapter is largely based on~\cite{Ferrari:2013pq}.

The simplest deformation we consider is the Coulomb branch deformation, which corresponds to turning on the vacuum expectation values of the scalar fields of the $\nn=4$ theory. This breaks both conformal invariance and R-symmetry but preserves sixteen supersymmetries. The resulting dual geometry asymptotically coincides with the usual $\AdSS$ background in the \<UV> but the metric and the Ramond-Ramond five-form field strength are modified in the \<IR> at the scales set by the scalar expectation values. Our field theory calculations yield a perfect match with the known near-horizon limit of the general multi-centered D3-brane solution, for both the metric and the Ramond-Ramond form.

The second case we consider is the non-commutative deformation \cite{Seiberg:1999vs,Douglas:2001ba}. It breaks conformal invariance but preserves both supersymmetry and R-symmetry. This model does not seem to have a \<UV> fixed point and, accordingly, the known supergravity dual \cite{Hashimoto:1999ut,Maldacena:1999mh} does not have a boundary in the \<UV> and it is likely that a purely field theoretic description does not exist. However, at sufficiently large distance scales, the model approaches the undeformed $\nn=4$ theory and the physical interpretation of both the field theory and its dual supergravity background becomes clear. The geometry we find is then fully consistent with the background proposed in \cite{Hashimoto:1999ut,Maldacena:1999mh}.

Finally, we investigate the so-called $\beta$-deformation \cite{Leigh:1995ep}. In its most general form \cite{Frolov:2005dj}, it breaks supersymmetry completely but preserves conformal invariance in the planar limit \cite{Ananth:2007px,Ananth:2006ac}\footnote{\label{fn:beta}\emph{Note added:} conformality of the theory when supersymmetry is broken is a subtle issue. At small 't Hooft coupling, the running of additional double-trace operators that are not inherited from the parent $\nn=4$ theory seems to break conformal invariance \cite{Fokken:2013aea,Fokken:2013mza,Fokken:2014soa}. This effect starts at second order in the deformation parameters. However, since the known dual supergravity background has an \<AdS> factor \cite{Frolov:2005dj}, conformal invariance seems to be restored in the supergravity regime. As explained further, our approach can probe the metric only to first order in the deformation and is hence insensitive to this breaking of conformal symmetry.}. The supergravity solution \cite{Lunin:2005jy,Frolov:2005dj} is known when the deformation parameters are small, which ensures that the $\alpha'$ corrections can be neglected. Again, our solution is fully consistent with supergravity, including for the Neveu-Schwarz and Ramond-Ramond three-form field strengths. Let us note that the form of the dilaton was already derived from instanton calculus in very interesting previous papers (see \cite{Georgiou:2006np,Durnford:2006nb} and related research in \cite{hep-th/9810243,hep-th/9901128,hep-th/9908035,hep-th/9908201,hep-th/9910118,hep-th/0604090,Akhmedov:1998pf}). 

The plan of the chapter is as follows. In \cref{framesec}, we briefly review how the general set-up, already explained in more details in \cref{sec:holoprobe}, can be applied for the considered cases and present our main results. In particular, we emphasize the new subtleties associated with the use of D-instantons in backgrounds that have a non-constant dilaton \cite{Ferrari:2013pi}. The details of the calculations are included in \cref{DerivSec}.

Besides the notations and some algebraic identities presented in~\cref{chap:notations}, also the appendices~\ref{chap:Myers} and~\ref{chap:sugrasols} are relevant for this chapter.
They contain respectively the expansion of Myers' action up to order five and a review of the supergravity solutions dual to the non-commutative and the $\beta$-deformed models.

\section{Set-up and main results}
\label{framesec}

Because we are dealing with deformations of $\AdSS$, the approach is extremely similar to the one presented in \cref{sec:holoprobe}.
We thus consider the path integral for a system of $N\gg1$ background 
D3-branes and $K$ probe D-instantons and we need to compute the probe effective action~$S_{\text{eff}}$ as in~\cref{eq:Zeff} from the microscopic action for the system as in~\cref{eq:Zmic}.
To do this, we need to integrate over all the extra degrees of freedom not present in the effective description of a probe in a closed string background and compare the resulting action with Myers' to read-off the background.
The construction of the effective action was explained in~\cref{sec:holoprobe} and we are now going to explain how to read-off the background once we have such an action at our disposal.

\subsection{\label{MDactionSec} On Myers' D-instanton action}

To analyse the action $\Seff$, we limit ourselves to the bosonic part, setting $\Psi=0$ in \eqref{eq:Zeff}. We then write the ten $K\times K$ matrices $Z_{M}$, $1\leq M\leq 10$, as
\begin{equation}\label{Zexp} Z_M=z_M\1+\ls^2 \epsilon_M\end{equation}
and expand $\Seff$ in powers of $\epsilon$,
\begin{equation}
  \Seff=\sum_{n\ge0}\Seff^{(n)}=\sum_{n\ge0}\frac{1}{n!}\ls^{2n}c_{M_1\cdots M_n}(z)\tr\epsilon_{M_1}\cdots\epsilon_{M_n} \, .
  \label{Myersexpansion}
\end{equation}
The coordinates $z_{M}$ correspond to a given ten-dimensional space-time point and we have introduced powers of the string length
\begin{equation}\label{lsdef} \ls^{2} = 2\pi\alpha'\end{equation}
for convenience. Myers' prescription for the non-Abelian D-instanton action yields the coefficients $c_{M_{1}\cdots M_{n}}$ in terms of the supergravity fields, see formula \eqref{Mexpand} in Appendix \ref{appB}. Many terms in \eqref{Mexpand} are actually redundant, being fixed by general consistency conditions \cite{Ferrari:2013pi}. In order to derive the full set of supergravity fields, it is enough to consider the following combinations,
\begin{align}
\label{c0} &c = - 2i\pi\tau = 2i\pi\bigl( C_{0}-ie^{-\phi}\bigr)\\
\label{c3} &c_{[MNP]}  = -\frac{12\pi}{\ls^{2}}\partial_{[M}( \tau B - C_{2})_{NP]}\\
\label{c4} &c_{[MN][PQ]} = -\frac{18\pi}{\ls^{4}}e^{-\phi} 
\bigl(G_{MP}G_{NQ}-G_{MQ}G_{NP}\bigr)\\
\label{c5}
&c_{[MNPQR]}  = -\frac{120 i \pi}{\ls^{4}}
\partial_{[M}\bigl( C_{4} + C_{2}\wedge B - \frac{1}{2}\tau B\wedge B
\bigr)_{NPQR]}\, .
\end{align}

Myers' action has two basic limitations. The first comes from the symmetrized trace prescription \cite{Tseytlin:1997csa,Tseytlin:1999dj} used to fix the ordering ambiguities due to the non-commu\-ting nature of the variables $Z$. This prescription is valid up to order five in the expansion \eqref{Myersexpansion} but is known to fail at higher orders \cite{Bain:1999hu,Hashimoto:1997gm}. This caveat will be of no concern to us, since equations \eqref{c0}--\eqref{c5} show that the expansion up to order five is sufficient to fix unambiguously all the supergravity fields. 

The second limitation comes from the fact that the formulas \eqref{c0}--\eqref{c5} are valid only to leading order in the small $\ls^{2}$, or supergravity, approximation. This implies that our microscopic calculations of $\Seff$, which do not rely on a small $\ls^{2}$ approximation, can be compared with Myers' only when $\ls^{2}\rightarrow 0$. When comparing our results with the known supergravity solutions, this restriction is harmless, since the solutions are themselves known at small $\ls^{2}$ only.

Let us point out, however, that some of the basic structural properties of the action, which are visible in the formulas
\eqref{c0}--\eqref{c5}, must be valid to all orders in $\ls^{2}$ because they are consequences of the general consistency conditions discussed in \cite{Ferrari:2013pi}. One of the most interesting properties is that the coefficients $c_{[MNP]}$ and $c_{[MNPQR]}$, viewed as the components of differential forms
\begin{align}\label{F3def} F^{(3)} &= \frac{1}{3!}c_{[MNP]}\,\d z^{M}\wedge\d z^{N}\wedge\d z^{P}\, ,\\ \label{F5def}
F^{(5)} &= \frac{1}{5!}c_{[MNPQR]}\,\d z^{M}\wedge\d z^{N}\wedge\d z^{P}\wedge\d z^{Q}\wedge\d z^{R}\, ,\end{align}
must always be closed,
\begin{equation}\label{F3F5closed} \d F^{(3)} = 0\, ,\quad \d F^{(5)}=0\, .\end{equation}
Locally, we can thus write
\begin{equation}\label{dc} F^{(3)} = -\frac{4\pi}{\ls^{2}}\,\d C^{(2)}\, ,\quad
F^{(5)} = -\frac{24 i \pi}{\ls^{4}}\,\d C^{(4)}\, .\end{equation}
Since the two- and four-form potentials $C^{(2)}$ and $C^{(4)}$ are well-defined to all order in $\ls^{2}$, formulas \eqref{c3} and \eqref{c5} can actually be used to \emph{define} the Ramond-Ramond and Neveu-Schwarz form fields to all order in $\ls^{2}$, 
\begin{equation}\label{C2C4NSRR} C^{(2)} = \tau B - C_{2}\, ,\quad C^{(4)} = C_{4} + C_{2}\wedge B -\frac{1}{2}\tau B\wedge B\, ,\end{equation}
modulo the general gauge transformations that are discussed in details in \cite{Ferrari:2013pi}. One of our main goal in the present chapter will be to compute the forms \eqref{F3def} and \eqref{F5def} for the Coulomb branch, non-commutative and $\beta$-deformations of the conformal $\nn=4$ gauge theory. As explained in the next Subsection we can then use \eqref{C2C4NSRR} to compare with supergravity in appropriate limits.

Other properties of the Myers action will not, however, be preserved by the $\ls^{2}$ corrections. For example, the only general constraint on the fourth order coefficient $c_{[MN][PQ]}$ is that it should have the same tensorial symmetries as the Riemann tensor. This does not imply a factorization in terms of a second rank symmetric tensor as in \eqref{c4} and thus such a factorization property is generically lost when $\ls^{2}$ corrections are included.

\subsection{\label{useSec}On the use of the non-Abelian D-instanton action}

There is one last crucial limitation associated with the use of D-instantons to derive the supergravity background \cite{Ferrari:2013pi}. Intuitively, this limitation is related to the fact that a D-instanton, sitting at a particular point, cannot be expected in general to probe the geometry of the full space-time manifold. This restriction is waived if the effective action, evaluated at $Z_{M}=z_{M}\1$, $\Seff(z\1)=Kc(z)$, does not depend on $z$, or, equivalently,
if the axion-dilaton $\tau$ is constant. This is the case for the $\nn=4$ gauge theory at any point on its Coulomb branch. However, for a generic background with non-constant axion-dilaton, the instantons are forced to sit at the critical points of $c(z)=-2i\pi\tau (z)$. This condition becomes strict when $N\rightarrow\infty$, being equivalent to the saddle-point approximation of the integral \eqref{eq:Zeff}.

An alternative way to understand the same limitation is to study the effect of general matrix coordinate redefinitions on the effective action. It is explained in \cite{Ferrari:2013pi} that, when $\d c$ is generic, one can actually gauge away the coefficients $c_{M_{1}\cdots M_{n}}$ for $n\geq 2$ in the expansion \eqref{Myersexpansion} by an allowed matrix transformation $Z\mapsto Z'$.

For the purposes of the present chapter, we shall deal with this difficulty by using a perturbative approach around the $\AdSS$ background on which the instantons can freely move. This is possible because the non-commutative and $\beta$-deformed models are continuous deformations of the $\nn=4$ gauge theory and thus the associated dual backgrounds will be themselves continuous deformations of the $\AdSS$ background. 

Let us denote by $\eta$ the deformation parameter; $\eta$ is the 
dimensionless ratio $\theta/\ls^{2}$ for the non-commutative theory discussed in section \ref{NCSec} or the combination $\lambda\gamma^{2}$ for the $\beta$-deformed theory studied in section \ref{BetaSec}. Let us also denote by $\smash{c^{*}_{M_{1}\cdots M_{n}}}$ the coefficients in the expansion \eqref{Myersexpansion} for the undeformed $\AdSS$ background. In our models, the gradient of the axion-dilaton and the corrections to the metric and five-form field strength turn out to be of order $\eta^{2}$. Hence,
\begin{align}\label{adpert} c(z) &= c^{*} + O(\eta^{2})\, ,\\
\label{metpert} c_{[MN][PQ]}(z) &= c_{[MN][PQ]}^{*}(z) + O(\eta^{2})\, ,\\
c_{[MNPQR]}(z) & = c_{[MNPQR]}^{*}(z) + O(\eta^{2})\, , \label{metpert2}
\end{align}
whereas the three-form field strengths are turned on at leading order,
\begin{align}\label{c3pert} c_{[MNP]}(z) = O(\eta)\, .\end{align}
The general variation of $c_{[MNP]}$ under an arbitrary redefinition of the matrix coordinates corresponds to a standard tensorial transformation under diffeomorphisms plus terms proportional to the gradient of $c$ \cite{Ferrari:2013pi} which, by \eqref{adpert}, are $O(\eta^{2})$. \emph{This means that the Neveu-Schwarz and Ramond-Ramond forms $B$ and $C_{2}$ are unambiguously fixed in terms of the microscopic calculation of the coefficient $c_{[MNP]}$ of the D-instanton effective action to leading order in the deformation parameter $\eta$.} 

Moreover, since the background derived from $\Seff$ unambiguously matches with the $\AdSS$ supergravity background \cite{Ferrari:2012nw} in the undeformed theory, we can always choose the same coordinate systems in both points of view at $\eta=0$. In the deformed $\eta\not=0$ models, the coordinate systems $z_{\text{mic}}$ and $z_{\text{\<SUGRA>}}$ used in the effective action $\Seff$ and in the supergravity solution respectively no longer necessarily agree, but the discrepancy must be of order $\eta$,
\begin{equation}\label{zzequal} z_{\text{mic}}=z_{\textsc{sugra}} + O(\eta)\, .\end{equation}
The associated ambiguity in the axion-dilaton field $c(z)$ is then of order
\begin{equation}\label{deltac} \delta c = \delta z_{M}\partial_{M} c =O(\eta\partial c) = O(\eta^{3})\, .\end{equation}
\emph{This means that the leading $O(\eta^{2})$ non-constant term in the axion-dilaton field, see \eqref{adpert}, is unambiguously fixed in terms of the microscopic calculation of $c(z)$.}

The conclusion is that, by using D-instantons, we have only access to the leading deformations of the $\AdSS$ background, through the $O(\eta)$ terms in $B$ and $C_{2}$ and the $O(\eta^{2})$ term in $\tau$. Beyond this order, the instantons can no longer probe the full space-time geometry due to the non-trivial dilaton profile. In particular, the backreaction on the metric and five-form cannot be obtained.

Of course, the above restrictions do not apply if we use particles or 
higher-dimensional branes, which can probe the geometry with their kinetic energy. Examples are worked out in \cref{chap:D4brane} and~\cite{Ferrari:2013wla}.

\subsection{\label{SetupExamples}The Examples}

We now present our main results, postponing the detailed derivations to the next section. It is convenient to separate the ten space-time coordinates $(z^{M})$ into four coordinates $(x_{\mu})$ parallel to the background branes and six transverse coordinates $(y_{A})=\vec y$.
The radial coordinate $r$ is defined by
\begin{equation}\label{rdef0}
  r^2=\vy^2\, .
\end{equation}

\subsubsection{The Coulomb branch}

Our first example is the Coulomb branch deformation of the conformal $\text{U}(N)$, $\nn=4$ gauge theory studied in \cite{Ferrari:2012nw}. This deformation is parameterized 
by the scalar expectation values as
\begin{equation}\label{vevCB}
\left<\varphi_A\right> = \ls^{-2}\, \diag (y_{1A},\ldots,y_{NA})\, ,\quad
1\leq A\leq 6\, .\end{equation}

The supergravity fields, derived from the expansion of the D-instanton effective action computed in section \ref{CBSec} by comparing with \eqref{c0}--\eqref{c5}, read
\begin{align}\label{taudefCB0}
	\tau &= \frac{4i\pi N}{\lambda}-\frac \vartheta {2\pi}  , \\
\label{metricCB0}	\d s^2 &= H^{-1/2} \d x_\mu \d x_\mu + H^{1/2} \Bigl( \d r^2 + r^2 \d \Omega_5^2 \Bigr) \, , \\\label{F5CB0}
	F_5 &= -\frac{N\ls^4}{\pi R^5}\Bigl( \frac{r^4}{R^4}\, y_{A} \frac{\partial H}{\partial y_{A}}\, \omega_{\Sfive} + i \frac{R^4}{r^4} \, y_{A}\frac{\partial H^{-1}}{\partial y_{A}} \omega_{\AdS} \Bigr) \, .
\end{align} 
We have denoted the metric on the unit round five-sphere by $\d \Omega_5^2$ and used the definitions
\begin{align}\label{defHCB}
  	H(\vec y)& =\frac 1N\sum_{f=1}^N \frac{R^4}{\bigl( \vec y-\vec y_{f}\bigr)^{4}}\, ,\\\label{omegaAdSdef}
	\omega_{\AdS}& = \frac{\vy^{2}y_{A}}{R^3}\,
\d x_{1}\wedge\cdots\wedge\d x_{4}\wedge\d y_{A}\, ,\\
\label{omegaS5def}
\omega_{\Sfive} &= \frac{1}{5!}\frac{R^5 y_{F}}{\vy^{6}}\epsilon_{ABCDEF}\,\d y_{A}\wedge\cdots\wedge\d y_{E}\, .
\end{align}
The radius $R$ is related to the string scale and the 't~Hooft coupling $\lambda$ as
\begin{equation}\label{Rlambdarel} R^{4}=\alpha'^{2}\lambda = \frac{\ls^{4}\lambda}{4\pi^{2}}\, .\end{equation}
The parameter $\vartheta$ is the bare theta angle. The solution \eqref{taudefCB0}, \eqref{metricCB0} and \eqref{F5CB0} matches perfectly the supergravity solution for the multi-centered D3-brane background (a detailed presentation of BPS brane supergravity solutions can be found e.g.\ in \cite{Stelle:1998xg}) in the standard Maldacena scaling limit.

Let us note that the axion-dilaton $\tau$ given by \eqref{taudefCB0} is a constant for the present solution. The D-instantons can thus move freely on the entire space-time geometry and the restriction discussed in \ref{useSec} does not apply. Moreover, the match between the microscopic calculation and the supergravity solution is found at finite $\ls^{2}$ or, equivalently, for 
any value of the 't~Hooft coupling. This suggests that, similarly to the undeformed $\text{AdS}_{5}\times
\text{S}^{5}$ background \cite{Berkovits:2004xu,Kallosh:1998qs,Banks:1998nr}, the near-horizon multi-centered D3-brane background could be exact, with vanishing $\ls^{2}$ corrections to both Myers' action and to the supergravity equations of motion.

Beyond the details of the solution, let us emphasize that general properties like the self-duality of the five-form field strength with respect to the metric \eqref{metricCB0},
\begin{equation}
	\star F_5 = -i F_5 \, ,
\end{equation}
or the quantization of the five-form flux in units of the D3-brane charge,
\begin{equation}\label{5formfluxSol1}
	\int _{\vy^{2}=r^{2}} F_5 = 4\pi^2 \ls^4 N(r) \, ,
\end{equation}
where $N(r)$ counts the number of D3-branes with $\vec y_f^{\,2} < r^2$,
which are fundamental consistency requirements from the point of view of the closed string theory, are highly non-trivial and rather mysterious consequences of the microscopic, field theoretic calculation of the effective action. 

\subsubsection{\label{NCsubex}The non-commutative deformation}

Our second example is the non-commutative deformation of the $\nn=4$ gauge theory. This deformation amounts to imposing non-trivial commutation relations among the space-time coordinates \cite{Seiberg:1999vs,Douglas:2001ba}. The most general deformation is parameterized by a real antisymmetric matrix $\theta_{\mu\nu}$, with
\begin{equation}
  [ x_\mu, x_\nu]=-i\theta_{\mu\nu} \, .
  \label{NCrelation}
\end{equation}
Up to an $\text{SO}(4)$ rotation, we may assume that the only non-vanishing components are $\theta_{12}=-\theta_{21}$ and $\theta_{34}=-\theta_{43}$, with corresponding self-dual and anti self-dual parts
\begin{equation}\label{sdeq} \theta_{12}^{\pm} = \theta_{34}^{\pm} = \frac{1}{2}\bigl(
\theta_{12}\pm\theta_{34}\bigr)\, ,\quad \theta_{\pm}^{2} = \theta^{\pm}_{\mu\nu}\theta^{\pm}_{\mu\nu} = (\theta_{12}\pm\theta_{34})^{2}\, .\end{equation}
As discussed in section \ref{NCSec}, it can be convenient for some purposes to make the rotation to imaginary Euclidean time $x^{4}\rightarrow ix^{4}$, in which case $\theta_{34}$ is imaginary and $(\theta_{\pm})^{*}=\theta_{\mp}$.

The large $N$ solution of the microscopic model, presented in details in section \ref{NCSec}, then yields an effective action \eqref{Myersexpansion} with
\begin{equation} \label{NCc} c=i\vartheta + \frac{8\pi^2 N}\lambda+N
 \left(\sqrt{1+4 \theta_{+}^{2}r^4/R^{8}}-1\right)+N\ln\biggl(\frac{\sqrt{1+4 \theta_{+}^{2}r^4/R^{8}}-1}{2\theta_{+}^{2}r^4/R^{8}}\biggr)\, . 
\end{equation}
Since the coefficient $c$ depends non-trivially on the transverse coordinates $\vec y$, the discussion of section \ref{useSec} implies that the physical information contained in the effective action is obtained by expanding in $\eta_{\pm}=\theta_{\pm}/\ls^{2}$ around the undeformed $\AdSS$ background. Precisely, \eqref{NCc} can be used to find the axion-dilaton $\tau=i c/(2\pi)$ up to terms of order $\eta^{3}$, giving the predictions
\begin{align}
  C_0&=\frac{\vartheta}{2\pi}-\frac{4i\pi N}{\lambda}\frac{\theta_{12}\theta_{34}}{\ls^4}\frac{r^4}{R^4} +O(\ls^{-2}\theta)^3 \, , \label{C0NCsol}\\
  e^{-\phi}&=\frac{4\pi N}{\lambda}\left[ 1+\left( \left(\frac{\theta_{12}}{\ls^2}\right)^2+\left(\frac{\theta_{34}}{\ls^2}\right)^2 \right)\frac{r^4}{R^4} \right] +O\bigl(\ls^{-2}\theta\bigr)^3 \, .
  \label{phiNCsol}
\end{align}

Moreover, our microscopic calculation yields a third order coefficient $c_{[MNP]}$ and thus a three-form $F^{(3)}$ of the form \eqref{dc}, with a two-form potential $C^{(2)}$ given by
\begin{equation}
  C^{(2)}=\frac{N\ls^2}{2i\pi\theta_{+}^{2}}\biggl[1- \sqrt{1+4\theta_{+}^{2}r^4/R^{8}}\biggr]\theta^+_{\mu\nu}\,\d x^\mu\wedge\d x^\nu \, .\label{NComega}
\end{equation}
From the discussion of section \ref{useSec}, we know that only the term linear in the deformation parameter is physical. By using \eqref{C2C4NSRR}, we explain in section \ref{NCSec} that this yields the predictions
\begin{align}
\begin{split}
  C_2 &=-\frac{r^4}{R^4}\biggl[ \Bigl( \frac{4i\pi N}{\lambda} \frac{\theta_{34}}{\ls^2} + \frac{\vartheta}{2\pi}\frac{\theta_{12}}{\ls^{2}}\Bigr)  
    {\d x_1\wedge\d x_2}\\
    &\hskip 3cm + \Bigl(\frac{4i\pi N}{\lambda}
    \frac{\theta_{12}}{\ls^2} + 
    \frac{\vartheta}{2\pi}\frac{\theta_{34}}{\ls^2}\Bigr)
  {\d x_3\wedge\d x_4} \biggr] +O\bigl(\ls^{-2}\theta\bigr)^2 \, ,
  \end{split}
   \label{C2NCsol} \\
  B& =\frac{r^4}{R^4}\left[ \frac{\theta_{12}}{\ls^2}{\d x_1\wedge\d x_2}+ \frac{\theta_{34}}{\ls^2}{\d x_3\wedge\d x_4}  \right] +
  O\bigl(\ls^{-2}\theta\bigr)^2 \label{BNCsol} \, .
\end{align}

We can now compare the above results with the supergravity solution. This solution was derived independently by Hashimoto and Itzhaki on the one hand \cite{Hashimoto:1999ut} and Maldacena and Russo on the other hand \cite{Maldacena:1999mh}.
As explained previously, to compare the supergravity and microscopic solutions, we must expand in the deformation parameters $\theta_{12}/\ls^{2}$ and $\theta_{34}/\ls^{2}$, which enter into the functions $\Delta_{12}$ and $\Delta_{34}$ defined in \eqref{SolNCDelta}. For the $C_{0}$ field, this expansion plays no r\^ole and indeed equations \eqref{C0NCsol} and \eqref{C0NCsugra} match. For the dilaton field,
we find a match between \eqref{phiNCsol} and \eqref{phiNCsugra} to quadratic order, consistently with our discussion in section \ref{useSec}.
For the $B$ and $C_{2}$ fields, to compare supergravity with \eqref{BNCsol} and \eqref{C0NCsol}, we must use the approximation $\Delta_{12}\simeq\Delta_{34}\simeq 1$ to keep the leading contribution in the deformation parameter only. We again find a perfect match with the microscopic calculation, in the regime where both can a priori be compared. 

As a final remark, let us note that the dimensionless expansion parameter governing the deformation with respect to the conformal $\nn=4$ model is not really $\eta\sim\theta/\ls^{2}$ but rather the combination 
\begin{equation}\label{expparamic} \eta_{\text{mic}}=\frac{\theta\, r^2}{R^4}\sim\frac{\theta}{\ls^{2}}\frac{r^{2}}{\ls^{2}\lambda}\end{equation}
in the microscopic formulas \eqref{NCc}, \eqref{NComega} and 
\begin{equation}\label{expparasugra}\eta_{\textsc{sugra}}= \frac{\theta}{\ls^{2}}\frac{r^{2}}{R^{2}}\sim
\frac{\theta}{\ls^{2}}\frac{r^{2}}{\ls^{2}\sqrt{\lambda}}\end{equation}
in the supergravity solution. In the microscopic formulas, $\lambda$ is a priori arbitrary, but the supergravity solution can be trusted only at large $\lambda$. The condition $\eta_{\textsc{sugra}}\ll 1$ thus automatically implies $\eta_{\text{mic}}\ll 1$ in the supergravity limit. However, 
the condition $\eta_{\textsc{sugra}}\ll 1$ cannot be satisfied for all $r$, 
even if we choose the deformation parameter $\theta/\ls^{2}$ to be arbitrarily small; we have to restrict ourselves to the region $r\ll \ls^{2}\lambda^{1/4}/\theta^{1/2}$, where the solution is indeed a small deformation of the $\AdSS$ background. This means that, even for infinitesimal $\theta$, the theory is completely changed in the \<UV>, a well-known difficulty associated with non-commutative field theories.

\subsubsection{The $\beta$-deformation}

Our last example is the $\beta$-deformed $\nn=4$ gauge theory. The most general deformation studied in section \ref{BetaSec} is parameterized by three real parameters $\gamma_{1}$, $\gamma_{2}$ and $\gamma_{3}$ and breaks all supersymmetries. Let us discuss here the slightly simpler $\nn=1$ preserving case $\gamma=\gamma_{1}=\gamma_{2}=\gamma_{3}$. In $\nn=1$ language, the $\nn=4$ multiplet decomposes into one vector multiplet and three chiral multiplets $\Phi_{1}$, $\Phi_{2}$ and $\Phi_{3}$. The $\beta$-deformation then simply amounts to replacing the $\nn=4$ preserving superpotential term $\tr [\Phi_{1},\Phi_{2}]\Phi_{3}$ by
$\tr (e^{i\pi\gamma}\Phi_{1}\Phi_{2}\Phi_{3}- e^{-i\pi\gamma}\Phi_{1}\Phi_{3}\Phi_{2})$.

To describe the solution of the model it is convenient to introduce the polar coordinates $(\rho_{i},\theta_{i})$, $1\leq i\leq 3$, defined in terms of the transverse coordinates $\vec y$ by
\begin{alignat}{3}\nonumber y_{1}&= \rho_{1}\cos\theta_{1}\, ,\quad &
y_{3} &=\rho_{2}\cos\theta_{2}\, , \quad & y_{5} &=\rho_{3}\cos\theta_{3}\, ,
\\\label{polardef0}
y_{2} &=\rho_{1}\sin\theta_{1}\, ,&
y_{4} &=\rho_{2}\sin\theta_{2}\, ,& y_{6} &=\rho_{3}\sin\theta_{3}\, ,
\end{alignat}
together with
\begin{equation}\label{ridef0} r_{i} =\frac{\rho_{i}}{\sqrt{\rho_{1}^{2}+\rho_{2}^{2}+\rho_{3}^{2}}}=\frac{\rho_{i}}{|\vec y|}\,,\end{equation}
which satisfy the constraint 
\begin{equation}\label{ricons} r_{1}^{2}+r_{2}^{2}+r_{3}^{2}=1\, .\end{equation}
We shall also use the spherical angles $(\theta,\phi)$ defined by
\begin{equation}\label{thetaphidef0} r_{1}=\sin\theta\cos\phi\, ,\quad r_{2}=\sin\theta\sin\phi\, ,\quad r_{3}=\cos\theta\, .\end{equation}
The large $N$ solution of the microscopic theory, derived in section \ref{BetaSec}, yields
\begin{equation}\label{Betac} c = \frac{8\pi^{2}N}{\lambda} + i\vartheta - N \ln\bigl(
1-4(\rr)\sin^{2}(\pi \gamma)\bigr)\, .\end{equation}
Expanding to second order in the deformation parameter $\gamma$ as required by the discussion in section \ref{useSec}, we obtain the prediction
\begin{equation}\label{phiBetasol} e^{-\phi} = \frac{4\pi N}{\lambda} 
\Bigl( 1+\frac{1}{2}\lambda\gamma^{2}\bigl(\rr\bigr) + O\bigl(\lambda\gamma^{4}\bigr)
\Bigr)\, .\end{equation}
Moreover, the two-form $C^{(2)}$ defined in \eqref{dc} is found to be
\begin{multline}\label{omegaBetasol}
C^{(2)} = \frac{4 N\ls^{2}}{\pi}\sin\bigl(2\pi\gamma\bigr)\biggl[ G_{1}\wedge\d\theta_{1}
+ G_{2}\wedge\d\theta_{2}+ G_{3}\wedge\d\theta_{3}\\
-\frac{i}{4}\frac{r_{1}^{2}r_{2}^{2}\d\theta_{1}\wedge\d\theta_{2}
+ r_{1}^{2}r_{3}^{2}\d\theta_{1}\wedge\d\theta_{3}
+r_{2}^{2}r_{3}^{2}\d\theta_{2}\wedge\d\theta_{3}}{1-4(\rr)\sin^{2}(\pi\gamma)}\biggr]\, ,
\end{multline}
with
\begin{align}\label{G1def0} \d G_{1} &= \frac{r_{1}r_{2}r_{3}\bigl(r_{1}^{2}
+ (r_{2}^{2}+r_{3}^{2})\cos (2\pi\gamma)\bigr)}{\bigl(1-4(\rr)\sin^{2}(\pi\gamma)\bigr)^{2}}\,\sin\theta\, \d\theta\wedge\d\phi\, ,\\
\label{G2def0} \d G_{2} &= \frac{r_{1}r_{2}r_{3}\bigl(r_{2}^{2}
+ (r_{1}^{2}+r_{3}^{2})\cos (2\pi\gamma)\bigr)}{\bigl(1-4(\rr)\sin^{2}(\pi\gamma)\bigr)^{2}}\,\sin\theta\, \d\theta\wedge\d\phi\, ,\\
\label{G3def0} \d G_{3}& = \frac{r_{1}r_{2}r_{3}\bigl(r_{3}^{2}
+ (r_{1}^{2}+r_{2}^{2})\cos (2\pi\gamma)\bigr)}{\bigl(1-4(\rr)\sin^{2}(\pi\gamma)\bigr)^{2}}\,\sin\theta\, \d\theta\wedge\d\phi\, .
\end{align}
To obtain a prediction for $B$ and $C_{2}$, we are instructed by the discussion in section \ref{useSec} to expand to linear order in the deformation parameter $\gamma$. In this limit,
\begin{equation}\label{approxGs}
\d G_{1}\simeq \d G_{2}\simeq\d G_{3} \simeq r_{1}r_{2}r_{3}\,
\sin\theta\,\d\theta\wedge\d\phi = \d\omega_{1}\end{equation}
and \eqref{C2C4NSRR} then yields
\begin{align} \label{C2Betasol} C_{2}&=-8 N\ls^{2}\gamma\omega_{1}\wedge
\bigl(\d\theta_{1}+\d\theta_{2}+\d\theta_{3}\bigr) + O\bigl(\gamma^{2}\bigr)\, ,
\\
\label{BBetasol} B & =-\frac{\ls^{2}\lambda}{2\pi}\,\gamma\bigl(
r_{1}^{2}r_{2}^{2}\d\theta_{1}\wedge\d\theta_{2}
+ r_{1}^{2}r_{3}^{2}\d\theta_{1}\wedge\d\theta_{3}
+r_{2}^{2}r_{3}^{2}\d\theta_{2}\wedge\d\theta_{3}\bigr) + O\bigl(\gamma^{2}\bigr)\, .
\end{align}

The supergravity dual of the $\beta$-deformed theory was studied by Lunin and Maldacena in \cite{Lunin:2005jy} (or, more generally when $\gamma_{1}$, $\gamma_{2}$ and $\gamma_{3}$ are distinct, by Frolov in \cite{Frolov:2005dj}, see section \ref{BetaSec}). This is reviewed in Appendix \ref{CCC}. The supergravity solution can be trusted as long as the two conditions
\begin{equation}\label{sugracondBeta} \lambda\gg 1\, ,\quad \lambda\gamma^{4}\ll 1\, ,\end{equation}
are satisfied. The discussion in section \ref{useSec} implies that supergravity can be compared with the above microscopic solution only when the background is a small perturbation of the undeformed $\AdSS$ solution. This occurs when $\lambda\gamma^{2}\ll 1$, in which case the functions $1/\sqrt{G}$ and $\sqrt{G}$ in equations \eqref{tbeta} and \eqref{Bbeta} can be simplified. This yields a perfect match with \eqref{phiBetasol}, \eqref{C2Betasol} and \eqref{BBetasol}.

\section{Derivation of the solutions\label{DerivSec}}

Our starting point is the microscopic probe action \cref{eq:SinstN4} for $K$ D-instantons in the undeformed conformal $\nn=4$ model. It reads
\begin{multline}\label{Sp1} S_{\text p}=K\Bigl(\frac{8\pi^{2}N}{\lambda}
 +i\vartheta\Bigr)+ \frac{4\pi^{2}N}{\lambda}\truK \Bigl\{
2iD_{\mu\nu}\bigl[X_{\mu},X_{\nu}\bigr] -
\bigl[X_{\mu},\phi_{A}\bigr]\bigl[X_{\mu},\phi_{A}\bigr]\\
-2  \Lambda^{\alpha}_{\ a}\sigma_{\mu\alpha\dot\alpha}
\bigl[X_{\mu},\bar\psi^{\dot\alpha a} \bigr]
- \bar\psi_{\dot\alpha}^{\ a}\Sigma_{Aab}
\bigl[\phi_{A},\bar\psi^{\dot\alpha b}\bigr]\Bigr\}\\\hskip 3cm
+ \frac{i}{2}
\tilde q^{\alpha}D_{\mu\nu}\sigma_{\mu\nu\alpha}^{\hphantom{\mu\nu\alpha}\beta}q_{\beta}+\frac{1}{2}\tilde q^{\alpha}\phi_{A}\phi_{A}q_{\alpha}-\frac{1}{2}
\tilde\chi^{a}\Sigma_{Aab}\phi_{A}\chi^{b}\\+ \frac{1}{\sqrt{2}}
\tilde q^{\alpha}\Lambda_{\alpha a}\chi^{a} + \frac{1}{\sqrt{2}}
\tilde\chi^{a}\Lambda^{\alpha}_{\ a}q_{\alpha} + \cdots \end{multline}
The $\cdots$ represent couplings with the local fields of the $\nn=4$ gauge theory living on the background D3-brane worldvolume. These terms are described in \cite{Ferrari:2012nw,Green:2000ke,Billo:2002hm} and enter crucially into the computation of the expectation value \eqref{eq:Zeff} of the superdeterminant $\mathscr D$, but most of them play no role\footnote{\label{fn:CB} The only couplings to the D3-branes that can contribute in this case are those involving fields that do not vanish at the classical level. This is the case for the scalars on the Coulomb branch and the relevant terms will be discussed in \cref{CBSec}.} when this determinant is not quantum corrected. As discussed in \cref{sec:holoprobe}, we can thus discard them (with the caveat of~\cref{fn:CB}) for our present purposes. 

The fields in the vector multiplet $(\phi_{A},\Lambda_{\alpha a},D_{\mu\nu})$ are auxiliary fields that can be easily integrated out from \eqref{Sp1} to yield the usual \<ADHM> constraints and measure on the instanton moduli space. However, keeping these variables is crucial to solve the model at large $N$. In particular, the action, as written in \eqref{Sp1}, is quadratic in the hypermultiplet fields, a property that would be lost if we integrate out the six scalars $\phi_{A}$. Instead, we can integrate exactly over the moduli $q,\tilde q,\chi, \tilde\chi$ which belong to the fundamental of $\uN$. This yields an effective action which is automatically proportional to $N$ and can thus be treated classically when $N\rightarrow\infty$.

The microscopic actions for the deformed theories that we study in the present chapter are simple modifications of \eqref{Sp1} and their large $N$ limit can be studied along the same lines. Since our goal is to obtain the bosonic effective action, we shall always set $\Lambda_{\alpha a}$ and $\bar\psi^{\dot\alpha a}$ to zero in the following.
Also recall that
\begin{equation}\label{defY} Y_{A}=\ls^{2}\phi_{A}\, ,\end{equation}
because the moduli $\phi_A$ play the role of the six transverse coordinates.

\subsection{The Coulomb branch deformation}
\label{CBSec}
\subsubsection{The microscopic action}

The Coulomb branch deformation amounts to turning on non-zero expectation value $\langle\varphi_{A}\rangle$ for the $\nn=4$ scalars. 
The microscopic action is then modified by making the replacement
\begin{equation}\label{CBmod} \phi_{Ai}^{\ \ j}\delta_{f}^{f'}\rightarrow
\phi_{Ai}^{\ \ j}\delta_{f}^{f'} -\langle\varphi_{Af}^{\ \ f'}\rangle
\delta_{i}^{j} = \phi_{Ai}^{\ \ j}\delta_{f}^{f'} - \ls^{-2}y_{fA}\delta_{f}^{f'}\delta_{i}^{j}\end{equation}
in the third line of \eqref{Sp1}. We have indicated all the $\uN$ and $\text{U}(K)$ indices explicitly for clarity. This modification is actually best understood as coming from the coupling of the scalar fields 
$\varphi_{A}$ to the moduli in the $\cdots$ part of the action \eqref{Sp1} that we have not written down explicitly.

\subsubsection{The effective action}

Integrating out $q,\tilde q,\chi,\tilde\chi$ yields the effective action
\begin{multline}\label{SeffboseCB} S_{\text{eff}}(X,Y,D)=K\Bigl(\frac{8\pi^{2}N}{\lambda}
 +i\vartheta\Bigr)\\+
\frac{4\pi^{2}N}{\ls^{4}\lambda}\truK \Bigl\{
2i\ls^{4} D_{\mu\nu}\bigl[X_{\mu},X_{\nu}\bigr] -
\bigl[X_{\mu},Y_{A}\bigr]\bigl[X_{\mu},Y_{A}\bigr]\Bigr\}+ \ln\Delta_{q,\tilde q} - \ln\Delta_{\chi,\tilde\chi}\, .
\end{multline}
The logarithm of the superdeterminant $\ln(\Delta_{q,\tilde q}/\Delta_{\chi,\tilde\chi})$ is the sum of the term obtained by integrating over the bosonic variables $q,\tilde q$,
\begin{equation}\label{Deltaqq} \ln\Delta_{q,\tilde q} = \sum_{f=1}^N \ln \det \Bigl(\big(Y_A-y_{f A}\big)^2\otimes \1_{2\times2} + i \ls ^4 D_{\mu\nu}\otimes \sigma_{\mu\nu} \Bigl) \end{equation}
and the term obtained by integrating over the fermionic variables $\chi,\tilde\chi$,
\begin{equation}\label{Deltacc} -\ln\Delta_{\chi,\tilde\chi} =-\sum_{f=1}^N
\ln \det \bigl( \Sigma_A \otimes \big( Y_A - y_{f A} \big) \bigl)\, .\end{equation}
This action is proportional to $N$ and thus can be treated classically at large $N$. In particular, the fluctuations of
$X$, $Y$ and $D$ are suppressed. The six matrices $Y_{A}$ are interpreted as the six coordinates for the emerging space (from the field theory point of view) transverse to the background D3-branes. Together with the four $X_{\mu}$s, 
they correspond to the ten matrix coordinates $Z_{M}$ in the non-Abelian D-instanton action \eqref{Myersexpansion}. Consequently, to compare \eqref{SeffboseCB} with \eqref{Myersexpansion}, we simply need to integrate out the additional variables $D_{\mu\nu}$ by solving the saddle-point equation
\begin{equation}\label{saddleCBD} \frac{\partial\Seff}{\partial D_{\mu\nu i}^{\ \ \ j}}
= 0\end{equation}
and plugging the solution $D_{\mu\nu}=\langle D_{\mu\nu}\rangle$ back into \eqref{SeffboseCB},
\begin{equation}\label{SeffNClast} \Seff (X,Y) = \Seff \bigl(X,Y,\langle D\rangle\bigr)\, .\end{equation}

Our goal is to expand $\Seff(X,Y)$ as in \eqref{Myersexpansion}, up to the fifth order and then use \eqref{c0}--\eqref{c5} to read off the supergravity background. This calculation is very similar to the one performed in \cite{Ferrari:2012nw}. We set
\begin{equation}\label{XYexp}
X_\mu = x_\mu \1 + \ls^2 \epsilon_\mu \, , \quad Y_A = y_A \1 + \ls^2 \epsilon_A
\end{equation}
and solve \eqref{saddleCBD} perturbatively in $\epsilon$. Using the standard notation $[\epsilon_\mu,\epsilon_\nu]^+$ for the self-dual part of the commutator (see \eqref{sddef}) and defining the function
\begin{equation}\label{defH}
	H(\vec y) =\frac 1N\sum_{f=1}^N \frac{R^4}{\bigl( \vec y-\vec y_{f}\bigr)^{4}}\, ,
\end{equation}
where $R$ is given by \eqref{Rlambdarel},
we obtain
\begin{equation}\label{Dorder3CB}
\langle D_{\mu\nu} \rangle= iH^{-1} \left[ \epsilon_\mu,\epsilon_\nu \right]^+ + \frac {i\ls^2}{2}\partial_A H^{-1}\left( \epsilon_{A} [\epsilon_\mu,\epsilon_\nu]^+ + [\epsilon_\mu,\epsilon_\nu]^+ \, \epsilon_{A}\right) +  O\bigl(\epsilon^{4}\bigr) \, .
\end{equation}
Let us note that since $\langle D \rangle$ solves the equation of motion \eqref{saddleCBD}, it enters into \eqref{SeffboseCB} at order $\langle D\rangle^{2}$ and thus the expansion \eqref{Dorder3CB} to third order in $\epsilon$ is sufficient to get the expansion of \eqref{SeffboseCB} to fifth order.

Plugging \eqref{Dorder3CB} into $\eqref{SeffboseCB}$, expanding the determinants by using the relation
\begin{equation}\label{detexp} \ln\det (M+\delta M) = \ln\det M + \sum_{n\geq 1}
\frac{(-1)^{n+1}}{n}\tr (M^{-1}\delta M)^{n}\end{equation}
and computing the resulting traces by using the identities \eqref{2sigmaid} and \eqref{s1}--\eqref{s5} in the Appendix, we find that the first, second and third order action in \eqref{Myersexpansion} vanish, due to many cancellations between the bosonic and fermionic contributions \eqref{Deltaqq} and \eqref{Deltacc},
\begin{equation}\label{Sefforder123CB}S_\eff ^{(1)}=S_\eff ^{(2)}=S_\eff ^{(3)}=0\, .\end{equation}
On the other hand, the action is non-trivial at the fourth and fifth orders, %
\begin{align}\label{Sefforder4CB}\begin{split}
\Seff^{(4)}&=-\frac{\ls^8}{2R^4}\truK\Bigl\{2[\epsilon_A,\epsilon_\mu][\epsilon_A,\epsilon_\mu]+H^{-1}[\epsilon_\mu,\epsilon_\nu][\epsilon_\mu,\epsilon_\nu] \\& \hskip 6cm+ \frac 12 H [\epsilon_A,\epsilon_B][\epsilon_A,\epsilon_B]\Bigr\}\, ,\end{split} \\
\label{Sefforder5CB}
\begin{split}
\Seff^{(5)} &=-\frac{\ls^{10}}{2R^4}\partial_A H^{-1} \truK \Bigl\{ \epsilon_A [\epsilon_\mu,\epsilon_\nu][\epsilon_\mu,\epsilon_\nu] + 2 \epsilon_{\mu\nu\rho\lambda} \epsilon_A \epsilon_\mu \epsilon_\nu \epsilon_\rho \epsilon_\lambda\\ &\hskip 1cm
-H^2 \epsilon_A [\epsilon_B,\epsilon_C][\epsilon_B,\epsilon_C] 
-\frac{2iH^2}{5}\epsilon_{ABCDEF}\epsilon_{B}\epsilon_{C}\epsilon_{D}\epsilon_{E}\epsilon_{F} \Bigr\}\,.
\end{split}
\end{align}

\subsubsection{The holographic background}

The results of the previous subsection are perfectly consistent with the general ideas explained in \cref{sec:holoprobe,framesec}. The effective action that we have obtained can be matched with the non-Abelian action for D-instantons embedded in a non-trivial ten-dimensional geometry, with background supergravity fields fixed by comparing 
\eqref{Sefforder123CB}, \eqref{Sefforder4CB} and \eqref{Sefforder5CB} with \eqref{Mexpand} or equivalently \eqref{c0}--\eqref{c5}.

The conditions $\Seff^{(1)}=\Seff^{(2)}=0$ imply that the axion-dilaton is a constant,
\begin{equation}\label{tauCB}
	\tau =  \frac{4i\pi N}{\lambda}-\frac \vartheta {2\pi}\, ,
\end{equation}
whereas $\Seff^{(3)}=0$ yields
\begin{equation}\label{3formCB} B=C_{2}=0\, .\end{equation}
On the other hand, the fourth order term \eqref{Sefforder4CB} allows to identify the coefficient $c_{[MN][PQ]}$ which turns out to be precisely of the required form \eqref{c4}, with a metric
\begin{equation}\label{MetricCB}
G_{\mu\nu}=H^{-1/2} \delta_{\mu\nu} \, , \quad G_{AB}=H^{1/2}\delta_{AB} \,, \quad G_{A\mu}=0
\end{equation}
which is equivalent to \eqref{metricCB0}. Finally, we get the completely antisymmetric coef\-fi\-cient $c_{[MNPQR]}$ from \eqref{Sefforder5CB}, which yields the five-form field strength by comparing with \eqref{c5} and using \eqref{3formCB},
\begin{equation}\label{5FormCB}
({F_5})_{ABCDE} = -\frac{N\ls^4}{\pi R^4} \partial_F H \epsilon_{ABCDEF} \,,\enspace
({F_5})_{A\mu_1\cdots \mu_4}=-\frac{iN\ls^4}{\pi R^4} \partial_A H^{-1} \epsilon_{\mu_1 \cdots \mu_4}\, ,
\end{equation}
and all the other independent components (not related to \eqref{5FormCB} by antisymmetry) vanishing. This is equivalent to the formula \eqref{F5CB0}.

\subsection{The non-commutative deformation}
\label{NCSec}
\subsubsection{The microscopic action}

The non-commutative deformation can be elegantly implemented by replacing all ordinary products $fg$ appearing in the microscopic action by the so-called Moyal $*$-product defined by
\begin{equation}\label{Moyaldef} f*g = e^{-\frac{i}{2}\theta_{\mu\nu}P^{f}_{\mu}P^{g}_{\nu}}\cdot (fg)\, ,\end{equation}
where $P_{\mu}^{f}$ and $P_{\mu}^{g}$ are the translation operators acting on $f$ and $g$ respectively and $\theta_{\mu\nu}$ is an arbitrary antisymmetric matrix \cite{Seiberg:1999vs,Douglas:2001ba}. The only moduli in \eqref{Sp1} transforming non-trivially under translations are the matrices $X_{\mu}$, with $P_{\mu}\cdot X_{\nu} = -i\delta_{\mu\nu}$. It is then easy to check that the only term affected by the use of the $*$-product is the commutator term
\begin{equation}\label{stareffectNC}\tr D_{\mu\nu} [X_{\mu},X_{\nu}] \rightarrow\tr D_{\mu\nu}\bigl( X_{\mu}*X_{\nu}-
X_{\nu}*X_{\mu}\bigr) =\tr D_{\mu\nu}\bigl( [X_{\mu},X_{\nu}] + i\theta_{\mu\nu}\bigr)\, .\end{equation}
This simple reasoning reproduces the well-known modification of the \<ADHM> construction in non-commutative gauge theories \cite{Nekrasov:1998ss}. Note that, in particular, the action only depends on the self-dual part 
$\theta_{\mu\nu}^{+}$ of the non-commutative parameters because the modulus $D_{\mu\nu}$ is itself self-dual.

\subsubsection{The effective action\label{effactNC}}

Integrating out $q,\tilde q,\chi,\tilde\chi$ from the microscopic action yields
\begin{multline}\label{SeffboseNC} S_{\text{eff}}(X,Y,D)=K\Bigl(\frac{8\pi^{2}N}{\lambda}
 +i\vartheta\Bigr) -\frac{4\pi^{2}N}{\ls^{4}\lambda}\tr
 \bigl[X_{\mu},Y_{A}\bigr]\bigl[X_{\mu},Y_{A}\bigr] \\ - 
N \ln \det \bigl( \Sigma_A \otimes Y_A \bigl)
 + \mathcal S\bigl([X_\mu,X_\nu]^+,\vec Y^2, D;\theta_+\bigr)\, .
\end{multline}
We have singled out the $D$-dependent piece in the action,
\begin{multline}\label{calSNC}
\mathcal S\bigl([X_\mu,X_\nu]^+,\vec Y^2,D\bigr)=\frac{8i\pi^{2}N}{\lambda}\tr D_{\mu\nu}\bigl(\bigl[X_{\mu},X_{\nu}\bigr]+i\theta_{\mu\nu}\bigr)^+ \\+ N\ln \det \bigl(\vec 
Y^{2}\otimes \1_{2} + i \ls ^4 D_{\mu\nu}\otimes \sigma_{\mu\nu} \bigl)\, .
\end{multline}
Let us note that the determinants appearing in \eqref{SeffboseNC} and \eqref{calSNC} are special cases of the determinants \eqref{Deltaqq} and \eqref{Deltacc} studied in the previous subsection. The crucial difference comes from the saddle-point equation \eqref{saddleCBD}, which now picks 
a new term in $\theta_{\mu\nu}$,
\begin{equation}
  \frac{\partial \mathcal S}{\partial D_{\mu\nu j}^{\ \ \ i}}=\frac{8\pi^2}\lambda\Bigl( [X_\mu,X_\nu]^{+\,j}_{\ i}+i\theta_{\mu\nu}^{+}
  \delta_{i}^{j} \Bigr)+\ls^4 
  \Bigl(\vec Y^{2}\otimes\1_{2}+i\ls^4D_{\rho\kappa}\otimes\sigma_{\rho\kappa}\Bigr)^{-1\ j\beta }_{i\alpha }\!\!\!\!\!\!\sigma_{\mu\nu\beta}^{\ \ 
\ \alpha}=0\, .
  \label{EOMDNC}
\end{equation}
This equation must be solved for $D_{\mu\nu}=\langle D_{\mu\nu}\rangle$, order by order in the expansion \eqref{XYexp}. 

By using \eqref{inv1sigma}, we find a quadratic equation for the zeroth order solution. Picking the root that behaves smoothly when $\theta_{\mu\nu}\rightarrow 0$ yields
\begin{equation}\label{solD0NC} \langle D_{\mu\nu}\rangle = \frac{\lambda}{8\pi^{2}\theta_{+}^{2}}\biggl( 1 - \sqrt{1+4\theta_{+}^{2}r^{4}/R^{8}}
\biggr) \theta_{\mu\nu}^{+} + O\bigl(\epsilon\bigr)\end{equation}
in terms of the transverse radial coordinate \eqref{rdef0} and the parameter $\theta_{+}$ defined in \eqref{sdeq}. Plugging this result into \eqref{calSNC} and \eqref{SeffboseNC} and computing the determinants using \eqref{detsigmaid} and \eqref{detsigmaa}, we get the zeroth order coefficient \eqref{NCc} for the effective action.

The first, second and completely symmetric third order coefficients in the expansion \eqref{Myersexpansion} of the effective action are fixed in terms of the derivatives of $c$ by consistency conditions \cite{Ferrari:2013pi}. To get further information, we thus need to compute the completely antisymmetric third order coefficient or equivalently the three-form $F^{(3)}$ defined in \eqref{F3def}. From \eqref{s3}, we see that the determinant in \eqref{SeffboseNC} cannot contribute to the completely antisymmetric coefficient. A priori, we thus simply need
to plug the solution of \eqref{EOMDNC} to the third order in $\epsilon$ into \eqref{calSNC}. However, the algebra to do this calculation explicitly is daunting. Very fortunately, the discussion can be greatly simplified by using the following argument.

The basic idea is to note that the $D$-dependent piece \eqref{calSNC} of the effective action and thus the saddle-point equation \eqref{EOMDNC} as well depend only on the combinations $\smash{\vec Y^{2}}$ and $[X_{\mu},X_{\nu}]^{+}=\ls^{4}[\epsilon_{\mu},\epsilon_{\nu}]^{+}$ of the matrices $Y_{A}$s and $X_{\mu}$s. The same must be true after plugging $D_{\mu\nu}=\langle D_{\mu\nu}\rangle$ into $\mathcal S$. If we define
\begin{equation}\label{epsrdef} \vec Y^{2} = r^{2} + \ls^{2}\epsilon_{r} = r^{2} + 2\ls^{2}\vec y\cdot\vec\epsilon + \ls^{4}\vec{\epsilon\,}^{2}\, ,\quad
\end{equation}
the expansion of $\mathcal S$ in powers of $\epsilon$ is then most conveniently written in terms of $[\epsilon_{\mu},\epsilon_{\nu}]^{+}$
and $\epsilon_{r}$. It will actually be useful to replace $[\epsilon_{\mu},\epsilon_{\nu}]^{+}$ by a completely general self-dual matrix $M_{\mu\nu}^{+}$ in \eqref{calSNC} and \eqref{EOMDNC}, which is not necessarily a commutator, and solve the equations in term of this more general matrix. We simply have to keep in mind that $M_{\mu\nu}^{+}$ will be identified with $[\epsilon_{\mu},\epsilon_{\nu}]^{+}$ at the end of the calculation and is thus of order $\epsilon^{2}$.
The most general single-trace expansion up to order three then reads
\begin{multline}\label{Scalexpand} \mathcal S
\bigl(M_{\mu\nu}^{+},r^{2}+\ls^{2}\epsilon_{r}\bigr) = K s(r^{2}) + \ls^{2}s'(r^{2})\tr\epsilon_{r}
+\frac{\ls^{4}}{2}s''(r^{2})\tr\epsilon_{r}^{2} + \frac{\ls^{6}}{6}s'''(r^{2})\tr
\epsilon_{r}^{3}\\ + \ls^{4}s_{\mu\nu}(r^{2})\tr M_{\mu\nu}^{+}
 + \ls^{6}s'_{\mu\nu}(r^{2})\tr\epsilon_{r}
M_{\mu\nu}^{+} + O\bigl(\epsilon^{4}\bigr)\, ,
\end{multline}
where the primes denote the derivatives with respect to $r^{2}$.
The zeroth order coefficient $s(r^{2})$ is determined by the zeroth order solution \eqref{solD0NC} or equivalently \eqref{NCc},
\begin{align}\label{s0sola} s(r^{2}) &= c - i\vartheta-\frac{8\pi^{2}N}{\lambda} + N\ln r^{4} \\\label{s0solb} &= N
 \Bigl(\sqrt{1+4 \theta_{+}^{2}r^4/R^{8}}-1\Bigr)+N\ln\biggl(\frac{\sqrt{1+4 \theta_{+}^{2}r^4/R^{8}}-1}{2\theta_{+}^{2}/R^{8}}\biggr)\, .
\end{align}
Since $\mathcal S$ does not depend on $r^{2}$ and $\epsilon_{r}$ independently but only through the combination $r^{2}+\ls^{2}\epsilon_{r}$, the expansion \eqref{Scalexpand} must be invariant under the simultaneous shifts \cite{Ferrari:2013pi}
\begin{equation}\label{shift1} r^{2}\rightarrow r^{2} + \ls^{2} a\, ,\quad \epsilon_{r}\rightarrow\epsilon_{r} - a\1\, ,\end{equation}
for any real number $a$. This fixes 
the terms in $\tr\epsilon_{r}$, $\tr\epsilon_{r}^{2}$ and $\tr\epsilon_{r}^{3}$ in terms of the derivatives of $s$ and the term in $\tr\epsilon_{r}M_{\mu\nu}^{+}$ in terms of $s'_{\mu\nu}$ as indicated. To fix $s_{\mu\nu}(r^{2})$, we can then use another shift symmetry, under
\begin{equation}\label{shift2} M_{\mu\nu}^{+}\rightarrow 
M_{\mu\nu}^{+} + i\xi_{\mu\nu}^{+}\, ,\quad
\theta_{\mu\nu}^{+}\rightarrow\theta_{\mu\nu}^{+}-\ls^{4}\xi_{\mu\nu}^{+}\, ,\end{equation}
for any self-dual $\xi_{\mu\nu}^{+}$. This symmetry comes
from the fact that only the combination $\ls^{4} M_{\mu\nu}^{+}+ i\theta_{\mu\nu}^{+}$ enters in the generalized versions of the equations \eqref{calSNC} and \eqref{EOMDNC}, in which $[X_{\mu},X_{\nu}]^{+}$ has been replaced by $\ls^{4}M_{\mu\nu}^{+}$. This replacement is useful precisely because it allows to consider the symmetry \eqref{shift2}, by waiving the tracelessness condition that any commutator must satisfy. The invariance of \eqref{Scalexpand} under \eqref{shift2} then yields
\begin{equation}\label{smunusol} s_{\mu\nu} = -i\frac{\partial s}{\partial\theta_{\mu\nu}^{+}} = \frac{iN}{\theta_{+}^{2}}\biggl[1-\sqrt{1+4 \theta_{+}^{2}r^4/R^{8}} \biggr]\theta_{\mu\nu}^{+}\, .
\end{equation}
Plugging this result in \eqref{Scalexpand} for $M_{\mu\nu}^{+}=[\epsilon_{\mu},\epsilon_{\nu}]^{+}$ and using \eqref{epsrdef} immediately yields the piece
\begin{equation}\label{piece} 2\ls^{6}s'_{\mu\nu}y_{A}\tr \epsilon_{A}[\epsilon_{\mu},\epsilon_{\nu}]\end{equation}
of the effective action contributing to the three-form $F^{(3)}$ in \eqref{F3def}, from which we obtain
\begin{equation}\label{F3NCsol} F^{(3)} = 4s'_{\mu\nu}y_{A}\d x^{\mu}\wedge\d
x^{\nu}\wedge\d y^{A} = \d\bigl[2 s_{\mu\nu}\d x^{\mu}\wedge\d x^{\nu}\bigr]\, .\end{equation}
This is equivalent to the formula \eqref{NComega} for the two-form $C^{(2)}$ defined in \eqref{dc}.

\subsubsection{The holographic background}

In this example, there is a non-trivial contribution \eqref{NCc} to the action at order $\epsilon^0$.
As we have extensively discussed in sections \ref{useSec} and \ref{NCsubex}, the physical content of this formula is obtained by expanding up to quadratic order in the deformation parameter $\theta_{+}$ and comparing with \eqref{c0}. This yields
\begin{equation}
  \tau= ie^{-\phi}-C_{0}= -\frac{\vartheta}{2\pi}+\frac{4i\pi N}\lambda\left( 1+
  \frac{\theta_{+}^{2}r^{4}}{2\ls^4 R^{4}}\right)+O\bigl(\ls^{-2}\theta\bigr)^3 \, .
  \label{NCtauSUGRA}
\end{equation}
To disentangle the dilaton and the axion fields from \eqref{NCtauSUGRA}, one has to be careful because the fields do not need to be real-valued in the Euclidean. It is thus convenient to rotate the $x_{4}$ coordinate to Minkowskian time which, from \eqref{NCrelation}, implies that $\theta_{34}$ is purely imaginary. After this rotation, the dilaton $\phi$ and the axion $C_0$ are real and we can then take the real and imaginary parts of \eqref{NCtauSUGRA} to get \eqref{C0NCsol} and \eqref{phiNCsol}.

Similarly, the action at third order yields \eqref{NComega} as we have shown. The physical content of this contribution is found by expanding to linear order in $\theta_{+}$, see sections \ref{useSec} and \ref{NCsubex}.
From \eqref{dc} and \eqref{c3}, this yields
\begin{align}
  \tau B-C_2&=\frac{4i\pi N}{\lambda}\frac{ r^4}{R^4} \frac{\theta^\sd_{\mu\nu}}{\ls^2} \d x_\mu\wedge\d x_\nu+O\bigl(\ls^{-2}\theta\bigr) \, .
  \label{BC2NC}
\end{align}
To disentangle the Neveu-Schwarz and Ramond-Ramond fields $B$ and $C_2$ from \eqref{BC2NC}, we again rotate to Minkowskian signature in which $x_4$ and $\theta_{34}$ are purely imaginary and the fields $B$ and $C_2$ are real. Taking the real and imaginary parts of \eqref{BC2NC} then yields \eqref{C2NCsol} and \eqref{BNCsol}. 

As a final remark, let us note that we have also computed the effective action to the fourth order. As mentioned in section \ref{useSec}, only the term linear in the deformation parameter $\theta$ is physical. Consistently with the supergravity solution, this linear term is found to vanish. At quadradic order in $\theta$, we find a coefficient $c_{[MN][PQ]}$ which does not factorize as in \eqref{c4}, as expected.

\subsection{The \texorpdfstring{$\beta$}{β}-deformation}
\label{BetaSec}
\subsubsection{The microscopic action}

In parallel with the case of the non-commutative theory, the $\beta$-deformation can be implemented\footnote{This is only true for single trace terms, see \cite{Fokken:2013aea}.} by replacing the ordinary products $fg$ appearing in the microscopic action by a $*$-product \cite{Lunin:2005jy}. Let us denote by $Q_{i}$, $1\leq i\leq 3$, the charges associated with the $\u_{1}\times\u_{2}\times\u_{3}$ subgroup of $\text{SO}(6)$ corresponding to the rotations in the 1-2, 3-4 and 5-6 planes in $\vec y$-space respectively. The charge assignments according to the $\text{SU}(4)$ quantum numbers is indicated in the Appendix \ref{NotAppSec}, Table \ref{chargesU1}. The $*$-product is then defined by
\begin{equation}\label{starbeta} f * g = e^{i\pi\epsilon_{ijk}\gamma_{i}Q_{j}^{f}Q_{k}^{g}}fg\, ,\end{equation}
where $\epsilon_{ijk}$ is the totally antisymmetric symbol, the charges $Q_{i}^{f}$ and $Q_{i}^{g}$ act on $f$ and $g$ respectively and $\gamma_{1}$, $\gamma_{2}$ and $\gamma_{3}$ are three deformation parameters that we shall assume to be real. When $\gamma_{1}=\gamma_{2}=\gamma_{3}$, $\nn=1$ supersymmetry is preserved, but supersymmetry is completely broken otherwise. In the supersymmetric case, the model is conformal in the planar limit \cite{Leigh:1995ep}. When supersymmetry is broken, the situation is less clear, see \cref{fn:beta} on page~\pageref{fn:beta}.

The only terms in \eqref{Sp1} that are affected when we use the $*$-product are the Yukawa couplings $\bar\psi[\phi,\bar\psi]$ and $\tilde\chi\phi\chi$. To compute the bosonic part of the effective action, we only need $\tilde\chi\phi\chi$. According to \eqref{starproductid}, the effect of the $*$-product on this term
is equivalent to replacing the matrices $\Sigma_{A}$ by deformed versions $\tilde\Sigma_{A}$,
\begin{equation}\label{starbetaeffect} \tilde\chi^{a}*\Sigma_{Aab}\phi_{A}*\chi^{b}
=\tilde\chi^{a}\tilde\Sigma_{Aab}\phi_{A}\chi^{b}\, .\end{equation}
The explicit formulas for the matrices $\tilde\Sigma_{A}$ are given in
\eqref{Sigmabetadef6D}.

\subsubsection{The effective action}

Integrating out $q$, $\tilde q$, $\chi$ and $\tilde\chi$ from the deformed microscopic action, we get
\begin{multline}\label{SeffboseBeta} S_{\text{eff}}(X,Y,D)=K\Bigl(\frac{8\pi^{2}N}{\lambda}
 +i\vartheta\Bigr)\\+
\frac{4\pi^{2}N}{\ls^{4}\lambda}\tr \Bigl\{
2i\ls^{4} D_{\mu\nu}\bigl[X_{\mu},X_{\nu}\bigr] -
\bigl[X_{\mu},Y_{A}\bigr]\bigl[X_{\mu},Y_{A}\bigr]\Bigr\}+ \ln\Delta_{q,\tilde q} - \ln\tilde\Delta_{\chi,\tilde\chi}\, ,
\end{multline}
where
\begin{align}\label{DeltaqqBeta} \ln\Delta_{q,\tilde q} &= N\ln \det \bigl(\vec 
Y^{2}\otimes \1_{2} + i \ls ^4 D_{\mu\nu}\otimes \sigma_{\mu\nu} \bigl)\, ,\\\label{DeltaccBeta}
\ln\tilde\Delta_{\chi,\tilde\chi} & =N
 \ln \det \bigl( \tilde\Sigma_A \otimes Y_A \bigl)\, .\end{align}
The dependence of $\Seff(X,Y,D)$ on $D_{\mu\nu}$ is exactly the same as in the undeformed model studied in \cite{Ferrari:2012nw}. The solution of the saddle-point equation \eqref{saddleCBD} is thus given by \eqref{Dorder3CB} for $\vec y_{f}=\vec 0$. In particular, when we write \eqref{XYexp}, $\langle D_{\mu\nu}\rangle$ is of order $\epsilon^{2}$ and will contribute to $\Seff$ only at order four or higher in $\epsilon$.

To leading order, \eqref{SeffboseBeta} yields
\begin{equation}\label{cBeta} c = \frac{8\pi^{2}N}{\lambda} + i\vartheta + 2N\ln{\vec y\,}^{2} - N\ln\det U\, ,\end{equation}
where the matrix $U$ is defined by
\begin{equation}\label{Udef} U = y_{A}\tilde\Sigma_{A}\, .\end{equation}
The determinant of $U$ can be computed straightforwardly in terms of the polar coordinates introduced in \eqref{polardef0},
\begin{equation}\label{detU} \det U = \rho_{1}^{4} + \rho_{2}^{4} + \rho_{3}^{4} + 2\cos(2\pi\gamma_{1})\rho_{2}^{2}\rho_{3}^{2}
+ 2\cos(2\pi\gamma_{2})\rho_{1}^{2}\rho_{3}^{2}
+ 2\cos(2\pi\gamma_{3})\rho_{1}^{2}\rho_{2}^{2}\, .\end{equation}
Plugging this result in \eqref{cBeta} and using the coordinates $r_{i}$ defined in \eqref{ridef0} yields
\begin{multline}\label{cfBeta} c = \frac{8\pi^{2}N}{\lambda} + i\vartheta
\\- N \ln\Bigl[ 1 - 4 \bigl(r_{2}^{2}r_{3}^{2}\sin^{2}(\pi\gamma_{1}) +
r_{1}^{2}r_{3}^{2}\sin^{2}(\pi\gamma_{2})
+r_{1}^{2}r_{2}^{2}\sin^{2}(\pi\gamma_{3})\bigr)\Bigr]\, .
\end{multline}
Let us note that this result was also obtained in the context of standard instanton calculus in \cite{Georgiou:2006np,Durnford:2006nb}.

The effective action at first and second order is fixed in terms of the derivatives of $c$. New information is found in the completely antisymmetric coefficient at order three, which yields the three-form $F^{(3)}$ defined in \eqref{F3def}. Expanding in $\epsilon$ using \eqref{detexp}, we see that
both determinants \eqref{DeltaqqBeta} and \eqref{DeltaccBeta} contribute to the third order action, but only \eqref{DeltaccBeta} yields a completely antisymmetric term. Explicitly, we get a nice and compact result,
\begin{equation}\label{F3Beta} F^{(3)} = -\frac{N}{3}\tr \bigl( 
U^{-1}\d U\wedge U^{-1}\d U\wedge U^{-1}\d U\bigr)\, .\end{equation}
In particular, this formula makes manifest the fact that $\d F^{(3)}=0$. However, 
the evaluation of the trace on the right-hand side is extremely tedious to perform manually, because the explicit expressions for the matrix $U$ and its inverse $U^{-1}$ are very complicated. We have thus implemented the calculation in Mathematica. The resulting formulas greatly simplify when using the coordinates defined in \eqref{polardef0}, \eqref{ridef0} and \eqref{thetaphidef0}. To linear order in the deformation parameters, which is all we need to compare with supergravity, we find, for the two-form potential defined in \eqref{dc},  
\begin{multline}\label{C2linearorder}
C^{(2)} = 8 N\ls^{2}\Bigl[ \omega_{1}\wedge\bigl(\gamma_{1}\d\theta_{1} + \gamma_{2}\d\theta_{2}+\gamma_{3}\d\theta_{3}\bigr) \\- \frac{i}{4}
\bigl( \gamma_{1}r_{2}^{2}r_{3}^{2}\,\d\theta_{2}\wedge\d\theta_{3}
+ \gamma_{2}r_{3}^{2}r_{1}^{2}\,\d\theta_{3}\wedge\d\theta_{1} +
\gamma_{3}r_{1}^{2}r_{2}^{2}\,\d\theta_{1}\wedge\d\theta_{2}\bigr)\Bigr] + 
O\bigl(\gamma^{2}\bigr)\, ,
\end{multline}
where the one-form $\omega_{1}$ is defined by the condition
\begin{equation}\label{domega1} \d \omega_{1} = r_{1}r_{2}r_{3}\sin\theta\,\d\theta\wedge\d\phi\, .\end{equation}
The exact result in the supersymmetry preserving case $\gamma_{1}=\gamma_{2}=\gamma_{3}=\gamma$ is given in \eqref{omegaBetasol}. Similar formulas can be obtained in other special cases, but they are not particularly illuminating. The general formula for arbitrary finite $\gamma_{i}$s is very complicated and we shall refrain from writing it down explicitly.

\subsubsection{The holographic background}

Expanding \eqref{cfBeta} to quadratic order in the deformation parameters and using \eqref{c0} yields
\begin{equation}\label{phiBeta} e^{-\phi} = \frac{4\pi N}{\lambda}\Bigl( 1 + \frac{1}{2}\lambda\bigl( \gamma_{1}r_{2}^{2}r_{3}^{2} + \gamma_{2}r_{3}^{2}r_{1}^{2} + \gamma_{3}r_{1}^{2}r_{2}^{2}\bigr) + O\bigl(\lambda\gamma^{4}\bigr)\Bigr)\, .\end{equation}
When the background is a small deformation of the undeformed $\AdSS$ solution, i.e.\ when $\lambda\gamma_{i}^{2}\ll 1$,
this is a perfect match with the supergravity solution \eqref{tbeta} and \eqref{Gdef}, consistently with the discussion in section \ref{useSec}. Similarly, \eqref{C2linearorder} and \eqref{C2C4NSRR} yield
\begin{align}
\label{BgenBeta}
 B & =\frac{\lambda}{4\pi N} \im C^{(2)} \nonumber \\
 &= -\frac{\ls^{2}\lambda}{2\pi}\bigl(
\gamma_{1}r_{2}^{2}r_{3}^{2}\,\d\theta_{2}\wedge\d\theta_{3}
+ \gamma_{2}r_{3}^{2}r_{1}^{2}\,\d\theta_{3}\wedge\d\theta_{1} +
\gamma_{3}r_{1}^{2}r_{2}^{2}\,\d\theta_{1}\wedge\d\theta_{2}\bigr) \nonumber \\
& \phantom{=-} + O\bigl(\gamma^{2}\bigr)\, ,\\
\label{C2genBeta}
\begin{split}
C_{2} & = -\re C^{(2)}-\frac{\vartheta}{2\pi} B\\& = 
-8 N\ls^{2} \omega_{1}\wedge\bigl(\gamma_{1}\d\theta_{1} + \gamma_{2}\d\theta_{2}+\gamma_{3}\d\theta_{3}\bigr)-\frac{\vartheta}{2\pi} B + 
O\bigl(\gamma^{2}\bigr)\, .\end{split}
\end{align}
After making the $\text{SL}(2,\mathbb R)$ transformation $C_{0}\rightarrow C_{0}+\frac{\vartheta}{2\pi}$, $C_{2}\rightarrow C_{2}- \frac{\vartheta}{2\pi} B$ to generalize the solution to an arbitrary bare $\vartheta$ angle, we find again a beautiful match with the supergravity background \eqref{Bbeta} and \eqref{C2beta} in the appropriate limit.

Actually, in the present case, it seems that the discussion of section \ref{useSec} can be slightly refined. Indeed, because the imaginary part of $c$ given by \eqref{cBeta} is a constant, it turns out that the general matrix coordinates redefinitions do not act on $\re F^{(3)}$ \cite{Ferrari:2013pi}. This three-form is thus unambiguously fixed by our microscopic calculations, even when the perturbation with respect to the undeformed conformal $\nn=4$ gauge theory is large. As a consequence, to compare with supergravity, we do not have to impose $\lambda\gamma_{i}^{2}$ to be small. The only relevant constraint is of course the validity of the supergravity solution itself, which is the weaker condition $\lambda\gamma_{i}^{4}\ll 1$ together with $\lambda\gg 1$. In this limit, we are allowed to expand the microscopic results as in \eqref{C2linearorder},
since $\gamma_{i}\ll 1$. However, we are not allowed to simplify the function $G$ defined by \eqref{Gdef} in the supergravity solution, because $\lambda\gamma_{i}^{2}$ may be large. Remarquably, we do find agreement with the microscopic prediction, because the real part of $C^{(2)}$ is related to the right-hand side of \eqref{C2beta} which does not depend on $G$.

\chapter{The enhançon mechanism from a fractional D-instanton probe}
\label{chap:enhancon}
  In this chapter, the $\nn=2$ field theory realized by D3-branes on the $\CC^2/\ZZ_2$ orbifold is studied. The
  dual supergravity solution exhibits a repulson singularity cured by the
  enhançon mechanism. By comparing the open and closed string descriptions of a probe D-instanton,
  it is possible to compute the exact non-perturbative profile of the supergravity
  twisted field, which determines the supergravity background. We then show how the
  non-trivial \<IR> physics of the field theory translates into the stringy effects that
  give rise to the enhançon mechanism and the associated excision procedure.
 
\section{The context}
After the original proposal by Maldacena for a duality between $\nn=4$ Yang-Mills theory and type \<IIB> superstrings on $\AdSS$ \cite{Maldacena:1997re}, a lot of work focused on the construction of string theory duals to more realistic field theories.
One of the directions that proved most fruitful consists in placing D3-branes on a singular Calabi-Yau threefold in order to break supersymmetry down to $\nn=1$.
The simplest example, the conifold, was studied by Klebanov and Witten in \cite{Klebanov:1998hh}.
The low-energy dynamics of D3-branes on the conifold is described by a conformal two-node quiver gauge theory, with gauge group $\SU(N)\times\SU(N)$.
Adding $M$ fractional branes to this setup, one can engineer a theory with unequal ranks for the two factors of the gauge group, which now have non-vanishing $\beta$-functions.
The corresponding supergravity dual was found by Klebanov and Strassler in \cite{Klebanov:2000hb}.
A remarkable aspect of the solution is that the Ramond-Ramond fluxes have a logarithmic dependence on the radial coordinate, which corresponds in the field theory to a cascade of Seiberg dualities.
A second remarkable aspect is that the conifold gets deformed in the \<IR>, corresponding to confinement in the gauge theory dual.

From the field theory point of view, a close cousin of the conifold is the $\CC^2/\ZZ_2$ orbifold: the field theory corresponding to D3-branes on this orbifold is also a two-node quiver gauge theory, which now preserves $\nn=2$ supersymmetry.
By giving appropriate mass terms to the two adjoint chiral multiplets, one can make this theory flow to the Klebanov-Witten one \cite{Klebanov:1998hh}.
For equal ranks of the gauge groups, the theory is again conformal \cite{Lawrence:1998ja} and the dual supergravity background is simply a $\ZZ_2$ orbifold of the five-sphere in the $\AdSS$ solution \cite{Kachru:1998ys}.
Similarly to what is done in the $\nn=1$ case, one can break conformality by taking the ranks to be different; this again corresponds to adding $M$ fractional D3-branes to the $N$ regular ones.
The supergravity dual was found by \cite{Bertolini:2000dk,Polchinski:2000mx} following \cite{Klebanov:1999rd}, and presents several puzzling features.
Firstly, like for its $\nn=1$ counterpart, the logarithmic dependence of the fluxes on the radial coordinate calls for a dual which is a cascading field theory.
But Seiberg duality is a purely $\nn=1$ phenomenon, which complicated early attempts towards a field theory interpretation \cite{Aharony:2000pp,Petrini:2001fk,Billo:2001vg}.
Eventually, the authors of \cite{Benini:2008ir} put forward a consistent picture for the mechanism responsible for the cascade in analogy with the baryonic root transition of $\nn=2$ \<SQCD> \cite{Argyres:1996eh}.
However, the main puzzle is that the supergravity solution has a singularity in the \<IR> of the repulson type \cite{Kallosh:1995yz,Cvetic:1995mx,Behrndt:1995tr}: there is a region where a probe experiences a repulsive force, which makes the solution unphysical.
That the solution is singular could be expected on general grounds: $\nn=2$ theories do not confine and correspondingly there is no $\nn=2$-preserving deformation of the $S^5/\ZZ_2$ space that could cure the singularity as happens for the Klebanov-Strassler solution.
The singularity must be resolved differently by string theory and it was argued that in holographic duals to $\nn=2$ theories, this happens through the enhançon mechanism \cite{Johnson:1999qt}.
At a finite value of the radial coordinate, the enhançon radius, the supergravity solution cannot be trusted anymore because some branes become tensionless, providing new light degrees of freedom that are not described by supergravity and can possibly be responsible for the resolution of the singularity.
Drawing inspiration from the behavior of roots of the Seiberg-Witten curve, the authors of \cite{Johnson:1999qt} argue that, inside the enhançon radius, the supergravity solution must be excised and replaced with a solution with constant fluxes, similarly to what happens inside a conducting material in Maxwell theory.
To the extent of the authors’ knowledge, this excision procedure has never been justified in full generality from a microscopic point of view, even if partial results have been obtained by focusing on limiting cases \cite{Cremonesi:2009hq}.

The present work aims to fill this gap.
We compute directly from the field theory the profile of the twisted supergravity field $\gamma$, which encodes the backreaction of the fractional branes and completely determines the supergravity solution once the configuration of regular branes is given.
This computation will be done with arbitrary values for the gauge theory couplings, which translates in the string theory dual to having arbitrary values for the string coupling $g_s$ and the string length $\ls=\sqrt{2\pi\alpha'}$, and for any point on the Coulomb branch of the theory.
We will prove that the twisted supergravity field $\gamma$ can be written in terms of field theory data as
\begin{equation}
  2\pi i\,\gamma(z) = 2\pi i\,\gamma^{(0)}  -\beta\,\int_1^{T_r(z)}
\frac{\mathrm{d}v}{\sqrt{(v^2-\alpha_1^2)(v^2-\alpha_2^2)}}\,,
  \label{eqn:resultgamma}
\end{equation}
where $z$ is a complex coordinate on the orbifold fixed plane, $\gamma^{(0)}$ is the asymptotic value of $\gamma$, $T_r$ is a ratio of polynomials encoding the choice of Coulomb branch vacuum and $\alpha_i$, $\beta$ are specific coupling-dependent constants.
All these quantities will be defined precisely in due course.
Choosing the particular vacua that have been studied from the supergravity side and taking the large $N$ limit of \eqref{eqn:resultgamma}, we can then derive that an enhançon mechanism takes place at a radius that perfectly matches the supergravity expectations: $\gamma$ is constant inside this radius, confirming the proposal of \cite{Johnson:1999qt}.

To make the proof of \eqref{eqn:resultgamma} possible, we draw on recent developments in two very different research lines.
The first of these is the use of D-brane probes to derive holographic string theory backgrounds from the field theory side.
Like in \cite{Ferrari:2012nw} and in \cref{chap:adsdeformed}, we consider a setup where the background branes are D3-branes and the probe is a small number of D-instantons (i.e.\ \Dmi{}-branes).
More specifically, the probe we will use is a single fractional \Dmi{}-brane.
The open-string realization of the \Dmi{}/D3 system in flat space can be straightforwardly generalized to the orbifold setting \cite{Argurio:2007vqa,Argurio:2008jm} by following the same procedure as for D3-branes \cite{Douglas:1996sw,Lawrence:1998ja}.
The action for \Dmi{}-branes in the presence of D3-branes (in a ``near-horizon'' limit) has a purely field theoretic intepretation\footnote{In this chapter, we will deal only with a fractional \Dmi{}-brane that sits on a quiver node also occupied by D3-branes and can directly be interpreted as a gauge theory instanton. In the case where the node is occupied by at most one D3-brane, that gauge group does not receive instanton corrections in field theory, and the \Dmi{}-brane corresponds to a ``stringy instanton''. Nevertheless, it turns out that also stringy instantons can be given a gauge theory interpretation in a suitable \<UV> completion \cite{Argurio:2012iw}.} as the \<ADHM> action for supersymmetric instantons \cite{Dorey:2002ik}.
The \Dmi{}-brane couples to the D3-branes through moduli that transform in an (anti-)fundamental representation of the four-dimensional gauge group and one can always integrate them out exactly.
However, the integration of the D3-brane fields involves the computation of a full-fledged non-chiral correlator in the four-dimensional gauge theory, which seems intractable in general.
In the conformal case, this correlator turns out to be trivial and one can recover the full supergravity background by matching the action for several \Dmi{}-branes with the non-Abelian probe brane action of \cite{Myers:1999ps,Taylor:1999pr}.
This is also the case for regular D3-branes on orbifold singularities whose field theory description is conformal.
Unfortunately, for the $\CC^2/\ZZ_2$ orbifold with fractional D3-branes which is the focus of the present chapter, the theory is non-conformal and one cannot reconstruct the full supergravity multiplet in this way.
We circumvent this difficulty by using a fractional \Dmi{}-brane as a probe instead of a regular one. 
This brane couples only to the twisted sector at the orbifold singularity, which captures the essential information on the background.
Applying this procedure in this case yields the following identity relating the twisted supergravity field $\gamma$ to a field theory correlator,
\begin{equation}
\gamma(z)=\gamma^{(0)}+\frac{i}{\pi}\left\langle\tr_{M}\log\left({z-Z_1}\right)\right\rangle
-\frac{i}{\pi}\left\langle\tr_{M}\log\left({z-Z_0}\right)\right\rangle\,,
\label{eqn:resultcorr}
\end{equation}
where $Z_0$ and $Z_1$ are the adjoint scalars of the two gauge groups normalized to have units of length.
This identity was derived in \cite{Billo:2012st} by computing string worldsheet diagrams, but we will rederive it much more straightforwardly.
The identity \eqref{eqn:resultcorr} also involves expectation values in the full gauge theory on the D3-branes and one might think naively that not much has been gained by focusing on the twisted sector.
There is a crucial difference however between these correlators and the correlator one is faced with in the untwisted sector: in \eqref{eqn:resultcorr}, only the chiral fields $Z_0$ and $Z_1$ enter.
This gives us more control and allows us to compute them explicitly by exploiting the impressive recent progress in the resummation of instanton corrections to $\nn=2$ quiver gauge theories \cite{Nekrasov:2012xe,Fucito:2012xc}.

The plan of the chapter is as follows.
In section~\ref{sec:review}, we review the supergravity background corresponding to D3-branes on the $\CC^2/\ZZ_2$ orbifold and explain the enhançon mechanism that has been conjectured to cure the \<IR> singularity.
In section~\ref{sec:micro}, we detail the microscopic model we start with, consisting of $N$ regular D3-branes, $M$ fractional D3-branes of each type and one fractional \Dmi{}-brane.
In section~\ref{sec:gamma}, we derive equation \eqref{eqn:resultgamma} by building the effective action for the \Dmi{}-brane and comparing it with the supergravity probe action.
The computation of the correlators in \eqref{eqn:resultcorr} is quite technical and we have chosen to present it separately in appendix~\ref{sec:appendix}.
In section~\ref{sec:enhancon}, we take the large $N$ limit of this result, showing explicitly that the enhançon mechanism takes place.
Finally, we conclude in section~\ref{sec:conc} by giving some perspectives on possible future work.

\section{D3-branes on the \texorpdfstring{$\CC^2/\ZZ_2$}{C²/Z₂} orbifold. A review}
\label{sec:review}

The $\CC^2/\ZZ_2$ orbifold is a representative of a larger family, the ADE
 orbifolds. These are built as $\CC^2/\Gamma_{\textrm{ADE}}$,
with $\Gamma_{\textrm{ADE}}$ being a discrete subgroup of $SU(2)$. The theories living on
D3-branes placed on these orbifolds are $\nn=2$ superconformal quiver gauge theories. The
Coulomb phase of these theories is non-conformal, and can be engineered in the string
picture by including fractional D3-branes. The model we are interested in, with $\Gamma_{\textrm{ADE}}=\ZZ_2$,
is also known as the affine $A_1$ quiver theory. In this section, we review what we have learnt about the workings of the
gauge/string duality in this example. Most of what we say can be found in \cite{Benini:2008ir},
where this model was thoroughly studied.

\subsection{A supergravity perspective}

Our setup is made up of a large number of parallel $N$ regular and $2M$ fractional D3-branes in\footnote{As we will later deal with instantons, it is more convenient to rotate to Euclidean signature from the start.}
$\RR^{4}\times\CC\times\CC^2/\ZZ_2$. We use coordinates $\left(x^{\mu},z,z^2,z^3\right)$
for this space, and the $\ZZ_2$ acts as $\left(z^2,z^3\right)\to\left(-z^2,-z^3\right)$.
The regular branes can probe the full transverse space $\CC\times\CC^2/\ZZ_2$,
while the fractional branes are constrained to live at the orbifold singularity,
which is the complex $z$-plane at the origin of $\CC^2$ in this case. There are two types of fractional branes, which
we will denote as type 0 and type 1, and we will consider $M$ branes of the first type,
and $M$ of the second type. A regular brane can be thought of as a bound
state of a type 0 and a type 1 fractional brane. For some purposes, it is useful to think of
the fractional D3-branes as wrapped D5-branes. Recall that the orbifold $\CC^2/\ZZ_2$ can
be seen as the singular limit of a smooth \<ALE> manifold (in our case it is the Eguchi-Hanson
space \cite{Eguchi:1978gw}) where a homologically non-trivial 2-cycle $\Sigma$ collapses.
The type 1 and type 0 fractional D3-branes correspond to D5-branes wrapped on $\Sigma$ and
$-\Sigma$ respectively, stabilized by certain background fluxes.

The presence of fractional branes induces the excitation of some of the twisted modes of
type \<IIB> string theory. Thinking of the fractional D3-branes as wrapped D5-branes, it is
easy to understand that the reduction of the potentials $C_2$ and $B_2$ on the exceptional
cycle $\Sigma$ will give rise to non-zero twisted scalars $c$ and $b$. These two fields can only
depend on $z,\bar z$, as the fractional D3-branes can only probe this plane, and
are conveniently combined to form the complex field:
\begin{equation}
\gamma= c+\left(C_0+i\,e^{-\Phi}\right)b=\frac{1}{2\pi\ls^2}\int_{\Sigma}\left(C_2+\frac{i}{g_s}\,B_2\right)\,,
\label{eqn:g.def}
\end{equation}
to which we will generically refer as the twisted supergravity field. In writing the last equality
we have taken into account that the axio-dilaton is constant,
$C_0+i\,e^{-\Phi}=\frac{i}{g_s}$, since it does not couple to D3-branes.
Such branes do source a $C_4$ potential, and of course backreact on the metric. Instead of
writing the expression for all these fields, which can be found for instance in
\cite{Bertolini:2000dk}, the point we want to emphasize here is that \emph{the full
type \<IIB> background follows\footnote{Essentially the metric and the RR potential $C_4$ are
determined by a warp factor $H(z,z^1,z^2)$, which is determined itself by solving a Poisson
equation sourced by the the regular and the fractional D3-branes. The contribution of the latter comes with
a $|\partial_z\gamma|^2$ factor. The position of the former must be specified as the only extra input.} once the twisted supergravity field $\gamma$ is known}.
Because of $\nn=2$ supersymmetry, $\gamma$ depends holomorphically on $z$, i.e.\
$\partial_{\bar z}\gamma=0$.

The profile of the twisted supergravity field is in turn solely determined by the positions
of the fractional D3-branes. This follows from its equation of motion, that can be derived
from the type \<IIB> supergravity action taking into account the twisted supergravity supermultiplet and the fractional D-brane sources:
\begin{equation}
\Delta\gamma=2i\sum_{j=1}^M\left(\delta^2(z-z_j)-\delta^2(z-\tilde{z}_j)\right)\,,
\label{eqn:g.eq}
\end{equation}
where the fractional branes of type 1 sit at positions $z_j$, and those of type 0 sit
at $\tilde{z}_j$. Notice that the profile of $\gamma$ is only sensitive to genuine fractional
D3-branes: if $z_i=\tilde z_j$ for some pair $(i,j)$, these two fractional branes form a regular
D3-brane and do not source $\gamma$ anymore, in agreement with the fact that $\gamma$ does not couple to regular branes.
It is easy to solve the two-dimensional Laplace equation \eqref{eqn:g.eq} to obtain:
\begin{equation}
\gamma=\frac{i}{\pi}\left(\sum_{j=1}^M\log(z-z_j)-\sum_{j=1}^M\log(z-\tilde{z}_j)\right)+\gamma^{(0)}\,.
\label{eqn:g.pert}
\end{equation}
The value of $\gamma^{(0)}$ is clearly the asymptotic value, as $z\to\infty$, of $\gamma$.
There is a preferred value of $b$ and $c$ for perturbative string theory:
 if we choose
$\gamma^{(0)}=\frac{i}{2g_s}\Leftrightarrow\lim_{z\to\infty}(c,b)=(0,\frac{1}{2})$,
the world-sheet propagating on this orbifold is a free \<CFT> \cite{Aspinwall:1996mn,Blum:1997fw}.
We will see shortly that this value is also special from the field theory point of view, and we will often make this choice for simplicity.

The take-home message is then that the supergravity background is determined by the way in
which we distribute the fractional D3-branes in the geometry, and this information is encoded
in the twisted field $\gamma$. The distribution of branes is naturally related to
the different vacua of the dual gauge theory, as we now explain.

\subsection{A field theory perspective}

When we look at our brane system from far away, that is at large $|z|$, we essentially see
a stack of $N+M$ regular D3-branes on the $Z_2$ orbifold, since the fact that the positions
of type 0 and type 1 fractional are \emph{a priori} different becomes irrelevant.
We effectively obtain a theory with only $N+M$ regular branes. The field theory dual to this setup is well-known
\cite{Kachru:1998ys}. It is the $\nn=2$ superconformal quiver theory with gauge group $\SU(N+M)_0\times \SU(N+M)_1$.
This theory has a rich moduli space of vacua, with both Coulomb and Higgs branches.
The Higgs branch corresponds to giving \VEV s to the bifundamentals of the quiver. 
As is well known, the field theory on the Higgs branch is not very interesting, since both the superpotential and the K\"ahler potential are not renormalized \cite{Argyres:1996eh}.
In the brane picture, (the mesonic part of) this branch has a nice geometrical interpretation:
it corresponds to the possible configurations of regular D3-branes occupying certain positions
in the transverse space $\CC\times\CC^2/\ZZ_2$. 
Notice that in the covering space $\CC\times\CC^2$, the branes have to be arranged in pairs of orbifold images $(z,\pm z^2,\pm z^3)$.
Another possibility is to have some D3-branes at the origin of $\CC^2$ which maps to the orbifold singularity of $\CC\times \CC^2/\ZZ_2$.
Those D3-branes do not need to be paired and can have arbitrary positions along the $\CC$ plane with coordinate~$z$; those are fractional branes.
The different configurations for fractional branes correspond in the field theory to the Coulomb branch, obtained by giving expectation values to the two adjoint fields. Denoting these
fields by $\varphi_0$, $\varphi_1$ (see figure~\ref{fig:Z2fracquiver}), at the perturbative level we
can identify their respective non-zero eigenvalues with the $\tilde{z}_j$, $z_j$ of \eqref{eqn:g.eq}.
There are also mixed branches, where both bifundamentals and adjoints acquire \VEV s.

We are interested in the \<IR> physics of the Coulomb branch. More precisely we will be mainly
concerned with the point that was dubbed ``enhançon vacuum'' in \cite{Benini:2008ir}. It
is classically defined by $\varphi_0$ having $M$ prescribed non-zero
eigenvalues, or equivalently by having $M$ fractional branes of type 0 sitting at the roots of
$\tilde{z}_j^M=-z_0^M$, where $|z_0|$ is an arbitrary \<UV> scale. Below this scale, we are
left with an effective theory describing $N$ regular branes plus $M$ fractional branes of type 1 sitting at $z=0$. The gauge group is Higgsed
down to $SU(N)_0\times SU(N+M)_1$ if we take into account that all the $U(1)$ factors are \<IR> free
and decouple. 
Such an effective theory is not conformal, as 
reflected by the running of $\gamma$ in equation \eqref{eqn:g.pert}, which for this vacuum
reads
\begin{equation}
\gamma=\frac{i}{\pi}\log\frac{z^M}{z^M+z_0^M}+\frac{i}{2g_s}\quad\underset{\textrm{large }M}{\rightsquigarrow}
\quad\gamma\approx\begin{cases}
\frac{iM}{\pi}\log\frac{z}{z_0\,e^{-\frac{\pi}{2g_sM}}} & \textrm{if }|z|<|z_0|\\
\frac{i}{2g_s} & \textrm{if }|z|>|z_0|\end{cases}
\label{eqn:g.enh}
\end{equation}
In order for the classical supergravity solution that follows from \eqref{eqn:g.enh} to be a good description of the
gauge theory, one should require as usual that $N$ and $M$ be large. Using the complexified gauge couplings
$\tau_a=\frac{\vartheta_a}{2\pi}+\frac{4\pi\,i}{g_a^2}$ for the two $\SU(N+M)_a$ factors, the holographic relations between the gauge couplings and the supergravity fields read
\begin{equation}
\tau_0+\tau_1=\frac{i}{g_s} \, , \qquad\qquad \tau_1=\gamma \, .
\label{eqn:holotaugs}
\end{equation}
The first relation is the standard holographic dictionary applied to the diagonal $\SU(N+M)$ gauge group; the second one can be shown by a fractional probe brane analysis \cite{Bertolini:2000dk}.
This implies the following relation between the bare gauge couplings and the asymptotic values of the dilaton and $\gamma$:
\begin{equation}
  8\pi g_s=g_a^2\,,\qquad \lambda_a=(N+M)g_a^2=8\pi g_s(N+M)\,,
  \label{eqn:holo1}
\end{equation}
defining the 't Hooft couplings $\lambda_a$.
In particular, we see that $g_0=g_1$, which can be traced back to the fact that we have chosen the special asymptotic value $\frac{i}{2 g_s}$ for the twisted supergravity field $\gamma$.
Notice that the relations \eqref{eqn:holotaugs} are not restricted to $z\to\infty$. Indeed the second 
one provides an exact match between the supergravity
running of $\gamma$ and the perturbative running of the gauge couplings (which is exhausted at one-loop).

Strictly speaking, the supergravity description is a faithful one for small $g_s$, large $N$ and $M$ so that
$g_s(N+M)\gg1$.
We take $g_s(N+M)$ to be a large, but finite, number.
Since below $|z_0|$ the gauge couplings run in opposite directions, at a certain scale, one of the gauge couplings blows up.
The supergravity approximation breaks down there, the second relation in \eqref{eqn:holotaugs} no longer holds, and a stringy resolution is needed.
There are several ways to proceed, related to different non-perturbative completions of the same perturbative physics.
Let us discuss this in a bit more detail below.

\subsection{Non-perturbative physics and the enhançon}
\label{ssec:enh}

With the amount of supersymmetry that we have, the perturbative series for
correlation functions of protected operators in the field theory truncate at one loop. Any other quantum correction must come from
instantons, i.e.\ with a pre-factor $e^{-l/g_a^2}$ ($l$ being a positive number). At large $N,M$ and
fixed 't Hooft coupling, $g_a^2\sim 1/(N+M)\to0$ and these corrections are exponentially suppressed.
This is why supergravity outside the enhançon matches exactly the one-loop field theory, although they are expected to be valid in opposite regimes of $\lambda_a$.
Nevertheless, it is known that non-perturbative corrections
can still contribute in the 't Hooft limit \cite{Douglas:1995nw}, as will occur in our model.
Such corrections are proportional to $e^{-l/\lambda_a}$, which does not have to be small.

Let us now follow the holographic RG flow of our theory from equation \eqref{eqn:g.enh}, assuming
that both $N$ and $M$ are large. We have a conformal theory above the scale
$|z_0|$. Below this scale, $\gamma$ starts to run, inducing a running of the couplings. Recall from
\eqref{eqn:g.def} that the imaginary part of $\gamma$ gives us the gauge coupling $1/g_1^2$ in field theory and the scalar $b$ in supergravity.
This scalar should be in the range $\left[0,1\right]$ in order to have a proper field theory interpretation with positive $g_1^2,g_0^2$.
When we reach the scale
\begin{equation}
\rho_1=|z_0|\,e^{-\frac{\pi}{2g_sM}}\,,
\label{eqn:rho_1}
\end{equation}
$\gamma$ vanishes, and so does $b$.
At this point, from equation \eqref{eqn:holotaugs}, we see that $\lambda_1$ diverges.
Past this point, we can no longer trust the supergravity solution \eqref{eqn:g.enh}.
A way to think about it is that probe fractional branes become tensionless at $\rho_1$ (the tension of such branes is proportional to $b$).
Potentially, a whole fauna of stringy phenomena, not captured by the supergravity approximation,
could arise.

Nothing dramatic happens for the supergravity solution at $\rho_1$ though, so one could think of pushing the
gauge/string duality and come up with a possible field-theoretic interpretation below this scale. This is what the
authors in \cite{Benini:2008ir} did. They proposed an interpretation of the solution for $|z|<\rho_1$
\emph{\`a la} Klebanov-Strassler: we must perform a Higgsing in the field theory, interpreted
as a large gauge transformation in the supergravity background\footnote{Notice that a large gauge
transformation is \emph{not} a gauge transformation. With it, we are changing the vacuum in the underlying field theory.
The fact that we have to perform this operation is not encoded in the supergravity background,
but it must be done by hand instead.} \cite{Benini:2007gx}.
This Higgsing is a strong coupling effect: it arises at a scale $\sim e^{-l/\lambda_a}$ where a gauge coupling blows up.
The non-trivial field theory vacuum responsible for the Higgsing is very similar to the baryonic root in $\nn=2$ \<SQCD> \cite{Argyres:1996eh}, it has hence been called a baryonic root transition.
The rank
of the gauge group with diverging coupling is reduced by $2M$ and the beta functions flip sign. The large gauge transformation shifts the twisted field of \eqref{eqn:g.enh}
in this region as $\gamma\to\gamma+\frac{i}{g_s}$.
If we keep going down the flow, we will hit another point where $g_0$ diverges, and the same operation
must be performed on the other gauge group. This can happen multiple times:  we say that the theory cascades. Apart from the
fact that here the Higgsings are not associated to Seiberg dualites since we have $\nn=2$ supersymmetry, there is a fundamental
difference with the Klebanov-Strassler case. In the latter, at the end of the cascade, the theory confines
(its dual counterpart is the deformation of the conifold). However, our $\nn=2$ model is not confining.
A different, but very interesting, phenomenon occurs. It has come to
be known as the enhançon, as originally named in \cite{Johnson:1999qt}. Let us discuss it from both sides
of the gauge/string duality.

From the supergravity point of view, we find that the background presents a singularity (where the metric
blows up) of a peculiar type: a repulson \cite{Kallosh:1995yz,Cvetic:1995mx,Behrndt:1995tr}. Close to it, there is a region of ``anti-gravity'',
characterized by a positive sign of $\partial_{|z|}g_{00}$. From the warp
factor of the supergravity solution, one can find that this anti-gravity region starts at around the scale
\begin{equation}
\rho_e= e^{-\frac{\pi N}{g_sM^2}}\rho_1\,.
\label{eqn:rho_e}
\end{equation}
Probe branes feel a repulsive potential below $\rho_e$, and cannot enter this region. This is supported
by a computation of the D3-brane Page charge, which gives \cite{Benini:2008ir}:
\begin{equation}
\int \left(F_5+B_2\wedge F_3\right)\propto\left(N+M\left[\frac{g_sM}{\pi}\log\frac{|z|}{\rho_1}\right]\right)\,,
\label{eqn:F5}
\end{equation}
where were are denoting by $[\cdot]$ the floor function. This shows that inside the region of
radius $\rho_e$ there is an unphysical negative D3 charge.
Even if we want to believe in the supergravity solution below the scale $\rho_1$, we can only trust it
down to the smaller scale $\rho_e$. The latter is the enhançon scale.
The standard lore in supergravity is that the $M$ fractional branes that were supposed to be at the origin expanded
to form a dense ring at the enhançon scale. Since inside this ring no branes are left, we should solve \eqref{eqn:g.eq} again
with this assumption. This obviously yields a constant $\gamma$ in this region. 
This correction \emph{by hand} of $\gamma$ is commonly known as the excision procedure.
Notice that since it is done manually, we could have chosen to perform the excision procedure already at the scale $\rho_1$, or any other scale in between where one of the gauge couplings diverges.
Different choices of where to perform the excision correspond to different choices of vacua (which only differ
non-perturbatively) in the field theory. Following the terminology of \cite{Benini:2008ir}, excising at $\rho_1$ ($\rho_e$)
corresponds to the enhançon (cascading) vacuum.

There is a field-theoretical phenomenon that takes place in the large $N$ limit of $\nn=2$ gauge theories, which
resembles very much the repulson singularity we just described. It was first noticed in \cite{Johnson:1999qt}
and is called enhançon mechanism for historical reasons (having to do with enhanced symmetries). Take for example $\nn=2$ \<SQCD> with gauge group $\SU(N)$. The \<IR> physics is controlled by the Seiberg-Witten (\<SW>) curve
\begin{equation}
y^2=\prod_{k=1}^N(x-\varphi_k)^2+4\Lambda^{2N}\,,
\label{eqn:SW}
\end{equation}
where $\Lambda$ is the strong coupling scale, and the $\varphi_k$ parameterize a point in the moduli space. At large $N$, for points
with $|\varphi_k|\gg\Lambda$, the branch cuts of $y(x)$ are very small and they are located near the classical
values $x=\varphi_k$. On the contrary, when $|\varphi_k|/\Lambda\to0$, the branch cuts become longer and remain at a finite distance
from the origin of the $x$-plane. They pile up at a ring of radius $2^{\frac{1}{N}}\Lambda$. If we consider the configuration
\begin{equation}
y^2=x^{2N-2}(x-\varphi)+4\Lambda^{2N}\,,
\label{eqn:SW.probe}
\end{equation}
that corresponds to a breaking $\SU(N)\to\SU(N-1)\times\U(1)$, and we track the two branch points associated to $\varphi$,
we see that for $|\varphi|\gg\Lambda$, they are close together and around $x=\varphi$. When $|\varphi|$ approaches $\Lambda$,
the branch points separate from each other, and melt into the ring of quantum roots. The branch points can never
penetrate inside this ring. Associating branch cuts with branes, clearly this resembles the enhançon phenomenon
found in supergravity.

However, to our knowledge, the connection between the \<SW> curve physics and the enhançon mechanism
has never been established in the literature in a completely top-down approach. This is of course a difficult problem, since its solution would involve
computing non-perturbative corrections to the supergravity background. Two important steps forward in this direction
have been taken, first by Cremonesi in \cite{Cremonesi:2009hq}, and more recently by the authors of \cite{Billo:2012st} (see also \cite{Martucci:2012jk}).
The former cleverly used the M-theory uplift of a brane configuration \cite{Witten:1997sc} corresponding to pure 
$\nn=2$ Yang-Mills to obtain  the non-perturbative corrections to the $\gamma$ profile.
The latter computed directly the corrections to the background by including
\Dmi{} branes in the configuration and resumming the string disc diagrams with any number of \Dmi{}-branes. They found a very compact expression for the (non-perturbatively) corrected profile of
$\gamma$,
\begin{equation}
\gamma=\gamma^{(0)}+\frac{i}{\pi}\left\langle\tr_{M}\log\frac{z-\ls^{-2}\varphi_1}{\mu}\right\rangle
-\frac{i}{\pi}\left\langle\tr_{M}\log\frac{z-\ls^{-2}\varphi_0}{\mu}\right\rangle\,,
\label{eqn:Lerdaetal}
\end{equation}
in terms of correlators of the quiver field theory. We will later arrive to this result in a simpler way without computing any string diagrams.
Moreover we will be able to evaluate explicitly these correlators.

Our goal is to to unravel the whole picture
that we have described hitherto from a purely microscopic description.

\section{The microscopic model}
\label{sec:micro}

In this section we detail the affine $A_1$ quiver theory governing the brane configuration on the $\CC^2/\ZZ_2$ orbifold,
paying special attention to the instanton sector that will be instrumental later on.
While the presentation we give makes use of string theory and D-branes, this is in no way necessary, as both the D3 and the \Dmi{}-branes' dynamics (in the ``near-horizon'' limit) can be described in field theory terms by gauge theories and instantons respectively.

\subsection{The four-dimensional gauge theory}

The field theory describing the dynamics (in the $\ls\to0$ field theory limit) of $N+M$ D3-branes of type~$0$ and $N+M$ D3-branes of type~$1$ is a four-dimensional superconformal $\nn=2$ gauge theory with gauge group $\U(N+M)_0\times\U(N+M)_1$ \cite{Douglas:1996sw,Lawrence:1998ja}.
This was explained in detail in \cref{sec:Dorbifolds}.
Recall that the orbifold group $\ZZ_2$ acts on $\CC^3$ in the following way,
\begin{equation}
  \gamma\cdot(z^1,z^2,z^3)=(z^1,-z^2,-z^3)
  \label{eqn:Z2action}
\end{equation}
where $\gamma$ is the non-trivial element of $\ZZ_2$.
This yields the $\CC\times\CC^2/\ZZ_2$ orbifold by identifying all points of $\CC^3$ with their image under \eqref{eqn:Z2action}.
The first coordinate is fixed under the orbifold action, hence the orbifold singularity is a complex plane $\CC$ (times the Euclidean space-time $\RR^4$).
This plane will play an important role in the rest of the chapter, and we will often write $z=z_1$ for conciseness.
The superpotential of this theory is given by \cref{eq:Z2superpotential} and allows for a rich classical moduli space of vacua.
We will be interested in the Coulomb branch, corresponding to giving \VEV s only to the adjoint fields $\varphi_0$ and $\varphi_1$, which are $\varphi_0^1$ and $\varphi_1^1$ in the notation of \cref{sec:Dorbifolds} (see \cref{eq:Z2Phi123}).
Classically, one can choose a gauge in which the adjoint fields are diagonal.
A point on the Coulomb branch is then parameterized by the expectation value of their diagonal matrix elements,
\begin{equation}
  \langle\varphi_0\rangle = \ls^{-2}\diag(\tilde z_1,\ldots,\tilde z_{N+M}) \, , \quad \langle\varphi_1\rangle = \ls^{-2}\diag(z_1,\ldots,z_{N+M}) \, ,
  \label{eqn:vevclass}
\end{equation}
where we have written the matrix elements in terms of quantities having dimension of length in order to interpret them as D3-brane positions, and we need to identify two configurations differing by a permutation of eigenvalues.
The D-brane interpretation of this vacuum configuration is the following.
As we have discussed in \cref{sec:Dinstantons}, the $\Phi^1$ coordinate corresponds to fluctuations of the D3-branes along the $\CC$ direction that is invariant under the orbifold action \eqref{eqn:Z2action}.
Once we orbifold, there are two independent $\U(N+M)$-adjoint scalars $\varphi_0$ and $\varphi_1$ on the Coulomb branch, which now describe the positions of two different stacks of D3-branes along the $\CC$ direction.
These two stacks correspond to the two different kinds of fractional branes.
Being free to choose \eqref{eqn:vevclass} arbitrarily then means that one can choose the position of the two types of fractional branes independently.
If the expectation values are completely generic, that is if no $z_i$ coincides with any $\tilde z_j$, the expectation values of the other fields have to vanish on $\mathcal M_{\text{cl}}$ and the fractional branes are stuck at the orbifold fixed locus.
When the positions of two branes of different type coincide, say $z_1=\tilde z_1$, one can see that a new branch of the classical moduli space opens up, the $(1,1)$ matrix element of the bifundamental fields are not required to vanish.
This corresponds to two fractional branes of different type forming a regular brane bound state.
This regular brane is then free to move away from the orbifold singularity.
Since we want a configuration of $N$ regular branes and $M$ fractional branes of each type, we need to have exactly $N$ pairs of eigenvalues of the two types coincide.
The vacuum that we want to consider, the enhançon vacuum, is further specified by the requirement of $\ZZ_M$ rotational symmetry and dependence on a single scale $|z_0|$,
\begin{equation}
  \langle\varphi_0\rangle=\ls^{-2}\diag(\underbrace{0,\ldots,0}_{\text{N times}},z_0\, \omega, z_0\, \omega^2,\ldots,z_0\, \omega^M) \, , \quad \langle\varphi_1\rangle=0 \, ,
  \label{eqn:vevenhanconvac}
\end{equation}
where $\omega$ is an $M$-th root of $-1$, $\omega^M=-1$.

At the quantum level, the $z_i$ and $\tilde z_j$ as defined by \eqref{eqn:vevclass} are not globally well-defined coordinates on the moduli space.
Instead, we need to use a set of independent gauge-invariant observables as coordinates of the Coulomb branch.
Rather than specifying those directly, we can instead encode the Coulomb branch vacuum in  a ratio of two polynomials of degree $M$, $T_r(z)=T_0(z)/T_1(z)$, where
\begin{equation}
  T_0(z)=\prod_{i=1}^{M+N}(z-\tilde z_i) \, , \qquad T_1(z)=\prod_{j=1}^{M+N}(z-z_j) \, .
  \label{T0T1def}
\end{equation}
Note that these $z_i$ and $\tilde z_j$ do \emph{not} coincide with the ones in \eqref{eqn:vevenhanconvac} in the quantum theory; they only agree perturbatively.
Instead, they are given by \VEV s of gauge invariant operators built from traces of $\varphi_0$ and $\varphi_1$.
From now on, we will assume that they are defined by \eqref{T0T1def} instead of \eqref{eqn:vevenhanconvac}.
Imposing the same constraints as on \eqref{eqn:vevenhanconvac} now requires \cite{Benini:2008ir}
\begin{equation}
  T_0(z)=z^N(z^M+z_0^M) \, , \qquad T_1(z)=z^{N+M} \, .
  \label{eqn:T0T1enhvac}
\end{equation}

\subsection{Adding a D-instanton probe}

Our goal is to derive the full non-perturbative profile of the twisted supergravity $\gamma$ from field theory data.
By \eqref{eqn:holotaugs}, at the perturbative level, $\gamma$ is related to the gauge coupling of one of the two $\U(N+M)$ gauge groups.
However, it is not so clear how to extend this relation to the non-perturbative level, since one needs to choose a regularization scheme to define a coupling and it is not clear \emph{a priori} which scheme is appropriate for the holographic interpretation of $\gamma$ as a twisted supergravity field.
A way out of this problem is to relate instead $\gamma$ to an observable in the field theory.
This observable will turn out to be intimately related to the effective action for a probe fractional \Dmi{}-brane, which we are going to construct from a \Dmi{}/D3-brane system along the lines of \cite{Ferrari:2012nw}.

We enrich the set-up of D3-branes on the $\CC^2/\ZZ_2$ orbifold that we discussed previously by adding a single fractional \Dmi{}-brane of type 1.
The combined system is described by a partition function of the schematic form
\begin{equation}
  \mathcal Z = \int \d\mu_{\text{D3}} \d\mu_{\text{\Dmi{}}} e^{-S_{\text{D3}}-S_{\text{\Dmi{}}}} \, .
  \label{eqn:ZD3D-1}
\end{equation}
In addition to the functional integration over the D3-brane fields weighted by the four-dimensional gauge theory action $S_{\text{D3}}$ discussed above, there is now also an (ordinary) integral over the fractional \Dmi{}-brane moduli, with an action $S_{\text{\Dmi{}}}$ that we now detail.
This action describes the low-energy dynamics of the -1/-1 strings starting and ending on the \Dmi{}-brane and the -1/3 strings with one endpoint on the \Dmi{} and the other on a D3-brane as well as their couplings to the D3-brane fields.
The -1/-1 strings are uncharged under the $\U(N+M)_0\times\U(N+M)_1$ gauge group, whereas the 3/-1 strings with their endpoint on the D3-branes of either type transform in a fundamental representation of the corresponding $\U(N+M)$ gauge group and have charge $-1$ under the \Dmi{}-brane $\U(1)$ gauge group.

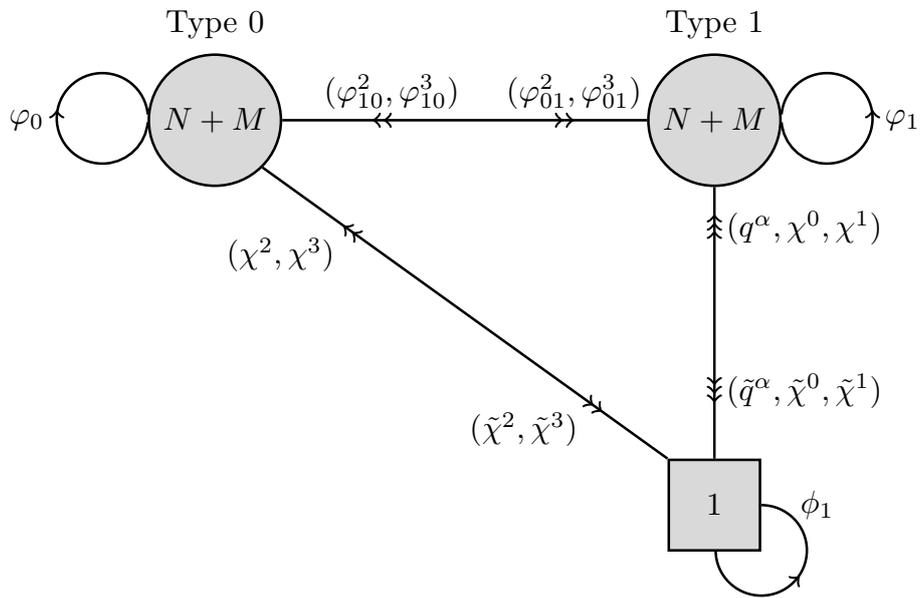
\begin{figure}
  \centering\usetikzlibrary{arrows,decorations.markings,positioning}
\pgfarrowsdeclaretriple{<<<}{>>>}{to}{to}
\pgfarrowsdeclaretriple{>>>}{<<<}{to reversed}{to reversed}
\begin{tikzpicture}
  [
  every path/.style={thick},
  group/.style={draw,fill=gray!30,minimum size=1cm},
  instanton/.style={rectangle,group},
  gauge/.style={circle,group}
  ]
  \node[instanton] (instanton 1) {$1$};
  \node[gauge] (gauge 1) [above=3 of instanton 1] {${N+M}$} node[above=\lineskip of gauge 1.north] {Type $1$};
  \node[gauge] (gauge 0) [left=4 of gauge 1] {${N+M}$} node[above=\lineskip of gauge 0.north] {Type $0$};
  \begin{scope}
    [decoration={markings,mark=at position 0.5 with {\arrow{<}}}]
    \draw [postaction={decorate}] (gauge 0.175) arc (5:355:0.5);
    \node [left=1 of gauge 0.west] {$\varphi_0$};
    \draw [postaction={decorate}] (instanton 1.east) arc (90:-180:0.5);
    \node [right=0.3 of instanton 1.east] {$\phi_1$};
    \draw [postaction={decorate}] (gauge 1.5) arc (175:-175:0.5);
    \node [right=1 of gauge 1.east] {$\varphi_1$};
  \end{scope}
  \begin{scope}
    [decoration={markings,mark=at position 0.3 with {\arrow{<<<}},mark=at position 0.9 with {\arrow{>>>}}}]
    \draw [postaction={decorate}] (instanton 1.north) -- (gauge 1.south) node [pos=0.25, right] {$(\tilde q^\alpha,\tilde \chi^0,\tilde \chi^1)$} node [pos=0.85, right] {$(q^\alpha,\chi^0,\chi^1)$};
  \end{scope}
  \begin{scope}
    [decoration={markings,mark=at position 0.3 with {\arrow{<<}},mark=at position 0.8 with {\arrow{>>}}}]
    \draw [postaction={decorate}] (gauge 0.east) -- (gauge 1.west) node [pos=0.3, above] {$(\varphi^2_{10},\varphi^3_{10})$} node [pos=0.8, above] {$(\varphi^2_{01},\varphi^3_{01})$};
  \end{scope}
  \begin{scope}
    [decoration={markings,mark=at position 0.2 with {\arrow{<<}},mark=at position 0.8 with {\arrow{>>}}}]
    \draw [postaction={decorate}] (instanton 1.north west) -- (gauge 0.south east) node [inner sep=15pt,pos=0.1,left] {$(\tilde\chi^2,\tilde\chi^3)$} node [inner sep=15pt,pos=0.7, left] {$(\chi^2,\chi^3)$};
  \end{scope}
\end{tikzpicture}
  \caption{The quiver of the $\CC^2/\ZZ_2$ orbifold with the \<UV> brane configuration that we consider: $N+M$ D3-branes of each type corresponding to a $\U(N+M)_0\times\U(N+M)_1$ gauge group and 1 \Dmi{}-brane of type~$1$.}
  \label{fig:Z2fracquiver}
\end{figure}

The action $S_{\text{\Dmi{}}}$ can be derived by a procedure similar to the one we followed for the four-dimensional gauge theory.
One starts with the action describing the \Dmi{}/D3 system in flat space \cite{Green:2000ke,Billo:2002hm} and one also embeds the $\ZZ_2$ orbifold group in the $\U(1)$ \Dmi{}-brane gauge group.
Since we are dealing with a fractional \Dmi{}-brane of type 1, the appropriate representation to take is not the regular representation as in \eqref{eq:Z2regrep}, but rather the non-trivial irreducible representation, $R_1(\gamma)=-1 \in \U(1)$.
One then needs to truncate the moduli to the modes invariant under the orbifold action, which has a form similar to \eqref{eq:Z2PhiI} but with one (both) regular representation(s) replaced by $R_1$ for -1/3 strings (-1/-1 strings), as required by the $\U(N+M)_0\times\U(N+M)_1\times\U(1)$ representation to which they belong.
The various fields and moduli surviving the orbifold projection, and hence present in this brane configuration, are summarized in the quiver diagram of figure~\ref{fig:Z2fracquiver}.
The modulus $\phi_1$ is a complex number that plays the role of position of the fractional \Dmi{}-brane on the orbifold fixed plane $\CC$.
For dimensional reasons, we will rather work with the modulus $z$, related to $\phi_1$ by a rescaling,
\begin{equation}
  z=\ls^2\,\phi_1\,.
  \label{eqn:z}
\end{equation}
The -1/3 strings with their endpoints on the D3-branes of type 1 provide an $\SU(2)$ doublet of bosonic moduli $q^\alpha$ and their two fermionic superpartners $\chi^0$ and $\chi^1$, all in the anti-fundamental representation of $\U(N+M)_1$.
Similarly, the 3/-1 strings provide the complex conjugate moduli $(\tilde q^\alpha, \tilde\chi^0,\tilde\chi^1)$ in the fundamental of $\U(N+M)_1$.
On the other hand, the strings stretched between the \Dmi{}-brane and the D3-branes of type 0 only provide fermionic moduli, $(\chi^2,\chi^3)$ and $(\tilde\chi^2,\tilde\chi^3)$, which are charged under $\U(N+M)_0$.
The action $S_{\text{\Dmi{}}}$ reads
\begin{multline}
  S_{\text{\Dmi{}}} = -2\pi i \tau_1+ \frac{1}{2}\tr \left\{ \frac{1}{2}(\tilde q^\alpha\phi^{\dagger}_1-\varphi^{\dagger}_1 \tilde q^\alpha)(\phi^1 q_\alpha -q_\alpha \varphi^1)  \right. \\
    \left. + \frac{1}{2}(\tilde q^\alpha\phi^1-\varphi^1 \tilde q^\alpha)(\phi^{\dagger}_1 q_\alpha -q_\alpha \varphi^{\dagger}_1)   -(\tilde\chi^{1}_{}\phi^\dagger_{1}-\varphi^\dagger_{1}\tilde\chi^{1}_{})\chi^0 + \tilde\chi^0(\phi^\dagger_{1}\chi^1-\chi^1\varphi^\dagger_{1}) \right. \\
    \left. + \tilde\chi^3(\phi^1\chi^2 + \chi^2 \varphi^0) - \tilde\chi^2(\phi^1\chi^3 + \chi^3 \varphi^0) \right\} + \cdots \, .
  \label{eqn:SD-1mic}
\end{multline}
In this expression, we have already taken the \<ADHM>/near-horizon limit \cite{Billo:2002hm,Ferrari:2012nw} and also dropped the terms involving the bifundamental fields since all their expectation values vanish on the Coulomb branch.
The fermionic neutral moduli $\psi$ (the superpartners of $\phi_1$) have not been written down explicitly either, since we are going to set them to $0$ anyway in the following.
The first constant term on the \RHS\ needs to be added in order to reproduce the instanton factor $e^{2\pi i\tau_1}$.

Let us comment on a subtlety concerning the gauge group of the theory in the near-horizon limit.
In this section, we have assumed it to be $\U(N+M)_0\times\U(N+M)_1\times\U(1)$, but the holographic dual describes only an $\SU(N+M)\times \SU(N+M)$ theory.
Indeed, the three commuting $\U(1)$ factors of the $\U(N+M)_0\times\U(N+M)_1\times\U(1)$ gauge group decouple.
The diagonal $\U(1)$ of the three gauge groups describes the movement of the full \Dmi{}/D3 system in the $\CC$-plane and, by translational invariance, no fields are charged under it.
The relative $\U(1)$ between the \Dmi{} and D3 gauge groups decouples since we will integrate out all the fields that are charged under it.
Finally, the anti-diagonal $\U(1)_B$ between the $\U(N+M)_0$ and $\U(N+M)_1$ groups is \<IR> free and becomes a global baryonic symmetry.

The relevant global symmetry group of our field-theoretic model will then be $\SU(2)_R\times\SU(2)_F\times\U(1)_A\times\U(1)_B$. The last factor emerges from the near-horizon limit as we explained, and the other three can be seen to be the commuting subgroups of the parent $\SU(4)_R$ that survive the orbifolding. Geometrically, both $\SU(2)_R$ and $\SU(2)_F$ correspond to rotations in $\CC^2/\ZZ_2$ (the latter acts holomorphically on $\CC^2$, contrarily to the former), while the $\U(1)_A$ rotates the $\CC$ factor.

\section{The twisted supergravity field}
\label{sec:gamma}

Our system of background fractional D3-branes plus a probe \Dmi{}-brane is described by the
microscopic model we spelled out in the previous section. This model is governed by an action
of the form $S_{\text{D3}}+S_{\text{\Dmi{}}}$, where $S_{\text{D3}}$ is an $\nn=2$ action with superpotential \eqref{eq:Z2superpotential} and $S_{\text{\Dmi{}}}$ is written in \eqref{eqn:SD-1mic}.
We are interested in obtaining the effective action $S_{\text{\Dmi{},eff}}$ for the probe in the holographic background, in the
spirit of \cite{Ferrari:2012nw}:
\begin{equation}
\int \d\mu_{\text{D3}} \d\mu_{\text{\Dmi{}}}\,e^{-S_{\text{D3}}-S_{\text{\Dmi{}}}}=\int \d z \d \bar z \d \psi\,e^{-S_{\text{\Dmi{},eff}}} \, ,
\label{eqn:S.eff}
\end{equation}
where $z$ is the modulus defined in \eqref{eqn:z} and $\psi$ is its fermionic superpartner.
On the \RHS\ we are interested only in the bosonic part of the effective action, hence we can safely set $\psi=0$ as anticipated in \eqref{eqn:SD-1mic}.
The bosonic part of $S_{\text{\Dmi{},eff}}$ is equal to the twisted supergravity field $\gamma$, up to numerical factors. One way
to see this is by thinking of the fractional probe \Dmi{}-brane as a D1-brane wrapped on the
exceptional cycle $\Sigma$. For zero world-sheet gauge field, the Euclidean action of such an object is:
\begin{multline}
\frac{1}{\ls^2}\left(\int_{\Sigma}\mathrm{d}^2\xi\,e^{-\Phi}\sqrt{\det\left[P\left(G+B_2\right)\right]}-
i\int_{\Sigma} P\left(C_0\,B_2+C_2\right)\right) \\
=-\frac{i}{\ls^2}\int_{\Sigma}\left(C_2+\left(C_0+i\,e^{-\Phi}\right)B_2\right)\,.
\end{multline}
Clearly, combining with the definition in \eqref{eqn:g.def}, we can write:
\begin{equation}
S_{\textrm{\Dmi{},eff}}=-2\pi i\,\gamma\,.
\label{eqn:S.g}
\end{equation}
We take the relation \eqref{eqn:S.g} as defining the twisted supergravity field outside the supergravity regime.
The first step to compute \eqref{eqn:S.eff} is to integrate out the fields that correspond to the degrees of freedom of the
\Dmi{}-D3 strings. In general this is done using large $N$ vector-model techniques (see 
\cite{Ferrari:2013pq,Ferrari:2013hg,Ferrari:2013wla} for examples and \cite{Ferrari:2013aba}
for a more general philosophy). In our case the integration can be done very simply since the action \eqref{eqn:SD-1mic} is quadratic in the moduli to be integrated out:
$q^{\alpha},\tilde{q}^{\alpha},\chi^0,\chi^1,\chi^2,\chi^3,\tilde{\chi}^0,\tilde{\chi}^1,
\tilde{\chi}^2,\tilde{\chi}^3$. Taking into account that the moduli with (without) a tilde are $(N+M)\times 1$ ($1\times(N+M)$) matrices, $\varphi_1$ and $\varphi_1^\dagger$ are adjoint fields with $(N+M)\times(N+M)$ components and $\phi,\phi^\dagger$ are $\CC$-number moduli, we can write the quadratic part of the action as 
\begin{equation}
S_{\text{\Dmi{}}}\supset\frac{1}{2}\left(q_{\alpha i}B\indices{^i_j}\tilde{q}^{\alpha j}+
\boldsymbol{\chi}_AF\indices{^A_B}\boldsymbol{\tilde{\chi}}^{B}\right)\,,
\end{equation}
where $\alpha=1,2$; $i,j$ go from $1$ to $N+M$ and $A,B$ go from $1$ to $4(N+M)$ because we have grouped the fermions as
\begin{align}
  \boldsymbol{\chi}&=\left(\chi^0_1,\ldots,\chi^0_{N+M},\chi^1_1,\ldots,\chi^2_1,\ldots,\chi^3_1,\ldots,\chi^3_{N+M}\right)\, ,\nonumber \\
  \boldsymbol{\tilde{\chi}}&=\left(\tilde{\chi}^{0 1},\ldots,\tilde{\chi}^{3 N+M}\right)^{\operatorname T} \, .
\label{eqn:chi}
\end{align}
The matrices $B$ and $F$ can be read from \eqref{eqn:SD-1mic}:
\begin{gather}
B\indices{^i_j}=\frac{1}{2}\left({\varphi_1}\indices{^i_k}{\varphi_1^\dagger}\indices{^k_j}+{\varphi_1^\dagger}\indices{^i_k}{\varphi_1}\indices{^k_j}\right)+\phi_1^\dagger\phi_1\delta\indices{^i_j}-\phi_1^\dagger{\varphi_1}\indices{^i_j}-{\varphi^\dagger}\indices{^i_j}\phi_1\,,\\
F=\left(\begin{array}{cccc}
0 & \phi_1^{\dagger}\mathbf{1}-\varphi_1^{\dagger} & 0 & 0\\
-\phi_1^{\dagger}\mathbf{1}+\varphi_1^{\dagger} & 0 & 0 & 0\\
0 & 0 & 0 & -\phi_1\mathbf{1}+\varphi_0 \\
0 & 0 & \phi_1\mathbf{1}-\varphi_0 & 0
\end{array}\right)\,,
\end{gather}
where we have written $F$ in $(N+M)\times(N+M)$ blocks and $\mathbf{1}$ represents the identity in each of these blocks. We notice that we can represent $B$ in matrix form as
\begin{equation}
B=\left(\phi_1\mathbf{1}-\varphi_1\right)\left(\phi_1^{\dagger}\mathbf{1}-\varphi_1^{\dagger}\right)+[\varphi_1,\varphi_1^\dagger]\,.
\label{eqn:B}
\end{equation}
As we did for the bifundamental fields, we can drop the last term $[\varphi_1,\varphi_1^\dagger]$ in \eqref{eqn:B} because it vanishes inside all correlators by the D-flatness condition for the gauge group $\SU(N+M)_1$ (with bifundamentals set to zero).
The result of the integration of the $q$ and $\chi$ moduli will be given by the ratio $\det F/\det B^2$ of the determinants of $F$ and $B$ (squared, because of the two indices of $q_\alpha$ and $\tilde q^\alpha$), which we can readily compute:
\begin{align}
\det\left(B\right)^2&=\det\left(\phi_1\mathbf{1}-\varphi_1\right)^2\det\left(\phi_1^{\dagger}\mathbf{1}-\varphi_1^{\dagger}\right)^2\,,\label{eqn:referee}\\
\det\left(F\right)&=\det\left(\phi_1\mathbf{1}-\varphi_0\right)^2\det\left(\phi_1^{\dagger}\mathbf{1}-\varphi_1^{\dagger}\right)^2\,.
\end{align}
Notice that when taking the quotient the dependence of the resulting expression on the D3 fields will be holomorphic, because the factors with $\phi_1^{\dagger}\mathbf{1}-\varphi_1^{\dagger}$ cancel between the bosonic and fermionic determinants. 
This will be key to performing the functional integral over the four-dimensional degrees of freedom. From
\eqref{eqn:S.eff}, taking into account also the first term in \eqref{eqn:SD-1mic}, we can write the integral to be performed as
\begin{multline}
\int\d\mu_{\text{D3}} \d\mu_{\text{\Dmi{}}}\,\frac{\det\left(\phi_1\mathbf{1}-\varphi_0\right)^2}{\det\left(\phi_1\mathbf{1}-\varphi_1\right)^2}
\,e^{2\pi i\tau_1-S_{\text{D3}}}\\
=\int \d\phi_1\d\bar\phi_1\d\psi\,e^{2\pi i\tau_1}
\left\langle\frac{\det\left(\phi_1\mathbf{1}-\varphi_0\right)^2}{\det\left(\phi_1\mathbf{1}-\varphi_1\right)^2}\right\rangle_{D3}\,.
\label{eqn:S.eff.2}
\end{multline}
If we use the identification \eqref{eqn:z} and we also rescale the fields $\varphi_i\to Z_i=\ls^2\varphi_i$,
we can write the following expression for $\gamma$ in terms of correlators in the quiver gauge theory:
\begin{equation}
e^{2\pi i\,\gamma}=e^{2\pi i\tau_1}\left\langle\frac{\det\left(z-Z_0\right)^2}{\det\left(z-Z_1\right)^2}\right\rangle=
e^{2\pi i\tau_1}\frac{\left\langle\det\left(z-Z_0\right)\right\rangle^2}{\left\langle\det\left(z-Z_1\right)\right\rangle^2}\,,
\label{eqn:g.corr}
\end{equation}
where we have used chiral factorization in the second equality, and we drop the explict $\mathbf{1}$ from now on.
This is a beautiful formula illustrating the emergence phenomenon, showing how the profile of the 
twisted supergravity field emerges from a ``microscopic'' quantity. What is even more striking is the
fact that we can compute these chiral correlators exactly. This is possible thanks to the recent remarkable
works \cite{Nekrasov:2012xe,Fucito:2012xc}, that extended the Seiberg-Witten technology to $\nn=2$
quivers.

Before writing the exact expression for the correlators, let us make contact with the formula
\eqref{eqn:Lerdaetal} obtained non-perturbatively  by string theory techniques \cite{Billo:2012st}. The computation of the $(N+M)\times (N+M)$ determinants of operators in \eqref{eqn:g.corr} entails a regularization scheme.
The natural way to define them is via the Fredholm determinant:
\begin{equation}
\det\left(z-Z\right)=\exp\left[\tr\log\left(z-Z\right)\right]\,.
\label{eqn:Fredholm}
\end{equation}
In this formula, both the exponential and the logarithm are to be understood as defined by their Taylor
series. When we act with a \VEV\ on the \LHS\ of \eqref{eqn:Fredholm}, because of chiral factorization,
on the \RHS\ we can act with the \VEV\ directly in the argument of the exponential.
So more explicitly for the case that concerns us, we write 
\begin{equation}
\left\langle\det\left(z-Z_a\right)\right\rangle^2=\exp\left[2\left\langle\tr\log\left(z-Z_a\right)\right\rangle\right]\,.
\label{eqn:Fredholm.VEV}
\end{equation}
Using \eqref{eqn:Fredholm.VEV} in \eqref{eqn:g.corr}, plus the fact that $\gamma^{(0)}$ equals the bare coupling $\tau_1$, we easily recover \eqref{eqn:Lerdaetal}.

This was to be expected, but the reader might be befuddled by the following puzzling aspect: the
computation of the twisted supergravity profile in \cite{Billo:2012st}, leading to \eqref{eqn:Lerdaetal},
involves resumming a series of string amplitudes encoding the interaction among D3 and \Dmi{} branes.
We are instead performing a simple Gaussian integration to arrive at the result.
The authors of \cite{Billo:2012st} essentially follow the opposite approach to ours.
They want to obtain non-perturbative corrections to the $\gamma$-profile \eqref{eqn:g.pert} by adding $k$ fractional \Dmi{}-branes to the D3-brane set-up (yielding a $\U(k)$ non-Abelian generalization of \eqref{eqn:SD-1mic}) and integrating them out.
On the one hand, these branes couple to $\gamma$. 
On the other hand, they can be interpreted as gauge theory instantons, relating in this way instanton corrections in gauge theory to corrections to the $\gamma$-profile.
Resumming the contributions for all values of $k$ then yields \eqref{eqn:Lerdaetal}.
On the contrary, our approach is to keep the \Dmi{}-brane and integrate out the D3-branes, yielding immediately the full gauge-theory correlator in \eqref{eqn:S.eff.2} with no need to make explicit nor resum the instanton series that contributes to it.

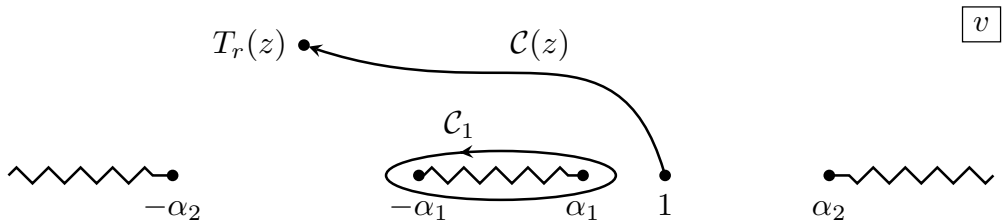
\begin{figure}
    \centering\usetikzlibrary{arrows,decorations.pathmorphing,positioning}
\begin{tikzpicture}
  [
  scale=1.8,
  every path/.style={thick},
  cut/.style={draw,decoration=zigzag, decorate},
  point/.style={shape=circle,draw,fill,scale=0.3},
  contour/.style={draw,->,>=stealth},
  ]
  \node (corner) at (3,1){} node[draw,thin] at (corner.south west) {$v$};
  \node[point] (-a2) at (-2,0) {} node [below=0.1 of -a2] {$\smash{-}\alpha_2$};
  \node[point] (a2) at (2,0) {} node [below=0.1 of a2] {$\alpha_2$};
  \node[point] (-a1) at (-0.5,0) {} node [below=0.1 of -a1] {$\smash{-}\alpha_1$};
  \node[point] (a1) at (0.5,0) {} node [below=0.1 of a1] {$\alpha_1$};
  \node[point] (one) at (1,0) {} node [below=0.1 of one] {$\smash{1}\vphantom{\alpha_1}$}; 
  \node[point] (z) at (-1.2,0.8) {} node [left=0 of z.west] {$T_r(z)$};
  \path[cut] (-3,0) -- (-a2);
  \path[cut] (3,0) -- (a2);
  \path[cut] (-a1) -- (a1);
  \path[contour] (one) .. controls (0.7,1) and (0,0.4) .. (z) node [above,midway] {$\mathcal C(z)$};
  \draw (0,0) ellipse (0.7 and 0.15);
  \path[contour] (110:0.7 and 0.15) arc(110:111:0.7 and 0.15) node [above=\lineskip] {$\mathcal C_1$};
\end{tikzpicture}
    \caption{
       The data specifying the integral \eqref{eqn:gammaintv} on the Riemann sphere $\CC\cup\{\infty\}$ with coordinate $v$.
       The integrand has two branch cuts: between $-\alpha_1$ and $\alpha_1$ and between $-\alpha_2$ and $\alpha_2$ through $\infty$.
       The ambiguity in the choice of $\mathcal C(z)$ is characterized by the integral \eqref{eqn:gammaCCp} along $\mathcal C_1$.
     }
    \label{fig:intgamma}
\end{figure}

Let us state now the final expression for the correlator in \eqref{eqn:g.corr}, leaving
all the details on how to extract it from \cite{Nekrasov:2012xe,Fucito:2012xc} for the appendix.
As usual when one deals with instantons, the result is more conveniently expressed in terms of the
variables:
\begin{equation}
\qq_a= e^{2\pi i\,\tau_a}=e^{-\frac{8\pi^2}{g_a^2}}\,e^{i\,\vartheta_a}\,,\qquad \qq=\qq_0\,\qq_1\,.
\label{eqn:qs}
\end{equation}
A contribution from a $k$-instanton of type $a$ comes with a factor $\qq_a^k$. Recalling
from the previous section that a point on the Coulomb branch of the quiver theory is specified by the
quotient of two monic polynomials $T_r=T_0/T_1$; in such a vacuum our correlator turns out to be:
\begin{equation}
2\pi i\left(\gamma-\gamma^{(0)}\right)=\beta(\qq_a)\int_z^{\infty}\mathrm{d}x
\,\frac{T'_r(x)}
{\sqrt{T_r(x)^2-\alpha_1(\qq_a)^2}\sqrt{T_r(x)^2-\alpha_2(\qq_a)^2}}\,,
\label{eqn:corr}
\end{equation}
where $\gamma^{(0)}=\tau_1$ and the precise definitions of $\alpha_1,\alpha_2$ and $\beta$ can be found in the appendix. For the discussion
that follows, it is enough to know that these quantities are well-behaved functions admitting a small
$\qq_a$ expansion:
\begin{equation}
\begin{aligned}
\beta(\qq_a)&=-\frac{i}{\sqrt{\qq_1}}\left(1+\qq_1-\qq_0+6\qq+\qq_0^2+\OO\left(\qq_a^3\right)\right)\,,\\
\alpha_1(\qq_a)&=2\sqrt{\qq_0}\left(1+\qq_1-\qq_0-6\qq+\qq_0^2+\OO\left(\qq_a^3\right)\right)\,,\\
\alpha_2(\qq_a)&=\frac{1}{2\sqrt{\qq_1}}\left(1+\qq_1-\qq_0+10\qq+\qq_0^2+\OO\left(\qq_a^3\right)\right)\,.
\end{aligned}
\label{eqn:baa}
\end{equation}
It is generally more convenient to change integration variables in \eqref{eqn:corr} to $v=T_r(x)$,
\begin{equation}
2\pi i\left(\gamma-\gamma^{(0)}\right)=-\beta\,\int_1^{T_r(z)}
\frac{\mathrm{d}v}{\sqrt{(v^2-\alpha_1^2)(v^2-\alpha_2^2)}}\,.
\label{eqn:gammaintv}
\end{equation}
The contour of integration in \eqref{eqn:gammaintv} and the branch cut structure of the integrand are represented in figure~\ref{fig:intgamma}.
The contour of integration $\mathcal C(z)$ must go from $v=1$ (corresponding to $z=\infty$) to $v=T_r(z)$ without crossing any branch cuts, but is otherwise arbitrary.
This does not fix $\gamma(z)$ unambiguously: one can choose a contour encircling the branch cut between $-\alpha_1$ and $\alpha_1$ an arbitrary number of times and the value of the integral will depend on this number.
If the contour $\mathcal C'(z)$ makes one more counter-clockwise circle around the cut than $\mathcal C(z)$, the difference in the resulting $\gamma$ functions is
\begin{equation}
  2\pi i (\gamma_{\mathcal C'}-\gamma_{\mathcal C})=-\beta\,\int_{\mathcal C_1}
\frac{\mathrm{d}v}{\sqrt{(v^2-\alpha_1^2)(v^2-\alpha_2^2)}} = 4\pi i\, .
  \label{eqn:gammaCCp}
\end{equation}
This has no physical significance as is obvious from \eqref{eqn:g.corr}. From
\eqref{eqn:g.def} we can also see that such a shift corresponds to a shift of $c$, or equivalently
a $4\pi$-shift of the $\vartheta$-angle in the field theory, which is of course not observable.
Similarly, the path can encircle the other branch cut an arbitrary number of times.
Since $\mathcal C_1$ can be continuously deformed into a path going around the branch cut between $\alpha_2$ and $-\alpha_2$ clockwise, this also corresponds to a non-observable $4\pi$ shift in the $\vartheta$-angle.

As a quick check of the formula \eqref{eqn:gammaintv}, we can recover the perturbative result \eqref{eqn:g.pert}.
What we have to do is to send $\qq_a\to0$, keeping only the leading order. Then
\begin{equation}
\beta\to-\frac{i}{\sqrt{\qq_1}}\,,\qquad\sqrt{v^2-\alpha_1^2}\to v\,,\qquad\sqrt{v^2-\alpha_2^2}\to\frac{i}{2\sqrt{\qq_1}}\,,
\end{equation}
which gives the trivial integral:
\begin{equation}
\pi i\left(\gamma-\gamma^{(0)}\right)=\int_1^{T_r(z)}\frac{\mathrm{d}v}{v}=
\log\left(T_r(z)\right)\, .
\end{equation}
When we use \eqref{T0T1def}, this is precisely the perturbative formula \eqref{eqn:g.pert} we were expecting. 

A less trivial check is to send only $\qq_0\to0$, but keep $\qq_1$ arbitrary.
This corresponds to suppressing the dynamics of the type~0 gauge group, which in this limit plays the role of a global flavor group.
Hence the theory one obtains is $\nn=2$ \<SQCD> with $2M$ flavors\footnote{Recall that $\gamma$ is completely insensitive to the $N$ regular branes.} on the Coulomb branch.
This is exactly the regime considered in \cite{Billo:2012st,Martucci:2012jk} and we can compare our formula \eqref{eqn:gammaintv} for $\gamma$ in this limit with theirs.
Setting $\qq_0=0$, the expansions \eqref{eqn:baa} truncate to
\begin{equation}
  \beta(0,\qq_1)=-\frac{i}{\sqrt{\qq_1}}(1+\qq_1) \, , \quad \alpha_1(0,\qq_1)=0, \, \quad \alpha_2(0,\qq_1)=\frac{1}{2\sqrt{\qq_1}}(1+\qq_1) \, .
  \label{eqn:baa.SQCD}
\end{equation}
The integral in \eqref{eqn:gammaintv} then reduces to
\begin{align}
  2\pi i\left(\gamma-\gamma^{(0)}\right)&=-\beta(0,\qq_1)\int_1^{T_r(z)}
  \frac{\mathrm{d}v}{v\sqrt{v^2-\alpha_2(\qq_1,0)^2}} \nonumber \\
  &=\frac{i\beta}{2\alpha_2}\left. \log \left( \frac{1-\sqrt{1-v^2/\alpha_2^{2}}}{1+\sqrt{1-v^2/\alpha_2^{2}}} \right) \right|_1^{T_r(z)}\, .
  \label{eqn:gammaSQCDcomp}
\end{align}
Using \eqref{eqn:baa.SQCD} and some elementary algebra, the lower bound contribution is found to be $-\log \qq_1=-2\pi i \gamma^{(0)}$, hence
\begin{equation}
  2\pi i\,\gamma(z)
  = \log \left( \frac{1-\sqrt{1-T_r(z)^2/\alpha_2^{2}}}{1+\sqrt{1-T_r(z)^2/\alpha_2^{2}}} \right) \, ,
  \label{eqn:gammaSQCD}
\end{equation}
in perfect agreement with the result of \cite{Billo:2012st,Martucci:2012jk}.

\section{Large \texorpdfstring{\textit{\textrm N}}{N} limit. The enhançon}
\label{sec:enhancon}

The expression \eqref{eqn:gammaintv} we wrote for the twisted supergravity field $\gamma$ (taking \eqref{eqn:g.corr} as its definition) is completely general, since
it has been derived from the field theory in full non-perturbative glory. In particular, it is valid all along the
RG flow for any point on the Coulomb branch, for any value of the couplings and any integer numbers $N,M$.
Looking at it from the string theory
perspective, it means that \eqref{eqn:gammaintv} contains all $g_s$ and $\alpha'$ corrections to the dynamics
of the brane array we are considering. However, for the time being we are only interested in using a small fraction
of this power. We consider small $g_s$ and large $N,M$, corresponding to the supergravity regime. For
convenience, we assume that the two bare gauge couplings are equal and that $N$ is proportional to $M$:
\begin{equation}
  \qq_0=\qq_1=e^{-\frac{\pi}{g_s}} \, , \quad N=p\,M\,,\quad p\in\mathbb{Q}\,.
\label{eqn:NM}
\end{equation}
As we discussed in section \ref{ssec:enh}, in this regime a curious phenomenon is taking place, that of the
enhançon. While the field-theoretical mechanism behind it is understood (recall it has to do with the impossibility
of bringing the roots of the Seiberg-Witten curve to the origin of the moduli space) and its effect in the
supergravity background (the need for an excision procedure below a certain scale) is also well-known, as far as we know
there is no fully general construction in the literature explaining the interplay of these aspects. We hope to
fill this gap here. The idea is to solve the integral \eqref{eqn:gammaintv} for different vacua, characterized by different functions
$T_r$, and analyze their large $M$ limit.

The large $M$ limit corresponds to taking $M\to\infty$ and $\qq_a\to0$, keeping the 't Hooft couplings $\lambda_a$ defined by \eqref{eqn:holo1} fixed, which translates to keeping
\begin{equation}
  \qq_a^{\frac{1}{M}}=e^{-\frac{8\pi^2(p+1)}{\lambda_a}}=e^{-\frac{\pi}{g_s M}} \quad\text{fixed} \, .
  \label{eqn:qalargeM}
\end{equation}
If we furthermore wanted to suppress the $\alpha'$ corrections and obtain two-derivative gravity we should take the limit $\lambda_a\to\infty$ in which \eqref{eqn:qalargeM} goes to~$1$.
We will however refrain from taking this limit, as it eliminates the separation between the scale $|z_0|$ at which the theory is Higgsed and the enhançon scale which, as we will see, is $\sim\qq_1^{l/M}|z_0|$ for some finite number $l$ that does not scale with $M$.

In the large large $M$ limit, $\alpha_1\to0$ by \eqref{eqn:baa}, and it seems that we can replace in \eqref{eqn:gammaintv} $v^2-\alpha_1^2$ by $v^2$.
This is not always true, depending on the value of the upper bound $T_r(z)$.
If $T_r(z)$ stays at a finite distance from $\pm\alpha_1$ in the large $M$ limit, one can choose an integration contour $\mathcal C(z)$ as in figure~\ref{fig:intgamma} that stays away from the branch points $\pm\alpha_1$ and this approximation is valid.
The computation of $\gamma$ then reduces to \eqref{eqn:gammaSQCD}, where one now has to take the large $M$ limit.
In other words, the large $M$ limit of this model reduces generically (in the sense we just discussed) to the large $M$ limit of $\nn=2$ \<SQCD> with $2M$ flavors, a result which was already anticipated by \cite{Benini:2008ir} from the study of the Seiberg-Witten curve, but that we have now shown directly on the twisted supergravity field.
Whether this condition on $T_r(z)$ is satisfied depends both on the Coulomb branch vacuum encoded by $T_r$ and the specific $z$ considered.
If it fails, one needs to do a more refined analysis, similarly to what was done in \cite{Ferrari:2001mg} for pure $\nn=2$ Yang-Mills theory.
For a given vacuum $T_r$, we will call the points that satisfy the condition ``\emph{ordinary points}'' and ``\emph{exceptional points}'' the ones that do not.

\subsection{The enhançon vacuum}

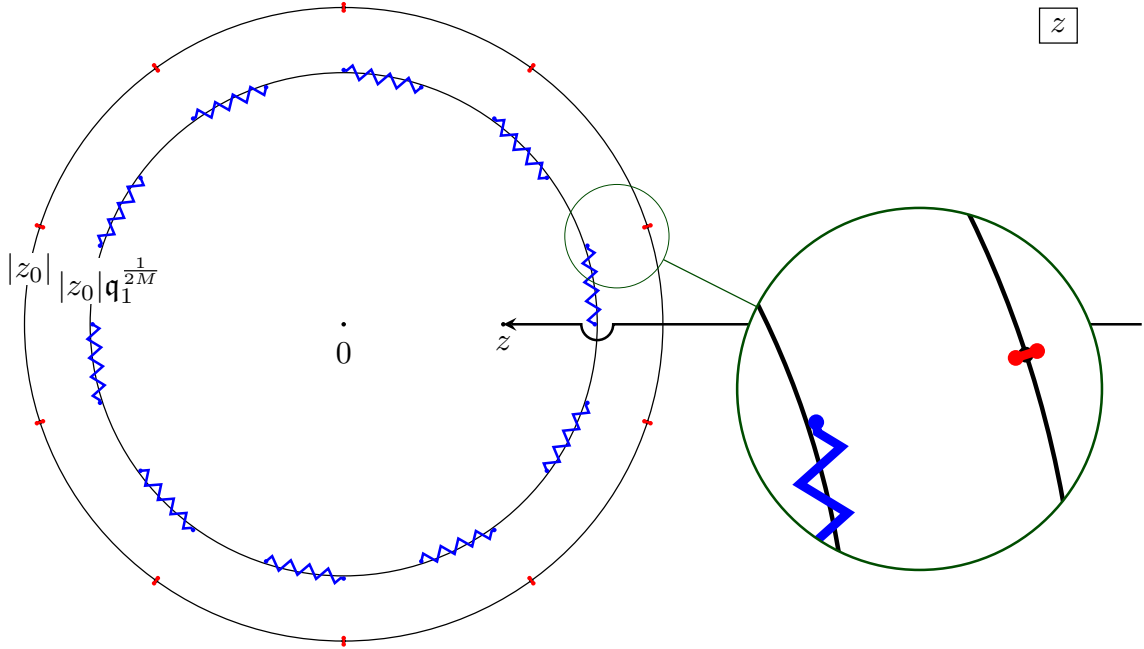
\begin{figure}[htb]
  \centering\usetikzlibrary{arrows,decorations.pathmorphing,decorations.markings,positioning,spy}
\newcommand*{\qscale}{0.1} 
\newcommand*{\numM}{10} 
\pgfmathsetmacro{\scaleaip}{0.990514} 
\pgfmathsetmacro{\scaleaim}{1.01059} 
\pgfmathsetmacro{\scaleaiip}{0.786793} 
\pgfmathsetmacro{\scaleaiim}{0.802742} 
\begin{tikzpicture}
  [
  scale=3.5,
  every path/.style={thick},
  point/.style={fill,circle,scale=0.15},
  coord/.style={thin,black,fill=white, inner sep=0},
  spy using outlines={circle, magnification=3.5, size=4cm, connect spies, every spy in node/.style={draw,fill=white}},
  ]
  \draw[coord] circle(1);
  \draw[coord] circle(0.794328);
  \node[coord,below=0.2] {$0$};
  \node[point] at (0,0) {};
  \node [draw,thin,below left,rectangle] at (2.3,1) {$z$};
  \foreach \x in {0,...,9} 
  {
    \draw[black] (18+36*\x:1) node[point] {}; 
    \draw[red] (18+36*\x:\scaleaip) node[point] {} -- (18+36*\x:\scaleaim) node[point] {};
    \draw[blue,decoration={zigzag,segment length=0.21cm,amplitude=2pt},decorate] (36*\x:\scaleaiip) node[point] {} -- (18+36*\x:\scaleaiim) node[point] {};
  }
 \draw[->,>=stealth] (2.5,0) -- (0.794328+0.05,0) arc (0:-180:0.05) -- (0.5,0) node[point] {} node[below] {$z$} ;
  \node[coord,xshift=5pt] at (170:0.794328) {$|z_0|{\mathfrak q}_1^{\frac{1}{2M}}$};
  \node[coord] at (170:1) {$|z_0|$};
  \spy[black!70!green] on (18:0.9*3.5) in node [below right] at (1.4,0.2);
\end{tikzpicture}
  \caption{The singularity structure in the $z$ plane of the integrand in \eqref{eqn:int.enh} for the enhançon vacuum with $M=10$, $\sqrt{\qq_1}=0.05$.
  Part of the figure has been enlarged for better visibility. The black dots are the roots of $T_0$ and $T_1$ corresponding to the classical positions of the fractional D3-branes. The red (blue) dots are the branch points at the scale $\sim|z_0|$ (at the enhançon radius) in the first (second) column of \eqref{eqn:branchpoints} and the very short red (zigzag blue) lines are the branch cuts joining them.
The integration path going from $\infty$ to $z$ is real except near the branch point at the enhançon radius.}
  \label{fig:enhancon}
\end{figure}

Let us first focus on arguably the simplest brane array: the one that corresponds classically to $M$ fractional
branes of type 0 distributed on a circle of radius $|z_0|$ and $M$ fractional branes of type 1 at the origin, where the
$N$ regular branes sit too. Of course, as we have already mentioned, this picture is corrected non-perturbatively,
where anyway it does not make sense to talk about brane positions. The way we characterize the configuration is by:
\begin{equation}
T_0=z^N\left(z^M+z_0^M\right)\,,\qquad T_1=z^{N+M}\,\quad\implies\quad T_r=1+\left(\frac{z_0}{z}\right)^M\,.
\label{eqn:enh.vac}
\end{equation}
Plugging this $T_r$ into the formula for $\gamma$ \eqref{eqn:gammaintv}, and using the rescaled variable $u=\frac{z_0}{z}$,
we can write
\begin{equation}
2\pi i\left(\gamma-\gamma^{(0)}\right)=-\beta\int_0^{\frac{z_0}{z}}\mathrm{d}u\,
\frac{M\,u^{M-1}}{\sqrt{(u^M+1)^2-\alpha_1^2}\,\sqrt{(u^M+1)^2-\alpha_2^2}}\,,
\label{eqn:int.enh}
\end{equation}
where we should recall that $\beta$, $\alpha_1$ and $\alpha_2$ depend on $\qq_a$. The contour of integration
must be chosen so as not to cross any branch cuts. Let us take for definiteness $z_0,z,\qq_a\in\RR$. Recall
that we want to work with small $g_s$, or equivalently small $\qq_a$, subject to the condition \eqref{eqn:qalargeM}. Given the
expansions in \eqref{eqn:baa}, we see that the branch points are located at
\begin{equation}
\begin{aligned}
&z=z_0\left|1+\alpha_1\right|^{-1/M}\omega^{k+\frac{1}{2}}\,, &&z=z_0\left|\alpha_2+1\right|^{-1/M}\omega^k\,,&&k=0\,\ldots M-1\,,\\
&z=z_0\left|1-\alpha_1\right|^{-1/M}\omega^{k+\frac{1}{2}}\,, &&z=z_0\left|\alpha_2-1\right|^{-1/M}\omega^{k+\frac{1}{2}}\,,&&k=0\,\ldots M-1\,,
\end{aligned}
\label{eqn:branchpoints}
\end{equation}
with $\omega=e^{\frac{2\pi i}{M}}$ an $M$-th root of unity. We take the branch cuts to link the branch points
sharing a column in \eqref{eqn:branchpoints}. These branch cuts are represented in figure~\ref{fig:enhancon}.
The first set of $2M$ branch
points are very close to the classical roots $z_0\omega^{k+\frac12}$. The length of these branch cuts is approximately
$\frac{2z_0\alpha_1}{M}$, which, using the identifications \eqref{eqn:holo1} and \eqref{eqn:qs}, is seen to be of order
$\OO\left(\frac{1}{M}\,e^{-\frac{4\pi^2(p+1)}{\lambda}M}\right)$. Such exponentially small branch cuts in the large $M$ limit can be associated with sharply localized D-branes. 
This matches precisely the expectations coming from supergravity, that is, to find $M$ D3-branes of type 0 at
the positions $z_0\,w^{k+\frac12}$. The second set of $2M$ branch points is quite different. They fill homogeneously a pair of circles of radius
\begin{equation}
|z_0|\left(\frac{1}{2\sqrt{\qq_1}}\pm1\right)^{-\frac{1}{M}}\approx|z_0|\,\qq_1^{\frac{1}{2M}}=|z_0|\,e^{-\frac{\pi}{2g_sM}}\,,
\label{eqn:enh.FT}
\end{equation}
and the distance between consecutive branch points is of order $\OO\left(\frac{1}{M}\right)$. Notice that the scale above
is exactly the enhançon scale \eqref{eqn:rho_1} arising from supergravity considerations. We will comment more on
this below. The length of the branch cuts is now of the same order as the separation between the cuts and the concept
of D3-brane is lost at this scale.

In view of \eqref{eqn:branchpoints}, we see that we can take the contour with $u$ real in \eqref{eqn:int.enh}. This
will not cross any branch cut. For small enough $z$, this path hits the branch point at $|z_0|\,e^{-\frac{\pi}{2g_sM}}$
but we can always go just below the cut as in figure~\ref{fig:enhancon}.

Let us now study the large $M$ limit of the integral \eqref{eqn:int.enh}.
The first step is to identify the exceptional points. By definition, they coincide with the branch points in the first column of \eqref{eqn:branchpoints} in the large $M$ limit and by consequence, the formula \eqref{eqn:gammaSQCD} does not apply to them.
Since in the large $M$ limit, the branch points densely fill the ring of radius $|z_0|$, the points with $|z|=|z_0|$ are exceptional.
All the points with fixed $z, |z|\neq|z_0|$ are then ordinary.
A more general way to construct an exceptional point is to scale its coordinate in the large $M$ limit, taking $z=z_b+w/M$ with $z_b$ a branch point in the first column of \eqref{eqn:branchpoints} and $w$ fixed in the large $M$ limit.
We will however not pursue this possibility here, since we are interested in the profile of $\gamma$ as a function of $z$ and the enhançon mechanism which happens at a scale \eqref{eqn:enh.FT}, well separated from $|z_0|$ at finite 't~Hooft coupling.

For ordinary points, the formula \eqref{eqn:gammaSQCD} is valid and we have
\begin{equation}
  2\pi i\, \gamma(z)=
  \log \left( \frac{1-\sqrt{1-T_r(z)^2/\alpha_2^{2}}}{1+\sqrt{1-T_r(z)^2/\alpha_2^{2}}} \right)\, , \quad T_r(z)=1+ \left( \frac{z_0}{z} \right)^M , \quad \frac{1}{\alpha_2^2}=4\qq_1 \, .
  \label{eqn:easy}
\end{equation}
Since $\alpha_2$ is very large, we can simplify further the expression for $\gamma$. But
for that, we have to be careful with the range of $z$. Let us distinguish three regions:

\paragraph{Region ${|z|>|z_0|}$ :}
This region corresponds to the \<UV> of the field theory, where we have a gauge group $\SU(N+M)\times\SU(N+M)$
and we expect conformality. Therefore $\gamma$ should not run. Here $|T_r(z)|\ll|\alpha_2|$, and we can approximate:
\begin{equation}
\sqrt{1-T_r(z)^2/\alpha_2^2}\approx 1 - \frac{T_r(z)^2}{2\alpha_2^2} \,.
\label{eqn:approx}
\end{equation}
Plugging this into \eqref{eqn:easy}, we obtain for the twisted field
\begin{equation}
\gamma=\gamma^{(0)}-\frac{i}{\pi}\log\left[1+\left(\frac{z_0}{z}\right)^M\right]\,.
\label{eqn:g.p.enh}
\end{equation}
For large $M$, we can neglect the second term inside the logarithm, and we obtain the desired result:
\begin{equation}
\gamma=\gamma^{(0)}=\frac{i}{2g_s}\,.
\end{equation}

\paragraph{Region ${|z_0|>|z|>\rho_1}$ :}
This is the region where $|T_r(z)|\ll|\alpha_2|$ still holds. Recall from \eqref{eqn:enh.FT}
that its boundary is at the scale $\rho_1$ of \eqref{eqn:rho_1}, where supergravity was predicting
the enhançon phenomenon. Given that the approximation is the same one used above, we can follow the
reasoning there, to arrive to the expression \eqref{eqn:g.p.enh}. In this case though, the term we 
have to neglect inside the logarithm is the first one. This gives
\begin{equation}
\gamma=\gamma^{(0)}+\frac{i\,M}{\pi}\log\frac{z}{z_0}\,.
\end{equation}
Again no surprises here. This is the expected result \eqref{eqn:g.enh}.

\paragraph{Region ${|z|<\rho_1}$ :}
This is the region where the supergravity background ceases to be trustable. It has been argued in the literature that one must
excise the supergravity background, leaving a constant $\gamma$ inside. Here we can \textit{prove} that this is indeed the
way the non-perturbative dynamics of the gauge theory translate onto the string side:

When we use that $|z|<\rho_1\implies |T_r(z)|\gg |\alpha_2|$ in this region (recall we are at large $M$), equation
\eqref{eqn:easy} becomes
\begin{equation}
2\pi i\,\gamma = - i\pi \, .
\label{eqn:jump}
\end{equation}
This corresponds to the result we would expect from an excision procedure, where the profile of $\gamma$
is frozen in the excised region to a constant value.
Notice however that $\gamma$ inside the enhançon is not its value at the enhançon radius as is often implied in the supergravity literature, but its real part jumps discontinuously in the large $M$ limit.
This cannot be seen of course in the supergravity analysis and is a purely stringy effect.
The jump of the real part of $\gamma$ has already been observed in \cite{Cremonesi:2009hq}.
\bigskip

Let us recapitulate the lessons learnt in this subsection about the enhançon vacuum. First of all, the $N$ regular
D3-branes just come along for the ride and play absolutely no role. This
just reflects the fact that they do not couple to $\gamma$. Second, the roots of the Seiberg-Witten curve
of the microscopic model enter in the game via the defining equation \eqref{eqn:int.enh}. The analysis of the
branch cuts allows us to discover that while at the first radius with branch cuts, at the scale $|z_0|$, there
are fractional D3-branes present (the branch cuts are exponentially short); at the second radius, at the scale
$\rho_1$, the branch cuts are longer and the brane interpretation no longer holds. Finally, we see that inside this
second region of branch cuts, i.e.\ inside the enhançon, the profile of the supergravity twisted field is constant. This agrees with the
\emph{ad hoc} procedure developed in the supergravity literature to cure the repulson singularity of the background, called
excision. Here everything follows from the quantum properties of the microscopic underlying theory.

\subsection{A generic vacuum}
\label{sec:genvac}

A natural question that comes to mind is how the picture we have obtained for the enhançon vacuum changes when
we consider different vacua. The answer is that, morally, nothing changes. The essential parts of the
discussion above apply as well for different (suitable) choices of $T_r$. Let us be a bit more precise.

Take generic polynomials $T_0$, $T_1$ as in \eqref{T0T1def}; $T_r$ is the quotient $T_0/T_1$. If $T_0$ and $T_1$ share any root, this root factors out of $T_r$ and plays no role.
Indeed, that would indicate the presence of a regular brane, which is invisible to $\gamma$.
Thus, we suppose that we have $M$ pairs of different roots, and we let $r$ of the roots of $T_1$ be zero, corresponding in the perturbative picture to $r$ fractional branes of type 1 sitting at the origin.
Let us look at the branch cuts of the integrand in \eqref{eqn:corr}.
The branch points coming from the first square root solve one of the equations
\begin{equation}
T_r=\frac{\prod_{i=1}^{M}\left(z-\tilde{z}_i\right)}
{\prod_{j=1}^{M}\left(z-z_j\right)}=\pm\alpha_1\,.
\end{equation}
Since $\alpha_1$ is very small, for finite $\tilde{z}_i$ the solutions to this equation are
$z\sim\tilde{z}_k\pm\alpha_1\frac{\prod_{i\neq k}(\tilde{z}_k-\tilde{z}_i)}{\prod_j(z-z_j)}\sim\tilde{z}_k$. The branch
cuts are exponentially short. The story for the branch cuts coming from the second square root is different:
\begin{equation}
T_r=\frac{\prod_{i=1}^{M}\left(z-\tilde{z}_i\right)}
{\prod_{j=1}^{M}\left(z-z_j\right)}=\pm\alpha_2\,.
\end{equation}
Given that $\alpha_2$ is very big, now we will have $M-r$ branch points of the form
$z\sim z_l\pm\alpha_2^{-1}\frac{\prod(\tilde{z}_k-\tilde{z}_i)}{\prod_{j\neq l}(z-z_j)}\sim z_l$, with $l\neq z$ and
short branch cuts associated. Regarding the remaining $r$ branch points, it is easy
to see that they will distribute homogeneously on a ring at the scale
\begin{equation}
|z|\approx \frac{\prod_{i=1}^{M}|\tilde{z}_i|^{\frac{1}{r}}}{\prod_{j=1}^{M-r}|z_j|^{\frac{1}{r}}}|4\qq_1|^{\frac{1}{2r}}=
\left(\frac{2\prod_{i=1}^{M}|\tilde{z}_i|}{\prod_{j=1}^{M-r}|z_j|}\right)^{\frac{1}{r}}\,e^{-\frac{\pi}{2g_s\,r}}\,.
\label{eqn:gen.enh}
\end{equation}
This distribution yields ``large'' branch cuts. 
In order to have a shell of branch cuts that will induce an enhançon mechanism, $r$ must be big.
Since we are taking $g_s$ to be small, we see that only if $r$ is of order $M$, the enhançon
phenomenon will be noticeable. In other words, as we already knew, the enhançon is a large $N$ phenomenon.

In the reasoning above, to obtain the formula \eqref{eqn:gen.enh} we assumed that the $\tilde{z}_i$ and the non-zero $z_j$ were finite,
meaning that they do not vanish in the $\qq_a\to0$ limit. But this condition is actually a bit too restrictive.
The approximations $\tilde{z}_k\pm\alpha_1\frac{\prod_{i\neq k}(\tilde{z}_k-\tilde{z}_i)}{\prod_j(z-z_j)}\sim\tilde{z}_k$
and $z_l\pm\alpha_2^{-1}\frac{\prod(\tilde{z}_k-\tilde{z}_i)}{\prod_{j\neq l}(z-z_j)}\sim z_l$ (by $\sim$ we
mean up to exponentially suppressed corrections $\OO(e^{-(\textrm{sth})M})$) still hold if the roots
$\tilde{z}_i,z_j$ contain $\qq_a$ factors in particular ways. This is the case of the cascading vacuum, where we distribute
$(2K+1)M$ fractional branes on $2K$ shells in order to trigger the baryonic root transitions at the scales where the
perturbative gauge couplings diverge, as discussed after \eqref{eqn:rho_1}. This distribution is characterized by the polynomials
\begin{align}
T_0&=\left(z^M+z_0^M\right)\prod_{i=0}^{K-1}\left(z^{2M}+\qq^{\frac32+2i}z_0^{2M}\right)\,,\nonumber \\
T_1&=z^M\prod_{j=0}^{K-1}\left(z^{2M}+\qq^{\frac12+2j}z_0^{2M}\right)\,.
\end{align}
It was indeed checked in \cite{Benini:2008ir} that for $T_0,T_1$ of the form above, the branch cuts associated to the solutions of $z^{2M}=-\qq^{\frac32+2i}z_0^{2M}$ and $z^{2M}=-\qq^{\frac12+2j}z_0^{2M}$ are exponentially short.

\medskip

In summary, we characterize a point on the Coulomb branch by two monic polynomials, $T_0$ with roots
$\tilde{z}_i$, and $T_1$ with roots $z_j$. We take $M$ of the roots $\tilde{z}_i$ to be the solutions of
$\tilde{z}_i^M=-z_0^M$ (recall this triggers the running below $|z_0|$),
and the rest of the roots to be freely distributed inside the circle of radius $|z_0|$.
When two roots $\tilde{z}_i$ and $z_j$ coincide, this signals the presence of regular branes, about which
we cannot say anything in our approach since they do not couple to $\gamma$. Otherwise, $\gamma(z)$ has small
branch cuts at the positions $\tilde{z}_i,z_j\neq0$. They can be interpreted as fractional branes of type 0 and
type 1 respectively. The roots $z_j=0$ cannot be interpreted as localized fractional branes. When we have a large
number $r$ of them, $r\sim\OO(M)$, the enhançon mechanism takes place at the scale \eqref{eqn:gen.enh}. As can be
seen from \eqref{eqn:gammaSQCD}, $\gamma(z)$ is constant inside the enhançon region (where $|T_r(z)|\ll|\alpha_2|$),
which matches the supergravity excision procedure. Notice that this general analysis does not include
exceptional points, neither the possibility of having roots that scale arbitrarily with $\qq_a$. Although our tools are
general enough to analyse these cases, we have not pursued this direction here.

\section{Outlook}
\label{sec:conc}

As we have already emphasized throughout the chapter, our main result is the computation, using field theory
 techniques (although often phrased in a stringy language), of the exact profile of the
twisted supergravity field \eqref{eqn:corr}. From the string point of view, this formula includes all $g_s$ and $\alpha'$
corrections. This allowed us to derive directly from field theory the enhançon mechanism proposed in the supergravity
literature.

Clearly, we have not fully exploited the power of the exact result \eqref{eqn:corr}.
Its validity for any value of $N$ in particular opens the possibility of studying $1/N$ corrections.
These corrections are expected to be important for the exceptional points that are very close to the branch points in the large $N$ limit \cite{Ferrari:2001mg}.
We have not delved either into the physical meaning of the curious ``imaginary'' jump of $\gamma$ at the enhançon radius, noted in \eqref{eqn:jump}.
Since such a jump is not observable in the classical supergravity regime, maybe our techniques could help shed some light on the nature of this stringy effect.
In addition, our results can be generalized in several directions that we believe merit further investigation.
One is the generalization to more general vacua, which fall outside the regime considered in section~\ref{sec:genvac}.
Among them we would like to point out the rather mysterious enhançon bearings of \cite{Benini:2008ir}, where some roots of the polynomial $T_0$ are put at a radius which sits inside the enhançon.
Another possibility is the extension of our findings to generic ADE orbifold singularities. We expect the field theory part of the computation to involve the same techniques we have used, albeit with more complicated integrals to evaluate; the physics should be richer since a more intrincate enhançon mechanism is expected with several enhançon radii.
Although the supergravity solutions have been discovered long ago \cite{Billo:2001vg}, as far as we know, the dual field theories have not been as explored in the literature as that of their $\CC^2/\ZZ_2$ counterpart.

\chapter{The D4-brane background from D0-brane probes}
\label{chap:D4brane}

In this chapter, we discuss how the approach presented in~\cref{sec:holoprobe} for the case of D3-branes probed by \Dmi{}-branes in type \<IIB> can be generalized in another direction, namely by considering D4-branes probed by D0-branes (also called D-particles).
This chapter is based on~\cite{Ferrari:2013hg}.
One could take more generally D$p$/D$(p+4)$ along the same lines; the related work~\cite{Ferrari:2013wla} considered the D1/D5 case, already reviewed in the thesis~\cite{antothese}.

Note that the point of view adopted in this chapter is a bit different: as in~\cite{Ferrari:2012nw}, the emphasis is on the field theory rather than the D-brane construction.
The moduli~$Y_A$ are not present from the start but are introduced in order to be able to integrate out the charged moduli by making the action depend quadratically on them.
The transverse dimensions to the D4-branes can then be considered to emerge from the quantum dynamics of the system.

\section{\label{lagsec} The quantum mechanical model}

The quantum mechanical system that we consider models the dynamics of $K$ D-particles interacting with $N$ coinciding D4-branes in type \<IIA> string theory. It is related by T-duality to the \Dmi{}/D3 system considered previously and can be interpreted as a $\text{U}(K)$ gauged  supersymmetric quantum mechanics on the \<ADHM> instanton moduli space. It can be obtained as in \cite{Aharony:1997th,Berkooz:1999iz,Ferrari:2012nw} from a scaling limit, associated with the near-horizon limit in the background of the D4-branes, of the dimensionally reduced $\text{U}(K)$ $\nn=1$ super Yang-Mills theory with one adjoint and $N$ fundamental hypermultiplets from six to one dimension.
This scaling limit is similar to the one for the \Dmi{}/D3 system discussed in~\cref{sec:holoprobe}, the only difference being that we need to fix the five-dimensional coupling~$g^2_5$, rather than the string coupling itself as in the D3-brane case \cite{Itzhaki:1998dd}.

For convenience, we work in Euclidean signature. The model preserves eight supercharges. It has a $\text{U}(K)$ gauge symmetry, a $\text{U}(N)$ flavor symmetry and a $\text{SU}(2)_{+}\times\text{SU}(2)_{-}\times\text{Spin}(5)$ global symmetry corresponding to $\text{SO}(4)\times\text{SO(5)}$ rotations transverse to the worldlines of the D-particles and preserving the configuration of the background D4-branes.
The fundamental degrees of freedom in the adjoint of $\text{U}(K)$, associated with D0/D0 strings, are the D4-brane worldvolume matrix space coordinates $X_{\mu}$ and a $\text{SU}(2)_{-}$ doublet of $\text{Spin}(5)$ spinor superpartners $\psi_{\dot\alpha}$.
The D0/D4 strings yield additional degrees of freedom $(q_{\alpha},\chi)$ and $(\tilde q^{\alpha},\tilde\chi)$ in the fundamental and anti-fundamental representations of $\text{U}(N)\times\text{U}(K)$. The bosonic $q_{\alpha}$ and $\tilde q^{\alpha}$ are doublets of $\text{SU}(2)_{+}$ and the fermionic $\chi$ and $\tilde\chi$ are $\text{Spin}(5)$ spinors.
In terms of these variables, the action contains complicated four-fermion terms and the \<ADHM> constraints must be imposed in the path integral.

The four-fermion terms can be greatly simplified by introducing a non-dynamical auxiliary $\text{SO}(5)$ vector $Y_{A}$ in the adjoint of $\text{U}(K)$, whereas the \<ADHM> constraints can be implemented by using adjoint Lagrange multipliers $(D_{\mu\nu},\Lambda^{\alpha})$, where
the bosonic $D_{\mu\nu}$ is self-dual and the fermionic $\Lambda^{\alpha}$ is a $\text{SU}(2)_{+}$ doublet of $\text{Spin}(5)$ spinors.
We shall see that these auxiliary variables play a crucial role, both at the technical level to solve the model \cite{Coleman:1980nk,ZinnJustin:1998cp,Ferrari:2000wq,Ferrari:2001jt,Ferrari:2002gy} and for the physical interpretation of the solution in terms of an emerging geometry, in a way akin to the case of D-instantons \cite{Ferrari:2012nw,hep-th/9810243,hep-th/9901128,Akhmedov:1998pf}.

We are going to focus on the bosonic part of the effective action for the D-particles. We can thus set the adjoint fermions $\psi_{\dot\alpha}$ and $\Lambda^{\alpha}$ to zero. In terms of the $\text{SO}(5)$ Dirac matrices $\Gamma_{A}$, charge conjugation matrix $C$ and $\mathfrak{su}(2)_{+}$ generators $\sigma_{\mu\nu}$,
the microscopic Lagrangian we start from then reads
\begin{multline}
 L=\frac12\nabla\tilde  q^{\alpha}\nabla q_{\alpha}
  +\frac{i}2\tilde\chi C\nabla\chi
 +\frac{i}2\tilde q^{\alpha}D_{\mu\nu}\sigma_{\mu\nu\alpha}^{\phantom{\mu\nu\alpha}\beta}q_{\beta} \\
  -\frac{i}{2\ls^2}\tilde\chi C\Gamma_{A}
  Y_{A}\chi
  +\frac1{2\ls^4}\tilde q^{\alpha}Y_AY_Aq_{\alpha}
  +\frac{\sqrt{2\pi}}{\gs\ls}\tr\Bigl(\! 1+\frac{1}{2} \nabla X_\mu \nabla X_\mu\\ -\frac1{2\ls^4}[Y_A,X_\mu][Y_A,X_\mu] 
  +i[X_\mu,X_\nu]D_{\mu\nu} \Bigr)  
   \, ,
  \label{Ldef}
\end{multline}
where $\nabla$ is the worldline covariant derivative. The non-trivial normalization of the trace term in \eqref{Ldef} is fixed by the D-particle mass in type \<IIA> string theory in terms of the string coupling $\gs$ and the string length $\ls$. 

Let us note that the full open-string description of the D0/D4 system includes a priori additional terms coupling the D4-brane worldvolume fields to the D-particle worldline variables. Similar terms for the \Dmi{}/D3 system have been discussed in \cite{Green:2000ke,Billo:2002hm}. A discussion of the effect of these terms, in a non-supersymmetric context, can be found in \cite{Ferrari:2013waa}. Presently, with eight supersymmetries, these terms are expected to be irrelevant for some of the contributions in the effective action, on which we focus \cite{Ferrari:2012nw,Ferrari:2013pq}. In particular, for the kinetic term (see \eqref{gsolu}), this is consistent with the non-renormalization theorem discussed in \cite{Diaconescu:1997ut}.

The Lagrangian \eqref{Ldef} is pre-geometric, in the sense that there is no dynamical variable associated with the motion of the D-particles in directions transverse to the D4-branes. The interactions between the D-particles and the D4s are described ``abstractly'' by the pre-geometric variables $(q,\chi,\tilde q,\tilde\chi)$. Our goal is to show that the strong quantum effects generated by these non-geometric interactions literally create five new dimensions of space in which the D-particles can move. Moreover, we are going to prove that the resulting ten-dimensional spacetime behaves classically at large $N$, is curved and supports a non-trivial dilaton and three-form field, precisely matching the near-horizon D4-brane type \<IIA> supergravity background \cite{Horowitz:1991cd,Itzhaki:1998dd}. 

\section{\label{solsec} The solution of the model}

The model \eqref{Ldef} can be solved at large $N$ because the interacting degrees of freedom $(q,\chi,\tilde q,\tilde\chi)$ carry only one $\text{U}(N)$ index and thus are vector-like variables. The leading large $N$ Feynman diagrams are then multi-loop bubble diagrams which can always be summed up exactly. The well-known technical trick to elegantly perform this sum \cite{Coleman:1980nk,ZinnJustin:1998cp,Ferrari:2000wq,Ferrari:2001jt,Ferrari:2002gy} is to rewrite the complicated interactions between the vector degrees of freedom by introducing auxiliary fields, in such a way that the vector variables only appear quadratically in the action. This is exactly what we have done when writing \eqref{Ldef} in terms of $Y_{A}$ and $D_{\mu\nu}$. One then integrates exactly over these variables to obtain a non-local effective action $S_{\text{eff}}$ for the auxiliary fields. This effective action is automatically proportional to $N$. It can thus be treated classically when $N$ is large. The tree diagrams of $S_{\text{eff}}$ reproduce the leading large $N$ bubble diagrams of the original action. 

In our case, fixing the worldline $\text{U}(K)$ gauge invariance such that $\nabla=\d_{t}$ is the ordinary time derivative, the effective action reads
\begin{equation} \label{Seff1} S_{\text{eff}}(X,Y,D)=\int\!\d t\, L_{\text{{tr}}} +N\bigl(\ln \Delta_B-\ln \Delta_F\bigr)\, ,\end{equation}
where $L_{\text{tr}}$ is the trace term in \eqref{Ldef} and
\begin{align}\label{bosdet}
  \Delta_B&=\det\bigl( -\d_{t}^2+\ls^{-4}Y_AY_A+iD_{\mu\nu}\otimes\sigma_{\mu\nu} \bigr) \, , \\ \label{ferdet}
  \Delta_F&= \det\bigl(-i\d_{t}+i\ls^{-2}Y_A\otimes\Gamma_A\bigr)
\end{align}
are bosonic and fermionic functional determinants obtained by integrating out $(q,\tilde q)$ and $(\chi,\tilde\chi)$ respectively. We now claim that the classical action \eqref{Seff1} describes the motion of the D-particles in a ten-dimensional spacetime with coordinates $X_{\mu}$ and $Y_{A}$. In other words, the auxiliary variables $Y_{A}$, which have acquired dynamics through the quantum loops of the vector-like variables, can be interpreted as the coordinates of the emerging five-dimensional space transverse to the D4-branes.

To prove that this interpretation is sensible, we first integrate out $D_{\mu\nu}$ which, at large $N$, can be done by solving the saddle point equation
\begin{equation}\label{Dsaddle} \delta S_{\text{eff}}/\delta D_{\mu\nu}(t) = 0\, .\end{equation}
This yields a new effective action
\begin{equation}\label{S10d} \tilde S_{\text{eff}}(X,Y) = S_{\text{eff}}\bigl(X,Y,\langle D\rangle\bigr)\end{equation}
which will be compared in the next section 
to the non-Abelian action for D-particles in a general type \<IIA> supergravity background. 

The action $\tilde S_{\text{eff}}$ can be most conveniently analyzed by expanding around time-independent diagonal configurations,
\begin{equation}
  X_\mu(t)=x_\mu\1_{K}+\ls^2\epsilon_\mu(t) \, , \quad Y_A(t)=y_A\1_{K}+\ls^2\epsilon_A(t) \, .
  \label{matrixexpansion}
\end{equation}
In this expansion, the determinants \eqref{bosdet} and \eqref{ferdet} can be computed by using the identity 
\begin{equation}
  \ln\det(M+\delta M)=\ln \det M-\sum_{k=1}^\infty \frac{(-1)^{k }}{k}\tr(M^{-1}\delta M)^k \, ,
  \label{Detexp}
\end{equation}
with 
\begin{equation}
  M=-\d_{t}^2+\ls^{-4}r^2  , \ \delta M=(2\ls^{-2}\vec y\cdot\vec\epsilon+\vec{\epsilon\,}^2)\1_2+i\left<D_{\mu\nu}\right>\sigma_{\mu\nu}
  \label{MB}
\end{equation}
in the bosonic case and
\begin{equation}
  M=-i\d_{t}+i\ls^{-2}y_A\Gamma_A \, , \quad \delta M=i\epsilon_A\Gamma_A 
  \label{MF}
\end{equation}
in the fermionic case. We use the notation
\begin{equation}\label{radialdef} r^{2} = y_{A}y_{A}={\vec y\,}^{2}\end{equation}
and $\vec y\cdot\vec\epsilon = y_{A}\epsilon_{A}$, etc. 
The corresponding bosonic and fermionic Green's functions read
\begin{align}
  G_B(t,t')& =\int \frac{\d\omega}{2\pi}\frac{e^{i\omega(t-t')}}{\omega^2+\ls^{-4}r^2} \, ,
  \label{defGB}\\
  G_F(t,t') &=\int \frac{\d\omega}{2\pi}\frac{e^{i\omega(t-t')}}{\omega^2+\ls^{-4}r^2}\bigl(\omega\1_4-i\ls^{-2}y_A\Gamma_A\bigr) \, .
  \label{GBdef}
\end{align}
The trace in \eqref{Detexp} involves both integrals over frequencies and traces over $\text{Spin}(5)$, $\text{SU}(2)_{+}$ and $\text{U}(K)$ indices.
At each order in the $\epsilon$ expansion, we can further expand in powers of the frequencies or, equivalently, in time derivatives.
The saddle-point equation \eqref{Dsaddle} can be solved similarly, both in the $\epsilon$ expansion \eqref{matrixexpansion} and in the derivative expansion.

If we write
\begin{equation}
  \tilde S_{\text{eff}}=\int\!\d t\,\sum_{p\ge0}L_{p} \, ,
  \label{Seffexp}
\end{equation}
where $L_p$ is of order $p$ in $\epsilon$, we find
\begin{equation}\label{L0L1sol} L_{0}= K\frac{\sqrt{2\pi}}{\gs\ls}\, ,\quad L_{1} = 0\, ,\end{equation}
which are simple consequences of supersymmetry. We have also computed $L_{2}$ and $L_{3}$ up to fourth order in derivatives,
\begin{align}\label{L2sol} L_{2} &=  \tr\Bigl( \frac{\sqrt{2\pi}\ls^3}{2g_s}\dot\epsilon_\mu\dot\epsilon_\mu+\frac{N\ls^6}{4r^3}\dot\epsilon_A\dot\epsilon_A \Bigr) + O\bigl(\ddot\epsilon^{2}\bigr)  \, , \\\label{L3sol}
  L_{3}&=-\frac{3N\ls^4}{4r^5}  \tr\bigl( \vec y\cdot\vec\epsilon\, \dot{\vec \epsilon}^{\,2} \bigr) + O\bigl(\epsilon\ddot\epsilon^2,\dot\epsilon^{2}\ddot\epsilon\bigr) \, , 
\end{align}
and $L_{4}$ and $L_{5}$ up to second order in derivatives,
\begin{align}\notag
 L_{4}&=-\tr \Bigl( \frac{\pi r^3}{g_s^2 N}[\epsilon_\mu,\epsilon_\nu][\epsilon_\mu,\epsilon_\nu]  +\frac{\ls^6N}{8r^3}[\epsilon_A,\epsilon_B][\epsilon_A,\epsilon_B]\\
 &\hskip 1.5cm+\frac{\sqrt{2\pi}\ls^3}{2g_s}[\epsilon_A,\epsilon_\mu][\epsilon_A,\epsilon_\mu] \Bigr) +O\bigl(\epsilon^{2}\dot\epsilon^2\bigr) 
 \, ,\label{L4} \\
  L_{5}&=-\frac{6\pi\ls^2r}{g_s^2N} \epsilon_{\mu\nu\rho\kappa}\tr\epsilon_\mu\epsilon_\nu\epsilon_\rho\epsilon_\kappa\vec y\cdot\vec\epsilon +\cdots+O\bigl(\epsilon^{3}\dot\epsilon^2\bigr) \mathrlap{.}
  \label{L5}
\end{align}
The $\cdots$ in \eqref{L5} are contributions to the action that are fixed in terms of \eqref{L4} by general consistency conditions \cite{Ferrari:2013pi}. We are now going to show that the terms \eqref{L0L1sol}--\eqref{L5} perfectly match with the expected form of the D-particle Lagrangian in a non-trivial background.

\section{\label{emergesec} The emergent geometry}

The non-Abelian action for D-particles in an arbitrary background can be computed using formulas in \cite{Myers:1999ps,Taylor:1999gq,Taylor:1999pr}, as reviewed in \cref{sec:Myers}, see \cite{Ferrari:2013pi} for more details. If we denote the space matrix coordinates by $Z^{i}$, $1\leq i\leq 9$, and expand $Z^{i} = z^{i} +\ls^{2} \epsilon^{i}$, then the Lagrangian, computed in the static gauge
\begin{equation}\label{staticgauge} x^{d}=x^{10} = t\, ,\end{equation}
can be conveniently written as a sum of terms with a fixed number of derivatives, 
\begin{multline}
\label{order0exp} L = \sum_{n\geq 0}\frac{1}{n!}\ls^{2n}
c_{\iin}^{(0)}(z,t)
\tr\epsilon^{i_{1}}\cdots\epsilon^{i_{n}} \\ +
\sum_{n\geq 0}\frac{1}{n!}\ls^{2(n+1)}
c^{(1)}_{\iin; k} (z,t) \tr\epsilon^{i_{1}}\cdots\epsilon^{i_{n}} \dot\epsilon^{k}\\ +
\ls^{4} c^{(2)}_{kl}(z,t)\tr\dot\epsilon^{k}\dot\epsilon^{l}+ \ls^{6}c^{(2)}_{i;kl}(z,t)\tr\epsilon^{i}\dot\epsilon^{k}\dot\epsilon^{l} + \cdots
\end{multline}
The $\cdots$ represent terms of higher orders in $\epsilon$ with two derivatives or with at least three derivatives.
The coefficients in \eqref{order0exp} can be expressed in terms of the type \<IIA> supergravity fields. Our goal is to find a match between \eqref{order0exp} and the corresponding terms in our microscopically computed Lagrangian \eqref{L0L1sol}--\eqref{L5}, with $\epsilon^{i} \equiv (\epsilon_{\mu},\epsilon_{A})$. 
We are seeking a static $\text{SO}(4)\times\text{SO}(5)$ preserving background which has vanishing Neveu-Schwarz $B$ field and Ramond-Ramond one-form.
The coefficients in \eqref{order0exp} can then be naturally expressed in terms of
the following combinations of $G_{MN}$, $1\leq M,N\leq 10$, and dilaton $\phi$,
\begin{equation}\label{gdef} g_{MN}  = e^{-2\phi}G_{MN}\, ,\quad 
\mathscr H_{ij} = \sqrt{g_{dd}}\biggl( \frac{g_{ij}}{g_{dd}} -
\frac{g_{di} g_{dj}}{g_{dd}^{2}}\biggr) \, .
\end{equation}
If $C$ and $\tilde C$ are the Ramond-Ramond three-form potential and its dual five-form respectively, $\d\tilde C=i*\d C$ in the Euclidean, the explicit formulas we need read~\cite{Ferrari:2013pi}
\begin{align}
\label{potD0} c^{(0)}  &= \frac{\sqrt{2\pi}}{\ls} \sqrt{g_{dd}}\, ,\quad
c^{(1)}_{i;k}  = \frac{\sqrt{2\pi}}{\ls}\partial_{i}\bigl(g_{dk}/\sqrt{g_{dd}}\bigr)\, ,
\\  
\label{pot2D0a} c^{(2)}_{kl}  &= \frac{\sqrt{2\pi}}{2\ls} \mathscr H_{kl} \, , \\
\label{potD0b} c^{(0)}_{[ijk]}  &= \frac{3\sqrt{2\pi}}{2\ls^{3}}
\partial_{[i} C_{jk]d}\, ,  \\
c^{(0)}_{[ij][kl]}  &= -\frac{9\sqrt{2\pi}}{\ls^{5}}
g_{dd}^{3/2}e^{4\phi} \bigl(\mathscr H_{ik}\mathscr H_{jl} - \mathscr H_{jk}\mathscr H_{il}\bigr) \, ,\label{potD0d} \\
c^{(0)}_{[ijklm]}  &= -\frac{60i\sqrt{2\pi}}{\ls^{5}}\partial_{[i}
 \tilde C_{jklm]d} \, . \label{potD0e}
\end{align}
Other combinations of coefficients either vanish, consistently with the vanishing of some of the background fields, or are expressed in terms of \eqref{potD0}--\eqref{potD0e} by solving the general consistency conditions discussed in \cite{Ferrari:2013pi}.

A single D-particle, $K=1$, probes only the metric $g_{MN}$. In this case, 
by comparing \eqref{potD0} and \eqref{pot2D0a} with \eqref{L0L1sol} and \eqref{L2sol} and by noting that non-vanishing constant components $g_{dk}$ would be inconsistent with $\text{SO(4)}\times\text{SO}(5)$, we find
\begin{equation}\label{gsolu} g_{MN}\, \d x^{M}\d x^{N} = \frac{1}{\gs^2}\Big(\d t^{2} + \d x_{\mu}\d x_{\mu} +
\frac{L^{3}}{r^{3}} {\d\vec y}^{\,2}\Big) \, ,\end{equation}
with
\begin{equation}\label{Length} L^{3} = \frac{N\gs\ls^{3}}{2\sqrt{2\pi}}\,\cdotp\end{equation}
%
Using the full non-Abelian action, $K>1$, we can get more information on the background. Indeed, the D-particles then couple to the Ramond-Ramond three-form through commutator terms. By comparing \eqref{potD0b} with \eqref{L3sol} and \eqref{potD0e} with \eqref{L5}, we can find $F_{4}=\d C$ or its dual $F_{6}$ unambiguously. Moreover, one can check that the double commutator term in the fourth order potential \eqref{L4} has precisely the correct structure to match with \eqref{potD0d}. Since $\mathscr H_{ij}$ is already known from the kinetic term \eqref{L2sol}, we can derive the dilaton profile from this term and then extract the string frame metric $\d s^{2}$ from 
\eqref{gsolu} and the first equation in \eqref{gdef}.
Overall, we get
\begin{align}
 & \d s^{2}=\frac{r^{3/2}}{L^{3/2}}(\d t^2+\d x_\mu\d x_\mu)+\frac{L^{3/2}}{r^{3/2}}\d{\vec y\,}^2 \, , \label{metric} \\
 & e^{\phi}=g_s \frac{r^{3/4}}{L^{3/4}} \, , \label{dilaton} \\
  &  F_{4} = \frac{L^{3}}{8\gs r^{5}}\epsilon_{ABCDE}\,y_{E}\,\d y_{A}\wedge\cdots\wedge\d y_{D}\, .
\end{align}
This background is in perfect agreement with the near-horizon D4-brane background \cite{Horowitz:1991cd,Itzhaki:1998dd}, including the relation between the supergravity length scale $L$ and string-theory parameters $g_s$ and $\ls$ and the correct normalization of the Ramond-Ramond form, consistently with the D4-brane charge in type \<IIA>.

\appendix
\part*{\addcontentsline{toc}{part}{Appendices}Appendices}
\chapter{\label{NotAppSec} Notations and conventions}
\label{chap:notations}
We work in Euclidean signature throughout this thesis and do not distinguish upper and lower vector indices in flat space.

The string length $\ls$ is related to $\alpha'$ by
\begin{equation}
  \ls^2=2\pi \alpha'
  \label{eq:lsalpha}
\end{equation}
in order to absorb some factors of $2\pi$.

\section{Indices and transformation laws}
See table~\ref{indices} for the \Dmi{}/D3 case of relevance for \cref{chap:instantons,chap:D-branes,chap:adsdeformed}.
The notations for \cref{chap:enhancon,chap:D4brane} are variations of this case and are discussed in the main text.

\begin{table}[hbt!]
  \caption{\label{indices}Conventions for the transformation laws of indices, fields and moduli of the \Dmi{}/D3 system. For maximum clarity, we have indicated all the indices associated to each field or modulus, whereas in the main text the gauge $\text{U}(N)$ and $\text{U}(K)$ indices are usually suppressed. The representations of $\text{Spin}(4)=\text{SU}(2)_{+}\times
\text{SU}(2)_{-}$ are indicated according to the spin in each $\text{SU}(2)$ factor. The $(1/2,1/2)$ of $\text{SU}(2)_{+}\times
\text{SU}(2)_{-}$ and the $\mathbf 6$ of $\text{SU}(4)=\text{Spin}(6)$ correspond to the fundamental representations of $\text{SO}(4)$ and $\text{SO}(6)$ respectively.}   
\centering
\begin{tabular}{@{}Ml@{\qquad\extracolsep{\fill}}McMcMcMc@{}}
\toprule
                                                               & \operatorname{Spin}(4) & \SU(4)          & \U(N)           & \U(K)\\
\midrule
\alpha, \beta, ... \ \text{(upper or lower)}                   & (1/2,0)                & \mathbf 1       & \mathbf 1       & \mathbf 1 \\
\dot\alpha, \dot\beta, ... \ \text{(upper or lower)}           & (0,1/2)                & \mathbf 1       & \mathbf 1       & \mathbf 1 \\
\mu, \nu, ...                                                  & (1/2,1/2)              & \mathbf 1       & \mathbf 1       & \mathbf 1 \\
a, b, ... \ \text{(lower)}                                     & (0,0)                  & \mathbf 4       & \mathbf 1       & \mathbf 1 \\
a, b, ... \ \text{(upper)}                                     & (0,0)                  & \mathbf{\bar 4} & \mathbf 1       & \mathbf 1 \\
A, B, ...                                                      & (0,0)                  & \mathbf{6}      & \mathbf 1       & \mathbf 1 \\
f, f', ... \ \text{(lower)}                                    & (0,0)                  & \mathbf 1       & \mathbf N       & \mathbf 1 \\
f, f', ... \ \text{(upper)}                                    & (0,0)                  & \mathbf 1       & \mathbf{\bar N} & \mathbf 1 \\
i, j, ... \ \text{(lower)}                                     & (0,0)                  & \mathbf 1       & \mathbf 1       & \mathbf K \\
i, j, ... \ \text{(upper)}                                     & (0,0)                  & \mathbf 1       & \mathbf 1       & \mathbf{\bar K}\\
A\indices{_\mu_f^{f'}}                                         & (1/2,1/2)              & \mathbf 1       & \textbf{Adj}    & \mathbf 1\\
\varphi\indices{_A_f^{f'}}                                     & (0,0)                  & \mathbf 6       & \textbf{Adj}    & \mathbf 1\\
\lambda\indices{_{\alpha a f}^{f'}}                            & (1/2,0)                & \mathbf 4       & \textbf{Adj}    & \mathbf 1\\
\bar\lambda\indices{^{\dot\alpha a}_{f}^{f'}}                  & (0,1/2)                & \mathbf{\bar 4} & \textbf{Adj}    & \mathbf 1\\
X\indices{_{\mu i}^{j}}                                        & (1/2,1/2)              & \mathbf 1       & \mathbf 1       & \textbf{Adj}\\
Y\indices{_{A i}^{j}}=\ls^{2}\phi\indices{_{A i}^{j}}          & (0,0)                  & \mathbf 6       & \mathbf 1       & \textbf{Adj} \\
\Lambda\indices{_{\alpha a i}^{j}}                             & (1/2,0)                & \mathbf 4       & \mathbf 1       & \textbf{Adj}\\
\bar\psi\indices{^{\dot\alpha a}_{i}^j}                        & (0,1/2)                & \mathbf{\bar 4} & \mathbf 1       & \textbf{Adj} \\
D\indices{_{\mu\nu i}^{j}}                                     & (1,0)                  & \mathbf 1       & \mathbf 1       & \textbf{Adj}\\
q\indices{_\alpha_i^f}                                         & (1/2,0)                & \mathbf 1       & \mathbf{\bar N} & \textbf{K}\\
\tilde q\indices{_\alpha_f^i}                                  & (1/2,0)                & \mathbf 1       & \mathbf N       & \mathbf{\bar K}\\
\chi\indices{^{a}_{i}^f}                                       & (0,0)                  & \mathbf{\bar 4} & \mathbf{\bar N} & \mathbf{K}\\
\tilde\chi\indices{^a_f^i}                                     & (0,0)                  & \mathbf{\bar 4} & \mathbf{N}      & \mathbf{\bar K} \\
\bottomrule
\end{tabular}

\end{table}
\section{Four-dimensional algebra}
With the standard Pauli matrices
\begin{equation}
  \tau_1=\begin{pmatrix}0 & 1\\1 & 0\end{pmatrix},\quad \tau_2=\begin{pmatrix}0 & -i\\i & 0\end{pmatrix}, \quad \tau_3=\begin{pmatrix}1 & 0\\0 & -1\end{pmatrix} \, ,
  \label{paulidef}
\end{equation}
we can define
\begin{equation}
  \sigma_{\mu\alpha\dot\alpha}=(\vec\tau,-i\1_{2})_{\alpha\dot\alpha}\,, \quad \bar\sigma_\mu^{\dot\alpha\alpha}=(-\vec\tau,-i\1_{2})^{\dot\alpha\alpha}
  \label{sigma1def}
\end{equation}
and
\begin{equation}
  \sigma_{\mu\nu}=\frac14(\sigma_{\mu}\bar\sigma_{\nu}-\sigma_{\nu}\bar\sigma_{\mu})\,, \quad 
  \bar\sigma_{\mu\nu}=\frac14(\bar\sigma_{\mu}\sigma_{\nu}-\bar\sigma_{\nu}\sigma_{\mu})\, .
  \label{sigma2def}
\end{equation}
The $\sigma_\mu$ and $\bar\sigma_\mu$ matrices obey the following Clifford algebra relations,
\begin{equation}
  \sigma_\mu \bar\sigma_\nu + \sigma_\nu \bar\sigma_\mu = -2 \delta_{\mu\nu} \1_2 \, , \quad
  \bar\sigma_\mu \sigma_\nu + \bar\sigma_\nu \sigma_\mu = -2 \delta_{\mu\nu} \1_2 \, .
  \label{eq:sigmaClifford}
\end{equation}

To raise and lower indices, we follow the \textcite{Wess:1992cp} convention.
In particular, $\epsilon^{12}=\epsilon^{\dot1\dot2}=1$, and the following identity holds,
\begin{equation}
  \epsilon^{\alpha\beta}\epsilon^{\dot\alpha\dot\beta}\sigma_{\mu\beta\dot\beta}=\bar\sigma_\mu^{\dot\alpha\alpha} \, .
  \label{form:sigmasigmabar}
\end{equation}

The following identity is very useful:
\begin{equation}
  \sigma_{\mu\nu}\sigma_{\rho\kappa}=\frac14(-\epsilon_{\mu\nu\rho\kappa}+\delta_{\nu\rho}\delta_{\mu\kappa}-\delta_{\mu\rho}\delta_{\nu\kappa})\1_{2} + \left( \delta_{\kappa [\nu}\sigma_{\mu]\rho}-\delta_{\rho[\nu} \sigma_{\mu]\kappa} \right) \, ,
  \label{2sigmaid}
\end{equation}
where $\epsilon_{\mu\nu\rho\sigma}$ is the completely antisymmetric tensor with $\epsilon_{1234}=+1$.

We denote by an upper ``+'' sign the projection of an antisymmetric tensor on its self-dual part,
\begin{equation}
  a^+_{\mu\nu}=\frac12(a_{\mu\nu}+\frac12\epsilon_{\mu\nu\rho\kappa}a_{\rho\kappa}) \, .
  \label{sddef}
\end{equation}
With these definitions $\sigma_{\mu\nu}$ is self-dual,
\begin{align}
  \sigma_{\mu\nu}=\sigma_{\mu\nu}^+ \, .
  \label{sigmasd}
\end{align}
Let us finally mention the following useful identities,
\begin{align}\det(\1_2+a_{\mu\nu}\sigma_{\mu\nu})& =1+a_+^2 \, ,
  \label{detsigmaid}\\
\bigl( \1_{2}+a_{\mu\nu}\sigma_{\mu\nu} \bigr)^{-1}&=
\frac{\1_{2}-a_{\rho\sigma}\sigma_{\rho\sigma}}{1+a_+^2}\, ,
  \label{inv1sigma}
\end{align}
where
\begin{equation}\label{defaplus}
a_+^2 = a^+_{\mu\nu}a^+_{\mu\nu}\, .
\end{equation}
\section{Six-dimensional algebra}
\subsection{Undeformed case}

We define
\begin{align}\nonumber
&\Sigma_{1}=
\begin{pmatrix} 0&-1&0&0\\1&0&0&0\\0&0&0&1\\0&0&-1&0\end{pmatrix} ,\
\Sigma_{2}=
\begin{pmatrix} 0&-i&0&0\\i&0&0&0\\0&0&0&-i\\0&0&i&0\end{pmatrix} ,\
\Sigma_{3}=
\begin{pmatrix} 0&0&-1&0\\0&0&0&-1\\1&0&0&0\\0&1&0&0\end{pmatrix} ,\\
\label{Sigmadef6D}
&\Sigma_{4}=
\begin{pmatrix} 0&0&-i&0\\0&0&0&i\\i&0&0&0\\0&-i&0&0\end{pmatrix} ,\
\Sigma_{5}=
\begin{pmatrix} 0&0&0&-1\\0&0&1&0\\0&-1&0&0\\1&0&0&0\end{pmatrix} ,\
\Sigma_{6}=
\begin{pmatrix} 0&0&0&-i\\0&0&-i&0\\0&i&0&0\\i&0&0&0\end{pmatrix}
\end{align}
and
\begin{equation}\label{Sigmabardef6D} \bar\Sigma_{A}= \Sigma_{A}^{\dagger}\, .\end{equation}
These matrices satisfy the algebra
\begin{equation}
\Sigma_A\bar\Sigma_B+\Sigma_B\bar\Sigma_A =2\delta_{AB}\1_4 
\end{equation}
as well as the relations
\begin{equation}\label{relSigma} \bar\Sigma_{A}^{ab}= \frac{1}{2}\epsilon^{abcd}
\Sigma_{Acd}\, ,\quad 
\Sigma_{Aab}= \frac{1}{2}\epsilon_{abcd}\bar\Sigma_{A}^{cd}\end{equation}
where the $\epsilon$s are completely antisymmetric symbols with $\epsilon_{1234}=\epsilon^{1234}=+1$.
Euclidean six-dimensional Dirac matrices, satisfying
\begin{equation}\label{Cliff6D} \bigl\{\Gamma_{A},\Gamma_{B}\bigr\} = 2\delta_{AB}\, ,\end{equation}
can then be defined by
\begin{equation}\label{gamma6Ddef} \Gamma_{A} = 
\begin{pmatrix}
0 & \Sigma_{A}\\ \bar\Sigma_{A} & 0
\end{pmatrix}\, .
\end{equation}

If $\vec v = (v_{A})_{1\leq A\leq 6}$ is a six-dimensional vector, one can check that
\begin{align}
\label{detsigmaa}\det(v_{A}\Sigma_A)&={\vec v\,}^4\,,\\
(v_{A}\Sigma_A)^{-1}&=\frac{v_{A}\bar \Sigma_A}{{\vec v\,}^{2}}\, \cdotp
\end{align}
In sections \ref{CBSec} and \ref{NCSec} of the main text, we have to compute the expansion of some determinants of the form
\begin{align}\label{appdetex} \ln\det\bigl(\Sigma_{A}\otimes (v_{A} + \ls^{2}\epsilon_{A})\bigr) &= \ln {\vec v\,}^4 + \sum_{k=1}^\infty\frac{(-1)^k}{k}\tr\left( \left(v_A \Sigma_A\right)^{-1}\Sigma_B \otimes \epsilon_{B} \right)^k  \nonumber \\
  &=\sum_{k=0}^{\infty}t^{(k)} \, .
\end{align}
Up to order five, this is done by using the trace formulas in \cite{Ferrari:2012nw}, which yield
\begin{align}
  \label{s1}t^{(1)} &= -\frac{4}{ v^{2}}\truK (\vec v\cdot\vec\epsilon\,)\,,\\
  \label{s2}t^{(2)} &= \frac2{{\vec v \,}^{4}}\truK \left[ 2(\vec v\cdot\vec\epsilon\,)^2 - {\vec v \,}^2 {\vec\epsilon\,}^2 \right] \,,\\
  \label{s3}t^{(3)} &= -\frac4{3 {\vec v \,}^{6}}\truK \left[ 4(\vec v\cdot\vec\epsilon\,)^3 - 3 {\vec v \,}^2 (\vec v\cdot\vec\epsilon\,){\vec\epsilon\,}^2 \right] \,,\\
  \label{s4}t^{(4)} &= \frac{8}{ {\vec v \,}^{8}} \truK \left[ (\vec v\cdot\vec\epsilon\,)^4 - {\vec v \,}^2 (\vec v\cdot\vec\epsilon\,)^2 {\vec\epsilon\,}^2 + \frac 14 {\vec v \,}^4 {\vec\epsilon\,}^4 - \frac18 {\vec v \,}^4 \epsilon_A\epsilon_B\epsilon_A\epsilon_B \right] \,,\\
  \label{s5}t^{(5)} &= -\frac{4}{{\vec v \,}^{10}} \truK \biggl[ \frac{16}5 (\vec v\cdot\vec\epsilon\,)^5 - 4{\vec v \,}^2 (\vec v\cdot\vec\epsilon\,)^3 {\vec\epsilon\,}^2 +
\frac i5 {\vec v \,}^4 v_A \epsilon_{A_1\cdots A_5A}\epsilon_{A_1}\cdots \epsilon_{A_5}
  \notag\\& + {\vec v \,}^4 \left( \vec v\cdot\vec \epsilon\, {\vec\epsilon\,}^4 - \vec v\cdot\vec \epsilon \, \epsilon_B\epsilon_C\epsilon_B\epsilon_C + \vec v\cdot\vec \epsilon \, \epsilon_B {\vec\epsilon\,}^2 \epsilon_B \right)  \biggr] \,.
\end{align}

Weyl spinors $\lambda_{a}$ and $\psi^{a}$ in the $\mathbf 4$ and $\mathbf{\bar 4}$ representations of the
rotation group $\text{Spin}(6)=\text{SU}(4)$ transform under a 
six-dimensional rotation parametrized by the antisymmetric matrix $\Omega$, $\delta x_{A}= -\Omega_{AB}x_{B}$, as
\begin{equation}\label{spinors6Dtl} \delta\lambda_{a} = -\frac{1}{2}\Omega_{AB}
\Sigma_{ABa}^{\ \ \ \ b}\lambda_{b}\, ,\quad
\delta\psi^{a} = -\frac{1}{2}\Omega_{AB}\bar\Sigma_{AB\ b}^{\ \ \ a}
\psi^{b}\, ,\end{equation}
where the generators of the rotation group are defined by
\begin{equation}\label{gen6Ddef} \Sigma_{AB}=\frac{1}{4}\bigl( \Sigma_{A}\bar
\Sigma_{B} - \Sigma_{B}\bar\Sigma_{A}\bigr)\, , \quad
\bar\Sigma_{AB}=\frac{1}{4}\bigl( \bar\Sigma_{A}
\Sigma_{B} - \bar\Sigma_{B}\Sigma_{A}\bigr)\, .\end{equation}
This yields in particular the charges under the $\u_{1}\times\u_{2}\times\u_{3}$ subgroup of $\text{SO}(6)$ corresponding to rotations in the 1-2, 3-4 and 5-6 planes respectively, see Table \ref{chargesU1}.

\subsection{\texorpdfstring{$\beta$}{β}-deformed case}

The $\u_{i}$ charges in Table \ref{chargesU1} are used to compute the $*$-product in section \ref{BetaSec}. In particular, deformed $\Sigma_{A}$ matrices can be defined by the identity
\begin{equation}\label{starproductid} \psi_{1}^{a}*\phi_{A}*\psi_{2}^{b}\,\Sigma_{Aab} = 
\psi_{1}^{a}\phi_{A}\psi_{2}^{b}\,\tilde\Sigma_{Aab}\, .\end{equation}
Explicitly, dropping the $0$s for readability, we have
\begin{spreadlines}{1em} 
  \begin{align}\nonumber
    \tilde\Sigma_{1}&=
    \left(\begin{smallmatrix}  &-i^{\gamma_{1}-\gamma_{2}}& & \\i^{-\gamma_{1}+
          \gamma_{2}}& & & \\ & & & i^{-\gamma_{1}-\gamma_{2}} \\ & &
        -i^{\gamma_{1}+\gamma_{2}}& \end{smallmatrix}\right)\, ,\\\nonumber
    \tilde\Sigma_{2} &=
    \left(\begin{smallmatrix}  &i^{\gamma_{1}-\gamma_{2}-1}& & \\i^{-\gamma_{1}+\gamma_{2}+1}& & & \\ & & &i^{-\gamma_{1}-\gamma_{2}-1}\\ & &
        i^{\gamma_{1}+\gamma_{2}+1}& \end{smallmatrix}\right)\, ,\\\nonumber
    \tilde\Sigma_{3}&=
    \left(\begin{smallmatrix}  & &-i^{-\gamma_{1}+\gamma_{3}}& \\ & & &
        -i^{\gamma_{1}+\gamma_{3}}\\i^{\gamma_{1}-\gamma_{3}}& & & \\ &i^{-\gamma_{1}-\gamma_{3}}& & \end{smallmatrix}\right)\, ,\\
    \label{Sigmabetadef6D}
    \tilde\Sigma_{4} & =
    \left(\begin{smallmatrix}  & &i^{-\gamma_{1}+\gamma_{3}-1}& \\ & & &i^{\gamma_{1}+\gamma_{3}+1}\\i^{\gamma_{1}-\gamma_{3}+1}& & & \\ &i^{-\gamma_{1}-\gamma_{3}-1}& & \end{smallmatrix}\right)\, ,\\\nonumber
    \tilde\Sigma_{5}& =
    \left(\begin{smallmatrix}  & & &-i^{\gamma_{2}-\gamma_{3}}\\ & &i^{-\gamma_{2}-\gamma_{3}}& \\ &-i^{\gamma_{2}+\gamma_{3}}& & \\i^{-\gamma_{2}+\gamma_{3}}& & & \end{smallmatrix}\right)\, ,\\\nonumber
    \tilde\Sigma_{6} &=
    \left(\begin{smallmatrix}  & & &i^{\gamma_{2}-\gamma_{3}-1}\\ & &i^{-\gamma_{2}-\gamma_{3}-1}& \\ &i^{\gamma_{2}+\gamma_{3}+1}& & \\i^{-\gamma_{2}+\gamma_{3}+1}& & & \end{smallmatrix}\right)\, .
  \end{align}
\end{spreadlines}
\begin{table}[hbt!]
  \caption{\label{chargesU1}Charges under $\u_{1}\times\u_{2}\times\u_{3}\subset\text{SO}(6)$. The spinors $\lambda_{a}$ and $\psi^{a}$ are arbitrary spinors in the $\mathbf 4$ and $\mathbf{\bar 4}$ representations of $\text{Spin}(6)$ respectively.}
  \hspace*{-20pt}\vbox{
\begin{equation}\nonumber
\begin{array}{@{}ccccrrrrrrrr@{}}
  \toprule
  \text{Group} & y_{1}+ i y_{2} & y_{3}+ i y_{4} & y_{5}+ i y_{6} & \lambda_{1} & \lambda_{2} & \lambda_{3} & \lambda_{4} & \psi^{1} & \psi^{2} & \psi^{3}
& \psi^{4}\\
\midrule
\u_{1}\ & 1 & 0 & 0 & \frac{1}{2} & \frac{1}{2} & -\frac{1}{2} &-\frac{1}{2}  & -\frac{1}{2} & -\frac{1}{2} & \frac{1}{2} & \frac{1}{2} 
\\ \addlinespace
 \u_{2}\ & 0 & 1 & 0 & \frac{1}{2} & -\frac{1}{2} & \frac{1}{2} &-\frac{1}{2}  &-\frac{1}{2}  &\frac{1}{2}  &-\frac{1}{2}  & \frac{1}{2} 
 \\ \addlinespace
 \u_{3}\ & 0 & 0 & 1 &\frac{1}{2}  &-\frac{1}{2}  &-\frac{1}{2}  &\frac{1}{2}  &-\frac{1}{2}  &\frac{1}{2}  &\frac{1}{2}  &-\frac{1}{2}  \\ \addlinespace
 \bottomrule
\end{array}
\end{equation}}
\end{table}

\chapter{Myers' non-abelian D-instanton action}
\label{appB}\label{chap:Myers}

Myers' non-abelian D-instanton action \cite{Myers:1999ps}, in the expansion \eqref{Myersexpansion} up to order five, is given in terms of the type \<IIB> supergravity fields by the following formulas \cite{Ferrari:2012nw},
\begin{align}\nonumber
S_{\text{eff}}^{(0)} & = -2i\pi K\tau\, ,\\\nonumber
S_{\text{eff}}^{(1)} & = -2i\pi \ls^{2}\partial_{M}\tau\tr \epsilon_{M}\, ,\\
\nonumber
S_{\text{eff}}^{(2)} & = -i\pi\ls^{4}\partial_{M}\partial_{N}\tau\tr
\epsilon_{M}\epsilon_{N}\, ,\\\nonumber
S_{\text{eff}}^{(3)} & =\bigl(-\frac{i\pi}{3}\ls^{6}\partial_{M}\partial_{N}\partial_{P}\tau - 
2\pi\ls^{4}\partial_{[M}(\tau B - C_{2})_{NP]}\bigr)\tr\epsilon_{M}\epsilon_{N}\epsilon_{P} \, ,\\\label{Mexpand}
S_{\text{eff}}^{(4)} & =\bigl( -\frac{i\pi}{12}\ls^{8}
\partial_{M}\partial_{N}\partial_{P}\partial_{Q}\tau
- \frac{3\pi}{2}\ls^{6}\partial_{M}
\partial_{[N}(\tau B - C_{2})_{PQ]}\\\nonumber
& \hskip 2cm-\pi\ls^{4}e^{-\Phi}(G_{MP}G_{NQ} - G_{MQ}G_{NP})\bigr)
\tr\epsilon_{M}\epsilon_{N}\epsilon_{P}\epsilon_{Q}\, ,
\\\nonumber
S_{\text{eff}}^{(5)} & = \Bigl(-\frac{i\pi}{60}\ls^{10}\partial_{M}\partial_{N}\partial_{P}\partial_{Q}\partial_{R}\tau 
- \frac{\pi}{3}\ls^{8}\partial_{P}\partial_{Q}\partial_{R}(\tau B - C_{2})_{MN}\\\nonumber &\hskip 3cm
- \pi\ls^{6}\partial_{R}\bigl(e^{-\Phi}
(G_{MP}G_{NQ} - G_{MQ}G_{NP})\bigr)\\\nonumber &\hskip 1cm
-i\pi\ls^{6}\partial_{[M}(C_{4}+ C_{2}\wedge B - \frac{\tau}{2}
B\wedge B)_{NPQR]}\Bigr)
\tr\epsilon_{M}\epsilon_{N}\epsilon_{P}\epsilon_{Q}\epsilon_{R}\, .
\end{align}

\chapter{Some type \<IIB>\ supergravity backgrounds}
\label{SolSUGRA}\label{chap:sugrasols}

We review in this appendix the known supergravity backgrounds dual to the non-com\-mu\-ta\-ti\-ve and $\beta$-deformed Euclidean $\nn=4$ super Yang-Mills theories studied in the main text. We use the standard relation between the radius $R$ and the 't~Hooft coupling $\lambda$,
\begin{equation}\label{Rlambdarelapp} R^{4}=\alpha'^{2}\lambda = \frac{\ls^{4}\lambda}{4\pi^{2}}\,\cdotp\end{equation}
The backgrounds are written at zero bare $\vartheta$ angle. The solutions at non-zero $\vartheta$ can be obtained by performing the $\text{SL}(2,\mathbb R)$ transformation $C_{0}\rightarrow C_{0} + \frac{\vartheta}{2\pi}$, $C_{2}\rightarrow C_{2}-
\frac{\vartheta}{2\pi} B$ and $C_{4}\rightarrow C_{4} + \frac{\vartheta}{4\pi}B\wedge B$, which automatically yields a new solution to the supergravity equations of motion.

\section{\label{App2}The dual to the non-commutative gauge theory}

The gravitational dual of the non-commutative deformation of the $\nn=4$ super Yang-Mills theory was derived by Hashimoto, Itzhaki, Maldacena and Russo in \cite{Hashimoto:1999ut,Maldacena:1999mh}.\footnote{Our formulas can be matched with those in \cite{Maldacena:1999mh} by making the replacements $R^2\rightarrow\alp R^2$, $\theta_{12}\rightarrow\tilde b' /(2\pi)$, $\theta_{34}\rightarrow\tilde b/(2\pi)$, $r\rightarrow\alp R^2 u$, $\lambda/(4\pi N)\rightarrow \hat g$ and $C_0\rightarrow-\chi$, $C_2\rightarrow-A$, $F_5\rightarrow-F$.} With non-vanishing non-commutative parameters $\theta_{12}=-\theta_{21}$ and $\theta_{34}=-\theta_{43}$, the solution for the string-frame metric and the other supergravity fields reads
\begin{align}
  \d s^2&=\frac{r^2}{R^2}\left[ \frac{ \d x_1^2+\d x_2^2 }{\Delta_{12}}+\frac{ \d x_3^2+\d x_4^2 }{\Delta_{34}} \right]+\frac{R^2}{r^2}\d r^2+R^2\d\Omega_5^2 \, , \label{metricNCsugra} \\
  e^{-\phi}&=\frac{4\pi N}{\lambda}\sqrt{\Delta_{12}\Delta_{34}} \, , \label{phiNCsugra} \\
  B&=\frac{r^4}{R^4}\left( \frac{\theta_{12}}{\ls^2}\frac{\d x_1\wedge\d x_2}{\Delta_{12}}+ \frac{\theta_{34}}{\ls^2}\frac{\d x_3\wedge\d x_4}{\Delta_{34}}  \right) \, , \label{BNCsugra} \\
  C_0&=-\frac{4i\pi N}{\lambda}\frac{\theta_{12}\theta_{34}}{\ls^4}\frac{r^4}{R^4} \, , \label{C0NCsugra} \\
  C_2&=-\frac{4i\pi N}{\lambda}\frac{r^4}{R^4}
  \biggl( \frac{\theta_{34}}{\ls^{2}}
    \frac{\d x_1\wedge\d x_2}{\Delta_{12}}+ 
    \frac{\theta_{12}}{\ls^{2}}
  \frac{\d x_3\wedge\d x_4}{\Delta_{34}} \biggr) \, ,
  \label{C2NCsugra} \\\label{C4NCsugra}
  C_4 &= \frac{16 \pi r^2}{R^3}\omega_4 -4i\pi\frac{r^6}{R^6}
\frac{\d x_1 \wedge \d x_2 \wedge \d x_3 \wedge \d x_4}{\Delta_{12}\Delta_{34}}\,,
\end{align}
where the functions $\Delta_{12}$ and $\Delta_{34}$ are defined by
\begin{equation}\label{SolNCDelta} 
\Delta_{12}=1+\left( \frac{\theta_{12}}{\ls^2} \right)^2\frac{r^4}{R^4} \, , \quad \Delta_{34}=1+\biggl( \frac{\theta_{34}}{\ls^2} \biggr)^2\frac{r^4}{R^4}\, \cdotp
\end{equation}
The $x_{1}$, $x_{2}$, $x_{3}$ and $x_{4}$ are the world-volume coordinates on which the gauge theory live, $r$ is the transverse radial coordinate, expressed in terms of the six transverse coordinates $\vec y = (y_{A})_{1\leq A\leq 6}$ as $r^2 = |\vec y|^2$, $\d\Omega_{5}^{2}$ is the metric on the five-dimensional round sphere of radius one and
$\omega_{4}$ is a four-form defined in terms of the volume form
\begin{equation}\label{defomega5}
	\omega_{\Sfive} = \frac{1}{5!}\frac{R^{5}y_{F}}{r^{6}}\epsilon_{ABCDEF}\,\d y_{A}\wedge\cdots\wedge\d y_{E}
\end{equation}
on $\Sfive$ of radius $R$ by
\begin{equation}\label{defgamma5}
	\d \omega_4 = \omega_{\Sfive}\, .
\end{equation}

The consistency of the supergravity approximation for the above solution requires as usual $\lambda\gg 1$.
In the far infrared region $r\ll R\ls/\sqrt{\theta}\sim \ls^{2}\lambda^{1/4}/\sqrt{\theta}$, the solution is a small deformation of the usual $\AdSS$ background and can be compared with the microscopic calculations presented in the main text. On the other hand, in the far ultraviolet region $r \gg R\ls/\sqrt \theta$, the metric \eqref{metricNCsugra} approximates another $\AdSS$ space, with a new radial coordinate $\tilde r=1/r$. Thus there is no conformal boundary at infinity, which signals that the non-commutative theory is not a standard \<UV>-complete quantum field theory.

\section{\label{CCC}The dual to the \texorpdfstring{$\beta$}{β}-deformed theory}

The gravitational dual of the $\beta$-deformed $\nn=4$ super Yang-Mills theory was derived by Lunin and Maldacena in \cite{Lunin:2005jy} in the $\nn=1$ supersymmetry preserving case $\gamma_{1}=\gamma_{2}=\gamma_{3}$ and generalized by Frolov in \cite{Frolov:2005dj} to arbitrary deformation parameters $\gamma_{1}$, $\gamma_{2}$ and $\gamma_{3}$. The solution for the string-frame metric and the other non-trivial supergravity fields reads
\begin{align}\label{metbeta}
\d s^{2} & = \frac{r^{2}}{R^{2}}\d x_{\mu}\d x_{\mu}
+\frac{R^{2}}{r^{2}}\d r^{2} + R^{2}\d\tilde\Omega_{5}^{2}\, ,\\
\label{tbeta}
e^{-\phi} &=\frac{4\pi N}{\lambda\sqrt{G}}\, ,\\\label{Bbeta}
B&=-\frac{\ls^{2}\lambda}{2\pi}\,   G\,
 \bigl(\gamma_{3}r_{1}^{2}r_{2}^{2}\d\theta_{1}\wedge
\d\theta_{2}+\gamma_{2}r_{1}^{2}r_{3}^{2}\d\theta_{3}\wedge
\d\theta_{1}+
\gamma_{1}r_{2}^{2}r_{3}^{2}\d\theta_{2}\wedge\d\theta_{3}\bigr)\, ,\\
\label{C2beta} C_{2}&= -8N\ls^{2}\,\omega_{1}\wedge\bigl(
\gamma_{1}\d\theta_{1}+\gamma_{2}\d\theta_{2}+\gamma_{3}\d\theta_{3}\bigr)
\, ,\\\label{C4beta}
C_{4} &= \frac{4N\ls^{4}}{\pi}\bigl(G\,\omega_{1}\wedge\d\theta_{1}\wedge\d\theta_{2}\wedge\d\theta_{3}-i\omega_{4}\bigr)\, .\end{align}
The coordinates $x_{\mu}$, $1\leq\mu\leq 4$, can be viewed as the world-volume coordinates of the background D3-branes. 
The coordinate $r$ is the usual transverse radial coordinate, expressed in terms of the six transverse coordinates $\vec y = (y_{A})_{1\leq A\leq 6}$ as $r^{2}=\vec y^{2}$. The coordinates $(r_{i},\theta_{i})_{1\leq i\leq 3}$ are defined by the relations
\begin{alignat}{3}\nonumber y_{1}&= \rho_{1}\cos\theta_{1}\, ,\quad &
y_{3} &=\rho_{2}\cos\theta_{2}\, , \quad & y_{5} &=\rho_{3}\cos\theta_{3}\, ,
\\\label{polardef}
y_{2} &=\rho_{1}\sin\theta_{1}\, ,&
y_{4} &=\rho_{2}\sin\theta_{2}\, ,& y_{6} &=\rho_{3}\sin\theta_{3}
\end{alignat}
and
\begin{equation}\label{ridef} r_{i} =\frac{\rho_{i}}{\sqrt{\rho_{1}^{2}+\rho_{2}^{2}+\rho_{3}^{2}}}=\frac{\rho_{i}}{|\vec y|}\,,\quad r_{1}^{2}+r_{2}^{2}+r_{3}^{2}=1\, .
\end{equation}
The function $G$ is given by 
\begin{equation}\label{Gdef} \frac{1}{G} = 1+\lambda\bigl(\gamma_{1}^{2}r_{2}^{2}r_{3}^{2}
+\gamma_{2}^{2}r_{1}^{2}r_{3}^{2}+\gamma_{3}^{2}r_{1}^{2}r_{2}^{2}\bigr)\, .\end{equation}
The metric \eqref{metbeta} describes an $\AdS\times\tilde{\text{S}}{}^{5}$ geometry for a deformed five-sphere $\tilde{\text{S}}{}^{5}$ endowed with the metric
\begin{equation}\label{defS5met} \d\tilde\Omega_{5}^{2}=\sum_{i=1}^{3}\bigl( \d r_{i}^{2} + G\, r_{i}^{2}\d\theta_{i}^{2}\bigr) + \lambda G\, r_{1}^{2}r_{2}^{2}r_{3}^{2}
\Bigl(\sum_{i=1}^{3}\gamma_{i}\d\theta_{i}\Bigr)^{2}\, .
\end{equation}
Defining the angles $\theta$ and $\phi$ by
\begin{equation}\label{thetaphidef} r_{1}=\sin\theta\cos\phi\, ,\quad r_{2}=\sin\theta\sin\phi\, ,\quad r_{3}=\cos\theta\, ,\end{equation}
the one-form $\omega_{1}$ in \eqref{C2beta} and \eqref{C4beta} satisfies
\begin{equation}\label{domeg1}\d\omega_{1} = r_{1}r_{2}r_{3}\, \sin\theta\,\d\theta\wedge\d\phi\end{equation}
and can be chosen to be
\begin{equation}\label{omeg1form} \omega_{1}=\frac{1}{4}\sin^{4}\theta\cos\phi\sin\phi\,\d\phi\, .\end{equation}
The four-form $\omega_{4}$ in \eqref{C4beta} satisfies
\begin{equation}\label{omega4cond}\d\omega_{4}= \omega_{\AdS}\, ,\end{equation}
where 
\begin{equation} \omega_{\AdS} = \frac{1}{R^{8}}\, r^{3}\d x_{1}\wedge\cdots\wedge\d x_{4}\wedge
\d r
\end{equation}
is the volume form on the unit radius $\AdS$ space. Explicitly, one can choose
\begin{equation}\label{omega4form}\omega_{4} = \frac{1}{4R^{8}}r^{4}\d x_{1}\wedge\cdots\wedge\d x_{4}\, .\end{equation}
Changes of $\omega_{1}$ and $\omega_{4}$ by exact forms correspond to a supergravity gauge transformation.

The supersymmetric $\beta$-deformed theory is conformal in the planar limit, which explains the fact that the $\AdS$ factor in the metric \eqref{metricNCsugra} is undeformed in this case. When the $\gamma_i$s are not all equal, conformal invariance appears to be broken in the perturbative regime \cite{Fokken:2013aea,Fokken:2013mza,Fokken:2014soa}.
As the metric \cref{metricNCsugra} still contains an $\AdS$ factor in this case, conformal invariance is restored in the supergravity regime. It requires, on top of the usual condition $\lambda\gg 1$, that $\gamma_{i}^{4}\lambda\ll 1$, as can be checked by evaluating the curvature of the deformed sphere \eqref{defS5met}. In particular, the $\gamma_{i}$s must be very small. This explains why the periodicity in the deformation parameters, $(\gamma_{1},\gamma_{2},\gamma_{3})\equiv (\gamma_{1}+n_{1},\gamma_{2}+n_{2},\gamma_{3}+n_{3})$ for any integers $n_{1}$, $n_{2}$, $n_{3}$, which is manifest in the microscopic theory and in particular in the effective action computed in section \ref{BetaSec}, cannot be seen in the supergravity solution. Finally, let us note that the background is a small deformation of the usual $\AdSS$ solution when $\gamma_{i}^{2}\lambda\ll 1$, a condition often used in the main text.

\chapter{Non-perturbative computation of the correlator}
\label{sec:appendix}\label{chap:correlator}

In this appendix we are going to derive the expression \eqref{eqn:corr} for the twisted supergravity field in terms of the two polynomials $T_0$ and $T_1$ specifying the Coulomb branch vacuum.
As explained in the main text, this requires to compute the correlator \eqref{eqn:g.corr}. 
This will be achieved by exploiting the recent results of \cite{Fucito:2012xc,Nekrasov:2012xe}, generalizing to $\nn=2$ quivers the microscopic approach to Seiberg-Witten theory \cite{Nekrasov:2002qd,Nekrasov:2003rj}.
We briefly review the main results of this work and explain in detail how they allow us to derive an explicit expression for the supergravity twisted field.

The main objects of study are the correlators
\begin{equation}
  y_a(z)=\exp \langle \tr \log (z-Z_a) \rangle \, ,
  \label{eqn:yadef}
\end{equation}
where the $Z_a, a=0,1$, are the adjoint fields (normalized to have dimension of length) of the two gauge groups, taken to be $\SU(N+M)_a$.
The functions $y_a(z)$ are generating functions for all correlators of the theory on the Coulomb branch as can be seen by Taylor expanding the logarithm in \eqref{eqn:yadef}
\begin{equation}
  y_a(z)=z^{N+M}\exp \left[-\sum_{k=2}^\infty \frac{1}{k z^k}\langle \tr Z_a^k \rangle \right] \, .
  \label{eqn:yaexp}
\end{equation}
There is no $k=1$ term in this expansion because we are dealing with $\SU(N+M)$ adjoint fields.
If the $Z_a$ were ordinary finite dimensional matrices instead of quantum fields, the Cayley-Hamilton theorem would express $\tr Z_a^k$ for $k> N+M$ in terms of the traces for $k=2,\ldots,N+M$.
These identities satisfied by the traces would ensure that all terms containing negative powers of $z$ in the expansion \eqref{eqn:yaexp} actually vanish and that $y_a$ is a polynomial, the characteristic polynomial of the matrix $Z_a$.
However, this needs not be the case in a quantum field theory as product of operators are not \emph{a priori} defined but require a choice of regularization scheme.
This regularization scheme will in general spoil the Cayley-Hamilton identities between the traces resulting in non-polynomial $y_a$.
In particular, this is the case for the most natural regularization scheme for instanton computations in $\nn=2$ theories obtained by turning on a non-commutative deformation and the $\Omega$-background.
Even though the $y_a$ are not polynomials, the main result of \cite{Fucito:2012xc,Nekrasov:2012xe} particularized to the case of the $\CC^2/\ZZ_2$ orbifold is that one can construct two functions of $y_0$ and $y_1$ which are actually polynomials of degree $N+M$, $\tilde T_0$ and $\tilde T_1$.
They read
\begin{align}
  \tilde T_0(z)&=\frac{y_0(z)}{\phi(\qq)}\thet3\left(\frac{y_1(z)^2}{\qq_1 y_0(z)^2} ; \qq^2 \right) \, , \label{eqn:tT0def} \\
  \tilde T_1(z)&=\left( \frac{\qq_1}{\qq_0} \right)^\frac{1}{4}\frac{y_0(z)}{\phi(\qq)}\thet2\left(\frac{y_1(z)^2}{\qq_1 y_0(z)^2} ; \qq^2 \right) \, , \label{eqn:tT1def}
\end{align}
where $\qq=\qq_0\qq_1$ and $\qq_a=e^{2\pi i\tau_a}$ are defined by the two holomorphic gauge couplings of the conformal theory.
We can then write down the following expansions of the $\qq$-Pochhamer symbol $\phi$ and the Jacobi $\theta$-functions,
\begin{align}
  \phi(\qq) &= \prod_{k=1}^\infty (1-\qq^k) \label{eqn:phidef} \, , \\
  \thet2(t; \qq) &= \sum_{n \in \ZZ+\frac{1}{2}} t^n \qq^{\frac{1}{2}n^2} \, , \label{eqn:th2def} \\
  \thet3(t; \qq) &= \sum_{n \in \ZZ} t^n \qq^{\frac{1}{2}n^2} \label{eqn:th3def} \, .
\end{align}
The polynomials $\tilde T_0$ and $\tilde T_1$ are not quite the same as the polynomials $T_0$ and $T_1$ used in the main text since they are not monic, i.e.\ the coefficients of the $z^{N+M}$ terms are not $1$, but rather
\begin{align}
  \tilde T_{0,0} &= \frac{1}{\phi(\qq)}\thet3\left(\frac{1}{\qq_1};\qq^2\right) \, , \label{eqn:T00} \\
  \tilde T_{1,0} &= \left( \frac{\qq_1}{\qq_0} \right)^{\frac{1}{4}}\frac{1}{\phi(\qq)}\thet2\left(\frac{1}{\qq_1};\qq^2\right) \, . \label{eqn:T10} 
\end{align}
Hence, we define the monic polynomials $T_0$ and $T_1$ by
\begin{equation}
  T_0(z)=\frac{\tilde T_0(z)}{\tilde T_{0,0}} \, , \quad T_1(z)=\frac{\tilde T_1(z)}{\tilde T_{1,0}} \, ,
  \label{eqn:defT0T1}
\end{equation}
which coincide with the polynomials used in the main text.

Plugging \eqref{eqn:yadef} into \eqref{eqn:g.corr}, we can express $\gamma$ as
\begin{equation}
  e^{2\pi i\gamma(z)}=\qq_1\frac{y_0(z)^2}{y_1(z)^2} \, .
  \label{eqn:gammay}
\end{equation}
The \RHS\ is the inverse of the argument of the $\theta$-functions in \eqref{eqn:tT0def}, \eqref{eqn:tT1def} and we thus need to invert those relations to obtain $\gamma$ in terms of $\tilde T_0$ and $\tilde T_1$.
For this, we use the following properties of the $\theta$-functions, which can easily be derived from the Fourier expansions \eqref{eqn:th2def} and \eqref{eqn:th3def}.
\begin{enumerate}
  \item \label{prop:1} The functions $\thet2(t;\qq)$ and $\thet3(t;\qq)$ are elliptic, i.e.\ holomorphic functions in $t$ associated to an elliptic curve $\mathcal E$ with complex structure $\qq=e^{2\pi i \tau}$.
    Defining also $t=e^{2\pi i u}$, this elliptic curve is the complex torus $\mathcal E=\CC/\Lambda$ where $\Lambda$ is the lattice $\Lambda=\{u \in \CC \vert u=m+n\tau , (m,n) \in \ZZ^2\}$.
    Holomorphicity of $\thet2$ and $\thet3$ is a consequence of the convergence of the series \eqref{eqn:th2def}, \eqref{eqn:th3def} for $|\qq|<1$ or equivalently $\im\tau>0$.

  \item \label{prop:2} The functions $\thet2(t;\qq)$ and $\thet3(t;\qq)$ enjoy periodicity properties: they are periodic under $u\to u+1$, being functions of $t$ only, and quasi-periodic under $u\to u+\tau$ or equivalently $t\to t\qq$,
    \begin{equation}
      \thet2(t\qq;\qq)= t^{-\frac{1}{2}}\qq^{-1}\thet2(t;\qq) \, , \quad   \thet3(t\qq;\qq)= t^{-\frac{1}{2}}\qq^{-1}\thet3(t;\qq) \, .
      \label{eqn:thperiod}
    \end{equation}

  \item \label{prop:3} The functions $\thet2(t;\qq)$ and $\thet3(t;\qq)$ each have a single simple zero on $\mathcal E$, 
    \begin{equation}
      \thet2(-1;\qq)=0 \, , \quad \thet3(-\qq^{\frac{1}{2}};\qq)=0 \, .
      \label{eqn:thzeroes}
    \end{equation}
\end{enumerate}
To invert \eqref{eqn:tT0def} and \eqref{eqn:tT1def}, we adopt the same strategy as \cite{Nekrasov:2012xe} and define
\begin{align}
  t^2&= \frac{y_1^2}{\qq_1 y_0^2} = e^{-2 \pi i \gamma} \, , \label{eqn:deft} \\
  T_r &= \frac{T_0}{T_1} = \frac{\tilde T_{1,0}}{\tilde T_{0,0}} \left( \frac{\qq_0}{\qq_1}\right)^{\frac{1}{4}} \frac{\thet3(t^2;\qq^2)}{\thet2(t^2;\qq^2)} \label{eqn:defTr} \, .
\end{align}
We can now use the three properties of the $\theta$-functions stated above to derive the following properties of $T_r$.
By property~\ref{prop:1}, $T_r$ is a meromorphic function.
By property~\ref{prop:2}, it is well-defined on $\mathcal E$ because the coefficients in the periodicity relations \eqref{eqn:thperiod} cancel in the ratio \eqref{eqn:defTr}.
By property~\ref{prop:3}, it has two simple poles at $t=\pm i$ and two simple zeroes at $t=\pm i\qq^{\frac{1}{2}}$.
Finally, it is an even function of $u$ since it is evaluated for $t^2$.
The fact that $T_r$ is an even meromorphic function on $\mathcal E$ with prescribed poles and zeroes allows us to rewrite it in a different way.
Indeed, the field of meromorphic functions on an elliptic curve is the field of fractions generated by the two elements $(X(t;\qq),Y(t;\qq))$ (subject to the relation \eqref{eqn:Weqn} to be discussed shortly), where
\begin{align}
  X(t;\qq)&=\wp(u; \tau) 
  \label{eqn:defX} \, , \\
  Y(t;\qq)&=2\pi i t \frac{\d X}{\d t} (t;\qq) = \wp'(u; \tau) \label{eqn:defY}
\end{align}
are the Weierstrass $\wp$-function and its derivative written in the more convenient $(t=e^{2\pi i u},\qq=e^{2 \pi i \tau})$ variables.
The function $\wp$ is an even meromorphic function on $\mathcal E$ with a double pole at the origin.
This implies that $\wp'$ is odd, and hence that $T_r$ can be written as a function of $\wp$ (or equivalently of $X$) only.
To match the poles and zeroes of \eqref{eqn:defTr}, the right combination is
\begin{equation}
  T_r = T_r^\infty \frac{X(t;\qq) - X_0}{X(t;\qq) - X_1} \, ,
  \label{eqn:TrX}
\end{equation}
where
\begin{gather}
  X_0=X(i\qq^{\frac{1}{2}};\qq) \, , \quad X_1=X(i;\qq) \, , \\
  T_r^\infty=\frac{\tilde T_{1,0}}{\tilde T_{0,0}} \left( \frac{\qq_0}{\qq_1}\right)^{\frac{1}{4}} \frac{\thet3(1;\qq^2)}{\thet2(1;\qq^2)}= \frac{\thet2(\qq_1^{-1};\qq^2)}{\thet3(\qq_1^{-1};\qq^2)}\frac{\thet3(1;\qq^2)}{\thet2(1;\qq^2)} \, ,
  \label{eqn:X0X1tTrinf}
\end{gather}
which are found by matching the zeroes at $t=\pm i\qq^{\frac{1}{2}}$, the poles at $t=\pm i$ and the value at $t=1$ (which is a pole of $X$) respectively.
We can now solve for $X$ in \eqref{eqn:TrX},
\begin{equation}
  X[T_r(z)](t;\qq)=\frac{T_r(z) X_1(\qq)- T_r^\infty(\qq_0,\qq_1)X_0(\qq)}{ T_r(z)- T_r^\infty(\qq_0,\qq_1)} \, ,
  \label{eqn:XTr}
\end{equation}
where we have spelled out the full parametric dependence of the different quantities involved.
We are now nearing the end of our journey through the land of elliptic functions: the \RHS\ is now $t$-independent and all that remains is to invert the relation between $t$ and $X$ to obtain $\gamma$ from \eqref{eqn:deft}.

This can be done by recalling that $X$ and $Y$ satisfy the following polynomial equation:
\begin{equation}
  Y(t;\qq)^2=4 X(t;\qq)^3 - g_2(\qq) X -g_3(\qq) \, ,
  \label{eqn:Weqn}
\end{equation}
which realizes the elliptic curve $\mathcal E$ as a projective variety inside $\mathbb P^2$.
The function $X(t;\qq) $ admits the Fourier expansion
\begin{equation}
  X(t;\qq)=-4 \pi^2 \left[ \frac{t}{(1-t)^2}+\frac{1}{12}+\sum_{k=1}^\infty k \frac{\qq^k}{1-\qq^k}(t^k + t^{-k} -2) \right] \, ,
  \label{eqn:Xseries}
\end{equation}
which coincides with the Weierstrass $\wp$-function by uniqueness.\footnote{It is meromorphic and periodic in $u$ with a double pole at $u=0$ of residue one; subtracting this pole gives a function which vanishes at zero. These properties define the Weierstrass $\wp$-function uniquely.}
The coefficients $g_2$ and $g_3$ are modular forms of weight 4 and 6 respectively.
Their Fourier expansions read
\begin{align}
  g_2(\qq) &=(-4\pi^2)^2 \left[ \frac{1}{12}+20\sum_{k=1}^\infty k^3 \frac{\qq^k}{1-\qq^k} \right] \, ,  \label{eqn:defg2} \\
  g_3(\qq) &=(-4\pi^2)^3\left[ -\frac{1}{216}+\frac{7}{3}\sum_{k=1}^\infty k^5 \frac{\qq^k}{1-\qq^k} \right] \, . \label{eqn:g3def}
\end{align}
The equation \eqref{eqn:Weqn} can be proven by showing that $Y^2-4X^3 + g_2 X + g_3$ is holomorphic and hence constant, and showing that this constant vanishes.

The equation \eqref{eqn:Weqn} plays a crucial role.
By plugging the explicit value \eqref{eqn:XTr} into \eqref{eqn:Weqn} and forgetting about the $t$-dependence of $Y$, we obtain the Seiberg-Witten curve of this model.
However, we are not interested in the Seiberg-Witten curve itself but in $t^2$.
It will be convenient to rewrite the equation \eqref{eqn:Weqn} in terms of its three roots $e_i(\qq)$:
\begin{equation}
  Y(t;\qq)^2=4\left[X(t;\qq)-e_1(\qq)\right]\left[X(t;\qq)-e_2(\qq)\right]\left[X(t;\qq)-e_3(\qq)\right] \, ,
  \label{eqn:Weqnroots}
\end{equation}
which are at 
\begin{equation}
  e_1(\qq)=X(-1;\qq), \, \quad e_2(\qq)=X(-\qq^{-\frac{1}{2}};\qq) \, , \quad e_3(\qq)=X(\qq^{\frac{1}{2}};\qq) \, .
  \label{eqn:eroots}
\end{equation}
Combining the equation \eqref{eqn:Weqnroots} with the definition \eqref{eqn:defY} of $Y$ and the value of $X$ in terms of $T_r$ \eqref{eqn:TrX}, we can write
\begin{equation}
   \frac{\d t}{t} = -2\pi i\, \frac{\d X[T_r]}{Y}=-\pi i\,  \frac{T_r'(z)\frac{\d X}{\d T_r}[T_r]}{\sqrt{\prod_{i=1}^3 \left( X[T_r(z)] -e_i \right)}} \, .
  \label{eqn:dt}
\end{equation}
The choice of branch for the square root must be fixed in order to match the perturbative result in the $z\to\infty$ corresponding to the UV of the theory.
The \RHS\ is independent of $t$, hence $t$ can be found by integrating this equation on a contour that does not cross any branch cuts,
\begin{equation}
  \log \frac{t(z)}{t_1}=-\pi i \int_{z^1}^{z} \frac{T_r'(x)\frac{\d X}{\d T_r}[T_r]}{\sqrt{\prod_{i=1}^3 \left( X[T_r(x)] -e_i \right)}} \, .
  \label{eqn:logt1t}
\end{equation}
To fix the lower bound, we use the relation \eqref{eqn:deft} between $t$ and $y_a$.
The large $z$ asymptotics of $y_a$ are $y_a(z)\sim z^{N+M}$ by \eqref{eqn:yaexp}, hence we have
\begin{equation}
  t_1=\lim_{z\to\infty} t(z)=\qq_1^{-\frac{1}{2}} \, .
  \label{eqn:tbound}
\end{equation}
Using the relation between $\gamma$ and $t$ \eqref{eqn:deft}, we finally obtain an explicit expression for $\gamma$,
\begin{equation}
  2\pi i\gamma(z)=2\pi i \tau_1 - 2\pi i\int^{\infty}_{z} \frac{T_r'(x)\frac{\d X}{\d T_r}[T_r]}{\sqrt{\prod_{i=1}^3 \left( X[T_r(x)] -e_i \right)}} \, .
  \label{eqn:gammaexpr}
\end{equation}

The integrand of \eqref{eqn:gammaexpr} can be massaged a bit in order to obtain a simpler expression.
First, one can evaluate $\frac{\d X}{\d T_r}$ from \eqref{eqn:XTr},
\begin{equation}
  \frac{\d X}{\d T_r}[T_r]=\frac{T_r^\infty(X_0-X_1)}{(T_r-T_r^\infty)^2} \, .
  \label{eqn:dXdTr}
\end{equation}
Plugging this into the integrand of \eqref{eqn:gammaexpr} and expanding $X[T_r]$ yields
\begin{equation}
  \frac{T_r'(x)\frac{\d X}{\d T_r}[T_r]}{\sqrt{\prod_{i=1}^3 \left( X[T_r(x)] -e_i \right)}}= - \frac{\beta}{2\pi i}\frac{T_r'(x)}{\sqrt{\prod_{j=0}^3(T_r(x) - E_j)}}\,,
  \label{eqn:intE}
\end{equation}
where
\begin{align}
  \beta &= - 2 \pi i \frac{T_r^\infty}{\sqrt{\prod_{i=1}^3(X_1-e_i)}} \, , \label{eqn:betadef} \\
  E_i &= T_r^\infty \frac{X_0-e_i}{X_1-e_i} \quad \text{for} \quad i=1,2,3 \, , \label{eqn:Eidef} \\
  E_0 &= T_r^\infty \, . \label{eqn:E0def}
\end{align}
Using identities relating the quantities $\wp-e_i$ to the Jacobi $\theta$-functions (see for instance \cite{NIST}), one can prove that $E_0=-E_1$ and $E_2=-E_3$. Hence by defining
\begin{equation}
  \alpha_1=E_2=-E_3 \, , \quad \alpha_2=E_0=-E_1 \, ,
  \label{eqn:alphadef}
\end{equation}
the \RHS\ of \eqref{eqn:intE} can then be further simplified to
\begin{equation}
  \frac{T_r'(x)\frac{\d X}{\d T_r}[T_r]}{\sqrt{\prod_{i=1}^3 \left( X[T_r(x)] -e_i \right)}}= - \frac{\beta}{2\pi i}\frac{T_r'(x)}{\sqrt{(T_r(x)^2 - \alpha_1^2)(T_r(x)^2 - \alpha_2^2)}} \, .
  \label{eqn:intisimple}
\end{equation}
This yields the formula \eqref{eqn:corr} quoted in the main text by plugging \eqref{eqn:intisimple} back into \eqref{eqn:gammaexpr}.

\backmatter
\printbibliography[heading=bibintoc]
\end{document}